\documentclass[review]{elsarticle}
\usepackage[colorlinks,citecolor=blue,linktoc=all,linkcolor=cyan]{hyperref}
\usepackage{graphicx}
\usepackage{subfig}
\usepackage{wrapfig}
\usepackage{epsfig}
\usepackage[T1]{fontenc}
\usepackage{dsfont} % use mathds instead of mathbb for outline fonts
\usepackage{mathrsfs} % provides mathscr without overwriting mathcal
\usepackage{slashed} % For Dirac slash notation.
\usepackage{amsmath}
\usepackage{amssymb}
\usepackage{amsbsy}
\usepackage{amsfonts}

\usepackage{multicol}
\usepackage{geometry} % For customizing page layout
\usepackage{graphics}
\usepackage{xcolor}
\usepackage{amsmath}
\usepackage[stable]{footmisc}

\usepackage[export]{adjustbox}

\usepackage{soul}
\setstcolor{blue}
\usepackage{booktabs} % For better table rules
\usepackage[utf8]{inputenc}
\usepackage{newunicodechar}
\usepackage[printonlyused]{acronym}
\newunicodechar{ }{~}

\usepackage{comment}

\newcommand{\br}{{\bf r}}

\newcommand{\bc}{\begin{comment}}
\newcommand{\ec}{\end{comment}}

\numberwithin{equation}{section}
\numberwithin{table}{section}
\numberwithin{figure}{section}

\journal{Progress in Particle and Nuclear Physics}

\usepackage{titlesec}
\usepackage{sectsty}
\titleformat{\section}{\normalfont\Large\bfseries}{\thesection}{1em}{}
\titleformat{\subsection}{\normalfont\large\bfseries}{\thesubsection}{1em}{}
\titleformat{\subsubsection}{\normalfont\normalsize\bfseries}{\thesubsubsection}{1em}{}
\begin{document}
	
	\begin{frontmatter}
		
		\title{Recent applications of the subtracted second RPA method}

		%authors, affiliations, corresponding author mention 
		\author[mymainaddress]{Danilo Gambacurta\corref{mycorrespondingauthor}} \cortext[mycorrespondingauthor]{Corresponding author}\ead{gambacurta@lns.infn.it}
		
		\author[mysecondaryaddress]{Marcella Grasso}

		\address[mymainaddress]{INFN-LNS, Laboratori Nazionali del Sud, 95123 Catania, Italy}
		\address[mysecondaryaddress]{Universit\'e Paris-Saclay, CNRS/IN2P3, IJCLab, 91405 Orsay, France}
		
		\begin{abstract}
			In this review, we discuss the most recent developments and applications of the Subtracted Second RPA (SSRPA), an extension of the Second RPA (SRPA), which overcomes its pathological issues encountered  within the Energy Density Functional theory. After recalling the formal properties of the SRPA and SSRPA, the anomalous behavior of SRPA is shown and discussed by presenting several applications with different kinds of nuclear interactions. The most recent pathology-free SSRPA studies are then presented both for charge-conserving and charge-exchange nuclear excitations. The comparison with experimental data is presented to assess and quantify the
			improvement introduced by the SSRPA with respect to the RPA and SRPA. The impact of beyond-mean-field correlations induced in SSRPA is also qualitatively estimated in connection with the modeling of the nuclear equation of state.
		We conclude by discussing the future perspectives of the SSRPA, focusing on its potential connections with some current experimental challenges and outlining necessary theoretical extensions and numerical developments.
		\end{abstract}
		
		
	\end{frontmatter}
	
	\newpage
	
	\thispagestyle{empty}
	\tableofcontents
	
	%to begin the line numbers: 
	%\linenumbers

	%beginning of the core of the manuscript
		\newpage
		
	\section{Introduction}\label{Sec:Intro}
	\subsection{Nuclear collective excitations: experimental and theoretical challenges }

	Giant Resonances (GRs) \cite{Harakeh_book}  are an emergent property of atomic nuclei arising from a coherent, collective motion of the individual  nucleons driven by the residual interaction. They show up as a concentration of transition strength within a specific energy range of the excitation spectrum. The interaction is defined as 'residual' relative to the nuclear ground-state mean-field (MF), where nucleons behave, to a first approximation, as independent particles.  While collective excitations are indeed a common feature of many-body systems, the complex nature of the nuclear force gives rise to a distinct variety of nuclear modes. These are primarily characterized by their spin, isospin, and angular momentum quantum numbers, though their intrinsic properties may also depend significantly on the excitation energy, nuclear mass, and shape. 	Their theoretical modeling requires a precise knowledge of the nuclear interaction, ideally derived within the underlying and more fundamental QCD theory. While  significant progress has been made in the last years in this direction, a unified description from the ground state to high-energy excitations is still far from being fully achieved.  Phenomenological effective interactions, though less microscopic and physically-grounded, often yield better quantitative agreement, being specifically tuned on selected nuclear data. For this reason, it is highly unlikely that a single phenomenological force can describe all types of collective excitations with equal accuracy.  Besides the accurate knowledge of the nuclear force, it is also crucial to use  nuclear models which include, in a hierarchically-structured manner, many-body correlations in the nuclear wave-functions.  Starting from the simplest single Slater determinant approximation, multi particle - multi hole configurations can be progressively added up, or perturbative and diagrammatic expansions employed.
	
	From the experimental point of view, the standard methodology 
 for investigating the intrinsic properties and dynamic behavior of atomic nuclei involves subjecting them to tunable external perturbations and analyzing their response afterwards. In this way, valuable insights into the energetic landscape, underlying symmetries, and fundamental interactions governing the system's evolution can be obtained. By correlating the applied perturbation with the observed response, one can reconstruct detailed information about the system's Hamiltonian, characterize the effective nucleon-nucleon interaction, identify the nature and energy of collective excitations, access and quantify damping mechanisms. The external perturbations often arise from interactions with various probes or through specific nuclear reactions including inelastic electron scattering, electromagnetic and hadronic probes, and Coulomb excitation. The key observables in these experiments typically include the excitation spectra, transition probabilities, angular distributions (which provide information on the multipolarity and spins of the excited states) and decay modes.

GRs are of fundamental importance, not only for understanding nuclear structure and dynamics but also for constraining the nuclear Equation of State (EoS), which governs the behavior of dense nuclear matter in extreme astrophysical environments such as neutron stars and core-collapse supernovae. The constraints imposed on the EoS propagate directly into astrophysical modeling, particularly regarding, for example, the neutron star structure and mass-radius relation and the key processes of nucleosynthesis. For example, beta-decay is a critical nuclear structure property governing r-process nucleosynthesis \cite{Mumpower2016}, an event considered to take place predominantly during the merger of neutron stars \cite{Perego2020,Amend2022}. The interplay between nuclear collective excitations and the EoS thus forms a crucial bridge between terrestrial nuclear physics experiments and the astrophysical understanding of neutron stars, supernovae, and the cosmic origin of elements in the multimessanger physics era.

 Unlike bound nuclear states with sharp energy levels, GRs are typically high-lying excitations embedded in the continuum, making them unstable and associated with finite lifetimes. Quantum mechanically, this instability is reflected in a broadening of their strength distribution, quantified by the resonance width ($\Gamma$), which is inversely related to the lifetime. The total width of a GR includes the various damping processes that dissipate the initial collective energy and is typically decomposed as:

\begin{equation}
	\Gamma =\Gamma^{\mathrm{Landau}}+ \Gamma^{\uparrow} + \Gamma^{\downarrow},
\end{equation}
where $\Gamma^{\mathrm{Landau}}$ accounts for Landau damping, $\Gamma^{\uparrow}$ is the escape width and $\Gamma^{\downarrow}$ the spreading width.

The Landau damping $\Gamma^{\mathrm{Landau}}$ is often dominant at low to moderate excitation energies. This mechanism involves the scattering of the collective mode of nucleons near the Fermi surface, mostly produced by one-particle one-hole ($1p-1h$) excitations. This process is a MF effect analogous to zero-sound damping in Fermi liquids and is naturally incorporated within the linear response theory. The escape width $\Gamma^{\uparrow}$ arises from the direct decay of the collective mode via particle emission into the continuum. This one-body mechanism involves the coupling of the collective excitation to unbound single-particle states, typically resulting in neutron or proton emission, and in gamma decay for electromagnetic modes. It is dominant in light and weakly bound nuclei, where separation energies are low and the coupling to the continuum is enhanced.% The magnitude of $\Gamma^{\uparrow}$ is sensitive to the asymptotic behavior of single-particle wave-functions, and proximity to particle emission thresholds. 
 The spreading width $\Gamma^{\downarrow}$ reflects the internal damping of the collective motion through its coupling to increasingly complex and higher-order excitations. This redistribution of strength leads to a fragmentation of the collective mode and is a key signature of its dissipation into the compound nucleus. Microscopically, this involves the coupling of the $1p-1h$ doorway states with $(2p-2h)$, $(3p-3h)$, or more generally $(np-nh)$ configurations. This coupling leads to a fragmentation of the collective strength over a broad spectrum of more complicated states, resulting in an irreversible loss of coherence of the original mode. The theoretical description of these mechanisms requires going beyond the linear response approximation, on which the majority of studies are based.

Recent advances in high-resolution $\gamma$-ray spectroscopy have significantly enhanced energy resolution and detection efficiency. These capabilities are essential for probing the fine structure and fragmentation of collective excitations across a wide energy range. In particular, they allow for a precise determination of strength distributions, enabling detailed insights into damping mechanisms and nuclear level densities. %High-resolution measurements are especially critical for resolving the complex, fragmented structure of GRs. 
The advent of next-generation Radioactive Ion Beam facilities has revolutionized the study of collective excitations in exotic and weakly bound nuclei. These facilities provide access to isotopes far from stability, characterized by extreme neutron-to-proton ratios and low binding energies. Moreover, radioactive ion beam experiments allow the study of how strength fragmentation and resonance widths evolve with increasing isospin asymmetry. High-resolution magnetic spectrometers, often used in light-ion inelastic scattering reactions—such as $(p,p')$, $(\alpha, \alpha'$), and $(d,d')$—enable selective excitation of specific multipolarities and isospin modes. These reactions, especially when combined with particle-gamma or particle-neutron coincidence techniques, allow for the detailed mapping of strength distributions and the identification of damping mechanisms. Inverse kinematics reactions further expand the  reach to short-lived exotic nuclei. When combined with high-resolution detectors and recoil particle identification, these setups allow for precise reconstruction of excitation energies and decay properties. %In weakly bound systems, widths are typically larger due to the increased probability of nucleon emission, underscoring the importance of these techniques for studying continuum coupling and resonance fragmentation. 
Coincidence measurements play a central role in disentangling complex decay pathways. Particle-gamma coincidence provides clear identification of excited states and their decay modes, while particle-neutron coincidence is essential in neutron-rich nuclei, where neutron emission is the dominant decay channel. These methods help differentiate between statistical and direct decay processes, offering a detailed understanding of the damping mechanisms and escape widths in weakly bound nuclear systems. 
These experimental advancements have significantly improved our ability to probe the microscopic structure and decay dynamics of collective excitations, particularly in nuclei near the drip lines, where the interplay between collectivity and continuum coupling produces new regimes of nuclear behavior.

A GR is characterized by three key observables that describe its distribution in the excitation energy spectrum.
The total strength $S_{\text{tot}}$ represents the integrated transition probability over the energy range of the resonance. It provides a quantitative measure of the collectivity and transition probability of the mode. A collective mode typically exhausts a large fraction of the relevant sum rule \cite{Harakeh_book}. The centroid energy $E_c$ denotes the average excitation energy at which the resonance is centered. It is typically defined using the moments of the strength distribution. It provides insight into the characteristic energy scale of the collective motion and depends on factors such as nuclear mass, isospin asymmetry, and the type of multipole mode.
The line shape of the resonance is often modeled using phenomenological Lorentzian or Gaussian shape profiles depending on the above discussed total width $\Gamma$. However, a central, and still open, question is whether GRs can be adequately described by single Lorentzian/Gaussian distributions with large widths or more complex structures are required. To address this, in the last years, high-resolution experiments have been conducted to investigate the detailed structure of the GR line shape. Investigating the role of individual components in nuclear decay processes is effectively achieved through coincidence experiments. The contribution of the spreading width can be quantitatively assessed by comparing experimental results with theoretical predictions, using models where different kinds of couplings are included. The underlying idea implies thus a hierarchy of widths and timescales, leading to a layered and hierarchically-ordered fragmentation of GR strength. This fragmentation suggests the existence of distinct lifetimes associated with each coupling step, spanning energy scales from the total width (of the order of MeV) to the widths of compound nuclear states (of the order of eV in heavy nuclei).

Various methodologies have been proposed to extract quantitative information regarding the characteristic scales of fine structure. The wavelet analysis is one of the most promising methods \cite{Shevchenko2008_wavelet}. When applied to experimental spectra, wavelet analysis yields quantitative insights into the fine structure by resolving its inherent scales. A parallel analysis of resonance strength distributions, derived from microscopic models that incorporate various resonance decay mechanisms, allows for deductions about the fundamental origins of this fine structure. In the last two decades, systematic investigations performed in various facilities have shown that the dominant mechanism underlying fine structure varies for different GRs \cite{FINE_VonNeumann2019_Review,FINE_Shevchenko2004_GQR,FINE_Shevchenko2009_GQR,FINE_Poltoratska2014_GDR,FINE_Usman2016_GQR,FINE_Kureba2018_GQR,FINE_Fearick2018_GDR,FINE_Jingo2018_GDR,FINE_Donaldson2020_GDR,FINE_Bahini2022_GMR,FINE_Carter2022_GDR,FINE_Bahini2023_GMR,FINE_Bahini2024_GMR}. The isoscalar giant quadrupole resonance (ISGQR) was the initial focus of systematic fine structure studies across the nuclear chart \cite{FINE_Shevchenko2004_GQR,FINE_Shevchenko2009_GQR,FINE_Usman2016_GQR,FINE_Kureba2018_GQR}. For the ISGQR, the crucial mechanism for spreading through the initial doorway state, specifically, the coupling between $1p-1h$ and $2p-2h$ states, has been established. In heavier nuclei, this process is primarily driven by coupling to low-lying surface vibrations, whereas in lighter nuclei, stochastic coupling gains increasing importance. 
%Fine structure persists even in highly deformed heavy nuclei, which might intuitively be expected to exhibit damping of spectral fluctuations due to extremely high level densities in the ISGQR excitation region. Recent findings even indicate that the fine structure, rather than the broader gross structure, directly evidences K-splitting of the ISGQR in deformed nuclei.
Experimental studies of fine structure have also been performed for the isovector giant dipole resonance (IVGDR) \cite{FINE_Poltoratska2014_GDR,FINE_Fearick2018_GDR,FINE_Jingo2018_GDR,FINE_Donaldson2020_GDR,FINE_Bahini2022_GMR,FINE_Carter2022_GDR}, by means of inelastic proton scattering experiments at energies of several hundred MeV by relativistic Coulomb excitation, which preferentially populates the IVGDR. The observed fine structure in the IVGDR primarily stems from the fragmentation of the $1p-1h$ strength, e.g., Landau damping, although there are also indications of the spreading width's relevance. 
The most recent studies, related to the isoscalar giant monopole resonance (ISGMR) \cite{FINE_Bahini2022_GMR,FINE_Carter2022_GDR,FINE_Bahini2023_GMR,FINE_Bahini2024_GMR}, seem to indicate that Landau damping is prominent in the medium-mass region. Conversely, the spreading width increases with mass number, making the largest contribution in heavy nuclei. The relative significance of these two contributions for the ISGMR lies between that of the IVGDR, where Landau damping dominates the spreading width even for heavy nuclei, and the ISGQR, where fine structure is almost entirely due to coupling to low-lying surface vibrations.

From a theoretical point of view, while MF or pure linear response approaches generally describe the total strength and corresponding centroid sufficiently well, the current challenge lies in accurately describing the total width and fine structure. The total width of a GR is experimentally accessible via high-resolution measurements and multipole decomposition analyses. However, disentangling its microscopic components demands detailed theoretical modeling. A comprehensive understanding of resonance widths provides crucial insights into the microscopic structure and dynamics of excited nuclei. Advances in experimental precision combined with the refinement of theoretical models, continue to enhance our understanding of nuclear damping mechanisms across the nuclear chart.

\begin{wrapfigure}{l}{0.5\textwidth}
	\centering
%	\vspace{2.5mm}
	\includegraphics[width=0.44\textwidth]{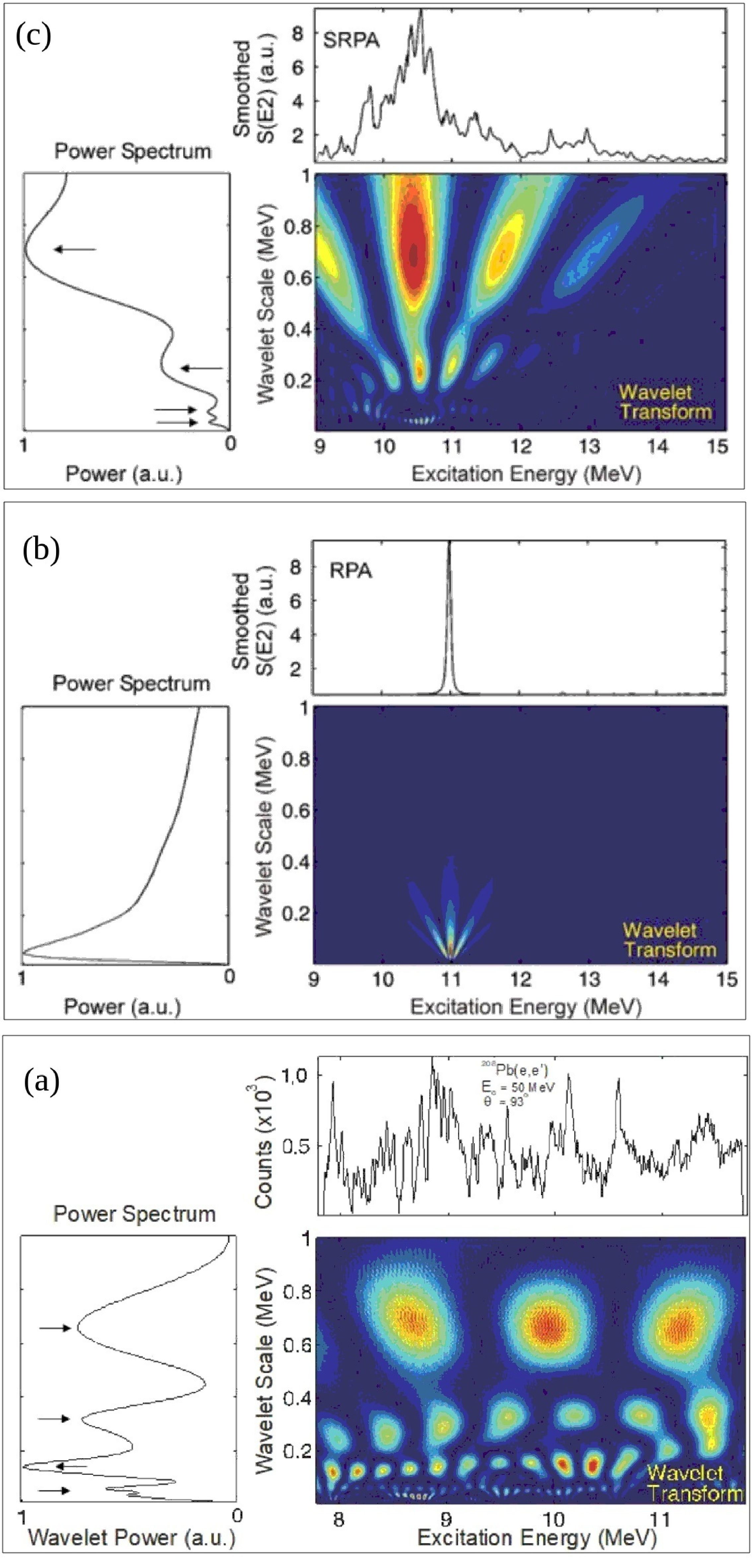}
	% 	\vspace{-1.5mm}
	\caption{Panel (a), top: ISGQR spectrum of the $^{208}$Pb(e, e') reaction. Right: squares of the wavelet coefficients from the CWT. Left: projection of the wavelet coefficients in arbitrary units. The results obtained, in RPA (panel (b)), and in SRPA (panel (a))  are shown. The arrows indicate characteristic scales.  Adapted from Ref. \cite{FINE_VonNeumann2019_Review}.}
	% 	\vspace{-7mm}
	\label{Fig:FINE_GQR_1}
\end{wrapfigure}

 Without aiming to be exhaustive, we recall some of the most advanced many-body methods that improve upon the MF framework, especially those that will be discussed later on. Each method introduces additional correlations and configuration spaces, aiming to capture the damping mechanisms and fragmentation phenomena characteristic of these collective modes.
The Random Phase Approximation (RPA), in its different formulations, (e.g., Time dependent Hartree-Fock (HF), linear response theory, Green function techniques, ...), can be considered the standard approach for the description of collective excitations. The Second RPA (SRPA) is its most natural extension, by including two-particle–two-hole ($2p-2h$) excitations in addition to the $1p-1h$ configurations included in RPA. This extension allows the model to describe more complex spreading mechanisms, particularly the coupling between the $1p-1h$ states and $2p-2h$ configurations. As a result, the SRPA can describe the spreading width and yield a more fragmented strength distribution. In self-consistent implementations based on effective interactions (Skyrme, Gogny, or relativistic functionals), the SRPA has shown promising results in reproducing the total width and the fine structure of GRs, especially in light and medium-mass nuclei.

\begin{wrapfigure}{l}{0.5\textwidth}
	\centering
	%	\vspace{2.5mm}
	%\vspace{-0.7 cm}
	\includegraphics[width=0.49\textwidth]{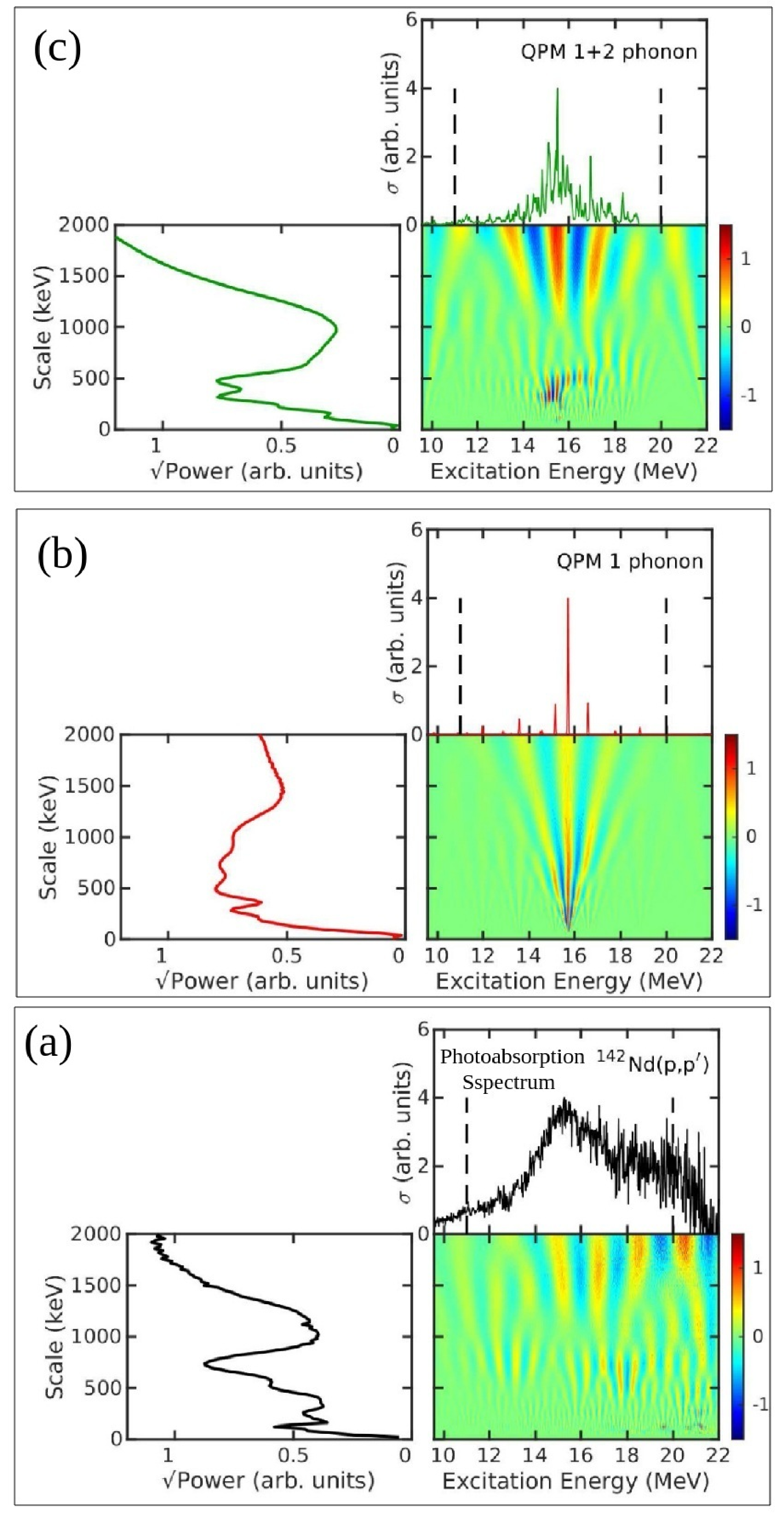}
	%	 	\vspace{-1.5 cm}
	\caption{Panel (a): CWT analysis of the photoabsorption spectrum for $^{142}$Nd. Top set (lower right): the real
		part of the CWT coefficients of the data. Top set (left column): The corresponding power spectrum for the excitation-energy
		region indicated by the vertical dashed lines.
		P(anel (b) and (c)), same as panel (a) but for the QPM 1 phonon
		and QPM 1 + 2 phonon calculations, respectively. Adapted from Ref. \cite{FINE_Donaldson2020_GDR}.}
	\label{Fig:FINE_GDR_1}
	%\vspace{-3 cm}
\end{wrapfigure}

The Particle-Vibration Coupling (PVC)\cite{Hamamoto1969,Baldo2015, Niu2015,Liu2024} and the  Relativistic Quasiparticle
Time-Blocking Approximation (RQTBA) \cite{Litvinova2008,Litvinova2010} are Green’s function methods introducing
dynamical correlations by allowing single-particle states to couple to collective phonons (e.g., low-lying surface vibrations or GRs).
They extend the response function beyond RPA by summing an infinite series of particle-hole-phonon coupling diagrams. These approaches capture the feedback of collective modes on the motion of individual nucleons, leading to a  microscopic description of the  spreading width, though they may be less efficient in describing the detailed fine structure of the strength function.
%The Particle-Vibration Coupling (PVC)\cite{Hamamoto1969,Baldo2015, Niu2015,Liu2024} model introduces dynamical correlations by allowing single-particle states to couple to collective phonons (e.g., low-lying surface vibrations or GRs). This approach captures the feedback of collective modes on the motion of individual nucleons, leading to a fragmentation of strength and a natural description of both the escape and spreading widths. The PVC model provides a microscopic explanation of damping in terms of virtual excitation of vibrational modes, thus reproducing the total width though it may be less accurate in describing the detailed fine structure of the strength function.

The Quasi-particle-Phonon Model (QPM) \cite{QPM,Soloviev1987} offers a fully microscopic framework based on the coupling of quasi-particles with phonons, including multi-phonon states. It allows for the systematic treatment of ground-state correlations and complex excited-state configurations involving $1p-1h$, $2p-2h$, and multi-phonon admixtures. The QPM is particularly successful in describing both the gross and fine structures of GRs, including the fragmentation patterns and detailed level densities. The QPM's ability to go beyond the linear response and include anharmonicities makes it suitable for modeling damping mechanisms and width evolution across isotopic chains. Most  applications are however performed by using semi-phenomenological frameworks with separable forces and adjusted parameters.

%The Relativistic Quasiparticle
%Time-Blocking Approximation (RQTBA)
%\cite{Litvinova2008,Litvinova2010} is a Green’s function method that allows for the inclusion of quasi-particle-vibration coupling in a time-dependent framework. It extends the response function beyond RPA by summing an infinite series of particle-hole-phonon coupling diagrams. The RQTBA provides a unified description of the escape width and the spreading width and can be implemented self-consistently using energy density functionals. By accounting for dynamic self-energy effects, it yields energy-dependent widths and accurately reproduces the spectral fragmentation of GRs.

\begin{wrapfigure}{l}{0.5\textwidth}
	\centering
	\vspace{2.5mm}
	\includegraphics[width=0.49\textwidth]{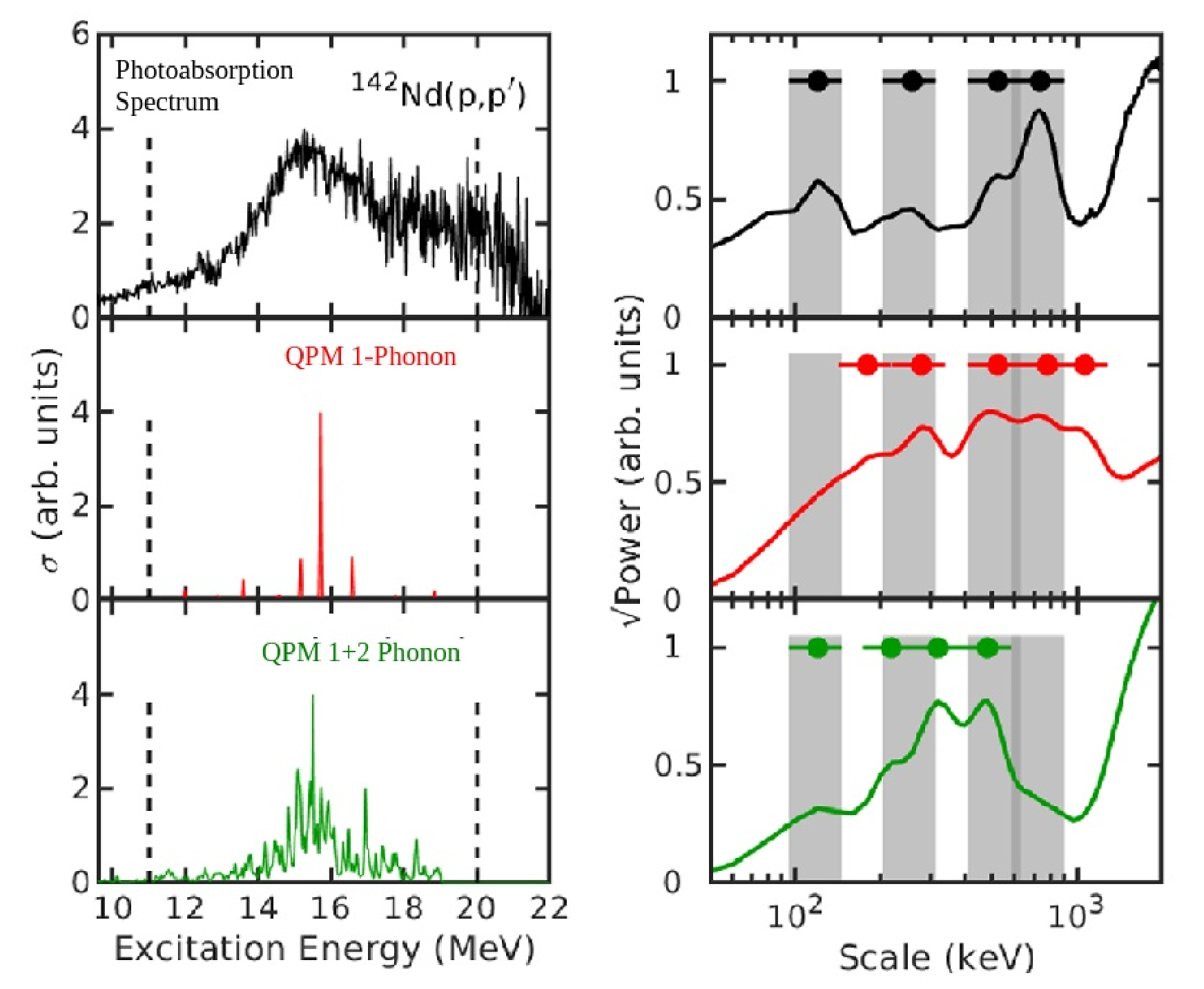}
	% 	\vspace{-1.5mm}
	\caption{Left column: Equivalent photoabsorption spectra for $^{142}$Nd and theoretical results. The top, middle, and bottom panels show the spectrum  of the experimental data, the QPM 1 phonon, and the QPM 1 + 2 phonon calculations, respectively. The right column displays the corresponding power spectra. The positions of the determined scales (filled circles with error bars) are compared across all panels using vertical gray bars to mark the experimental scales. Adapted from Ref. \cite{FINE_Donaldson2020_GDR}.
		}
	% 	\vspace{-7mm}
	\label{Fig:FINE_GDR_2}
\end{wrapfigure}

%%%The nuclear time-dependent density functional theory (TDDFT) \cite{Lacroix2004,Bulgac2013,Nakatsukasa2016} is a key tool for modeling nuclear dynamics describing the complex system's time evolution through single-(quasi)particle equations. However, the basic TDDFT provides only quasi-classical equations, severely underestimating quantum fluctuations. Efforts to include these fluctuations fall into deterministic approaches \cite{Sliusarenko2015,Lackner2015,Dietrich2010}) and stochastic methods \cite{Reinhard1992, Lacroix2006}. 
%%%% Efforts to include these fluctuations fall into deterministic approaches (e.g., truncated Bogolyubov-Born-Green-Kirkwood-Yvon hierarchy \cite{Sliusarenko2015}, reduced density matrix models \cite{Lackner2015}, transport theorie based on Fokker-Planck approximation \cite{Dietrich2010}) and stochastic methods \cite{Reinhard1992, Lacroix2006}. 
%%% The multi configurational TDDFT recently implemented \cite{Marevic2023,Marevic2024} offers a fully quantum alternative, where quantum fluctuations are introduced by tracking the time evolution of a mixing function derived from a general superposition of adiabatic many-body configurations.

The \textit{Ab Initio} methods, though more suited for ground state properties, have made substantial progress in describing collective modes, in the Coupled-Cluster \cite{Bacca2013,Sobczyk2021,Sobczyk2024, Marino2025}, Self-consistent Green’s functions \cite{Raimondi2019}, the projected generator
coordinate method \cite{Porro2024a,Porro2024b} and quasi-particle
finite-amplitude method \cite{Beaujeault2023}. While not originally designed to describe GRs, they have started to access resonant behavior and damping features by incorporating multi-particle–multi-hole excitations and continuum coupling explicitly. These methods promise a more fundamental understanding of collective excitations, although their applicability to heavy nuclei and high excitation energies remains a significant computational and theoretical challenge.

As previously mentioned, the fine structure of GRs has been a major focus in recent years. This research has  investigated the role of various damping mechanisms specifically to each type of GR and different nuclear mass regions. The wavelet analysis of the measured ISGQR spectra (upper frame) from the $^{208}$Pb$(e, e')$ reaction is presented in panel (c) of Figure \ref{Fig:FINE_GQR_1}, using the continuous wavelet transform (CWT) \cite{Shevchenko2008_wavelet}. The squared wavelet coefficients quantifying the intensity of structures resolved by the wavelets are displayed in the lower-right frame, with their magnitudes color-coded from red (large) to blue (small). Maxima at specific wavelet scale values can be observed across the ISGQR excitation energy region.
%	 The pattern of alternating maxima and minima along these lines arises from the oscillatory nature of the wavelet function. 
	The resulting power spectrum is shown in the bottom-left frame, where peaks indicate characteristic scales associated with structural features in the ISGQR energy region. The comparison with  microscopic calculations \cite{Lacroix2000} are shown in panel (a) and (b), for the  SRPA and RPA, respectively.  In the RPA model, which accounts only for $1p-1h$ transitions, the ISGQR strength is mostly concentrated in a single state. Consequently, the wavelet analysis reveals no significant characteristic scales, apart from a trivial one arising from the Gaussian folding (FWHM = 50 keV) introduced to simulate experimental resolution. In contrast, within the SRPA, the coupling to $2p-2h$ configurations leads to fragmentation of the ISGQR strength over many states, thereby giving rise to fine structure. The corresponding CWT and power spectrum display multiple characteristic scales. This clearly demonstrates the role of $2p-2h$ couplings in generating fine structure and the associated energy scales observed in experimental spectra. %Additionally, a broader characteristic scale at 2.1 MeV, corresponding to the total width of the resonance, is also identified.

In Figure \ref{Fig:FINE_GDR_1}, a similar comparison is shown, for the
experimental photo-absorption spectrum of the IVGDR in $^{142}$Nd
(panel (c)) and two examples of the corresponding model calculations, in the QPM model \cite{QPM,Soloviev1987}, including up to one-phonon (panel (b)) and two-phonons (panel (a)).
Although the inclusions of two-phonons produce a more fragmented distributions as in the ISGQR case, a different general behavior is also observed. A deeper insight can be obtained by considering the results shown in Figure \ref{Fig:FINE_GDR_2}. The figure shows the photoabsorption cross section for the $^{142}$Nd(p, p') reaction, along with its associated power spectrum, displayed in the top right panel. %The vertical dashed lines in the panels on the left side of the figure mark the excitation energy range from 11 to 20 MeV over which the wavelet coefficients were integrated. 
Characteristic energy scales are identified from the peaks and inflection points in the power spectrum and are shown as filled circles. The associated error bars represent one standard deviation of the respective width-like scale. The experimental energy scales are also shown as vertical gray bars. The middle and bottom panels present theoretical predictions from the QPM: the 1-phonon calculation (red) and the 1 + 2-phonon calculation (green), respectively. In the QPM 1-phonon calculation, each excitation includes a single dominant doorway state near the experimental peak, surrounded by more widely spaced, weaker doorway states.
\begin{figure}
	\centering
	\includegraphics[width=.8\linewidth]{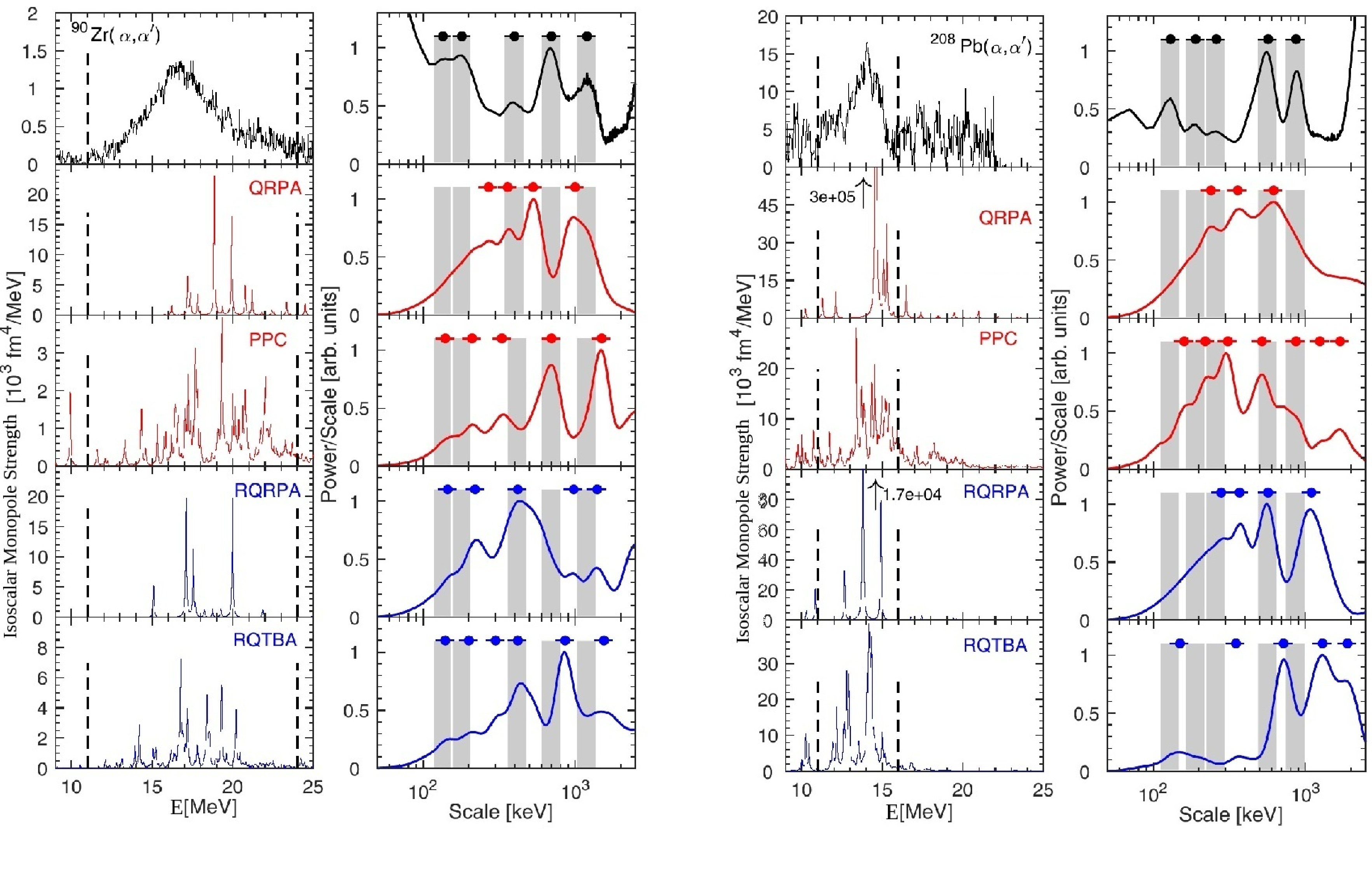}
	\caption{Wavelet Analysis of ISGMR strength in $^{90}$Zr (left side) and $^{208}$Pb (right side). The first row shows the experimentally measured strength followed by four different model predictions. The right column presents the corresponding power spectra determined by summing wavelet coefficients (dashed lines). Scale positions are shown by filled circles with errors. Experimental scales are represented by vertical gray bars. Adapted from Ref. \cite{FINE_Bahini2024_GMR}}.
	\label{Fig:FINE_GMR_1}
	% 	\vspace{2.1cm}
\end{figure}
The predicted power spectra reproduce energy scales that closely match those observed experimentally. The overall shape of the IVGDR is well captured by the QPM 1 + 2-phonon calculation, with the predicted maxima aligning well with those in the experimental spectrum. The model also reproduces similar energy scales in the range of several hundred keV, consistent with the 1-phonon results. One can see that additional low-energy scales absent in the 1-phonon predictions emerge in the 1 + 2-phonon case, indicating that these features arise from the inclusion of more complex configurations. %These results are consistent with previous analyses of IVGDR fine structure in lighter spherical nuclei \cite{FINE_Jingo2018_GDR} and in $^{208}$Pb \cite{FINE_Poltoratska2014_GDR}. 
In the theoretical calculations, the observed scales are primarily generated by the fragmentation of the $1p-1h$ strength, associated with Landau damping.

Figure \ref{Fig:FINE_GMR_1} presents the results of the wavelet analysis for the ISGMR, shown for $^{90}$Zr (left panels) and $^{208}$Pb (right panels).
Theoretical calculations include both relativistic and non-relativistic Quasi-particle RPA (QRPA), as well as extensions beyond QRPA that account for correlated $2p-2h$ states via the Skyrme-QPM model  and the RQTBA based on covariant Energy Density Functionals (EDFs). In general, QRPA-level models tend to overestimate the ISGMR centroid energy, while the inclusion of complex configurations leads to a downward energy shift, bringing theoretical predictions into better agreement with experimental data. This inclusion also results in increased fragmentation of the strength, with the effect being notably stronger in PPC than in RQTBA. Characteristic energy scales extracted from the power spectra are shown as filled circles: black for experimental data (and gray bars on the right-hand panels), red for QRPA and PPC, and blue for RQRPA and RQTBA. The characteristic scales derived from the fine structure are significantly modified when moving from QRPA to models including two-phonon or 2q $\otimes$ phonon configurations.
For $^{90}$Zr, the inclusion of complex configurations significantly affects the CWT results: the number of identified scales increases from 4 to 5 in PPC, and from 5 to 6 in RQTBA. Both PPC and RQTBA reproduce all experimentally observed scales below 1 MeV.  Interestingly, these two low-energy scales also appear in RQRPA, making this the only case where the lowest experimental scale is predicted to originate from Landau damping. In all other models, it is attributed to the spreading width. A larger scale above 1 MeV, consistent with the experimental one at $\approx$ 1 MeV, is found in both the PPC and RQTBA models, though slightly overestimated. In the case of $^{208}$Pb,
 % due to contamination from residual ISGDR strength, the comparison of strength distributions and wavelet power spectra is limited to the 11–17 MeV range. QRPA and RQRPA overestimate the experimental centroid by about 1 MeV and 0.5 MeV, respectively. The impact of 2p–2h coupling is modest, resulting in shifts of only 300–400 keV. 
 both PPC and RQTBA reasonably reproduce the total width. The PPC model quantitatively matches all experimental scale values within the typical uncertainty. The RQTBA calculation reproduces both the smallest and the largest of the observed scales (below 1 MeV). The results seem to indicate that, for the ISGMR, Landau damping plays a prominent role in medium-mass nuclei, while the spreading width becomes increasingly significant with higher mass numbers, dominating in heavy nuclei. This trend positions the ISGMR between the IVGDR, where Landau damping dominates even in heavy systems, and the ISGQR, in which fine structure is primarily attributed to coupling with low-lying surface vibrations, except possibly in lighter nuclei. However, more systematic experimental studies, in particular in the ISGMR case, are needed to confirm these findings, as well as the comparison with other beyond-mean-field (BMF) models.
 
% Beside the need of an overall description of the strength distributions in a quite large (several MeV) energy range, it might happen also that a very detailed knowledge within a much narrow region is needed. This is for example the case of the beta-decay, whose half-life is determined by the strength distribution within the so called beta-window. Also in this case, the role of complex configurations can be of paramount importance in reproducing experimental data and in addressing the quenching problem \cite{Engel_2017}.

As just shown in the previous examples, the study of the fragmentation, total width and characteristic scales, requires the support and guidance of reliable microscopic calculations. Indeed, from a theoretical point of view, describing the total width and fine structure of GRs requires capabilities beyond the simple MF and small-amplitude approximations like the RPA. Advanced many-body approaches have been developed to capture the fragmentation, spreading width, and continuum coupling, which are essential for an accurate and microscopic description of GRs. Each of these many-body approaches addresses specific limitations of the MF or standard RPA and contributes to a more accurate microscopic understanding of the width and fine structure of nuclear GRs. Their ongoing development and benchmarking against high-resolution experimental data continue to refine our knowledge of nuclear collective dynamics. The main focus of the present review is the SRPA method. We will specifically explore its more recent applications, especially those within the EDF framework that have been significantly enhanced by the subtraction procedure, a crucial improvement when employing effective interactions.

\subsection{The SRPA: past and present }
The SRPA, being a natural extension of the commonly used RPA, has been applied for several studies in the past, already starting from the 80's, see, for example, Refs \cite{Lacroix2000, Schwesinger1984,Drozdz1986,Takayanagi1988,Brink1993,Nishizaki1994,Ayik1999} . The primary goal of these works was to study the spreading width and dissipation mechanisms of GRs. 
This review will not discuss these studies which have been also presented in other review papers, see for example Ref. \cite{Drozdz1990}.
These studies, being pioneering, were compromised by two specific primary limitations. First, they were constrained by the then-limited computing power. Consequently, severe truncation in the model spaces and strong approximations in evaluating the beyond RPA matrix elements were employed. Fully self-consistent calculations were very rare already at RPA level, therefore those SRPA studies were performed by using single particle basis  phenomenologically tuned to the nucleus of interest and very often by using very simplified residual interactions. However, despite these simplifications, the anomalous, almost unphysical, shift of the SRPA nuclear response with respect to RPA was already observed and identified as pathological (see for example discussion at page 786 of Ref. \cite{Brink1993} and page 4 of Ref. \cite{Ayik1999}). This pathological issue was "cured" by neglecting systematically the real part of the self-energy correction induced in SRPA, which was responsible for the red-shift, and considering only the imaginary part producing the fragmentation of the strength. Though effective, this strategy was not the optimal one as inconsistencies might be artificially introduced %(the real and imaginary part of the self-energy are linked by a dispersive relation)
 and, moreover, this choice made the implementation of the SRPA approach even less consistent and robust.
Therefore the price paid for overcoming the RPA was the loss of the internal consistency that constitutes one of its key advantages within the EDF framework.
 
In the last 15-years, due also to the significant advances in computational power, the so-called large-scale SRPA calculations started to be performed. They employed model spaces as large as those used in RPA and fully incorporated all SRPA matrix elements. These calculations provided thus more systematic results, clearly illustrating their dependence on numerical cutoffs, the employed interaction, and various simplifying approximations compared to the full solution. Section \ref{Sec:Applications_SRPA} is entirely dedicated to the presentation and discussion of these SRPA calculations. In these studies, it was possible to study quantitatively and more accurately not only the expected broadening and fragmentation of the SRPA strength distribution, but also the systematic shift towards lower energy with respect to the RPA case. This red-shift is of the order of several MeV, independently of the employed interaction, microscopic (see Section \ref{Sec:Applications_SRPA_UCOM}) or phenomenological ones (see Sections \ref{Sec:Applications_SRPA_Skyrme} and \ref{Sec:Applications_SRPA_Gogny}), and it has been also very recently confirmed within a relativistic approach (see Section \ref{Sec:Applications_SSRPA_Relativistic}).
Moreover, stability issues were also observed by carefully scrutinizing the SRPA solutions (see Section \ref{Sec:Applications_SRPA_Stability}) caused by the SRPA's failure to satisfy the positive semi-definite condition, unlike the self-consistent RPA, where this property is assured (see Section \ref{Sec:Applications_SRPA_Stability}). 

The way out to this pathological behavior was provided by the so-called subtraction procedure, introduced by Tsalyev and discussed in Section \ref{Sec:FormalPart_SSRPA}. This procedure is indeed valid for any beyond RPA approach and it turns out to be a powerful regularization technique able to remove double counting and ultra-violet divergencies occurring in SRPA. The SRPA approach, when upgraded by the subtraction procedure and thus called Subtracted SRPA (SSRPA), proves to be pathology-free. This new framework retains the benefits of the original SRPA while preserving the advantages of a full self-consistent scheme in which all couplings from the underlying nuclear interaction are accounted for without arbitrary or uncontrolled approximations (both the real and imaginary part of the self-energy correction are taken into account).
As a matter of fact, the SSRPA allows for a safe and predictive implementation of a beyond RPA approach within the EDF framework. Sections \ref{Sec:Applications_SSRPA_CC} and \ref{Sec:Applications_SSRPA_CE} are dedicated to present recent applications of the SSRPA approach.

%\subsection{Comparison between SRPA and other BMF models}

Among the various extensions,  the
SRPA represents the most direct and conceptually transparent step to
beyond the RPA description. By explicitly including $2p$-$2h$
configurations in the excitation space, the SRPA is capable of
describing the fragmentation of collective states and producing a realistic
spreading width of GRs that can not be obtained within RPA. When formulated consistently from the same
EDF or Hamiltonian, SRPA and SSRPA preserve important sum
rules (discussed in Section \ref{Sec:FormalPart_SRPA_Moments}) and provide a fully microscopic account of correlations beyond the MF approximation. Its main drawbacks, however, lie in the rapid growth of the $2p-2h$ configuration space. Compared to other beyond RPA models, such as, for example, the PVC, RTQBA or QPM, where BMF correlations are introduced through collective phonons, in the SSRPA case the $2p-2h$ configurations are mostly selected in function of their unperturbed energy. The possibility to selectively choose
collective phonons depending on their energy, collectivity and multipolarity is an effective and efficient way to include many body correlations, which also generally results in less computationally demanding calculations. On the other side, while the selection of some phonons, though guided by physical arguments, is anyway arbitrary and not always doubtless, in the SRPA/SSRPA the dependence of the results can be studied just by considering their convergence with respect to the energy cut on the included $2p-2h$ configurations. Moreover, while the use of collective phonons is very efficient in the description of the width of GRs, the inclusion of the $2p-2h$ configurations, due to the higher density, generally better provides the fine structure of the strength distribution. 
The high computational cost of the SSRPA can be reduced by using approximated schemes (as for example the diagonal approximation) whose effect has to be carefully studied case by case. Alternatively, high performance computing techniques based on parallel computation and iterative eigensolvers have been employed to perform realistic and systematic calculations in the most recent SSRPA applications. Finally, the SSRPA approach is likely to be more suited for the study of double excitations, as for example double GRs \cite{Harakeh_book}, or double-beta decays \cite{Engel1999} compared to other models where nuclear excitations keep mainly a one-phonon nature.

	\newpage
	\section{Formal part}\label{Sec:FormalPart}
	\subsection{Introduction}\label{Sec:FormalPart_intro}
	The RPA is probably the most widely used method for a microscopic study of small-amplitude excitations in many kinds of many-body systems. It was first introduced in the context of collective oscillations of electron gas \cite{Bohm1953}.
	There are many different ways to derive the RPA: diagrammatic techniques\cite{GellMann1957}, time-dependent HF \cite{Rowe1966,RS.80,Negele1982}, Green's-function methods \cite{Fetter, Dickoff2004,Dickhoff2008}, the Equations of Motion (EoM) method \cite{RS.80,Rowe1968}, (see, for example, Ref. \cite{Chen2017} and \cite{Co2023} for a recent review in the context of electronic systems and nuclear physics, respectively). Out of the many possible approaches, we use here the EoM method, which formulates the many-body dynamics in terms of excitation operators within the second quantization frame. A key advantage of the EoM method is its direct path to the derivation of the  SRPA equations. This is achieved by introducing a more general type of excitation operators than those used in the standard RPA. Alternatively, the SRPA equations can be derived within a variational approach, where the ground state wave-function explicitly takes into account $2p-2h$ configurations \cite{Providencia1965}. 
	%\textcolor{red}{There should be also a paper within Green's function or in the coordinate space!!! } ).
%
	\subsection{The method of the equations of motion}\label{Sec:FormalPart_EOM}
The EoM  starts from the Heisenberg equation for an operator 
\begin{equation}
	i \hbar \frac{d Q}{d t}= \big[ Q,H\big]
\end{equation}
where $H$ is the nuclear Hamiltonian. By choosing $Q$ appropriately, for example an operator that creates excitations expressed as a linear combination of particle– hole excitations, one can derive the excitation spectrum and transition amplitudes of the system. Basically, the EoM approach finds collective excited states as eigenmodes of the commutator with the Hamiltonian. Deriving the RPA equations within the EoM  corresponds to summing the same infinite series of "ring" (particle-hole bubble) diagrams in perturbation theory \cite{Fetter}, or to linearize the time-dependent HF equations \cite{Rowe1966,RS.80}. %As mentioned above, one of the main advantages of the EoM is that extensions of the RPA can be obtained by including higher-order configurations in the Q operators. 

 Let's consider a set of eigenstates $\nu$ of the Hamiltonian $H$ 
\begin{displaymath}
H|\nu\big\rangle=\mathcal{E}_{\nu}\mid \nu\big\rangle
\end{displaymath}
where $\mid0\big\rangle$ is the ground state with energy $\mathcal{E}_{0}$. The present derivation is valid in the case of a Hamiltonian, e.g., with no density-dependent terms. Extensions to those cases can be found in Ref. \cite{Yannouleas1987, Gambacurta2011a} and will be discussed in Section \ref{Sec:FormalPart_SRPA_Rear}. The phonon operators $Q's$ are such that
\begin{equation}
Q_{\nu}^{\dag}\mid0\big\rangle=\mid\nu \big\rangle,
 \end{equation}
\begin{equation}
	Q_{\nu}\mid0\big\rangle=0,
	\label{Eq:vacuum} 
\end{equation}
the latter implying that the ground state is the vacuum of the phonon operators.
By means of simple algebraic steps, see for example \cite{RS.80}, it is easy to show that, for an arbitrary operator $\delta Q$, the following set of equations holds 
\begin{equation}\label{Eq:EqOfMot}
 \langle0 \mid \big[ \delta Q, [H,Q_{\nu}^{\dag}]\big]\mid0\rangle= \omega_{\nu} \langle0\mid[ \delta
 Q,
 Q_{\nu}^{\dag}]\mid 0\rangle
\end{equation} 
where $  \omega_{\nu}=\mathcal{E}_{\nu}-\mathcal{E}_{0} $
 is the excitation energy of the excited state $\nu$. Eq. (\ref{Eq:EqOfMot}) provides the equations of motion. As no approximations have been made, solving this equation is equivalent to solving the Schr\"{o}dinger equation.
 
 In order to derive equations that can be actually solved, one needs to resort to some approximations. The first one, which is common to both the RPA and SRPA, is due to the fact that the ground state introduced in Eq. (\ref{Eq:vacuum}), as a vacuum of the phonons, is indeed not known. The standard approximation is to replace this state with the HF one, i.e. a Slater determinant, which within a MF framework is naturally introduced to describe the ground state of the system.
 This replacement is also referred to as quasi-boson- approximation (QBA), as the commutator between $1p-1h$ operators is approximated as a bosonic commutation relation, implying that the pair (creation and annihilation) fermionic operators act as a bosonic operator\cite{RS.80,Rowe1968}. The approximation is valid in the limit where the occupation numbers correspond (or are close) to those of the uncorrelated MF ground state where particles are fully occupied below the Fermi level and empty above it.
 %The QBA, besides introducing a formal inconsistency, also introduces quantitative approximations that can be more or less severe depending on the several factors. 
 What truly distinguishes RPA and SRPA is the second approximation: the explicit choice of phonon operators employed to describe the excited states. This will be discussed in more detail below.

	\subsection{The HF approximation and the particle-hole representation}
	\label{Sec:HF}
	The HF approximation and the particle-hole representation are crucial concepts in understanding the behavior of many-body systems like atomic nuclei. The core idea of HF is to approximate the exact many-body wave function of the nucleus by a single Slater determinant. This Slater determinant is constructed from a set of single-particle wave functions, where each nucleon occupies one of these single-particle states. Each nucleon moves independently in an average field, called the MF, created by all the other nucleons.
	Once a MF of the nucleus is obtained from the HF approximation, one can use the particle-hole representation to describe excited states of the nucleus. In the HF ground state, the nucleons occupy the lowest available single-particle energy levels, where the highest occupied level is called the Fermi level. These states are referred to as "hole" states. A "particle" refers to a nucleon that has been excited to a state above the Fermi level. Excited states of the nucleus can be created by promoting a nucleon from a hole state to particle state. Mathematically, the particle-hole representation can be formulated using creation and annihilation operators in second quantization. 
	In the context of atomic nuclei, the particle-hole representation is essential for understanding nuclear excitations, such as collective vibrations and rotations. For example, a collective vibration can be described as a coherent superposition of many particle-hole states. The energy and properties of these excited states can be calculated by considering the interaction between the particles and holes, which is generally treated as residual interactions beyond the MF.
	
	\subsection{The m-scheme and J-scheme}
		\label{Sec:J_M_Scheme}
	The RPA and SRPA can be derived within two different schemes, which are related by a unitary transformation, the so-called m-scheme and J-scheme, each of them offering some advantages and disadvantages. In the m-scheme, the RPA/SRPA configurations are built using particle-hole creation and annihilation operators defined with respect to the m-quantum numbers, e.g., the z-axis component of the total angular momentum. The RPA/SRPA phonon operators, and the resulting eigenmodes correspond to states with a specific total M-projection and given parity, but they are generally a mixture of different total angular momentum J values. To obtain states with good J, one therefore needs to perform calculations for all possible M values and then using transition operators with a definite J-value to obtain the corresponding strength functions, or, alternatively, resorting to projection techniques. In the 	J-scheme, the RPA/SRPA configurations are built directly using particle-hole operators that are coupled to a specific total angular momentum J, projection M and parity $\pi$. This coupling is achieved using Clebsch-Gordan coefficients. As a result, the RPA/SRPA matrix has a block structure, each block belonging to a given set of quantum numbers. The eigenvectors and excitation energies obtained, directly correspond to states with good J and M. From a computational point of view, the m-scheme generally leads to a larger basis size compared to the J-scheme. However, the matrix elements in the m-scheme are simpler to calculate as they avoid the need for the angular momentum recoupling coefficients. The J-scheme uses a smaller basis size because the angular momentum coupling reduces the number of independent states. However, the calculation of the matrix elements is more involved. The J-scheme explicitly incorporates rotational symmetry by constructing states with good J. This can be an advantage for spherical nuclei, while in deformed systems, where the rotational symmetry is broken, the m-scheme is the usually adopted one. In the SRPA case, as the number of $2p-2h$ configurations is much larger than the $1p-1h$ ones, the J-coupled scheme is currently mandatory for realistic calculations.

	\subsection{The RPA and SRPA in the m-scheme}\label{Sec:FormalPart_RPA_SRPA_Mscheme} 
	%
%	Starting from the HF picture, where we have a ground state defined by occupied and unoccupied single-particle states, the RPA allows us to describe collective excitations as coherent superpositions of $1p-1h$ states. The method of the equations of motion provides the framework for deriving and solving the RPA equations. The HF approximation gives us a ground state $|HF\rangle$ which serves as our reference state or "vacuum", assumed to be also the best approximation of the true correlated ground states. 
		 In the RPA, the excited states are assumed to be created by the excitation operators $O^\dagger_\nu$ which are linear combinations of $1p-1h$ creation operators $(a^\dagger_a a_h)$ and one-hole one-particle ($1h-1p$) creation operators $(a^\dagger_h a_p)$. %Here, 'p' denotes a single-particle state above the Fermi level (particle), and 'h' denotes a state below the Fermi level (hole).
	 Each index "i" denotes a single particle states defined by its quantum numbers ($n_i l_i j_i m_i$), where 
	 $n_i$ is the principal quantum number identifying the energy of the state, $l_i$ defines orbital angular momentum,   
	 $j_i$ specifies the total angular momentum and $m_i$ is the magnetic quantum number given by the projection on the z-axis of the total angular momentum.
	 
	 The phonon operator
	\begin{equation}
		\label{Eq:OpRPA}
			Q^\dagger_\nu = \sum_{ph} X^\nu_{ph} a^\dagger_p a_h - Y^\nu_{ph} a^\dagger_h a_p,
	\end{equation}
 is  a superposition of $1p-1h$ and $1h-1p$ configurations  having a definite m-projection $m_\nu=m_p+m_h$ and parity $\pi_\nu=(-1)^{(l_p+l_h)}$. 
	The coefficients $X^\nu_{ph}$ and $Y^\nu_{ph}$ are amplitudes that determine the contribution of each particle-hole configuration to the excited state $|\nu\rangle$. The presence of the $Y^\nu_{ph}$ terms introduces ground state correlations beyond the simple HF picture. %If $Y^\nu_{ph}$ were zero, we would essentially be dealing with simpler scheme, where only $1p-1h$ excitations are taken into account, known as Tamm-Dancoff approximation \cite{RS.80}.
	
%	Considering the commutator of the Hamiltonian $H$ with the excitation operator:
%	$$[H, Q^\dagger_\nu] = \hbar \omega_\nu Q^\dagger_\nu$$
%	one can see that the excited state is an eigenstate of the Hamiltonian when acting on the excitation operator applied to the ground state, with the excitation energy $\hbar \omega_\nu$ as the eigenvalue.
	
 Substituting in Eq. (\ref{Eq:EqOfMot}) the $Q$ operators (\ref{Eq:OpRPA}), using $\delta Q=\{a^\dagger_p a_h,a^\dagger_h a_p\}$ and then taking the expectation value with respect to the HF ground state leads to a set of linear equations for the amplitudes $X^\nu_{ph}$ and $Y^\nu_{ph}$. The expectation values of the Hamiltonian, which typically includes a MF part (from HF) and a residual interaction $V_{res}$ (the part of the nucleon-nucleon interaction not captured by the HF-MF), can be computed using the HF properties (e.g., $a^\dagger_h | HF \rangle = 0$ and $a_p | HF \rangle = 0$).
	
	The resulting RPA matrix equations are
	\begin{equation}
		\label{Eq:RPAmat}
		\begin{pmatrix}
			\mathcal{A}_{11} & \mathcal{B}_{11} \\
			-\mathcal{B}_{11}^* & -\mathcal{A}_{11}^*
		\end{pmatrix}
		\begin{pmatrix}
			X^\nu_{1} \\
			Y^\nu_{1}
		\end{pmatrix}
		= \hbar \omega_\nu
		\begin{pmatrix}
			X^\nu_{1} \\
			Y^\nu_{1}
		\end{pmatrix}
	\end{equation}
	
	where the index $1$ stands for the  $1p-1h$ (and $1h-1p$) configurations and the matrices $\mathcal{A}_{11}$ and $\mathcal{B}_{11}$ are defined as :
	\begin{equation}
		\label{Eq:RPAmatA_M}
		\mathcal{A}_{11'} \equiv \mathcal{A}_{ph, p'h'} = \langle HF | [a^\dagger_p a_h, [H, a^\dagger_{h'} a_{p'}]] | HF \rangle
	\end{equation}
	\begin{equation}
		\label{Eq:RPAmatB_M}
	\mathcal{B}_{11'} \equiv	\mathcal{B}_{ph, p'h'} = \langle HF | [a^\dagger_p a_h, [H, a^\dagger_{p'} a_{h'}]] | HF \rangle.
	\end{equation}
	The explicit expression of the RPA matrices is given in Appendix \ref{Sec:AppMScheme}.
%Consider the Hamiltonian $H = \sum_i \epsilon_i a^\dagger_i a_i + \frac{1}{4} \sum_{tqrs} V_{tqrs} a^\dagger_t a^\dagger_q a_s a_r$, where $\epsilon_i$ are the single-particle energies obtained from the HF calculation and $V_{tqrs}$ are the antisymmetrized matrix elements ($V_{tqrs}=\langle tq\mid \hat{V}\mid rs \rangle-\langle tq\mid \hat{V}\mid sr \rangle$)of the residual two-body interaction $\hat{V}$, one gets:
%\begin{equation}
%\mathcal{A}_{ph, p'h'} = (\epsilon_p - \epsilon_h) \delta_{pp'} \delta_{hh'} + {V}_{ph'hp'}
%%\langle ph' | V_{res} | p'h' \rangle
%\end{equation}	
%\begin{equation}
%\mathcal{B}_{ph, p'h'} = {V}_{pp'hh'}.
%%\langle pp' | V_{res} | hh' \rangle	
%\end{equation}
%%where ${V}_{ijkl}={V}_{ijkl}-{V}_{ijlk}$.
 Solving the matrix eigenvalue problem yields the excitation energies $\hbar \omega_\nu$ (the eigenvalues) and the corresponding amplitudes $X^\nu_{ph}$ and $Y^\nu_{ph}$ (the eigenvectors) for each excited state.
 Imposing the orthonormalization of the RPA states
 \begin{equation}
 	\langle \nu \mid \nu' \rangle=	\langle 0 \mid Q_\nu Q^\dagger_{\nu'}\mid 0 \rangle=\langle 0 \mid \big[ Q_\nu, Q^\dagger_{\nu'}\big]\mid 0 \rangle\simeq
 	\langle HF \mid \big[ Q_\nu, Q^\dagger_{\nu'}\big]\mid HF \rangle =\delta_{\nu \nu'}
 	\label{Eq:ortho}
 	 \end{equation}
 	 one gets that the RPA states are normalized such as:
 	\begin{equation}
 			\label{Eq:norm_RPA_m}
 		 		\sum_{ph} \left( |X_{ph}^{\nu}|^2 - |Y_{ph}^{\nu}|^2 \right) = 1.
 		 	\end{equation} 

In	SRPA the 
excitation operators $Q^+_{\nu}$ are a superposition of $1p-1h$ and $2p-2h$ configurations:

\begin{equation}\label{Eq:srpa_op_m}
 Q_{\nu}^{\dagger}=\sum_{ph}(X_{ph}^{\nu}a_{p}^{\dag}a_{h}-Y_{ph}^
{\nu}a_{h}^{\dag}a_{p})
+ \sum_{p<p',h<h'}(X_{php'h'}^{\nu}
a_{p}^{\dag}a_{h}a_{p'}^{\dag}a_{h'}-Y_{php'h'}^{\nu}a_{h}^{\dag}a_{p}
a_{h'}^{\dag}a_{p'}).
\end{equation}
The $1p-1h$ configurations are built in such a way that the total angular projection $m_\nu$ and parity $\pi_\nu$ are given by $m_\nu=m_p+m_h$ and $\pi_\nu=(-1)^{(l_p+l_h)}$. For the $2p-2h$ configurations, we have in a similar way that $m_\nu=m_p+m_h+m_{p'}+m_{h'}$ and $\pi_\nu=(-1)^{(l_p+l_h+l_{p'}+l_{h'})}$.
Inserting Eq. (\ref{Eq:srpa_op_m}) in the equations of motion (\ref{Eq:EqOfMot}) one gets a set of 
equations
\begin{equation}\label{eq_srpa}
\left(\begin{array}{cc}
 \mathcal{A} & \mathcal{B} \\
 -\mathcal{B}^{*} & -\mathcal{A}^{*} \\
\end{array}\right)
\left(%
\begin{array}{c}
 \mathcal{X}^{\nu} \\
 \mathcal{Y}^{\nu} \\
\end{array}%
\right)=\omega_{\nu}
\left(%
\begin{array}{c}
 \mathcal{X}^{\nu} \\
 \mathcal{Y}^{\nu} \\
\end{array}%
\right),
\end{equation}
where:
\begin{displaymath}
\mathcal{A}=\left(\begin{array}{cc}
 \mathcal{A}_{11} & \mathcal{A}_{12} \\
 \mathcal{A}_{21} & \mathcal{A}_{22} \\
\end{array}\right),
% % \end{displaymath}
% \begin{displaymath}
\mathcal{B}=\left(\begin{array}{cc}
 \mathcal{B}_{11} & \mathcal{B}_{12} \\
 \mathcal{B}_{21} & \mathcal{B}_{22} \\
\end{array}\right),
%\end{displaymath}
%\begin{displaymath}
\mathcal{X}^{\nu}=\left(\begin{array}{cc}
 X_{1}^{\nu} \\
 X_{2}^{\nu} \\
\end{array}\right),
~~~~\mathcal{Y}^{\nu}=\left(\begin{array}{cc}
 Y_{1}^{\nu} \\
 Y_{2}^{\nu} \\
\end{array}\right).
\end{displaymath}

The indices $1$ and $2$ are a short-hand notation for the $1p-1h$ and $2p-2h$ configurations, respectively.
 $\mathcal{A}_{11}$ and $\mathcal{B}_{11}$ are the usual RPA matrices, $\mathcal{A}_{12}$ and $\mathcal{B}_{12}$ are the matrices coupling 
$1p-1h$ with $2p-2h$ configurations ($\mathcal{A}_{21}= \mathcal{A}_{12}^T$ and $\mathcal{B}_{21}= \mathcal{B}_{12}^T$) and $\mathcal{A}_{22}$ and $\mathcal{B}_{22}$ are the matrices
coupling $2p-2h$ configurations among themselves.
They are defined as
\begin{eqnarray}\label{a12}
	\mathcal{A}_{12}=\mathcal{A}_{ph,p_1p_2h_1h_2}&=&\big\langle HF |
	\big[a_{h}^{\dag}a_{p},[H,a_{p_1}^{\dag}a_{p_2}^{\dag}a_{h_2}a_{h_1}
	]\big]| HF \big\rangle 
\end{eqnarray}
\begin{eqnarray}\label{a22}
	\mathcal{A}_{22}=\mathcal{A}_{p_1h_1p_2h_2,p'_1h'_1p'_2h'_2}&=&\big\langle HF |\big[a_{h_1}^{\dag}a_{h_2}^{\dag}a_{p_1}a_{p_2},[H,
	a_{p'_2}^{\dag}a_{p'_1}^{\dag}a_{h'_2}a_{h'_1} ]\big]| HF \big\rangle,
\end{eqnarray}
the $\mathcal{B}_{12}$ is given by replacing $a_{h}^{\dag}a_{p}$ with $a_{p}^{\dag}a_{h}$ in Eq. (\ref{a12})
and the $\mathcal{B}_{22}$ is given by replacing $a_{h_1}^{\dag}a_{h_2}^{\dag}a_{p_1}a_{p_2}$ with $a_{p_2}^{\dag}a_{p_1}^{\dag}a_{h_2}a_{h_1}$ in Eq. (\ref{a22}).
%\begin{eqnarray}\label{b12}
%	\mathcal{B}_{12}=\mathcal{B}_{ph,p_1p_2h_1h_2}&=&\big\langle HF |
%	\big[a_{p}^{\dag}a_{h},[H,a_{p_1}^{\dag}a_{p_2}^{\dag}a_{h_2}a_{h_1}
%	]\big]| HF \big\rangle 
%\end{eqnarray}
%\begin{eqnarray}\label{b22}
%	\mathcal{B}_{22}=\mathcal{B}_{p_1h_1p_2h_2,p'_1h'_1p'_2h'_2}&=&\big\langle HF |\big[a_{p_2}^{\dag}a_{p_1}^{\dag}a_{h_2}a_{h_1},[H,
%	a_{p'_2}^{\dag}a_{p'_1}^{\dag}a_{h'_2}a_{h'_1} ]\big]| HF \big\rangle
%\end{eqnarray}
If QBA is used and the Hamiltonian contains only a one-body and (no density-dependent) two-body terms, it can be shown 
that $\mathcal{B}_{12}$, $\mathcal{B}_{21}$ and $\mathcal{B}_{22}$ are zero.  	The explicit expression of the matrix elements (\ref{a12}) and (\ref{a22}) is given in Appendix \ref{Sec:AppMScheme}.
%If QBA is used and the Hamiltonian contains only a one-body and (no density-dependent) two-body terms, it can be shown 
%that $\mathcal{B}_{12}$, $\mathcal{B}_{21}$ and $\mathcal{B}_{22}$ are zero. The other matrix elements are equal to: 
%
%\begin{eqnarray}\label{a12}
%\mathcal{A}_{12}=\mathcal{A}_{ph,p_1p_2h_1h_2}&=&\big\langle HF |
%\big[a_{h}^{\dag}a_{p},[H,a_{p_1}^{\dag}a_{p_2}^{\dag}a_{h_2}a_{h_1}
% ]\big]| HF \big\rangle =\chi(h_1,h_2){V}_{h_1pp_1p_2}\delta_{hh_2}-\chi(p_1,p_2){V}_{h_1h_1p_1h}\delta_{pp_2}
%\end{eqnarray}
%\begin{eqnarray}\label{a22}
%\mathcal{A}_{22}=\mathcal{A}_{p_1h_1p_2h_2,p'_1h'_1p'_2h'_2}&=&\big\langle HF |\big[a_{h_1}^{\dag}a_{h_2}^{\dag}a_{p_1}a_{p_2},[H,
%a_{p'_2}^{\dag}a_{p'_1}^{\dag}a_{h'_2}a_{h'_1} ]\big]| HF \big\rangle=\nonumber \\
% &=&(\epsilon_{p_1}+\epsilon_{p_2}-\epsilon_{h_1}-\epsilon_{h_2})\chi(p_1,p_2)\chi(h_1,h_2)\delta_{h_1h'_1}\delta_{p_1p'_1}\delta_{h_2h'_2}\delta_{p_2p'_2}+
%\chi(h_1,h_2){V}_{p_1p_2p'_1p'_2}\delta_{h_1h'_1}\delta_{h_2h'_2}+\nonumber\\&&
%\chi(p_1,p_2){V}_{h_1h_2h'_1h'_2}\delta_{p_1p'_1}\delta_{p_2p'_2}+
%\chi(p_1,p_2)\chi(h_1,h_2)\chi(p'_1,p'_2)\chi(h'_1,h'_2){V}_{p_1h'_1h_1p'_1}\delta_{h_2h'_2}\delta_{p_2p'_2}
%\end{eqnarray}
%
%where $\chi(ij)$ is the antisymmetrizer for the indices $i$, $j$.

 By using Eq. (\ref{Eq:ortho}) one gets for the SRPA states the normalization condition:
\begin{equation}
	\label{Eq:norm_SRPA_m}
	\sum_{ph} \left( |X_{ph}^{\nu}|^2 - |Y_{ph}^{\nu}|^2 \right)+ 	\sum_{p<p',h<h'} \left( |X_{php'h'}^{\nu}|^2 - |Y_{php'h'}^{\nu}|^2 \right)= 1.
\end{equation} 

%
%Expressions (\ref{a12}) and (\ref{a22}) are valid in cases where the interaction is not density dependent. Rearrangement 
%terms should be included in the case of density-dependent forces \cite{Gambacurta2011a}.

%
 \subsection{The RPA and SRPA in the J-scheme}\label{Sec:FormalPart_RPA_SRPA_Jscheme}

In the J-scheme, the excited states are described by phonon operators $Q_{\nu}^{\dagger}$ having definite angular momentum, projection and parity $JM\pi$. Excited states are expanded in $1p-1h$ and $2p-2h$ configurations, coupled to $JM\pi$. We use here the notation
$p$ (or $h$) representing all the quantum numbers of a particle 
(hole) state except the magnetic quantum number $m_p$ $(m_h)$. 

The RPA phonon operators are 
\begin{eqnarray} 
	Q_{\nu JM}^{\dagger} &=& 
	\sum_{ph} X_{ph}^{\nu JM} {A^{JM^{\dagger}}_{ph}}
	- (-1)^{J+M}\sum_{ph} Y_{ph}^{\nu ;JM} A^{J\, -M}_{ph}
		\label{E:OpRPA_J} 
\end{eqnarray} 
where $A^{JM^{\dagger}}_{ph}$ is a $1p-1h$ operator with quantum numbers $JM$: 
\begin{equation} 
	A^{JM^{\dagger}}_{ph} = \sum_{m_p,m_h} (-1)^{j_h-m_h}\langle j_p m_p j_h -m_h | JM_J \rangle
	a_{pm_p}^{\dagger} a_{hm_h}. 
\end{equation} 
The RPA equations have the same structure as in Eq. (\ref{Eq:RPAmat})
%\begin{equation}
%	\label{Eq:RPAmat}
%	\begin{pmatrix}
%		\mathcal{A}_{11} & \mathcal{B}_{11} \\
%		-\mathcal{B}_{11}^* & -\mathcal{A}_{11}^*
%	\end{pmatrix}
%	\begin{pmatrix}
%		X^\nu \\
%		Y^\nu
%	\end{pmatrix}
%	= \hbar \omega_\nu
%	\begin{pmatrix}
%		X^\nu \\
%		Y^\nu
%	\end{pmatrix}
%\end{equation}
%
%where the subscript $1$ stands for the $1p-1h$ configurations. T
where he $A_{11}$ and $B_{11}$ matrices in the angular momentum-coupled forms are given by in Eqs (\ref{Eq:RPA_A_J}) and (\ref{Eq:RPA_B_J}).

By using Eq. (\ref{Eq:ortho}) one gets the normalization in the RPA case:
\begin{equation}
	\label{Eq:norm_RPA_J}
	\sum_{ph} \left( |X_{ph}^{\nu JM}|^2 - |Y_{ph}^{\nu JM}|^2 \right) = 1.
\end{equation}

The SRPA operators are then defined as

\begin{eqnarray} 
Q_{\nu JM}^{\dagger} &=& 
\sum_{ph} X_{ph}^{\nu JM} {\mathcal{A}^{JM^{\dagger}}_{ph}}
- (-1)^{J+M}\sum_{ph} Y_{ph}^{\nu ;JM} \mathcal{A}^{J\, -M}_{ph}
\nonumber \\ 
&& 
 + \!\! \sum_{p_1\leq p_2,h_1\leq h_2;J_p,J_h} \!\! 
 \mathcal{X}_{p_1h_1p_2h_2J_pJ_h}^{\nu JM} {\mathcal{A}^{JM^{\dagger}}_{p_1h_1p_2h_2J_pJ_h}}
 - (-1)^{J+M}
 \mathcal{Y}_{p_1h_1p_2h_2J_pJ_h}^{\nu JM} \mathcal{A}^{J\, -M}_{p_1h_1p_2h_2J_pJ_h}
, 
\label{E:OpSRPA_J} 
\end{eqnarray} 
where ${\mathcal{A}^{JM^{\dagger}}_{p_1h_1p_2h_2J_pJ_h}}$ %=a^{\dagger}_pa_ha^{\dagger}_{p'}a_{h'}$
creates a $2p-2h$ state, coupled to the given quantum numbers:
\begin{eqnarray} 
\mathcal{A}^{JM^{\dagger}}_{p_1h_1p_2h_2J_pJ_h} &=& \hspace{-7mm} %\sum_{m_{p_1}m_{p_2}m_{h_1}m_{h_2}M_pM_h}
\sum_{m'sM_pM_h}
% \hspace{-9mm} 
 \langle j_{p_1}m_{p_1}j_{p_2}m_{p_2} | J_pM_p\rangle %\times
% \nonumber \\ & & \times
 \langle j_{h_1}m_{h_1}j_{h_2}m_{h_2} | J_hM_h\rangle 
% \nonumber \\ & & \times
 %\times
 (-1)^{J_j-M_h} 
 \langle J_pM_pJ_h-M_h | JM\rangle
 \nonumber \\ & & \times
 (1+\delta_{p_1p_2})^{-1/2}(1+\delta_{h_1h_2})^{-1/2} 
% \nonumber \\ & & \times
 \times
 a_{p_1,m_{p_1}}^{\dagger} 
 a_{p_2,m_{p_2}}^{\dagger} 
 a_{h_1,m_{h_1}} 
 a_{h_2,m_{h_2}} 
\end{eqnarray} 
obtained by coupling the two hole- and two particle- operators to a total $J_h$ and $J_p$, respectively, giving then the total angular momentum $J$. 

The SRPA equations are formally equivalent to Eq. (\ref{eq_srpa} ) and the corresponding matrix elements are given in Eqs (\ref{Eq:SRPA_A12_J}) and (\ref{Eq:SRPA_A22_J}).
% 
%The SRPA equations are then:
%\begin {equation} 
%\left( \begin{array}{cc|cc} 
%\mathcal{A}_{11} & \mathcal{A}_{12} & \mathcal{B}_{11} & \mathcal{B}_{12} \\
%\mathcal{A}_{21} & \mathcal{A}_{22} & \mathcal{B}_{21} & \mathcal{B}_{22} \\ \hline 
%-\mathcal{B}_{11}^{\ast} & -\mathcal{B}_{12}^{\ast} & -\mathcal{A}_{11} & -\mathcal{A}^{\ast}_{12} \\
% -\mathcal{B}_{21}^{\ast} & -\mathcal{B}_{22}^{\ast} & -\mathcal{A}_{21}^{\ast} & -\mathcal{A}^{\ast}_{22} \\ 
%\end{array} 
%\right) 
%\left( 
%\begin{array}{c} 
%X^{\lambda} 
%\\ 
%\mathcal{X}^{\lambda} 
%\\ 
%\hline 
%Y^{\nu} 
%\\ 
%\mathcal{Y}^{\lambda} 
%\end{array} 
%\right) 
%= \omega_{\nu}
%\left( 
%\begin{array}{c} 
%X^{\nu} 
%\\ 
%\mathcal{X}^{\lambda} 
%\\ 
%\hline 
%Y^{\nu} 
%\\ 
%\mathcal{Y}^{\lambda} 
%\end{array} 
%\right) 
%\label{eq_srpa} 
%. \end{equation} 
%If the Hamiltonian contains only a one- and two-body operators, then $\mathcal{B}_{21}=\mathcal{B}_{12}=\mathcal{B}_{22}=0$ as the reference state is the HF ground state. In the case of effective interaction containing also density dependent terms, rearrangement terms can appear in the $\mathcal{B}_{21}$ and $\mathcal{B}_{12}$ matrices \cite{Gambacurta2011a}.

 The normalization for the SRPA states reads:
\begin{equation}
	\label{Eq:norm_SRPA_J}
	\sum_{ph} \left( |X_{ph}^{\nu JM}|^2 - |Y_{ph}^{\nu JM}|^2 \right) +\sum_{p<p',h<h'} \left( |\mathcal{X}_{p_1h_1p_2h_2J_pJ_h}^{\nu JM}|^2 - |\mathcal{Y}_{p_1h_1p_2h_2J_pJ_h}^{\nu JM}|^2 \right)
	 = 1.
\end{equation} 
 
\subsection{Tamm-Dancoff approximations}
\label{Sec:Tamm-Dancoff} 
The Tamm-Dancoff (TD) \cite{RS.80} and Second TD (STD) \cite{Minato2016} approximations are obtained by neglecting the backward $Y'$s components in the phonon operators, e.g., 
	\begin{equation}
	\label{Eq:OpTammDancoff}
	Q^\dagger_\nu = \sum_{ph} X^\nu_{ph} a^\dagger_p a_h 
\end{equation} 
and
\begin{equation}\label{Eq:srpa_TammDancoff}
	Q_{\nu}^{\dagger}=\sum_{ph}X_{ph}^{\nu}a_{p}^{\dag}a_{h} +\sum_{p<p',h<h'}X_{php'h'}^{\nu}
	a_{p}^{\dag}a_{h}a_{p'}^{\dag}a_{h'}
\end{equation}
in the TD and STD, respectively.
The corresponding equations of motion are thus
	\begin{equation}\label{eq_tammdancoff}
	\mathcal{A} \mathcal{X}^{\nu}=\omega_{\nu} \mathcal{X}^{\nu}
\end{equation}
where
\begin{displaymath}
	\mathcal{A}=A_{11},
	\mathcal{X}^{\nu}=
		X_{1}^{\nu} \\
\end{displaymath}
in the TD case, and
\begin{displaymath}
	\mathcal{A}=\left(\begin{array}{cc}
		A_{11} & A_{12} \\
		A_{21} & A_{22} \\
	\end{array}\right),
	\mathcal{X}^{\nu}=\left(\begin{array}{cc}
		X_{1}^{\nu} \\
		X_{2}^{\nu} \\
	\end{array}\right),
\end{displaymath}
 in  the STD one. The TD (STD) equations are thus nothing but that the diagonlization of the Hamiltonian in the $1p-1h$ ($1p-1h$ plus $2p-2h$) space on top of the HF state. Ground state correlations inherently incorporated within the RPA are thus lost, which results also in technical/physical drawbacks, as for example the treatment of spurious components associated to broken symmetries\cite{RS.80}.
 Usually TD and STD calculations are performed only for numerical reasons, as the matrix to be handled and diagonalized has half the size of the RPA (or SRPA) matrix.
 
\subsection{Diagonal approximation in SRPA} 
\label{Sec:Diagonal_SRPA}
By projecting the SRPA equations of motion (\ref{eq_srpa}) on to the $1p-1h$ subspace \cite{Drozdz1986,Providencia1965}, e.g., by eliminating the $2p-2h$ $\mathcal{X}$ and $\mathcal{Y}$ amplitudes, it can be shown that the SRPA problem can be recast in an energy-dependent RPA-like problem 
\begin{equation} 
	\left( \begin{array}{cc} 
		\mathcal{A}_{1,1'}(\mathcal{\omega}_{\nu} ) & \mathcal{B}_{1,1'} \\ -\mathcal{B}_{1,1'}^{\ast} & -\mathcal{A}_{1,1'}^{\ast}(\mathcal{\omega}_{\nu} )
	\end{array} \right) 
	\left( \begin{array}{cc} 
		X^{\lambda} \\ Y^{\lambda} 
	\end{array} \right) 
	= \omega_\nu
	\left( \begin{array}{cc} 
		X^{\lambda} \\ Y^{\lambda} 
	\end{array} \right) 
	\label{E:SRPAred1} 
\end{equation} 

 where the $\mathcal{A}_{11}$ matrix depends on the excitation energies
\begin{equation}
	\label{E:Full} 
	\mathcal{A}_{1,1'}(\omega)=\mathcal{A}_{1,1'}+\sum_{2,2'}\mathcal{A}_{1,2}(\omega +i\eta -\mathcal{A}_{2,2'})^{-1}{\mathcal{A}}_{2',1'}
\end{equation}
through the coupling of the $1p-1h$ configurations to the $2p-2h$ ones. In order to calculate this energy-dependent part one has to invert the $A_{22}$ matrix  whose dimensions are generally very large, requiring thus a strong, often prohibitive, numerical effort. However, if the terms depending on the residual interaction are neglected, resorting to the so-called diagonal approximation, the inversion is algebraic.
If the coupling among the $2p-2h$ states is neglected, $A_{22}$ becomes diagonal and its elements are determined by the
unperturbed $2p-2h$ energies (diagonal approximation),
\begin{equation} 
	\mathcal{A}_{p_1h_1p_2h_2,p_1'h_1'p_2'h_2'} = 
	\delta_{p_1p_1'}\delta_{h_1h_1'}\delta_{p_2p_2'}\delta_{h_2h_2'}
	(e_{p_1}+e_{p_2}-e_{h_1}-e_{h_2}). 
	\label{Eq:Diagonal} 
\end{equation} 
the inversion is trivial and we have 
\begin{eqnarray} 
\mathcal{A}_{1,1'}(\omega)=	\mathcal{A}_{php'h'}(\omega ) &=& \mathcal{A}_{php'h'} 
	+ \sum_{p_1p_2h_1h_2} 
	\frac{\mathcal{A}_{ph;p_1p_2h_1h_2}\mathcal{A}^T_{p_1p_2h_1h_2;p'h'}}{\omega - (e_{p_1} + e_{p_2} - e_{h_1}-e_{h_2}) + i\eta_2} 
	\label{Eq:SRPAred2} 
\end{eqnarray} 
where the finite constant $\eta_2>0$ is generally employed in applications to smooth the poles of the function.

 \subsection{Rearrangement terms in SRPA}\label{Sec:FormalPart_SRPA_Rear}

 It's well established that when employing density-dependent forces, such as Skyrme or Gogny interactions, the residual interaction employed in constructing the RPA matrices  includes rearrangement terms. These terms arise from the derivative of the MF Hamiltonian with respect to the density, which corresponds to the second derivative of the energy-density functional. As the density-dependence of the interaction directly modifies the MF, rearrangement effects also appear in the single-particle energies. Extending this treatment to the SRPA case is not straightforward. In Ref. \cite{Gambacurta2011a}, the methodology originally proposed in Ref. \cite{Providencia1965} was adapted and generalized for the SRPA framework. In this specific formulation, the ground state of the system is described as a superposition of $1p-1h$ and $2p-2h$ configurations, all built upon the HF state. The coefficients associated with these configurations are used as variational parameters. The SRPA equations of motion are then derived by minimizing the expectation value of the Hamiltonian with respect to these parameters. The derivation, when reduced to the RPA limit, accurately recovers the correct RPA rearrangement terms. %, which are precisely defined by the second derivative of the EDF with respect to the density. 
  In the SRPA case, a first kind of rearrangement terms appears in the matrix $A_{22}$ leading to the correct definition of the single-particle energies, consistently with RPA.
 The same kind of terms appears also in the $A_{12}$ matrix and finally gives a vanishing contribution owing to the HF condition \cite{RS.80}. It is found that no other 
 rearrangement terms appear in the residual interaction of the 
 matrices $A_{22}$ and $B_{22}$. 
 The matrix $B_{22}$ is equal to zero for a two-body density-dependent interaction. 
 In the matrices $A_{12}$ and $B_{12}$, rearrangement terms of the residual 
 interaction are found and they are not given by applying the RPA prescription. The explicit derivation and expression of the so obtained rearrangement terms can be found in Appendix \ref{Sec:Rear} and their impact on the SRPA results are shown in Section \ref{Sec:ApplicationPart_SRPA_Rear}.

	 \subsection{Subtracted SRPA }\label{Sec:FormalPart_SSRPA}
	\subsubsection{Motivation within the EDF theory}
The EDF approach is a powerful method for describing the properties of many-body systems, and particularly atomic nuclei. In this context, where knowledge of the in-medium nucleon-nucleon interaction is still limited, the EDF framework proves especially useful.
The precise definition of a density-dependent interaction, particularly in cases lacking antisymmetric two-body matrix elements, presents a challenge. In practice, such interactions serve as a basis for constructing the expression of the total energy of the system and the subsequent variation of which yields the MF equations. The standard way to implement this scheme involves considering the energy as a functional of various densities, such as the number density and spin density \cite{DFT}.
% This methodology, termed Density Functional Theory (DFT), has found widespread application in condensed matter and atomic physics \cite{DFT}.

The  theoretical framework is based on the Hohenberg-Kohn theorem \cite{HohenbergKohn} and the Kohn-Sham procedure \cite{KohnSham}. While these principles require adaptation for nuclear physics to accommodate densities defined with respect to the nuclear center of mass, they retain their validity and applicability \cite{Engle2007}. For the sake of simplification, we limit our discussion here to the intrinsic one-body density, $\rho(r)$, neglecting other potential density and current dependencies within the functional. The key quantity is the functional E$[\rho]$, representing the expectation value of the underlying nuclear Hamiltonian as a function of the density $\rho$. The system's ground-state energy and density are then determined by minimizing E$[\rho]$. The Hohenberg-Kohn theorem establishes the universality of the energy-density functional E$[\rho]$. Specifically, the theorem states that under the influence of an additional local Hermitian operator, $\lambda Q(r)$ , where $\lambda$ is an arbitrary constant, E$[\rho]$ undergoes a well-defined modification:
\begin{equation}
\label{eq:univ}
E[\rho] \longrightarrow E_\lambda[\rho] = E[\rho] + \lambda \! \int \! d\br \, Q(\br) \rho(\br).
\end{equation}

The Kohn-Sham procedure ensures that any energy-density functional can be expressed in terms of single-particle orbitals, $\varphi_i(\mathbf{r})$, representing the system as non-interacting particles subject to a density-dependent external potential. The ground-state energy and density are then obtained by solving the  equations for these orbitals, derived from minimizing the functional with respect to the orbital wave functions. %Skyrme and Gogny energy-density functionals, are typical example of nuclear EDF, and they can be interpreted as approximations to the exact Kohn-Sham functional.

Consider now a small perturbation, characterized by the parameter $\lambda$. The system's density will change from the unperturbed ground-state density, $\rho_0$, to a perturbed density, $\rho_\lambda$, given by:

\begin{equation}
\label{eq:pert-dens}
\rho_\lambda = \rho_0 + \lambda \int d\mathbf{r}' \, R(\omega=0,\mathbf{r},\mathbf{r}') Q(\mathbf{r}'),
\end{equation}

where $R(\omega=0)$ represents the static response function of the underlying Hamiltonian. Due to the universality of the energy functional, it exactly reproduces $\rho_\lambda$ when modified according to Eq. \eqref{eq:univ}. Within the Kohn-Sham framework, the functional yields a MF effective Hamiltonian, which, nevertheless, accurately reproduces exact energies and densities. Consequently, the response function $R(\omega=0)$ is given by the RPA, corresponding to the small-amplitude limit of time-dependent MF theory \cite{RS.80}:
\begin{equation}
\label{eq:resp-zero}
R(\omega=0) = R^{RPA}_{KS} \,
\end{equation}
where, $R^{RPA}_{KS}$ denotes the RPA response function calculated using the Kohn-Sham representation of the EDF and the corresponding ground-state Slater determinant. Therefore, if the Skyrme functional were the exact Kohn-Sham functional, the Skyrme-RPA would yield the exact zero-frequency response function. Consequently, any modification to the response function must vanish in the static limit. This requirement, often referred to as "avoiding double counting," was highlighted  in Ref. \cite{Tselyaev2013}.

A time-dependent version of the Hohenberg-Kohn theorem, known as the Runge-Gross theorem \cite{RungeGross}, a time-dependent Kohn-Sham procedure, demonstrates that the full response function at any frequency $\omega$, satisfies:

\begin{align}
\label{eq:resp-eq}
R(\omega,\mathbf{r},\mathbf{r}') = R^0_{KS}(\omega,\mathbf{r},\mathbf{r}')+ \int d\mathbf{r}_1 d\mathbf{r}_2 \, R^0_{KS}(\omega,\mathbf{r}, \mathbf{r}_1) V(\omega,\mathbf{r}_1,\mathbf{r}_2) R(\omega,\mathbf{r}_2,\mathbf{r}') \, %\nonumber
\end{align}

where $R^0_{KS}$ is the bare Kohn-Sham (MF) response and $V(\omega)$ is a frequency-dependent effective interaction derived from the time-dependent energy-density functional, $\mathcal{E}[\rho(t),t]$. The approximation

\begin{equation}
\label{eq:adiab-approx}
V(\omega,\mathbf{r}_1,\mathbf{r}_2) \longrightarrow \frac{\delta^2 E[\rho]}{\delta \rho(\mathbf{r}_1) \delta \rho (\mathbf{r}_2)}\bigg|_{\rho_0} \equiv V^{RPA}(\mathbf{r}_1,\mathbf{r}_2) \,
\end{equation}

implies that the solution $R$ is equivalent to $R^{RPA}_{KS}$, which is independent of $\omega$. It amounts to assuming that $\mathcal{E}[\rho(t),t] = E[\rho(t)]$, i.e., that the time-dependent energy is simply the ground-state functional evaluated at the time-dependent density. This approximation is called the adiabatic limit. To extend beyond this limit, one must introduce an $\omega$ dependence into the effective two-body interaction. This is precisely what the SRPA does. However, given that $R^{RPA}_{KS}$ is correct (as correct as the Skyrme functional is) within the adiabatic limit, the SRPA must be modified to recover the RPA response at $\omega=0$.

%While several modifications to the SRPA's $V(\omega)$ could achieve this, The response function must also respect other constraints. Specifically, the quantity $\int d\mathbf{r} d\mathbf{r}' Q(\mathbf{r}) R(\omega,\mathbf{r},\mathbf{r}') Q(\mathbf{r}')$ must possess real and positive residues at poles on the positive real axis, satisfying the stability condition, ensuring thus a physical strength function. One straightforward method to guarantee this is the subtraction method. 
Defining the energy-dependent difference between the SRPA and RPA effective interactions as $W(\omega)$:

\begin{equation}
\label{eq:U}
W(\omega) \equiv V^{SRPA}(\omega) - V^{RPA}(\omega) \,
\end{equation}

the subtraction procedure involves the replacement:

\begin{equation}
\label{eq:subtraction-dft}
V^{SRPA}(\omega) \longrightarrow V^{SRPA}(\omega) - W(0) \,.
\end{equation}

This ensures that $V^{SRPA}(0) = V^{RPA}(0)$ after the substitution, as required, and thus, the SSRPA reduces to the RPA in the zero-frequency limit.

%Below, the matrix implementation of the subtraction procedure in the SRPA method is discussed.

\subsubsection{Subtraction method in SRPA}

As discussed in Section \ref{Sec:Diagonal_SRPA}, by projecting the SRPA equations in the $1p-1h$ space, the SRPA equations can be written as energy dependent RPA-like equations:
\begin{equation}\label{eq_srpa_proj}
	\left(\begin{array}{cc}
		 A_{11^{\prime}}(\omega) & B_{11^{\prime}}(\omega) \\
		 -B_{11^{\prime}}(\omega) & A_{11^{\prime}}(\omega) \\
		\end{array}\right)
		\left(%
		\begin{array}{c}
		 \mathcal{X}^{\nu} \\
		 \mathcal{Y}^{\nu} \\
		\end{array}%
		\right)=\omega_{\nu}
		\left(%
		\begin{array}{c}
		 \mathcal{X}^{\nu} \\
		 \mathcal{Y}^{\nu} \\
		\end{array}%
		\right),
		\end{equation}
where
\begin{eqnarray}
\label{arpa}
A_{11^{\prime}} (\omega) = A_{11^{\prime}}+\sum_{2,2^{\prime}} A_{12} (\omega + i \eta - A_{22^{\prime}})^{-1}
A_{2^{\prime}1^{\prime}} -
\sum_{2,2^{\prime}} B_{12} (\omega + i \eta + A_{22^{\prime}})^{-1}
B_{2^{\prime}1^{\prime}} \,\\
B_{11^{\prime}} (\omega) = B_{11^{\prime}} + \sum_{2,2^{\prime}} A_{12} (\omega + i \eta - A_{22^{\prime}})^{-1}
B_{2^{\prime}1^{\prime}} -
\sum_{2,2^{\prime}} B_{12} (\omega + i \eta + A_{22^{\prime}})^{-1}
A_{2^{\prime}1^{\prime}}\,. \nonumber
\label{brpa}
\end{eqnarray}
assuming that all the matrix blocks are real. These expressions are just the analogs of the
$\omega$--dependent interaction $V(\omega)$ discussed in the previous Section. Without so--called rearrangement terms, e.g., when no density-dependent forces are employed, $B_{12}$ and
$B_{21}$ would vanish. Thus, there would be no correction to $B_{11^{\prime}}$, which
would simply become the corresponding RPA matrix without any energy- dependent correction.

Let us denote by $\mathcal{A}_{11^{\prime}}(\omega)$ and $\mathcal{B}_{11^{\prime}}(\omega)$ the
energy--dependent corrections to $A_{11^\prime}$, and $B_{11^\prime}$:

\begin{eqnarray}
\mathcal{A}_{11^{\prime}} (\omega) = \sum_{2,2^{\prime}} A_{12} (\omega + i \eta - A_{22^{\prime}})^{-1}
A_{2^{\prime}1^{\prime}} -
\sum_{2,2^{\prime}} B_{12} (\omega + i \eta + A_{22^{\prime}})^{-1}
B_{2^{\prime}1^{\prime}}\, \\
\mathcal{B}_{11^{\prime}} (\omega) = \sum_{2,2^{\prime}} A_{12} (\omega + i \eta - A_{22^{\prime}})^{-1}
B_{2^{\prime}1^{\prime}} -
\sum_{2,2^{\prime}} B_{12} (\omega + i \eta + A_{22^{\prime}})^{-1}
A_{2^{\prime}1^{\prime}}\,. \nonumber
\end{eqnarray}

The subtraction procedure amounts to correct the RPA--like
matrices by subtracting from $A_{11^{\prime}}(\omega)$ and
$B_{11^{\prime}}(\omega)$ the static parts, e.g., their contribution at $\omega=0$, $\mathcal{A}_{11^{\prime}}(0)$ and
$\mathcal{B}_{11^{\prime}}(0)$:
\begin{equation}
\mathcal{A}^S_{11^{\prime}} (\omega)= A_{11^{\prime}} (\omega) - \mathcal{A}_{11^{\prime}}(0),
\label{sub1}
\end{equation}
\begin{equation}
B^S_{11^{\prime}} (\omega)= B_{11^{\prime}} (\omega) - \mathcal{B}_{11^{\prime}}(0).
\label{sub2}
\end{equation}
$\mathcal{A}^S_{11^{\prime}} (\omega)$ and $B^S_{11^{\prime}} (\omega)$ are then
substituted for $A_{11^{\prime}} (\omega) $ and $B_{11^{\prime}} (\omega)$ in
the energy--dependent RPA--like equations. One can go back to the original 
energy--independent equations in the coupled $1p-1h$ and $2p-2h$ spaces, having for the $A$ and $B$ matrices:
\begin{eqnarray}
\label{eq:absf}
\mathcal{A}^S=\left(\begin{array}{cc}
 A_{11'}+ \sum_{2,2'} A_{12} (A_{22'})^{-1}A_{2'1'} + \sum_{2,2'} B_{12} (A_{22'})^{-1} B_{2'1'} & A_{12} \\
 & \\
 A_{21} & A_{22'} \\
\end{array}\right)\, \\[.3cm]
\mathcal{B}^S=\left(\begin{array}{cc}
 B_{11'} + \sum_{2,2'} A_{12} (A_{22'})^{-1} B_{2'1'} + \sum_{2,2'} B_{12} (A_{22'})^{-1} A_{2'1'} & B_{12} \\
 & \\
 B_{21} & 0\\ %B_{22'} \\
\end{array}\right)\,. 
\end{eqnarray}
From a numerical point of view, the most demanding task in performing the subtraction is the inversion of the $A_{22'}$ matrix. In Section \ref{Sec:Applications_SSRPA_O16}, we will compare the full calculations, i.e., when the $A_{22'}$ is fully calculated and then inverted, with those when the matrix is assumed to be diagonal allowing for an exact algebraic inversion. We will see that the full calculation allows us to restore the RPA inverse energy--weighted moment $m_{-1}$, meaning that the SSRPA $m_{-1}$ value is found to be the same as the RPA, while some deviations are found when the diagonal approximation is used. %We also note that as the $m_{-1}$ is generally larger in the standard SRPA than in the RPA, the $m_{-1}$ restoration produces a shift energy upwards with respect to the ordinary SRPA of the multipole-response. 
The restoration of the inverse energy--weighted moment, implies that the static polarizability, is identical in the subtracted SRPA and the RPA, providing a new sum-rule benchmark for the SSRPA calculations (see also discussion at the end of the next Section \ref{Sec:FormalPart_SRPA_Moments}).
Section \ref{Sec:Applications_SSRPA_CC} and \ref{Sec:Applications_SSRPA_CE} are dedicated to the most recent application of the SSRPA framework.
 
\subsection{Moments and Sum Rules}
\label{Sec:FormalPart_SRPA_Moments}
Let's consider an arbitrary one-body operator $F$ exciting the nucleus from its ground state $\mid 0 \rangle$ to its eigenstate $\mid i \rangle$. The strength function
\begin{equation}
	S(\omega)= \sum_i \big|\langle i\mid F \mid 0 \rangle \big|^2 \delta (\omega - \omega_i)
\end{equation}
defines the response of the nuclear system to the action of the operator $F$, where $\omega_i$ is the excitation energy of the state $\mid i \rangle$.
One can then define the $k$-order moment $m_k$ as
\begin{equation}
	m_k=\int_0^\infty S(\omega) \omega^k d \omega= \sum_i \big|\langle i\mid F \mid 0 \rangle \big|^2 \omega^k_i.
\end{equation}
For operators such as $\langle 0\mid F \mid 0 \rangle=0$ and considering $k\geq0$, one has

\begin{equation}
	m_k=\langle 0\mid F^\dagger (H-E_0)^k F\mid 0 \rangle.
\end{equation}
The evaluation of the moments depends thus on the knowledge of the ground state or of the complete set of excited state of the system.

Let's now consider their evaluation and comparison within the RPA and SRPA \cite{Adachi1988}.\\
The standard RPA sum rule reads as
$$
m_k^{\mathrm{RPA}}(F) = \frac{1}{2} F^\dagger (C_{11})^k \mathcal{N} F
$$

where 
$$
\mathcal{N} = 
\begin{pmatrix}
	I & 0 \\
	0& -I 
\end{pmatrix},
C_{11} = 
\begin{pmatrix}
	A_{11} & B_{11} \\
	-B_{11}^* & -A_{11}^*
\end{pmatrix}.
$$

 In SRPA one has that

$$
m_k^{\mathrm{SRPA}}(F, F) = \frac{1}{2} F^\dagger ( \mathcal{S})^k \mathcal{N} F
$$
where 

$$
\mathcal{S} = 
\begin{pmatrix}
	C_{11} & C_{12} \\
	C_{21} & C_{22}
\end{pmatrix}
$$
with

$$
C_{11} = 
\begin{pmatrix}
	A_{11} & B_{11} \\
	-B_{11}^* & -A_{11}^*
\end{pmatrix},
C_{12} = 
\begin{pmatrix}
	A_{12} & 0 \\
	0 & -A_{12}^*
\end{pmatrix},
C_{22} = 
\begin{pmatrix}
	A_{22} & 0 \\
	0 & -A_{22}^*
\end{pmatrix},
C_{21} =C_{12}^T .
$$
Since the $F$ operator has only $1p-1h$ and $1h-1p$ components, (e.g. two-body components are zero) the following relations hold:
$$
m_0^{\mathrm{SRPA}}(F) = \frac{1}{2} F^\dagger \mathcal{N} F = m_0^{\mathrm{RPA}}(F)
$$

\begin{equation}
	\label{Eq:m1_RPA_SRPA}
	m_1^{SRPA}(F) = \frac{1}{2} F^\dagger C_{11} \mathcal{N} F = m_1^{RPA}(F)
\end{equation}

%\begin{aligned*}
\begin{equation}
		m_3^{\mathrm{SRPA}}(F) = \frac{1}{2} F^\dagger C_{11}^3 \mathcal{N} F +\frac{1}{2} F^\dagger \left( C_{12} C_{22} C_{21} + C_{11} C_{12} C_{21} + C_{12} C_{21} C_{11} \right) \mathcal{N} F \\
	= m_3^{\mathrm{RPA}}(G, F) + \delta m_3(G, F)
\end{equation}

%\end{aligned*}
The SRPA theory does not alter the non-energy-weighted sum rule $m_0$ and the Energy-Weighted Sum Rule
 (EWSR) $m_1$ with respect the standard RPA values, independently of
the nature of the one-body transition operators. On the contrary, the $m_{3}$ moment is modified in SRPA. 

The RPA sum rules satisfy the following relations: 

\begin{equation}
	m_0^{\mathrm{RPA}}(F) = \langle \mathrm{HF} | [F^\dagger, F] | \mathrm{HF} \rangle
\end{equation}

\begin{equation}
	m_1^{\mathrm{RPA}}(F, F) = \frac{1}{2} \langle \mathrm{HF} | \big[F^\dagger, [H, F]\big] | \mathrm{HF} \rangle
\end{equation}

\begin{equation}
m_3^{\mathrm{RPA}}(F) = \frac{1}{2} \langle \mathrm{HF} | \Big[\big[F^\dagger, H], [H, [H, F]\big]\Big] | \mathrm{HF} \rangle
\end{equation}
Special attention has to be paid in the evaluation of the commutators involving the Hamiltonian in case of density-dependent interactions, as discussed in Ref. \cite{Lipparini1989}.

%---
%

The Thouless theorem \cite{THOULESS1960} holds in the RPA case, determining that for a one-body operator
\begin{equation}
	\label{Eq:EWSR_RPA}
	m_1^{RPA}=\sum_i \big|\langle i\mid F \mid 0 \rangle \big|^2 \omega_i=\langle HF \mid \Big [ F^\dagger, \big[ F,H \big] \Big] \mid HF \rangle
\end{equation}
meaning that evaluation of the double commutator
in the HF ground state is equivalent to calculating the sum over the RPA eigenstates. Thanks to Eq. (\ref{Eq:m1_RPA_SRPA}), the same is true when the sum is performed over the SRPA eigenstates. Eq. (\ref{Eq:EWSR_RPA}) provides therefore a stringent benchmark for both RPA and SRPA approximations.
However, some remarks are in order. 
One of the main implication of the Thouless theorem is that, in a fully self-consistent\footnote{The exact definition and meaning of self-consistency  will be discussed in Section \ref{Sec:Applications_SRPA_Stability}.} HF plus RPA scheme, the  EWSR is preserved, and the spurious mode associated to broken symmetries should 
separate out from the physical spectrum and have zero energy\cite{THOULESS1960,RS.80}. 
This is a consequence of Eq. (\ref{Eq:EWSR_RPA}) and of the fact that the RPA matrix is semi-positive definite if the HF solution is variationally obtained, e.g., it corresponds to the energy minimum.
In the SRPA case, although the first moment is the same as in RPA, the previous condition is not fulfilled anymore by the SRPA matrix. The coupling with the $2p-2h$ configurations can indeed induce stability issues in the SRPA matrix, and therefore the Thouless theorem does not hold anymore. Special attention should be thus payed to the possible mixing with spurious components. These issues will be extensively discussed in Section \ref{Sec:Applications_SRPA_Stability}, also by means of numerical examples.  

In the SSRPA case, because of the subtraction procedure, the first moment $m_1$ is not anymore equal to the RPA one.
Therefore, Eq. (\ref{Eq:m1_RPA_SRPA}) can not be used anymore as a benchmark for SSRPA calculations. However, we note that, for the SRPA inverse energy-weighted sum rule, we have:
\begin{equation}
m_{-1}^{\mathrm{SRPA}}(F) = \frac{1}{2} F^\dagger (C_{11} - C_{12} C_{22}^{-1} C_{21})^{-1} \mathcal{N} F
\end{equation}

where 

$$
C_{11} - C_{12} C_{22}^{-1} C_{21}=\begin{pmatrix}
	A_{11} -A_{12}A_{22}^{-1}A_{21} & B_{11} \\
	- B_{11}^*& A_{11} -(A_{12}A_{22}^{-1}A_{21} )^* 
\end{pmatrix}.
$$

In SRPA, the contribution from $2p-2h$ space brings in the inverse EWSR $m_{-1}$ an additional energy-dependent term $-A_{12}A_{22}^{-1}A_{21}$ to $A_{11}$.
Comparing it with the correction term in Eq. (\ref{eq:absf}), where the contribution of the $B$ matrix is zero for non-density dependent forces, one can easily recognize that the correction induced in SRPA to the $m_{-1}$ evaluated at zero-energy, is exactly the term that is subtracted in the SSRPA. Therefore, the subtraction procedure does reestablish the RPA $m_{-1}$ value and  the inverse energy sum rule can be used as a benchmark test to judge the robustness and reliability of SSRPA calculations. Numerical tests will be presented and discussed  Section \ref{Sec:Applications_SSRPA_CC}.

	\newpage
	\section{Applications of the SRPA }\label{Sec:Applications_SRPA}
%	\subsection{Introduction}
In this Section, we present different SRPA large-scale calculations, meaning that large $2p-2h$ spaces are considered large enough to obtain convergent results, as compared to previous SRPA applications, where severe truncations had to be adopted. SRPA calculations performed using different kinds of nuclear interaction are presented, discussing their similarities and differences. The SRPA instabilities due to the fact that the SRPA matrix is not necessarily positive semi-definite, (even in a fully self-consistent framework, as it is the case for the RPA), will be discussed. The common difficulties encountered, independently of the kind of employed interaction, will be underlined, motivating thus the need of the subtraction procedure, to obtain stable and pathology-free results to be safely compared with experimental data. 
 \subsection{SRPA calculations based on the UCOM method}\label{Sec:Applications_SRPA_UCOM}
The Unitary Correlation Operator Method (UCOM) \cite{ROTH2004} is a many body technique to derive effective nucleon-nucleon interactions to be employed in microscopic nuclear theories. It starts with a realistic nucleon-nucleon interaction and explicitly treats short-range correlations, resulting in a softened, phase-shift equivalent interaction that we will indicate as $V_{UCOM}$. In the applications here discussed, $V_{UCOM}$ is derived from the Argonne V18 potential through a unitary transformation whose parameters are fixed using data from Helium-4 \cite{ROTH2005}.
%Previous applications have shown that $V_{UCOM}$, while successful in describing binding energies in perturbation theory (PT) \cite{ROTH2005}, underestimates charge radii, suggesting the need for a three-body term. HF calculations with $V_{UCOM}$ also underestimate binding energies and overestimate single-particle level spacing.
 RPA calculations using $V_{UCOM}$ \cite{Paar2006} provided reasonable agreement with experimental ISGMR energies but overestimated IVGDR and ISGQR energies by several MeV. The effect of $2p-2h$ configurations introduced within the UCOM-SRPA has been studied in several papers \cite{Papakonstantinou2009,Papakonstantinou2010,Papakonstantinou2014}, we report below the main results and conclusions.

\begin{figure}[h]
	\centering
	\includegraphics[width=0.8\textwidth]{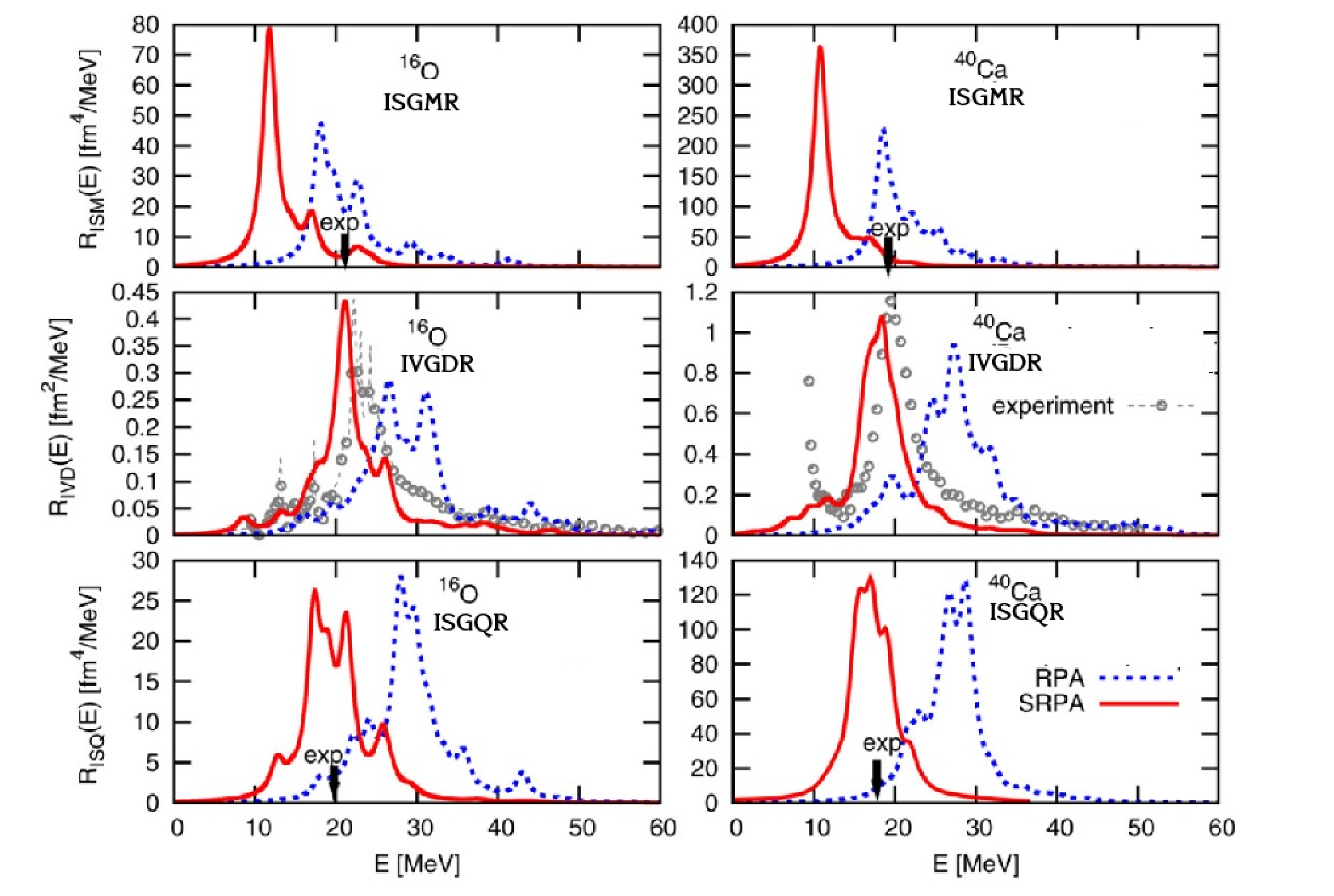}
	\caption{The RPA (blue dashed lines) and SRPA (full red lines) strength distributions for the ISGMR (top), IVGDR (middle) and ISGQR (bottom) cases in $^{16}$O (left) and $^{40}$Ca (right), compared with experiment (points, arrows). The calculated distributions have been folded with a Lorentzian with a width of 2 MeV. The experimental centroids of the ISGMR and the ISGQR were taken from Refs. \cite{Lui2001} ($^{16}$O) and \cite{Youngblood2001} ($^{40}$Ca). Photoabsorption cross sections are from Refs. \cite{Ahrens1975,LeBrun1987} ($^{16}$O) and \cite{Veyssiere1974}
		($^{40}$Ca). Adapted from \cite{Papakonstantinou2009}.}
	\label{Fig:Papa_1}
\end{figure} 
Figure \ref{Fig:Papa_1} shows the calculated ISGMR, IVGDR and ISGQR strength distributions for $^{16}$O and $^{40}$Ca, smoothed with a 2 MeV width Lorentzian. The SRPA calculations systematically show significantly lower centroid energies compared to the standard RPA results. This substantial difference arises from SRPA's explicit inclusion of couplings between $1p-1h$ and $2p-2h$ configurations, which re-normalizes the HF single-particle states, bringing their energies closer together and consequently lowering the $1p-1h$ excitation energies. In the upper panels of Figure \ref{Fig:Papa_1} the ISGMR distributions
are plotted. One can clearly see that the corresponding experimental centroid  energy is quite well reproduced at RPA level, while it is underestimated within SRPA. A possible explanation might be that three-body components of the nuclear interaction were not taken into account. 
 \begin{wrapfigure}{l}{0.5\textwidth}
	\centering
	%	\vspace{-3.5mm}
	\includegraphics[width=0.49\textwidth]{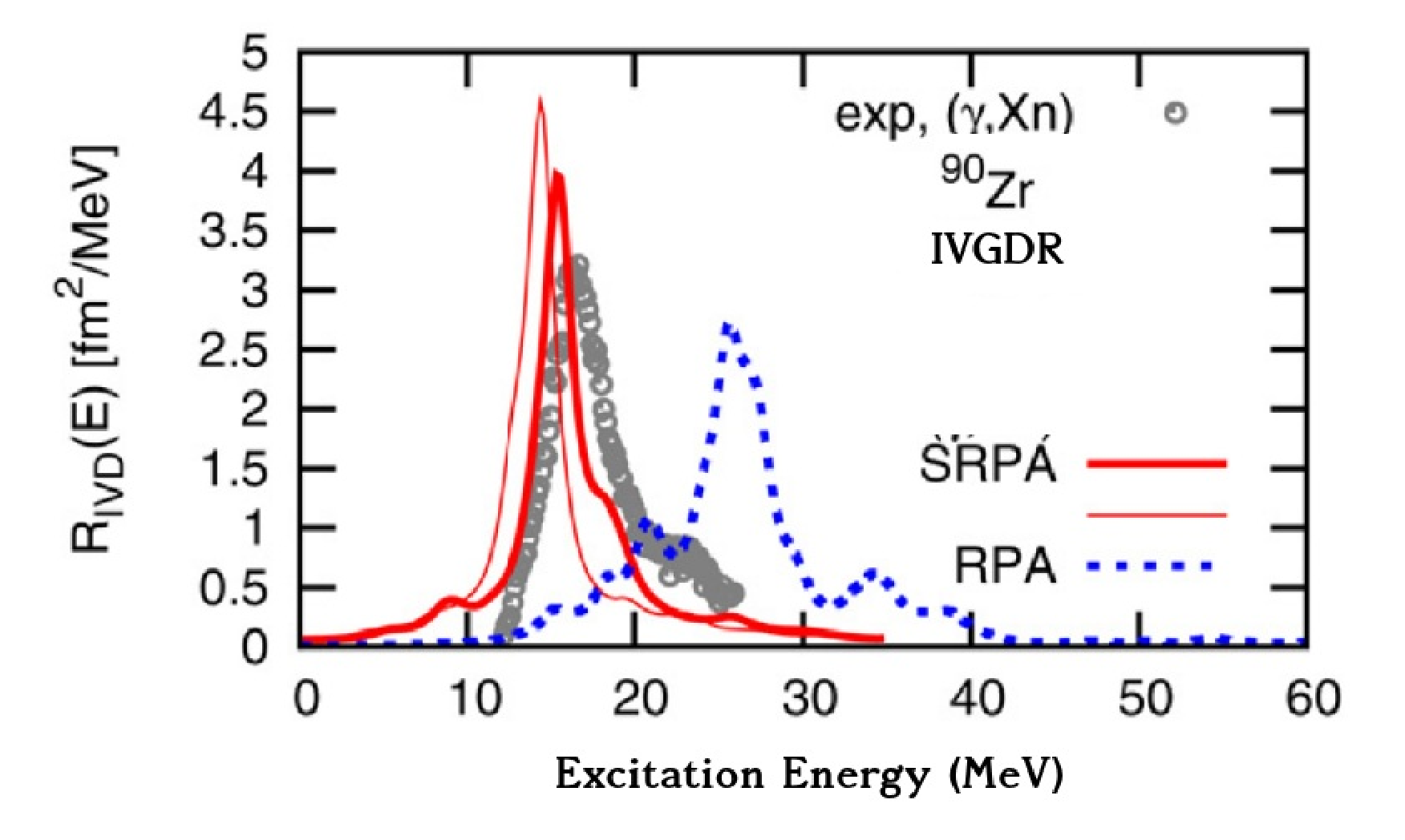}
	\vspace{-1.5mm}
	\caption{IVGDR transition strength distribution of $^{90}$Zr in RPA (dashed blue-line) and SRPA (solid red lines) compared with $(\gamma , Xn)$ data from \cite{Lepretre1971}. Figure adapted from \cite{Papakonstantinou2009}.}
	%	\vspace{-7mm}
	\label{Fig:Papa_3}
\end{wrapfigure} 
Comparing the dipole results with experimental data (middle panels), we see that SRPA provides a more realistic description of the IVGDR than RPA, though the centroid energy is slightly underestimated. This trend appears to continue in heavier nuclei. For example in Figure \ref{Fig:Papa_3}, the dipole transition strength distribution of $^{90}$Zr in RPA (dashed blue-line) and SRPA (solid red lines) is compared with  data from Ref. \cite{Lepretre1971}. SRPA results obtained with two different single-particle bases are shown, showing a weak sensitivity of the corresponding results. The isoscalar quadrupole strength distributions (lower panels of Figure \ref{Fig:Papa_1} ) demonstrate a very good agreement between SRPA calculations and experimental ISGQR centroids. This suggests that the coupling to higher-order configurations in SRPA improves the description of the nucleon effective mass, through self-energy corrections.

\begin{figure}
\includegraphics[width=.45\linewidth]{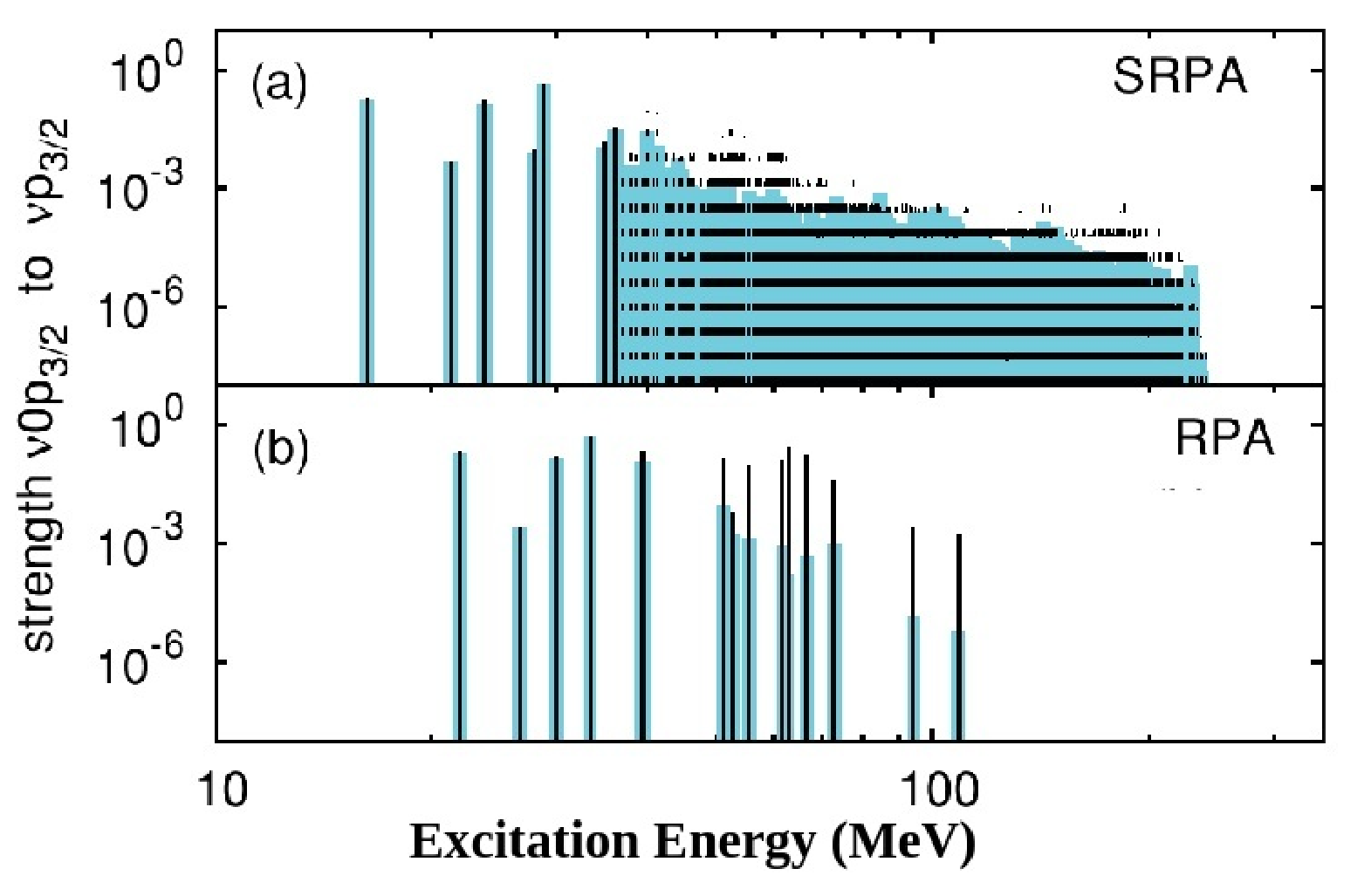}\hfill
\includegraphics[width=.55\linewidth]{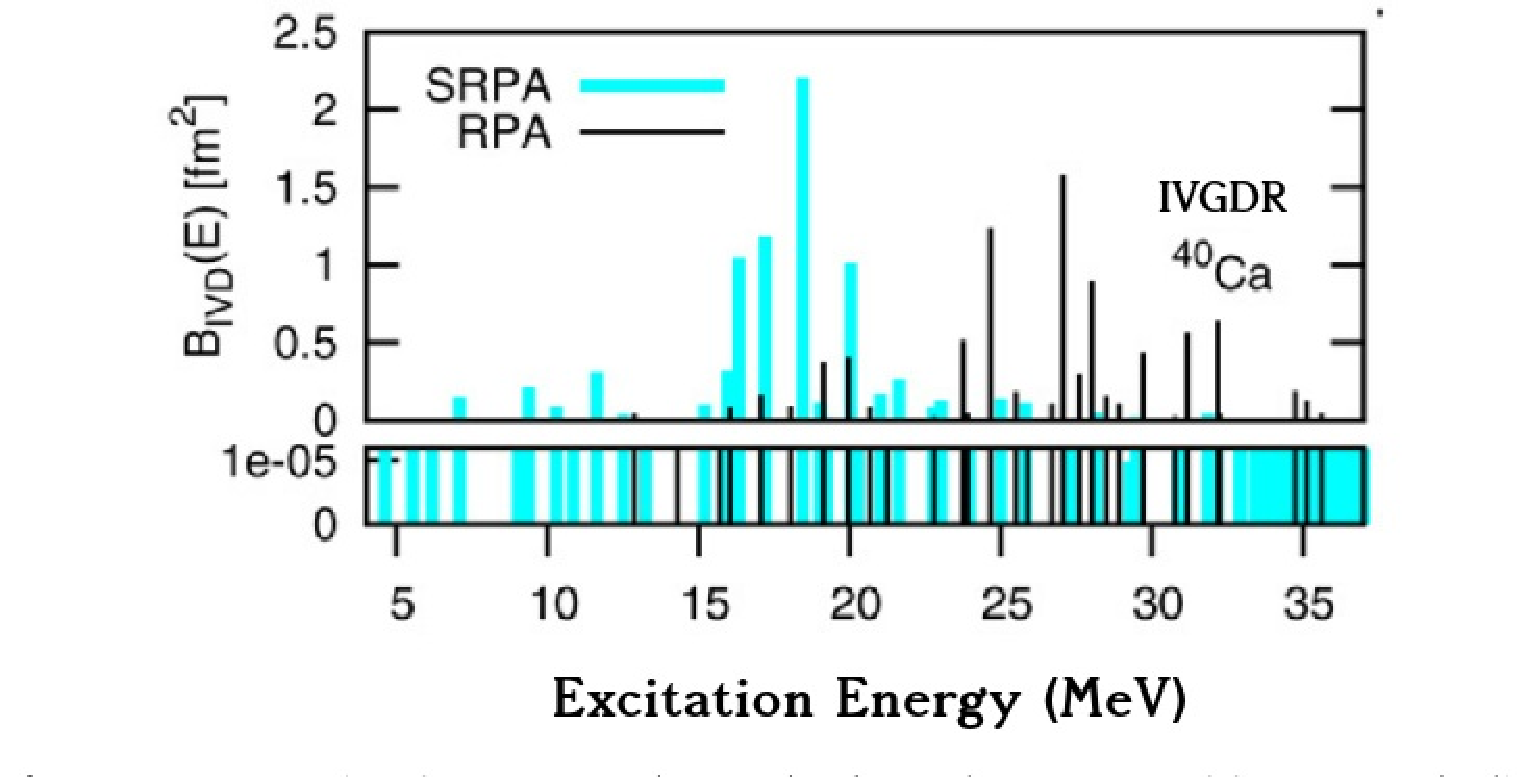}
\caption{Left: Fragmentation of the $\mid (\nu p_{3/2})(\nu 0p_{3/2})^{-1}\rangle_{0^+}$ configurations in the 0$^+$ response of $^{16}$O in (a) SRPA, (b) RPA. Thin dark bars show
the spectroscopic strength defined in Eq. (\ref{Eq:Sph}), thicker bars denote
the distribution of $\mid (\nu 1p_{3/2})(\nu 0p_{3/2})^{-1}\rangle_{0^+}$ (one particle shell only). Right: IVGDR transition strength distribution of $^{40}$Ca in RPA (black bars) and SRPA (cyan bars) are shown. Left  Figure adapted from Ref. \cite{Papakonstantinou2010}, right one from Ref. \cite{Papakonstantinou2009}.}
\label{Fig:Papa_2}
\end{figure}

 Figure \ref{Fig:Papa_1} also reveals that SRPA seems to predict narrower resonance widths than RPA. This counterintuitive result is attributed to the overall compression of the particle-hole spectrum in SRPA, which counteracts the expected increase in width due to increased fragmentation. While compression occurs, SRPA does introduce fragmentation, as shown in Figure \ref{Fig:Papa_2}. On the left side of this figure, we show the RPA and SRPA strength distributions
 \begin{equation}
 S_{ph}(E)=\sum_\nu (\mid X_{ph}\mid^2-\mid Y_{ph}\mid^2) \delta(E-E_\nu)
 \label{Eq:Sph}
 \end{equation}
 of the ph configurations $\mid (\nu p_{3/2})(\nu 0p_{3/2})^{-1};0^+\rangle$ contributing to the monopole strength in $^{16}$O. One can see that, already at RPA level a fragmentation is visible, in particular concerning a cluster of configurations at energies around 30 MeV and a second one around 60 MeV. In SRPA, a even more fragmented distribution appears, as well as shifted to lower energies, related to an effective compression of the single-particle spectrum induced by the $2p-2h$ configurations. The reduction of the width observed in UCOM-SRPA is related to the compression of the particle-hole spectrum, and it is not a general feature of SRPA. The SRPA width emerges from the interplay between the $2p-2h$ level density near the resonance centroid and the residual coupling between the $1p-1h$ and $2p-2h$ configurations. Consequently, this effect can be interaction-dependent. For instance, Skyrme-based calculations typically yield widths larger than RPA. A more detailed quantitative comparison will be presented in Sections \ref{Sec:Applications_SSRPA_Dipole_Ca}, \ref{Sec:Applications_SSRPA_Quadrupole} and \ref{Sec:Applications_SSRPA_GT}.

The fragmentation introduced in the SRPA can be seen also on the right side of Figure \ref{Fig:Papa_2}, where the IVGDR transition strength distribution of $^{40}$Ca in RPA (black bars) and SRPA (cyan bars) are shown. In the lower panel the y-scale is different, to make more visible the large amount SRPA states which are much more than in RPA.

\begin{figure}
\resizebox{\textwidth}{!}{\includegraphics{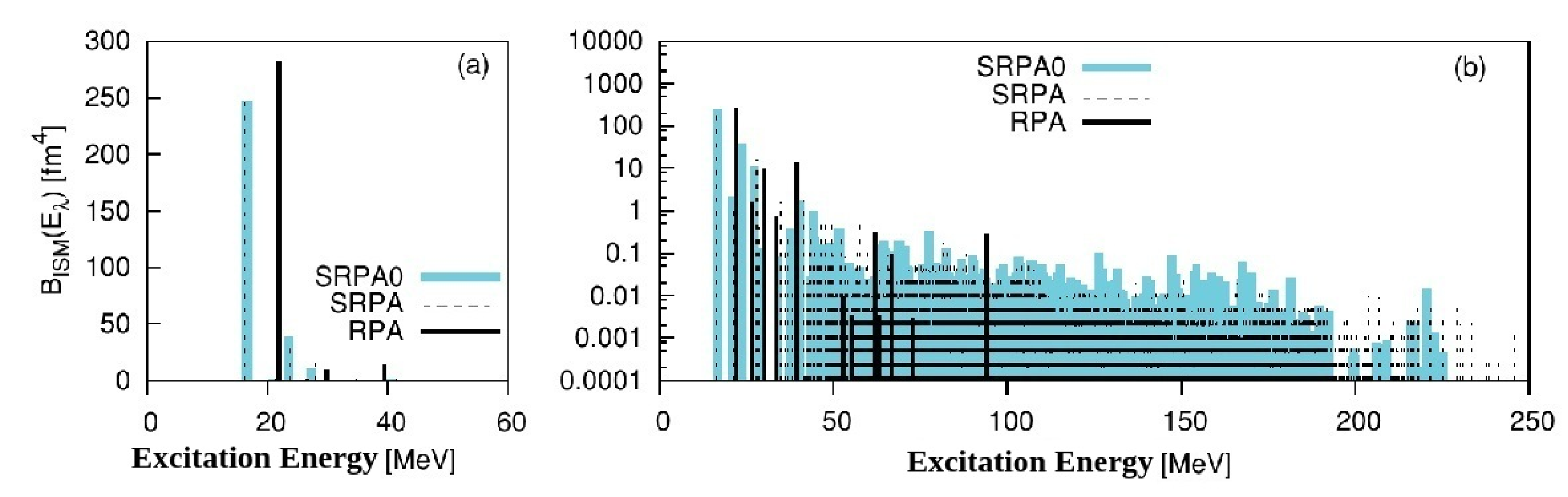}}
\caption{Isoscalar monopole response of $^{16}$O, calculated within the RPA, SRPA and SRPA in the diagonal approximation (SRPA0) as a function of the excitation energy, in linear (panel (a)) and logarithmic scale (panel (b)). From Ref. \cite{Papakonstantinou2010}.}
\label{Fig:Papa_4}
\end{figure}

\begin{figure}
\includegraphics[width=.55\linewidth]{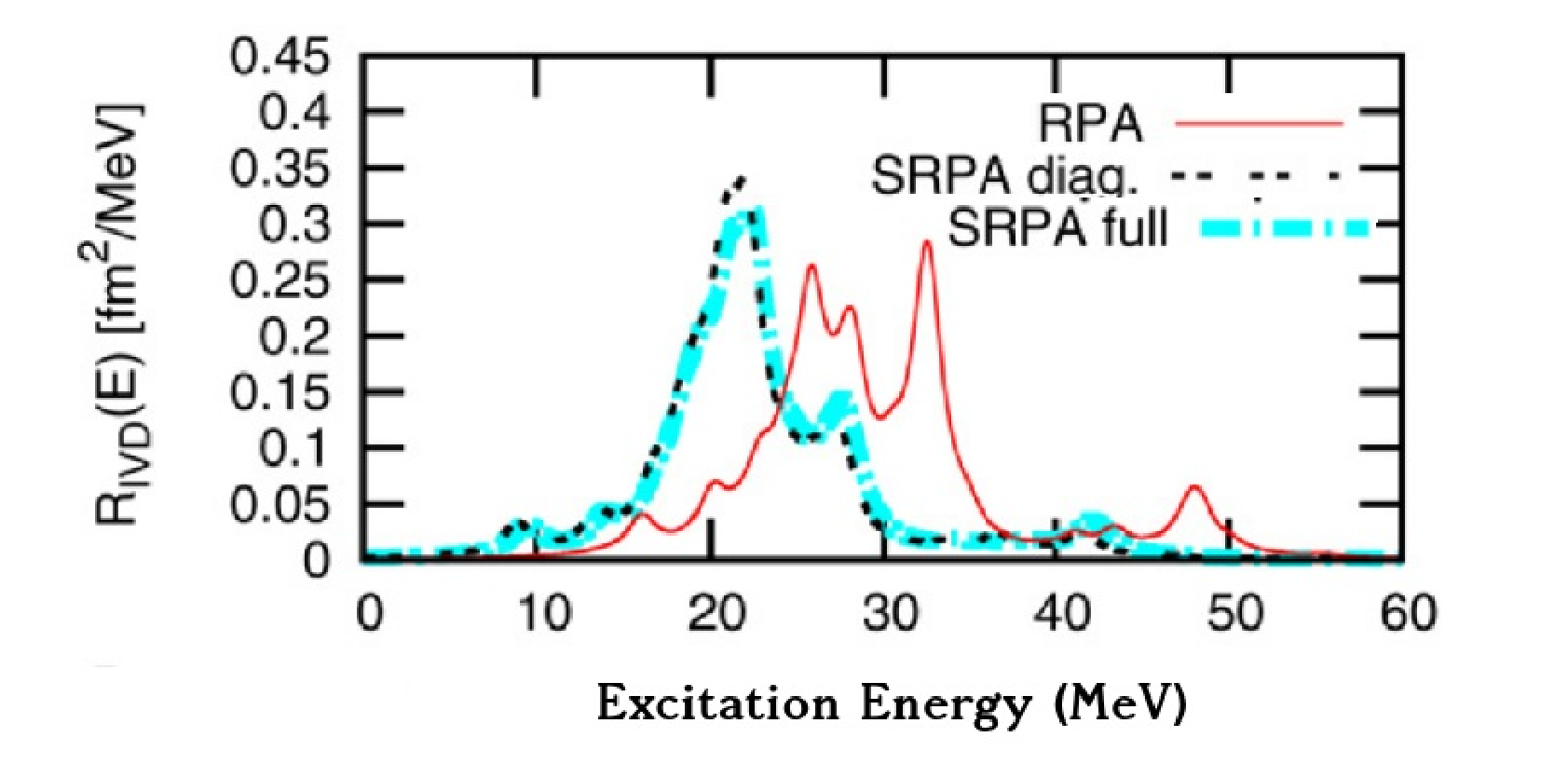}\hfill
\includegraphics[width=.4\linewidth]{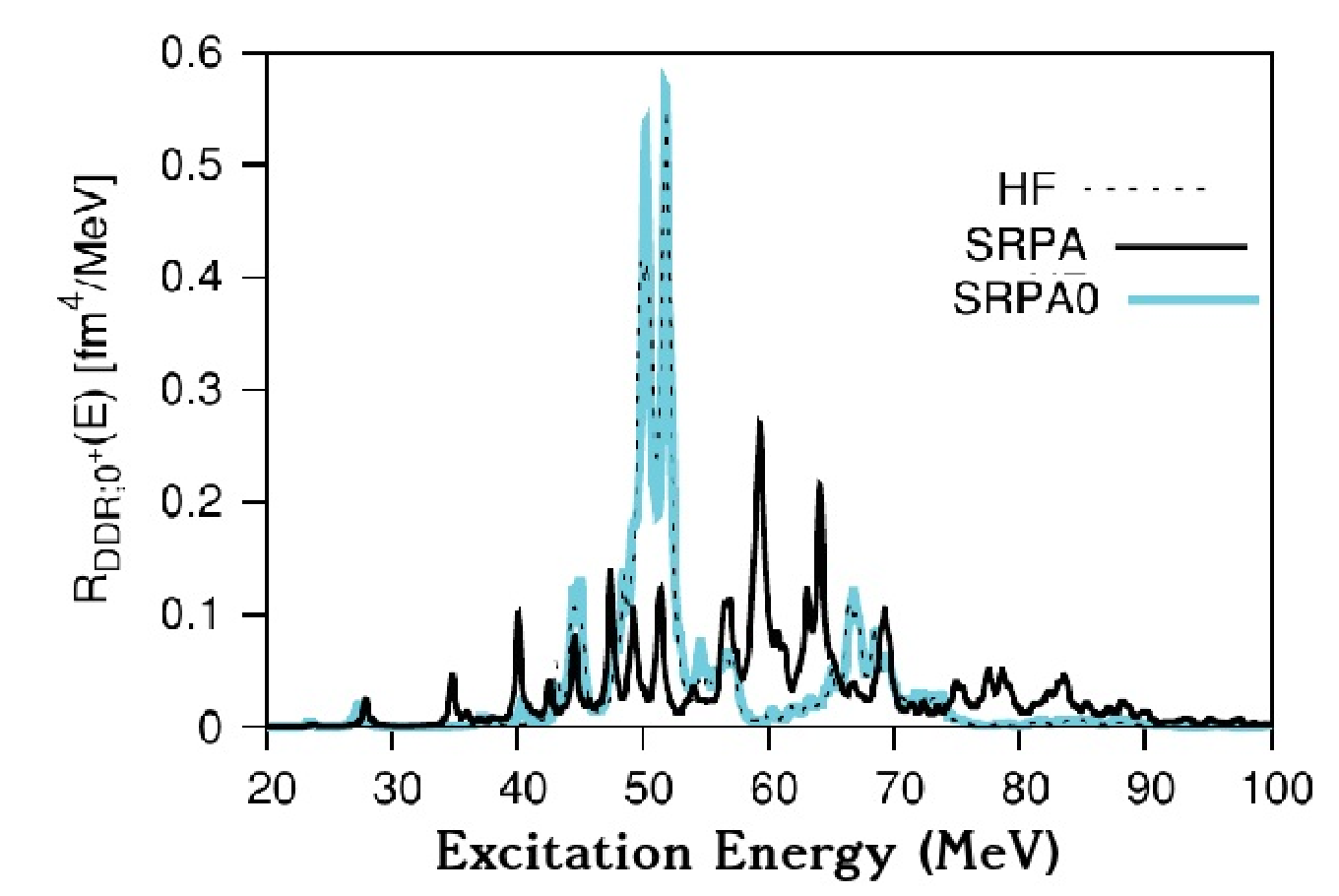}
\caption{Left: Isovector dipole response of $^{16}$O obtained in RPA (full solid red line), SRPA (dash-dotted cyan line) and SRPA in diagonal approximation (dashed black line).) Right: Monopole component of the double dipole
resonance of $^{16}$O, obtained in SRPA (solid thick black line) and SRPA0 (diagonal approximation, solid thin cyan line). The unperturbed response is also shown as a reference (black dashed line).
Left figure adapted from Ref. \cite{Papakonstantinou2009}, right one adapted from Ref. \cite{Papakonstantinou2010}.}
\label{Fig:Papa_5}
\end{figure}

It is interesting to see the effect of the diagonal approximation with the $V_{UCOM}$ interaction.
In Figure \ref{Fig:Papa_4}, the isoscalar monopole response of $^{16}$O, calculated within the RPA, SRPA and SRPA in the diagonal approximation (SRPA0) is plotted as a function of the excitation energy, in linear (panel (a)) and logarithmic scale (panel (b)). One can see that the effect of the residual interaction on the $2p-2h$ space is quite small, leading to very similar results whether the $2p-2h$ configurations are treated as unperturbed or not. The same is found for the dipole response of $^{16}$O, left side of Figure \ref{Fig:Papa_5}. In UCOM‑SRPA, the underlying  unitary transformation that softens short‑range correlations, leads to  weak off‑diagonal couplings among $2p-2h$ configurations. As a result, the diagonal approximation works well.

\begin{figure}
\includegraphics[width=.47\linewidth]{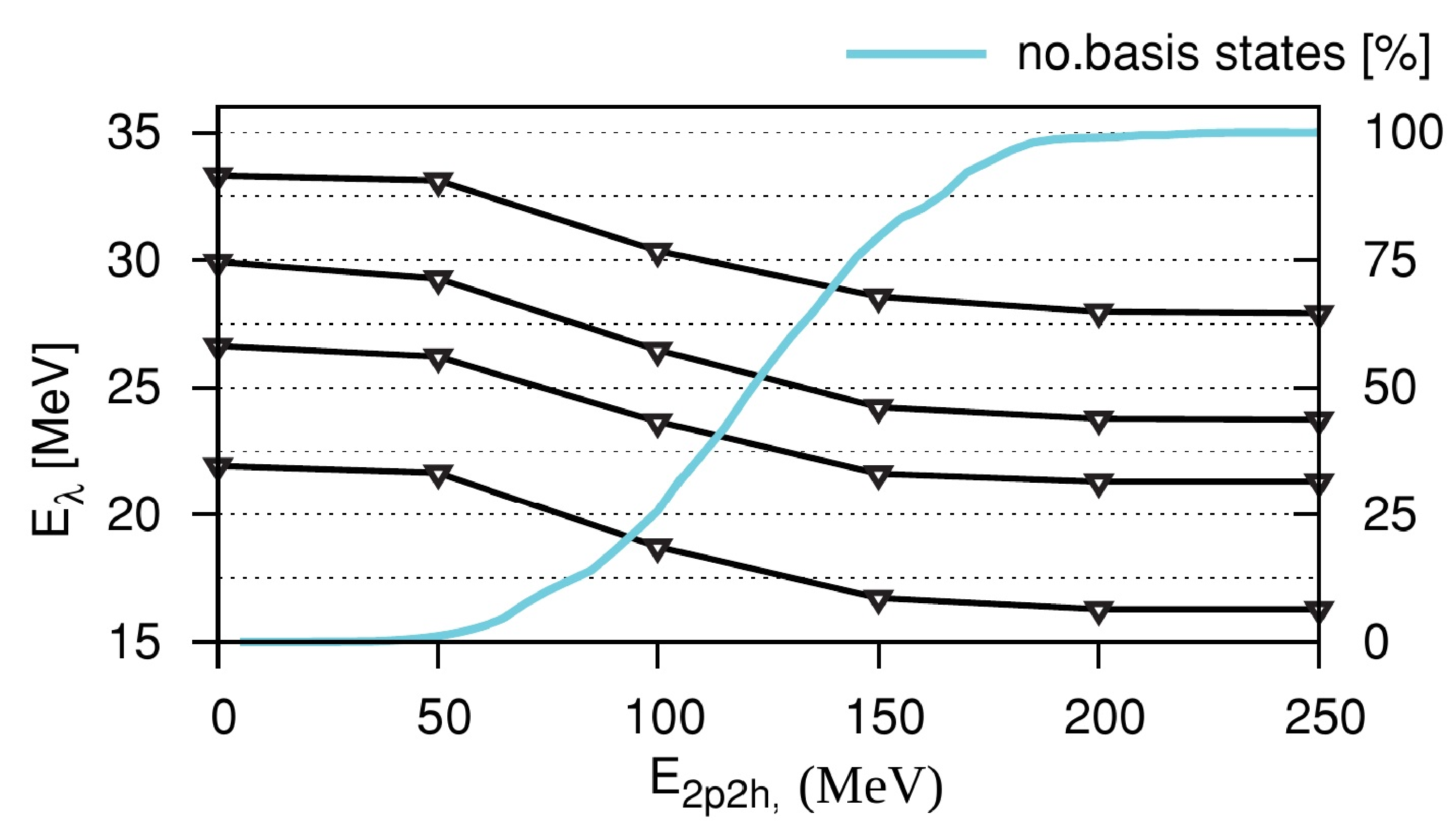}\hfill
\includegraphics[width=.52\linewidth]{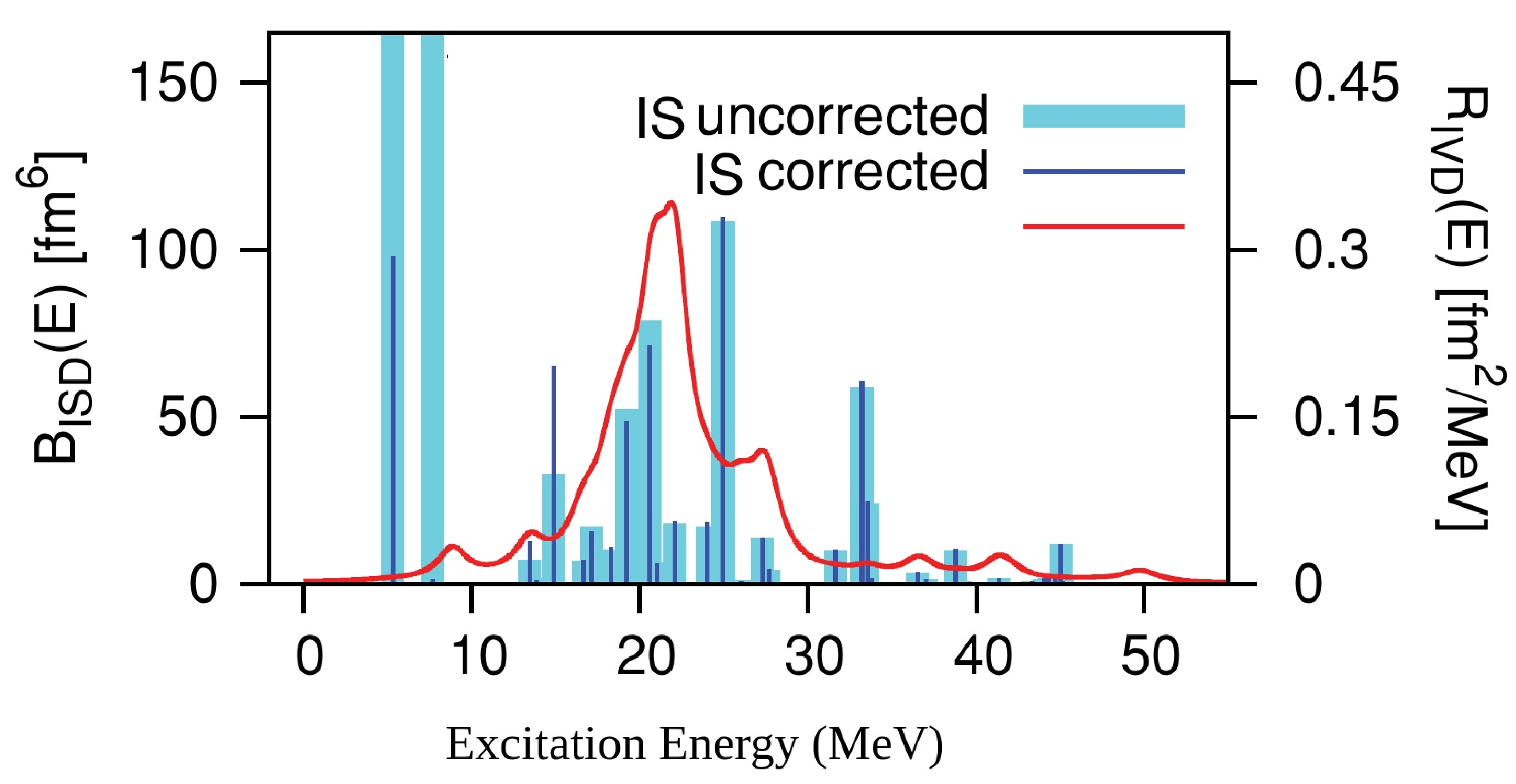}
\caption{Left: Energies of the four $0^+$ lowest eigenstates (axis on the left)
and percentage of $2p-2h$ states(axis on the right) as a function of the energy cutoff E$_{2p2h}$. Right: Isoscalar (IS) and isovector (IV) dipole strength distributions of $^{16}$O. The left y-axis refers to the isoscalar
 strength and the right one to the folded isovector distributions.
% The IS strength is shown for the uncorrected (corrected) IS dipole transition operator in thick cyan solid bars (thin blue solid bars). 
Adapted from Ref. \cite{Papakonstantinou2010}.}
\label{Fig:Papa_6}
\end{figure}

However, the impact of the residual interaction is important in the case of the response to a two-body transition operator. This can be seen, for the monopole component of the double dipole resonance two-body operator $D=[T^{IVD}_{E1}T^{IVD}_{E1}]_{0^+}$, which is shown on the right side of Figure \ref{Fig:Papa_5}, $T^{IVD}_{E1}$ being the standard one-body isovector dipole operator (\ref{Eq:Op-J1-IV-CM}). In this case, one can see that the diagonal approximation SRPA0 leads to results that are very close to the unperturbed ones and differs significantly from the full SRPA result. 
As a general feature, one can see that going from RPA to SRPA, a shift towards lower energy is observed. On the left side of Figure \ref{Fig:Papa_6}, the convergence of a few $0^+$ states lying at different energies with respect to the energy cutoff of on the $2p-2h$ configurations is shown. One can see that convergence is generally reached at an energy cutoff of approximately 200 MeV.

The treatment of the spurious mode associated with center-of-mass motion deserves special attention. The RPA calculation, being self-consistent, yields a spurious state near zero energy, leaving the rest of the spectrum uncontaminated \cite{Paar2006}. When higher order (than $1p-1h$ ones) configurations are included, this is not guaranteed anymore. In Ref. \cite{Tohyama2004}, it was shown that, when $2p-2h$ configurations are included, the single spurious state  appears at zero energy only if all the single-particle amplitudes are included in the phonon operators, thus including not only particle-hole and hole-particle amplitudes, but also particle-particle and hole-hole. Since SRPA does not include them, contaminations are possible. To assess this, one can analyze the isoscalar  dipole response, comparing the center mass corrected transition operator (\ref{Eq:Op-J1-isos-r3-corrected}) 
%$\propto r^3 - \frac{5}{3}\langle r^2 \rangle r$ % ($\propto r^3$).  
 with its uncorrected form (\ref{Eq:Op-J1-isos-r3}).  An example is presented on the right side of Figure \ref{Fig:Papa_6} for $^{16}$O, showing that only the low-energy region of the dipole spectrum is significantly affected by the operator choice.
 The spectrum around and beyond the  IVGDR peak remains largely unaffected. When a significantly consistent number of $2p-2h$ configurations is included, the spurious RPA state moves away from zero, potentially appearing at imaginary or even negative energies.  Furthermore, it can fragment, with spurious admixtures contaminating the rest of the spectrum, or additional eigenstates appearing nearby. A detailed analysis of the spurious mode problem in the SRPA was recently presented in Ref. \cite{Knapp2023}, demonstrating that center-of-mass motion contamination can affect all multipolarities, not just the dipole channel. One possible solution would be to implement Gram-Schmidt orthogonalization to construct a center-of-mass-spurious-free $2p-2h$ basis \cite{Knapp2023}. This has not yet been implemented in SRPA calculations.

%  \vspace{-1cm}

% \newpage
\subsection{SRPA with the Skyrme interaction}
\label{Sec:Applications_SRPA_Skyrme}
\subsubsection{Charge-conserving excitations within the SRPA}
\label{Sec:App_First_SRPA}
The first large-scale Skyrme-SRPA calculations were performed
 in Ref. \cite{Gambacurta2010}. The Skyrme SGII interaction parametrization \cite{SGII} was employed. These applications were performed without taking into account
the Coulomb and spin-orbit terms of the residual interaction. Therefore the calculations were not fully self-consistent, leading to EWSR violations of up to 5\% in the worst cases. The single particle space was chosen to be sufficiently large to ensure EWSR stability.
More precisely, $1p-1h$ configurations with unperturbed energies up to 100 MeV were included, while the convergence with respect to the energy cutoff E$_{cut}$ on $2p-2h$ configurations was studied.

Concerning the treatment of the rearrangement terms in SRPA, two distinct approximations were explored. The first one simply omits rearrangement terms in beyond-RPA matrix elements (results labeled as SRPA). The second approximation employs the standard RPA prescription for calculating the rearrangement terms also in beyond-RPA matrix elements, (results labeled as SRPA$^*$).
%This choice leverages the existing RPA formalism while recognizing that it may not be fully consistent within the SRPA context. 
While these approximations offered valuable insights into the SRPA framework, they represented interim solutions. A more rigorous and consistent treatment of the residual interaction within SRPA with density-dependent forces was derived in a subsequent paper \cite{Gambacurta2011a}, by using a variational derivation and it is discussed in Sections \ref{Sec:FormalPart_SRPA_Rear} and \ref{Sec:ApplicationPart_SRPA_Rear}.

We discuss the strength distributions for the doubly magic nucleus $^{16}$O for various multipolarities. For the monopole and quadrupole case, the transition operators (\ref{Eq:Op-J02-isos}) and (\ref{Eq:Op-J02-isov}) are used, while in the isovector dipole case  the transition operator (\ref{Eq:Op-J1-IV-CM}) is employed.
%and used are
%\begin{equation}\label{isos-oper}
% \mathcal{B}_{\lambda}^{IS}=\sum r_i^n Y_{\lambda 0}(\hat{r}_i)~~~
% \end{equation}
% \begin{equation}\label{isov-oper}
% \mathcal{B}_{\lambda}^{IV}=\sum r_i^n Y_{\lambda 0}(\hat{r}_i)\tau_z(i)
%\end{equation}
%in the isoscalar and isovector channel, respectively, where $n=\lambda$ except
%for $\lambda=0$ where $n=2$. For the isovector dipole case, the standard operator
%\begin{equation}\label{isov-oper-dipole}
% F^{IVD}=\sum_{p=1}^{Z} \frac{N}{A} r_p Y_{1 0}(\hat{r}_p) -\sum_{n=1}^{N} \frac{Z}{A} r_n Y_{1 0}(\hat{r}_n)
%\end{equation}
%is employed.
% The corresponding continuous strength distributions are generated by folding the discrete spectra with a 1 MeV width Lorentzian.
\begin{table}
 \begin{center}
\begin{tabular}{|c|c|c|}
\hline
 &$m_1 (T=0)$&$m_1 (T=1)$ \\
\hline
$\omega_{max}$ &~~~~~SRPA&SRPA\\
\hline
 40& 626.4381& 115.4153 \\
\hline
 50 & 648.9699& 147.8026 \\
\hline
 60 & 661.0194& 182.7364\\
\hline
 70 & 664.3803& 193.7896\\
\hline
 80 & 669.7185& 197.6874\\
\hline
 90 & 671.4575& 200.6472\\
\hline
 100 & 671.6515& 201.2473\\
\hline
 110 & 671.6515& 201.2473\\
\hline
RPA&671.6516&201.2494\\
\hline
\end{tabular}
\hspace{3cm}
\begin{tabular}{|c|c|c|c|c|c}
\hline
 &$m_1 (T=0)$&$m_1 (T=1)$ \\
\hline
$\omega_{max}$ &~~~~~SRPA$^{*}$&SRPA$^{*}$\\
\hline
 40& 616.6714 & 119.7808 \\
\hline
 50 & 650.1523& 155.1027 \\
\hline
 60 & 663.9445& 181.1740\\
\hline
 70 & 670.0688& 185.2818\\
\hline
 80 & 670.2844 & 197.3268\\
\hline
 90 & 671.5175 & 200.6471\\
\hline
 100 & 671.6515 & 201.2493\\
\hline
 110 & 671.6515& 201.2493\\
\hline
% \hline
RPA&671.6516&201.2494\\
\hline
\end{tabular}
\end{center}
\caption{Evolution of the monopole isoscalar and isovector first moments
obtained in SRPA, second and third columns respectively, as a function of the
$\omega_{max}$ parameter in MeV (see Eq. \ref{m1srpa}). In the last row, the
corresponding RPA values are reported with $\omega_{max}=100$ MeV. Left (right) table refers to result without (with) rearrangement terms. From Ref. \cite{Gambacurta2010}.}
\label{Tab:1}
\end{table}%
As discussed in Section \ref{Sec:FormalPart_SRPA_Moments}, the EWSR are satisfied in SRPA, and the first moment is identical in RPA and SRPA. Table \ref{Tab:1} presents, for the monopole case, the isoscalar and isovector values (second and third columns, respectively) of the SRPA first moment
\begin{equation}\label{m1srpa}
m_1=\sum_{\nu}^{\omega_{max}}\omega_\nu |\langle \nu \mid T\mid 0 \rangle |^2
\end{equation}
\begin{figure}
	\includegraphics[width=.5\linewidth]{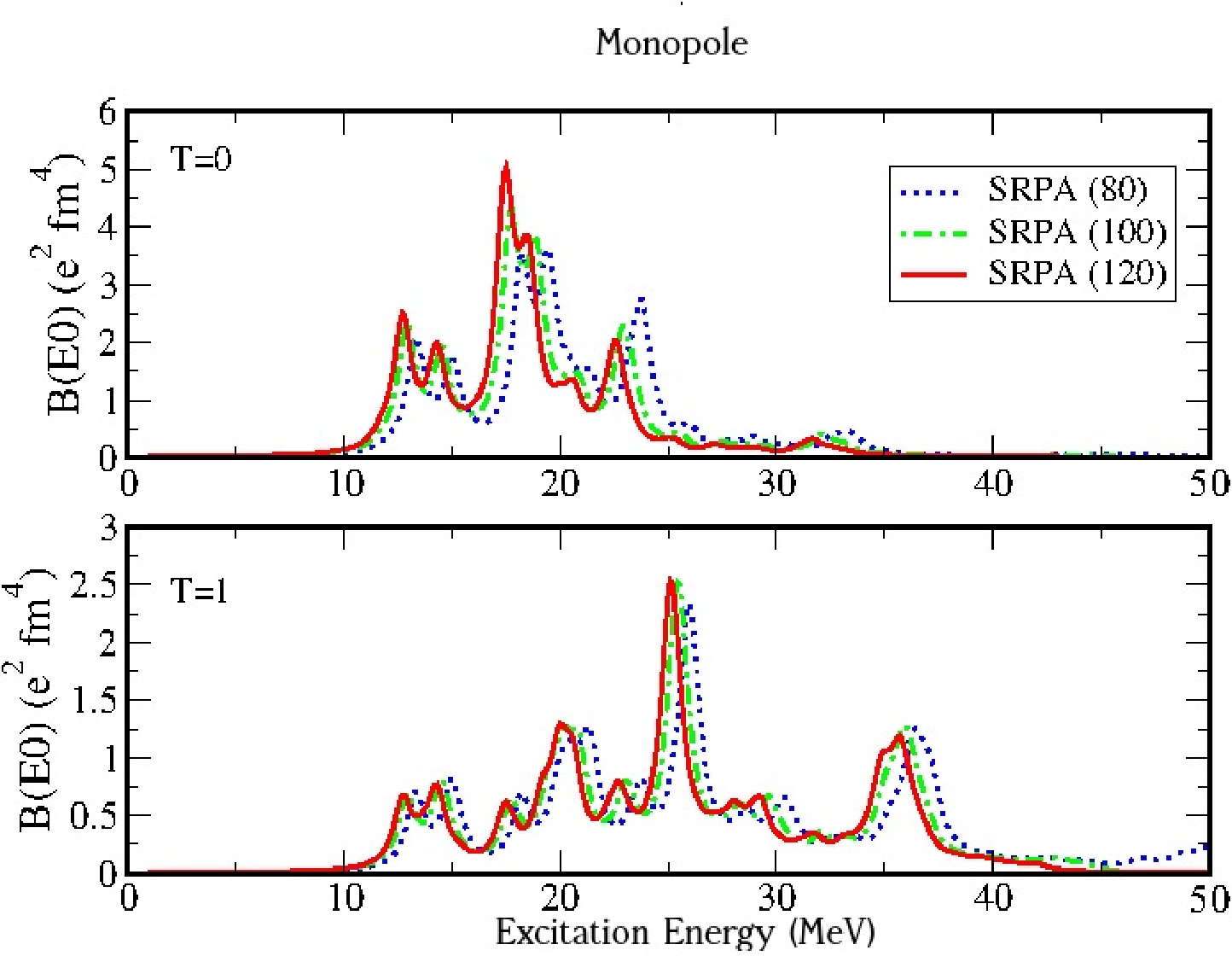}\hfill
	\includegraphics[width=.5\linewidth]{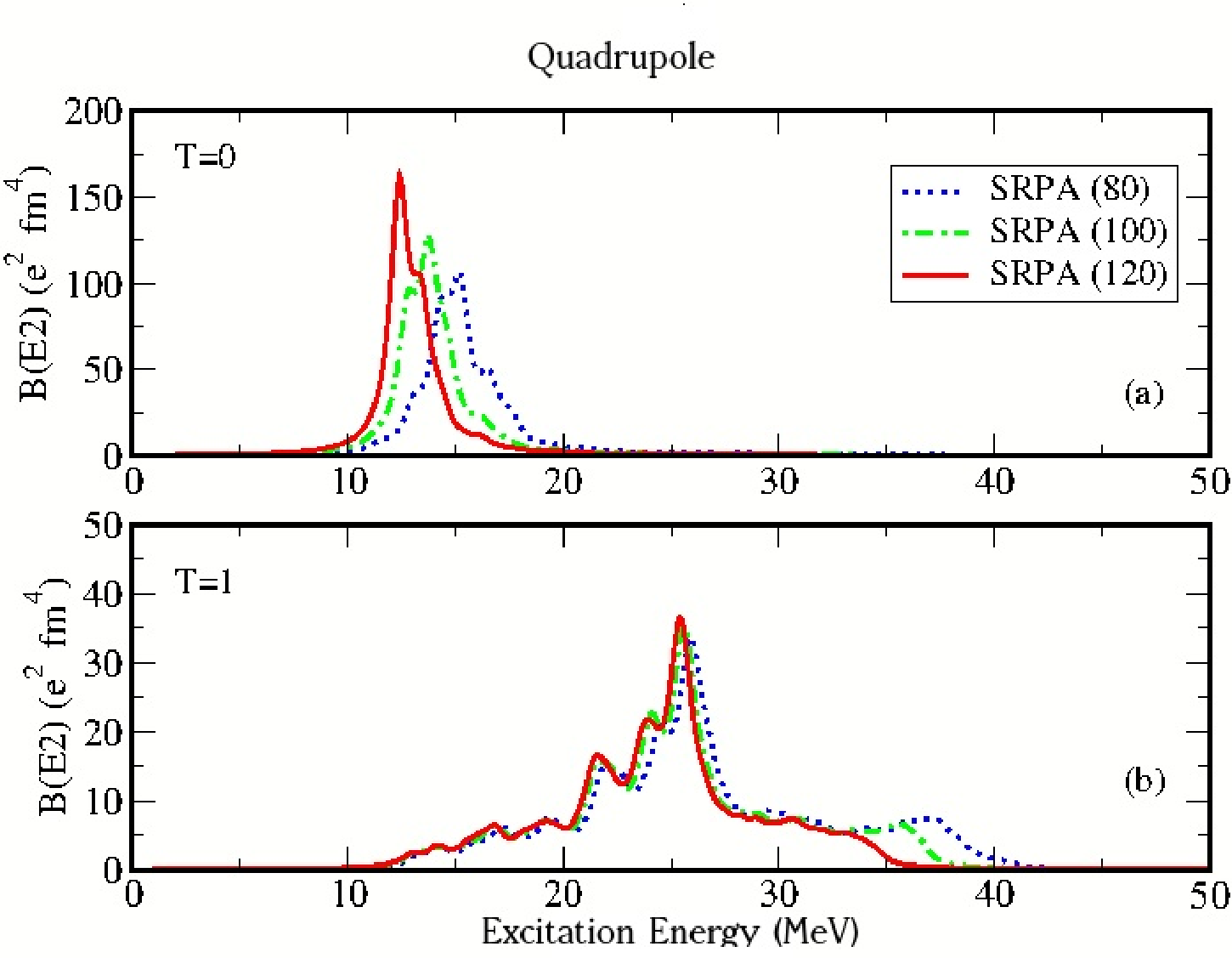}
	\caption{Isoscalar (upper panel) and isovector (lower panel) strength distributions for
		monopole (left side) and quadrupole (right side) states obtained in SRPA for increasing values of the energy cutoff on the $2p-2h$ configurations, indicated in MeV in parenthesis. Adapted from Ref. \cite{Gambacurta2010}.}
	\label{Fig:evolution-J0-J2}
\end{figure}
including all states with excitation energies below $\omega_{max}$ (shown in the first column). The last row reports the corresponding RPA values with $\omega_{max}$ = 100 MeV. As $\omega_{max}$ increases, the SRPA first moment values approach the RPA values.
A similar behavior is observed when rearrangement terms are included as defined in RPA (right table). However, a different distribution with respect to $\omega_{max}$ is found (see also right side of Figure. \ref{Fig:LogAndRearr}).
In Figure \ref{Fig:evolution-J0-J2}, the isoscalar (top) and isovector (bottom) monopole (left side) and quadrupole (right side) strength distributions are shown for different values of the energy cutoff on the $2p-2h$ configurations $E_{cut}$ (indicated in MeV in parenthesis in the figures). We can see that, increasing the cutoff, the strength distribution progressively decreases, and it reaches stability for $E_{cut}\simeq$ 120 MeV.

\begin{figure}
	\includegraphics[width=.5\linewidth]{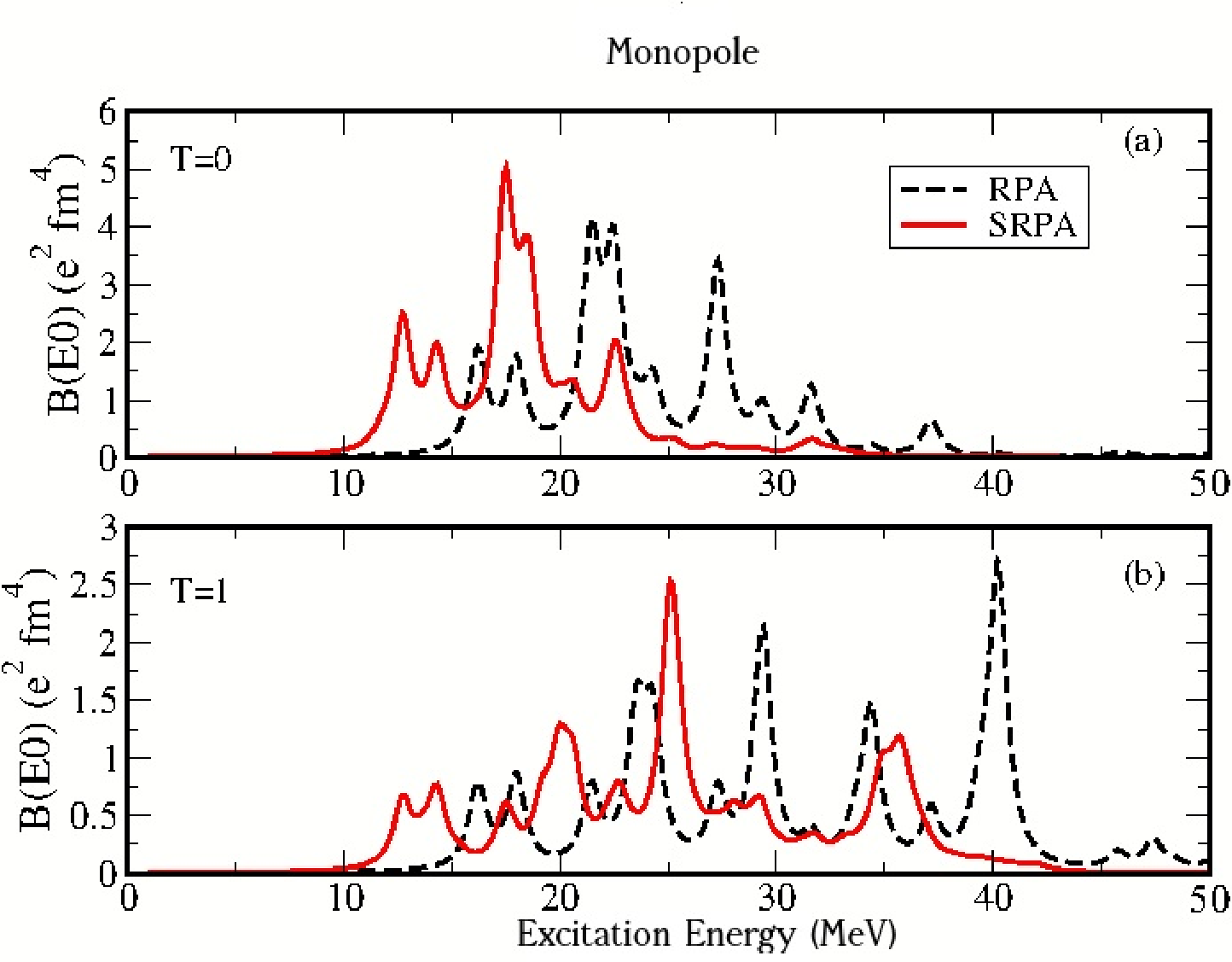}\hfill
	\includegraphics[width=.5\linewidth]{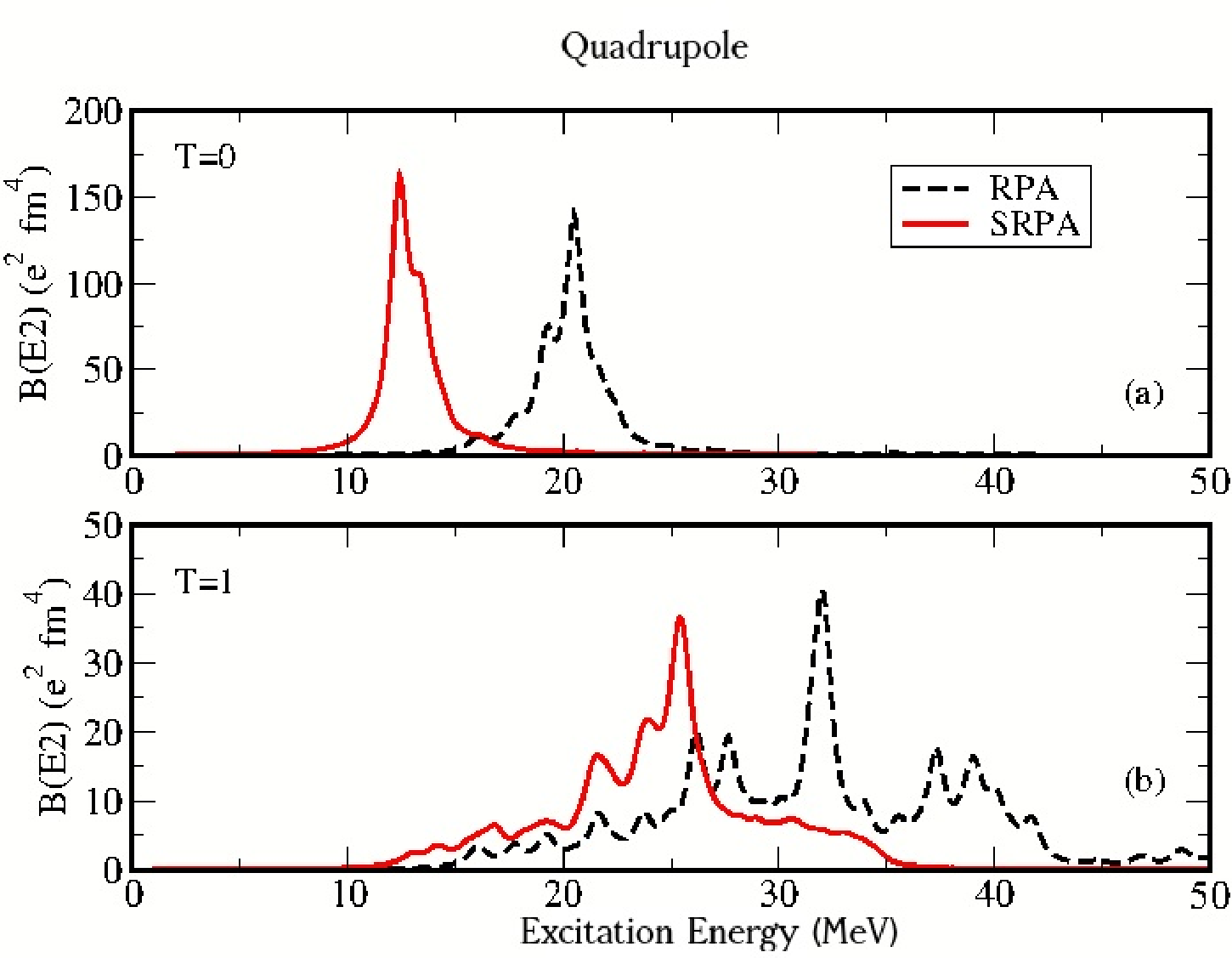}
	\caption{RPA, dashed (black) lines and SRPA, full (red) lines, for the isoscalar (upper panel) and isovector (lower panel) monopole (left side) and quadrupole (right side) strength distributions. Adapted from Ref. \cite{Gambacurta2010}.}
	\label{Fig:J0-J2}
\end{figure}

\begin{figure}
	\includegraphics[width=.5\linewidth]{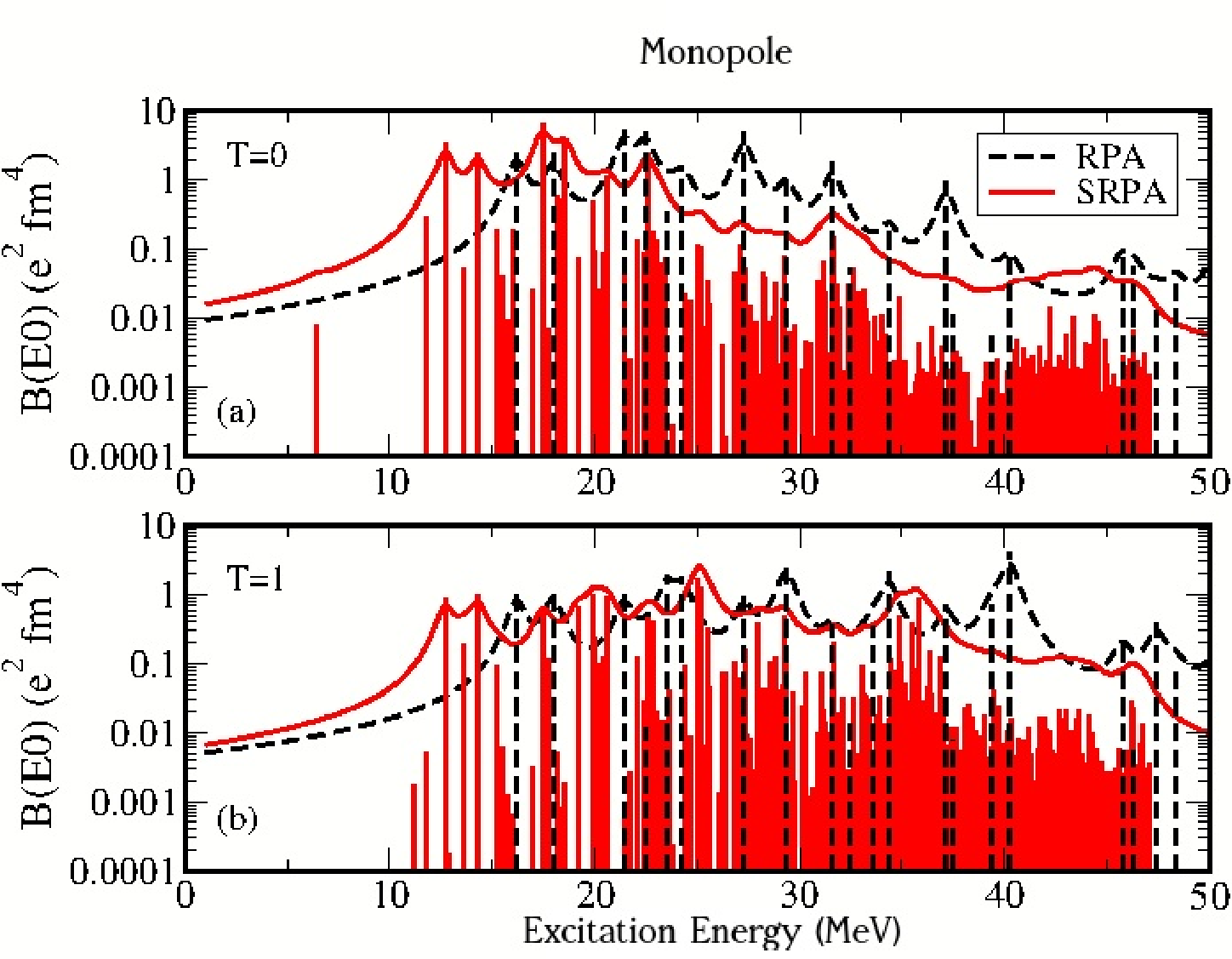}\hfill
	\includegraphics[width=.5\linewidth]{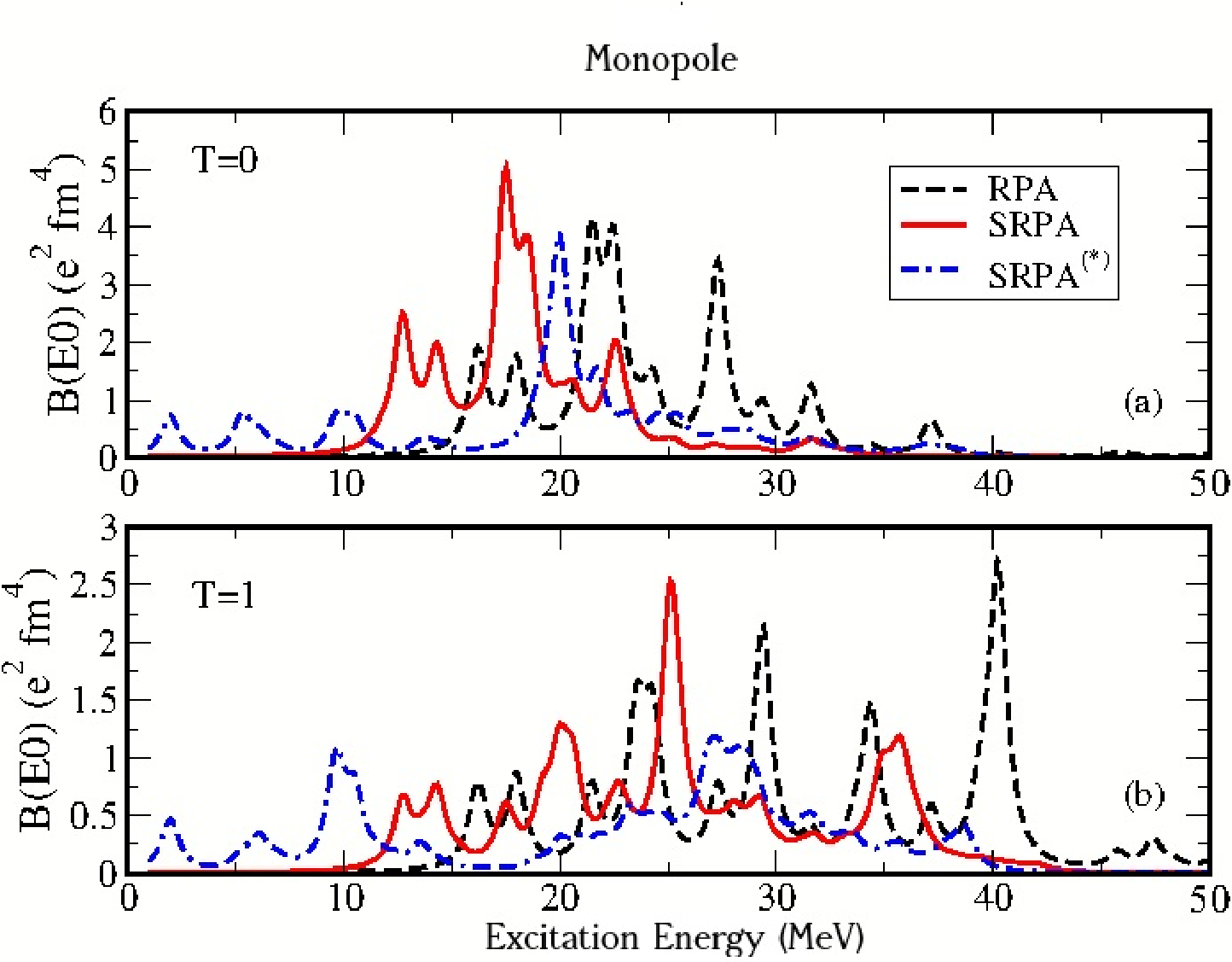}
	\caption{Left:Isoscalar (upper panel) and isovector (lower panel) strength distributions for monopole distribution but using a logarithmic scale in the
		ordinate. Right: Comparison between RPA, dashed (black) lines, SRPA in full (red) lines and SRPA with rearrangements terms, dot-dashed (blue) lines of the isoscalar (upper panel) and isovector (lower panel) monopole strength distributions. Adapted from Ref. \cite{Gambacurta2010}.}
	\label{Fig:LogAndRearr}
\end{figure}

In Figure \ref{Fig:J0-J2} the RPA (dashed black lines) and SRPA (solid red lines) results for the isoscalar (upper panels) and isovector (lower panels) monopole (left side) and quadrupole (right side) strength distributions are compared. SRPA calculations include all $2p-2h$ configurations with unperturbed energies up to E$_{cut}$ = 120 MeV. A common feature observed in all the cases is the pronounced shift of the SRPA strength distributions towards lower energies compared to their RPA counterparts. While the overall envelope of the strength distribution is mostly preserved, meaning the general shape and relative intensities of the peaks remain similar, this systematic red-shift is a significant effect.

Figure \ref{Fig:LogAndRearr} presents the monopole response, offering a detailed look at the impact of including complex configurations in the description of the spectra. The left panel employs a logarithmic scale for the strength axis, highlighting the richer fine structure of the SRPA response. This detailed structure, characterized by a dense distribution of discrete energy levels, is a direct consequence of the $2p-2h$ configurations considered in the SRPA calculation, significantly enriching the spectrum.
The right panel of Figure \ref{Fig:LogAndRearr} shows SRPA results obtained with and without the explicit inclusion of the rearrangement terms in the beyond-RPA matrix elements. The isoscalar (top panel) and isovector (bottom panel) monopole cases are shown, showing similar results. The residual interaction appears more repulsive with rearrangement terms, resulting in a smaller energy shift to lower energies compared to RPA, and a significant amount of strength appears at very low energy (below 10 MeV).
\begin{figure}
	\includegraphics[width=.5\linewidth]{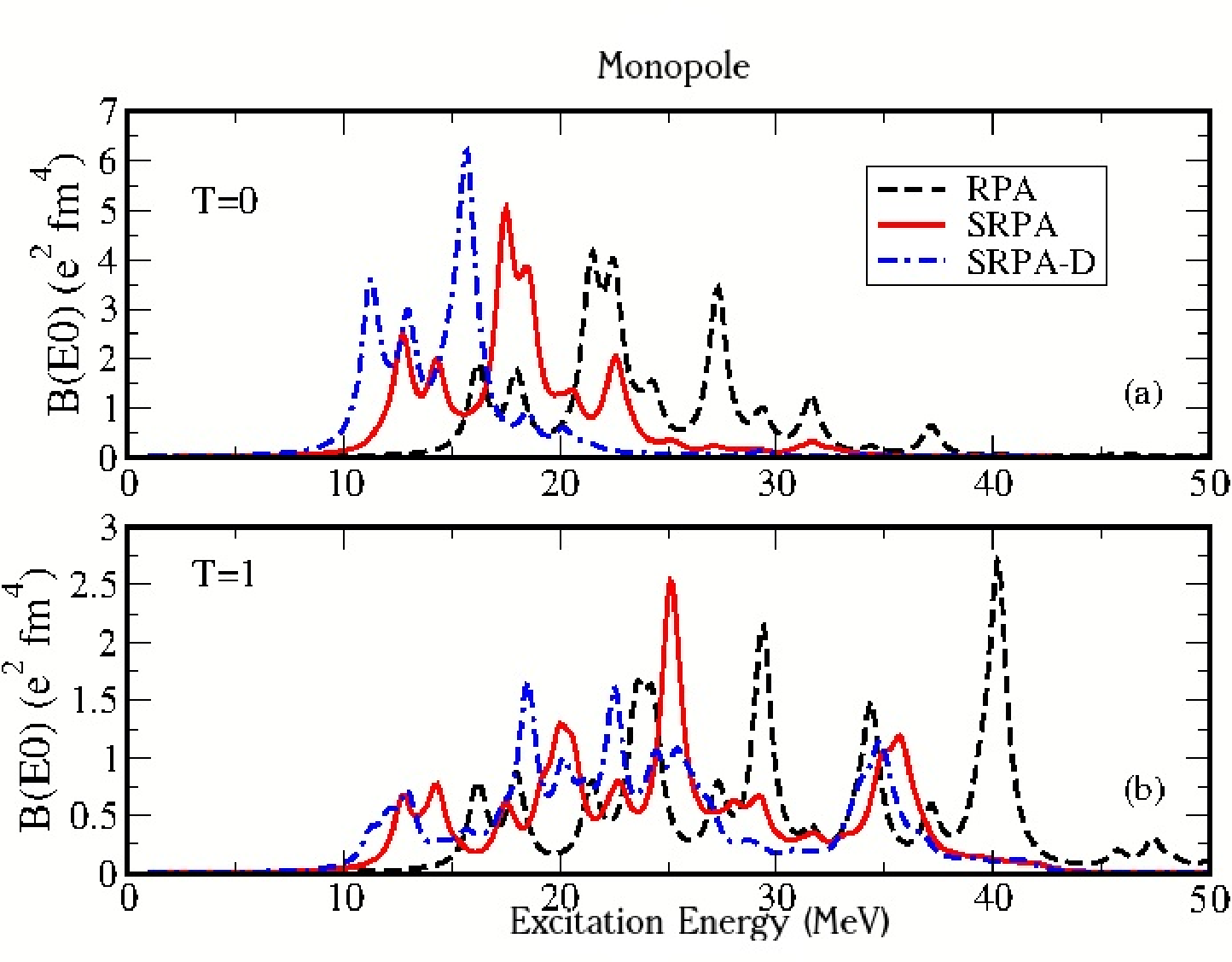}\hfill
	\includegraphics[width=.5\linewidth]{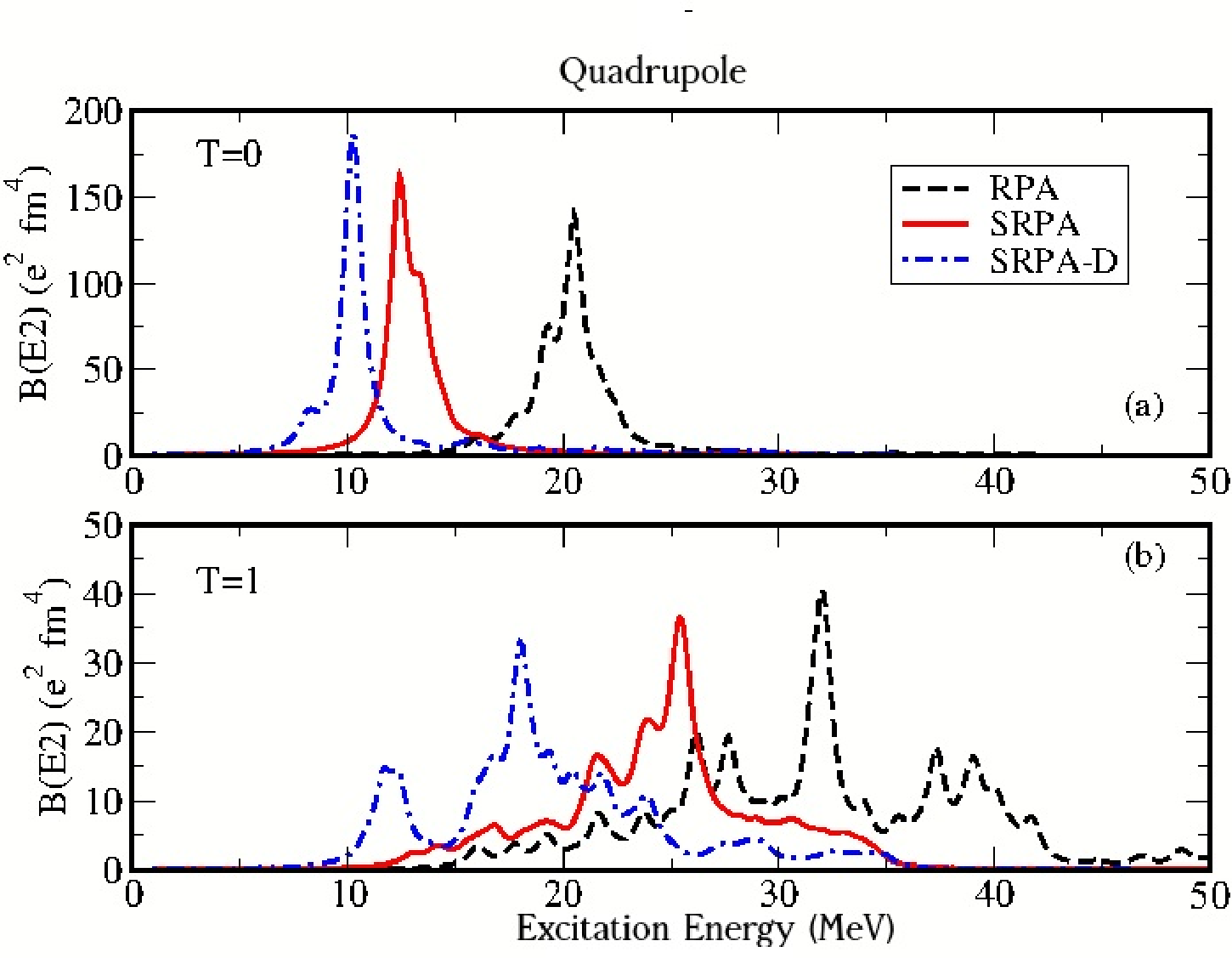}
	\caption{Comparison between RPA, dashed (black) lines, SRPA in full (red) lines and SRPA
		in the diagonal approximation, dot-dashed (blue) lines. The isoscalar (upper
		panel) and isovector (lower panel) monopole (left side) and quadrupole (right side) distributions are
		shown. Adapted from Ref. \cite{Gambacurta2010}.}
	\label{Fig:Diagonal}
\end{figure}

In Figure \ref{Fig:Diagonal}, the RPA (dashed black lines) and SRPA (solid red lines) results for the isoscalar (upper panel) and isovector (lower panel) monopole (left side) and quadrupole (right side) strength distributions are compared with the SRPA ones obtained when the diagonal approximation (\ref{Eq:Diagonal}) is used  (dot-dashed blue lines). One can see that in this case, at variance with the UCOM-SRPA results (see Figure \ref{Fig:Papa_4}), the impact of the residual interaction among the $2p-2h$ configurations is quite strong. These differences point out to the fact that strength of the residual interaction in the $2p-2h$ space depends on the structure of the underlying interaction and should be carefully analised case by case.
\begin{figure}
\includegraphics[width=.5\linewidth]{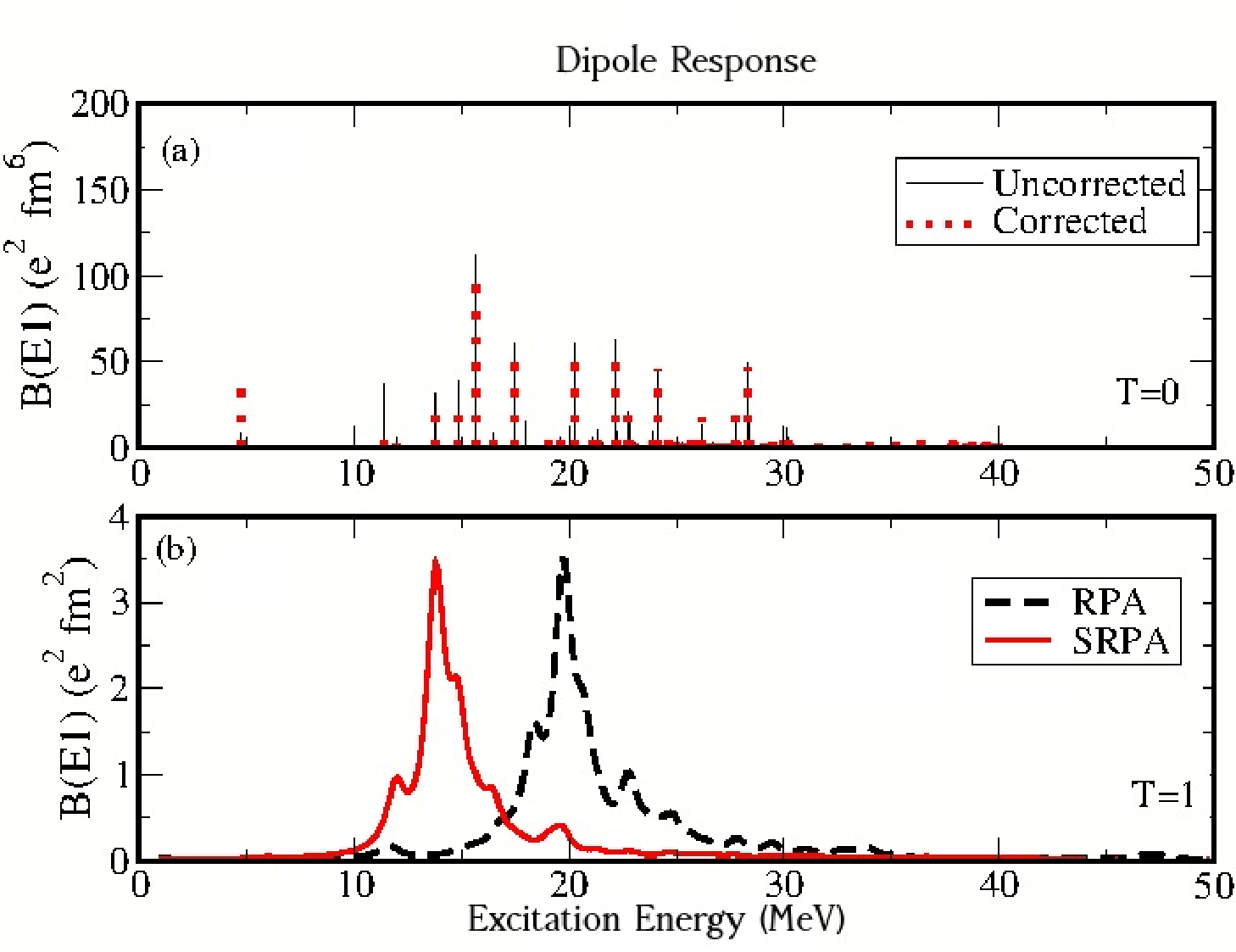}\hfill
\includegraphics[width=.5\linewidth]{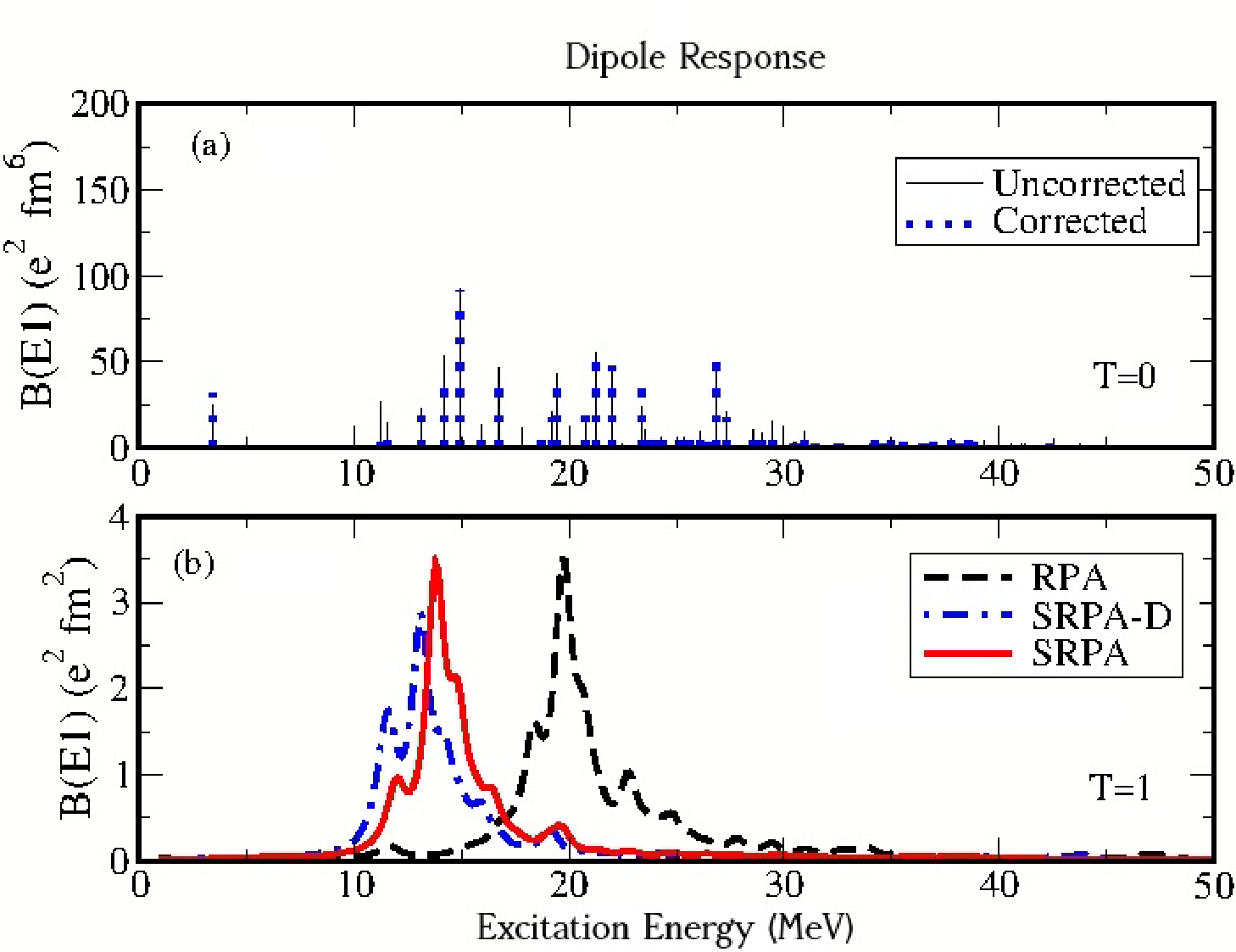}
\caption{Left side:  SRPA isoscalar dipole strength distribution using the transition operator (\ref{Eq:Op-J1-isos-r3})
	in full (black) lines and its corrected in dotted (red) lines form (\ref{Eq:Op-J1-isos-r3-corrected}) 
%	 $\sim r^3 - \frac{5}{3} \langle r^2 \rangle r$ 
	 in order to take into account center of mass corrections (upper panel);  RPA, dashed (black) line and SRPA in full (red) line isovector dipole strength distribution using the standard dipole transition operator (\ref{Eq:Op-J1-IV-CM}) (lower panel) . 
%	  of radial form $(\sim r)$. 
	  Right side:  isoscalar dipole strength distributions obtained within the SRPA in the diagonal approximation using a transition operator  (\ref{Eq:Op-J1-isos-r3})
%	   $(\sim r^3)$
	   in full (black) line using the transition operator (\ref{Eq:Op-J1-isos-r3-corrected})
	   to take into account center of mass corrections in dotted (blue) lines (upper panel);
	  % $\sim r^3 - \frac{5}{3} \langle r^2 \rangle r$ 
	    RPA, dashed (black) line, SRPA in the diagonal approximation, dot-dashed (blue) line and full SRPA in full (red) line isovector dipole strength distributions using the standard dipole transition operator (\ref{Eq:Op-J1-IV-CM}) (lower panel). Adapted from Ref. \cite{Gambacurta2010}. }
\label{Fig:Dipole}
\end{figure}

\begin{figure}
\includegraphics[width=.3\linewidth]{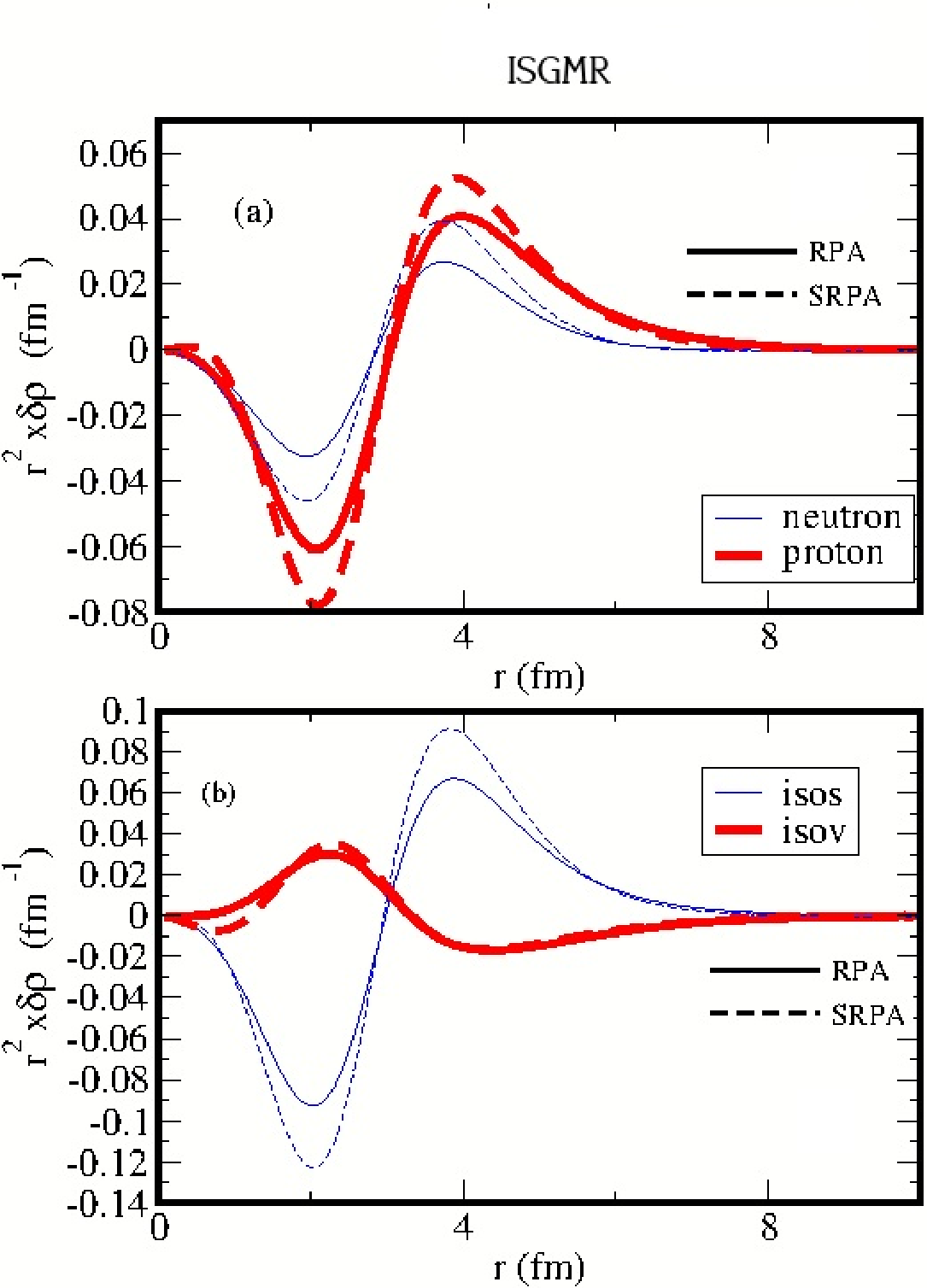}\hfill
\includegraphics[width=.3\linewidth]{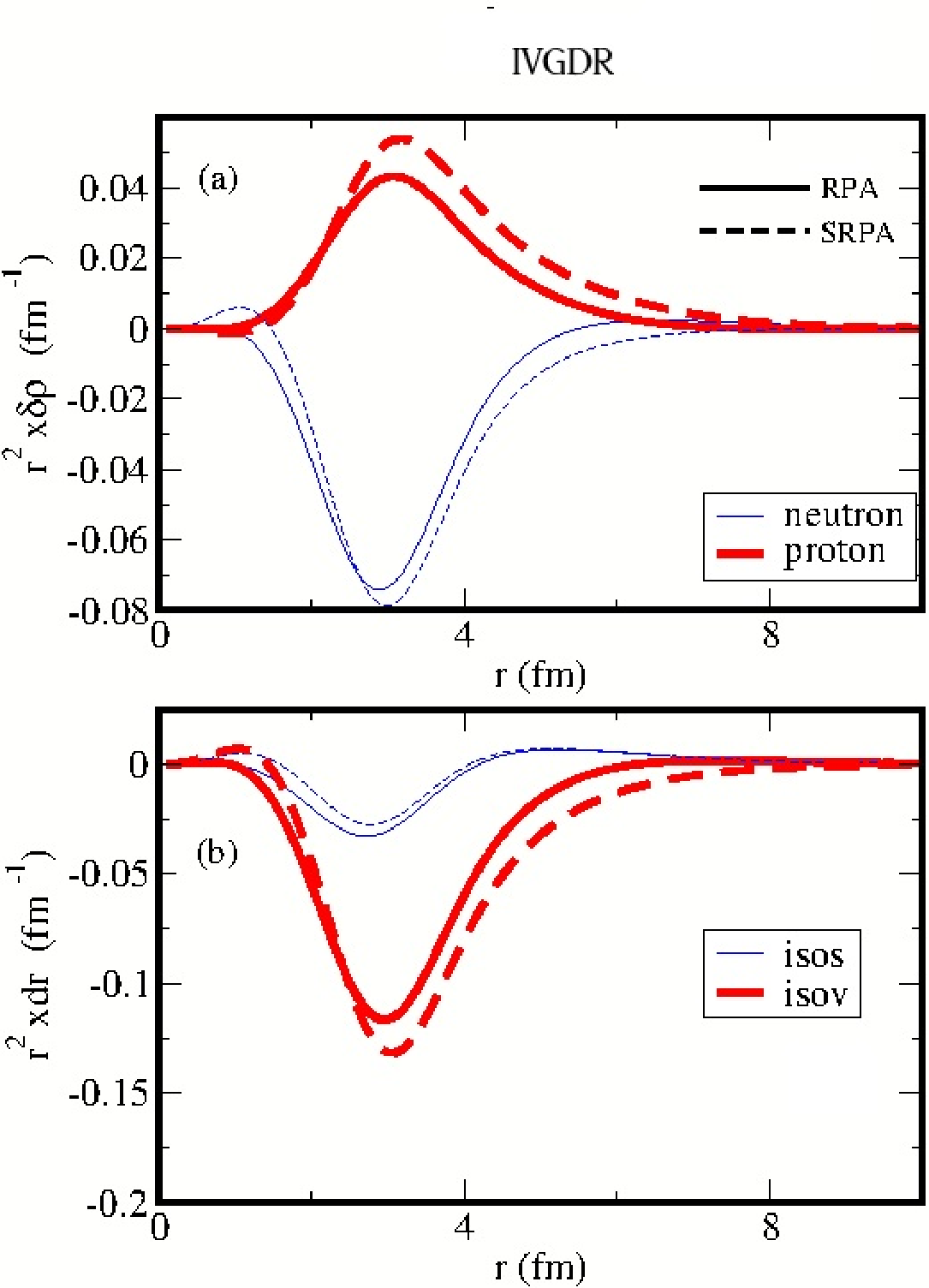}\hfill
\includegraphics[width=.3\linewidth]{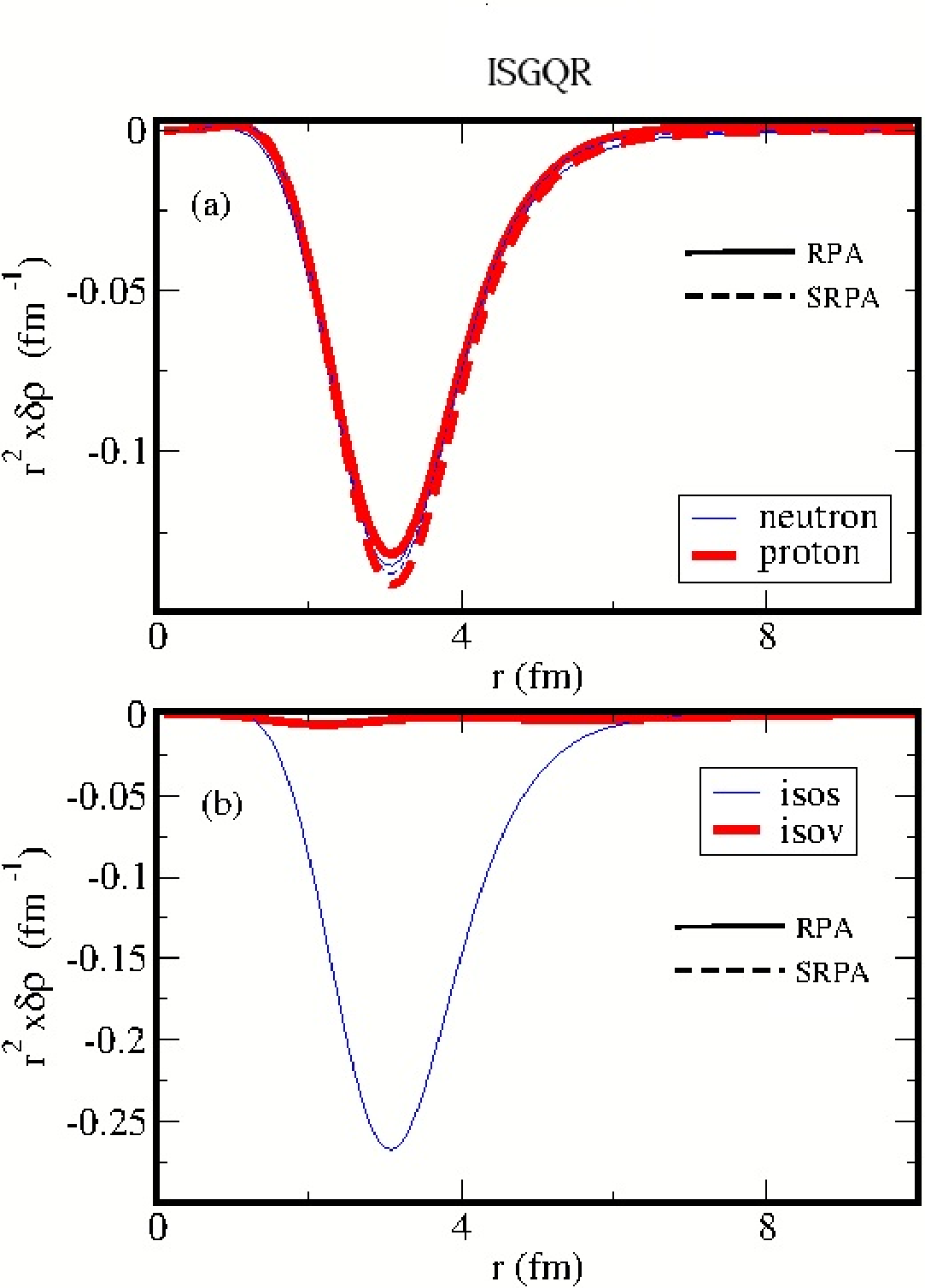}
\caption{Comparison between RPA (full lines) and SRPA (dotted lines) transition densities for the isoscalar monopole (left), isovector dipole (center) and isoscalar quadrupole (right) states. Panel (a) neutron (thin lines) and proton (bold lines). Panels (b) the isoscalar (thin lines) and isovector (bold lines) ones. Adapted from Ref. \cite{Gambacurta2010}. }
\label{Fig:TDR}
\end{figure}

We now consider the case of the dipole response. The Thouless theorem on the EWSR \cite{THOULESS1960}, valid in both RPA and SRPA \cite{Yannouleas1987}, guarantees that spurious excitations, arising from broken symmetries, should be decoupled and orthogonal to the physical spectrum. 
\begin{comment}
As detailed in Ref. \cite{Tohyama2004}, the presence of an approximate ground state and/or the inclusion of $2p-2h$ configurations requires the use of all single-particle amplitudes in constructing the elementary configurations to ensure that both single and double spurious modes reside at zero energy.	content...
\end{comment}
 In a fully self-consistent RPA (where the same interaction governs both the HF and RPA levels), the center-of-mass motion, a manifestation of translational invariance, appears precisely at zero energy, cleanly separated from the physical spectrum. However, the present RPA approach is not fully self-consistent, due to the omission of the Coulomb and spin-orbit terms in the residual interaction, resulting in the spurious state lying at approximately 1 MeV and exhausting over 96\% of the isoscalar EWSR. In SRPA, the coupling with $2p-2h$ configurations shifts the spurious state to imaginary energy. To investigate the potential mixing of spurious components with physical states, the isoscalar dipole strength distribution is analyzed using both the  transition operator (\ref{Eq:Op-J1-isos-r3}) and (\ref{Eq:Op-J1-isos-r3-corrected}) to take into account center of mass corrections. The results, presented in the upper panels of Figure \ref{Fig:Dipole}, show some discrepancies, particularly at low energies, suggesting some mixing. However, in the high-energy region corresponding to the IVGDR, the influence of spurious components is negligible. The lower panels of Figure \ref{Fig:Dipole} displays the isovector dipole results for the standard isovector operator (\ref{Eq:Op-J1-IV-CM}). Consistent with observations in the monopole and quadrupole cases, a significant shift of the strength distribution to lower energies is evident in SRPA compared to RPA. On the right side of Figure \ref{Fig:Dipole}, the results obtained using the diagonal approximation are presented. The upper panel compares isoscalar strength distributions calculated with the corrected (\ref{Eq:Op-J1-isos-r3-corrected}) and uncorrected transition operators (\ref{Eq:Op-J1-isos-r3}), again showing only significant differences at low energy. The lower panel compares isovector distributions from RPA, diagonal SRPA, and full SRPA calculations. The discrepancies between diagonal and full SRPA are smaller in the dipole channel than in the monopole and quadrupole cases, though not negligible.

We have seen that, as general trend, though a general shift towards lower energy of the strength distributions is observed in SRPA with respect to the original RPA ones, the overall structure of these distributions is maintained, suggesting that the intrinsic properties of the most collective states are not strongly modified. In Figure \ref{Fig:TDR}, we show the transition densities of the most collective states for the isoscalar monopole (left), isovector dipole (center) and isoscalar quadrupole (right) case, obtained in RPA and SRPA. One can see that indeed, the general features of the most collective states is indeed unchanged moving from RPA to SRPA.

\subsubsection{Charge-exchange excitations within the STD}

The effect of $2p-2h$ configurations within the Skyrme-EDF approach was also studied for charge-exchange excitations in Ref. \cite{Minato2016} within the STDA, (see Section \ref{Sec:Tamm-Dancoff}). In particular, the GT excitation in $^{24}$O, $^{34}$Si and $^{48}$Ca was studied, focusing on the role of the tensor force.
The SGII\cite{SGII} Skyrme  interaction for the central and spin-orbit forces \cite{SGII} was used and Te1 for the tensor force
\cite{Bai2011}. 

\begin{figure}
	\includegraphics[width=.32\linewidth]{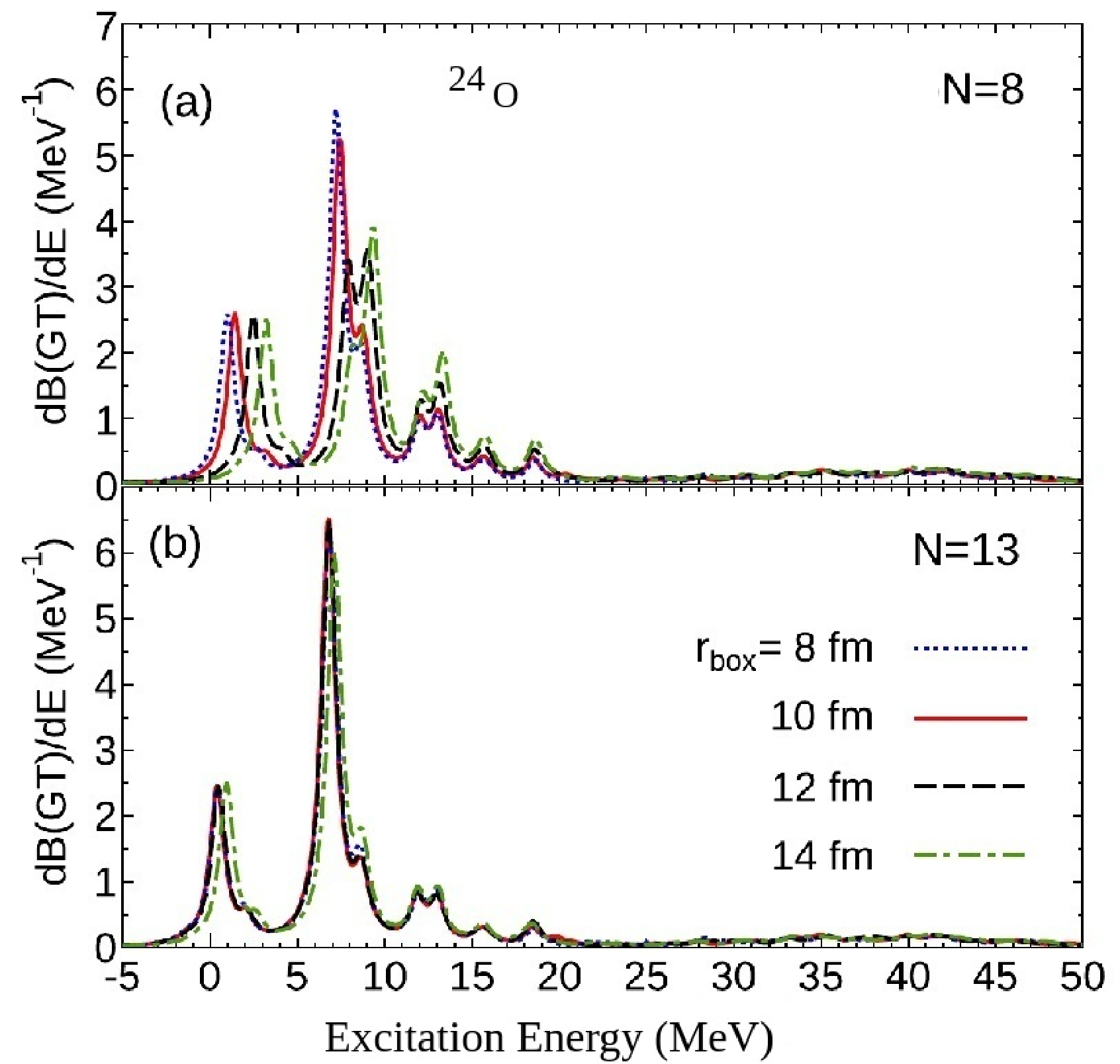}\hfill
	\includegraphics[width=.32\linewidth]{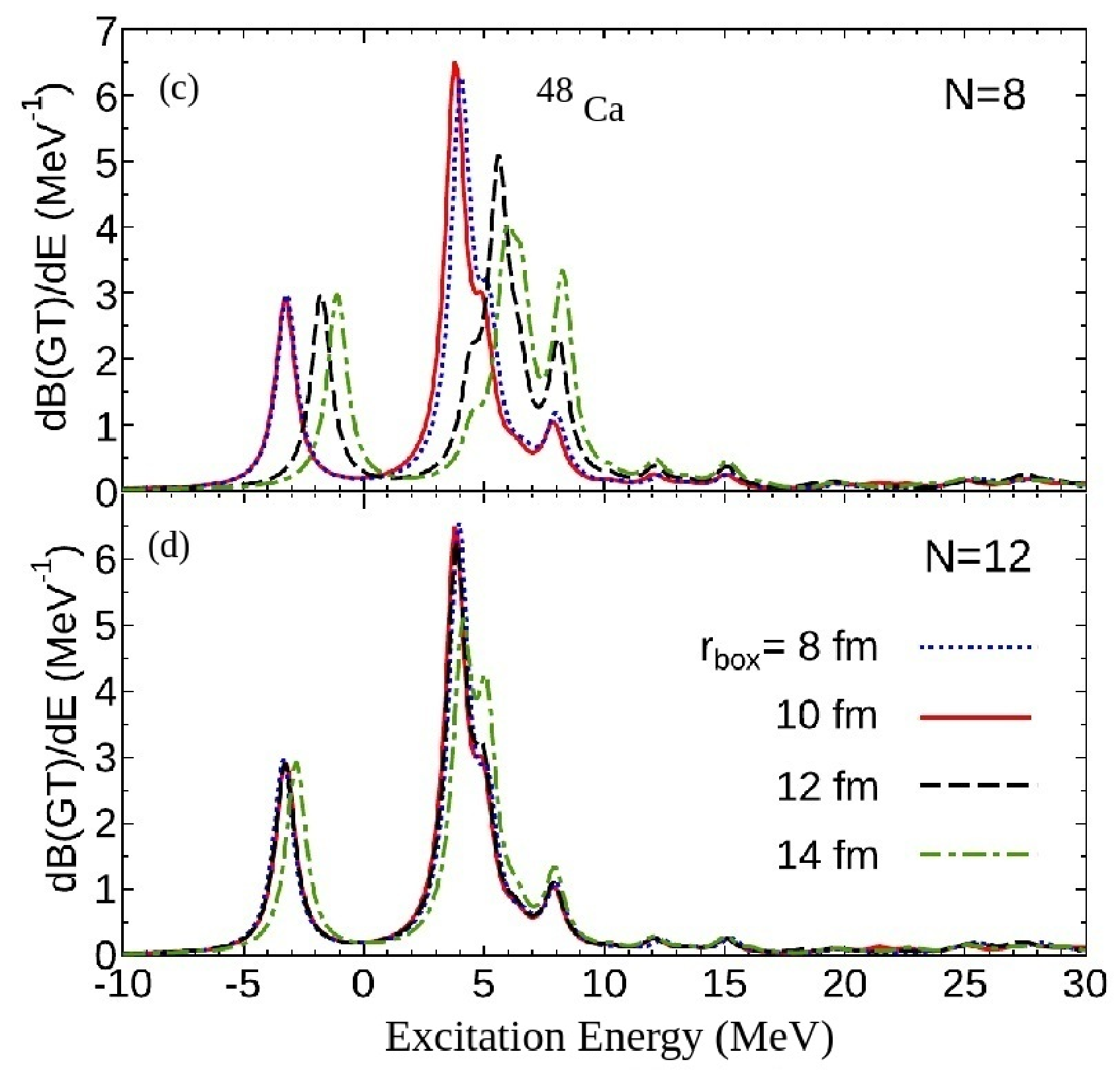}\hfill
	\includegraphics[width=.275\linewidth]{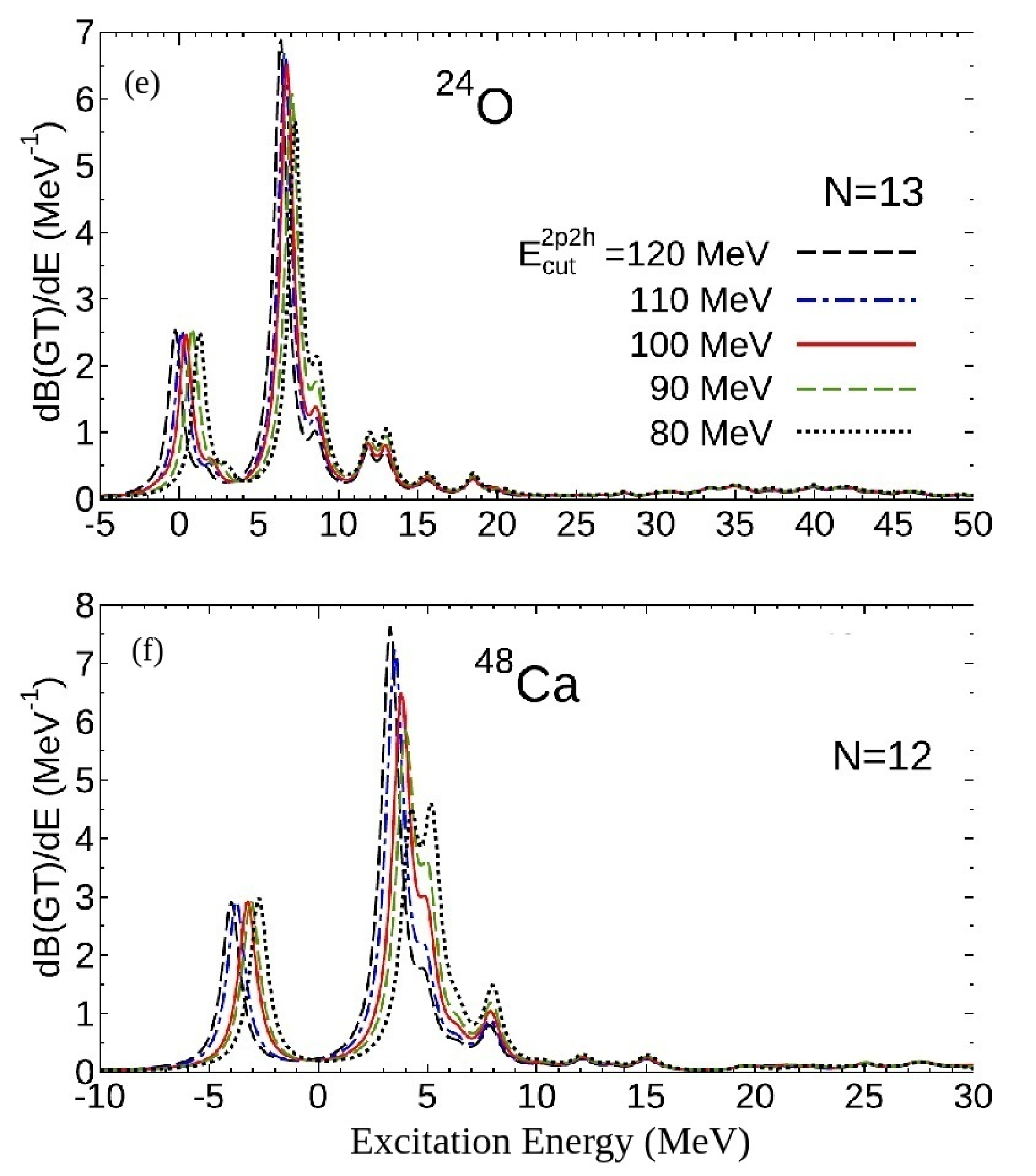}
	\caption{Left and center: GT strength distribution obtained in STDA 
		for various box boundary conditions  r$_{box}$.
		%		The strengths are smeared by the Lorentzian function with a 1 MeV width and a cutoff of 100 MeV is used for the $1p-1h$ configurations.
		Left: $^{24}$O for (a) N = 8 and (b) N = 13. Center side: $^{48}$Ca for (c) N = 8 and (d) N = 12. Right: GT strength distributions of $^{24}$O (e) and $^{48}$Ca (f) obtained in STD  with different cutoff energies on the $2p-2h$ configurations E$_{cut}^{2p-2h}$. Adapted  from Ref. \cite{Minato2016}.}
	\label{Fig:Minato-1}
\end{figure}

In Figure \ref{Fig:Minato-1}, the sensitivity of the results on the boundary box size  ($r_{\rm box}$) and a quantum number $N = 2n + l$, where $n$ and $l$ represent the number of nodes and the orbital angular momentum of the single-particle wave functions, respectively, is studied for $^{24}$O (panels (a) and (b)) and $^{48}$Ca (panels (c) and (d)). The STDA results in diagonal approximation are shown, with a fixed energy cutoff on the $2p-2h$ configurations=100 MeV. The strengths are smoothed using a 1 MeV width Lorentzian function. From the left side of the Figure,  one can see that the strength distribution for $^{24}$O does not converge for N=8, left side upper panel (a), while for $N=13$, lower panel (b), the results are rather insensitive to the various $r_{\rm box}$ values. However, as $N$ is increased, the strength distributions for different $r_{\rm box}$ values converge closer together. A similar behavior is observed for $^{48}$Ca, the results converging for $N = 12$.
\begin{figure}
	\vspace{-0.5 cm}
\includegraphics[width=.33\linewidth]{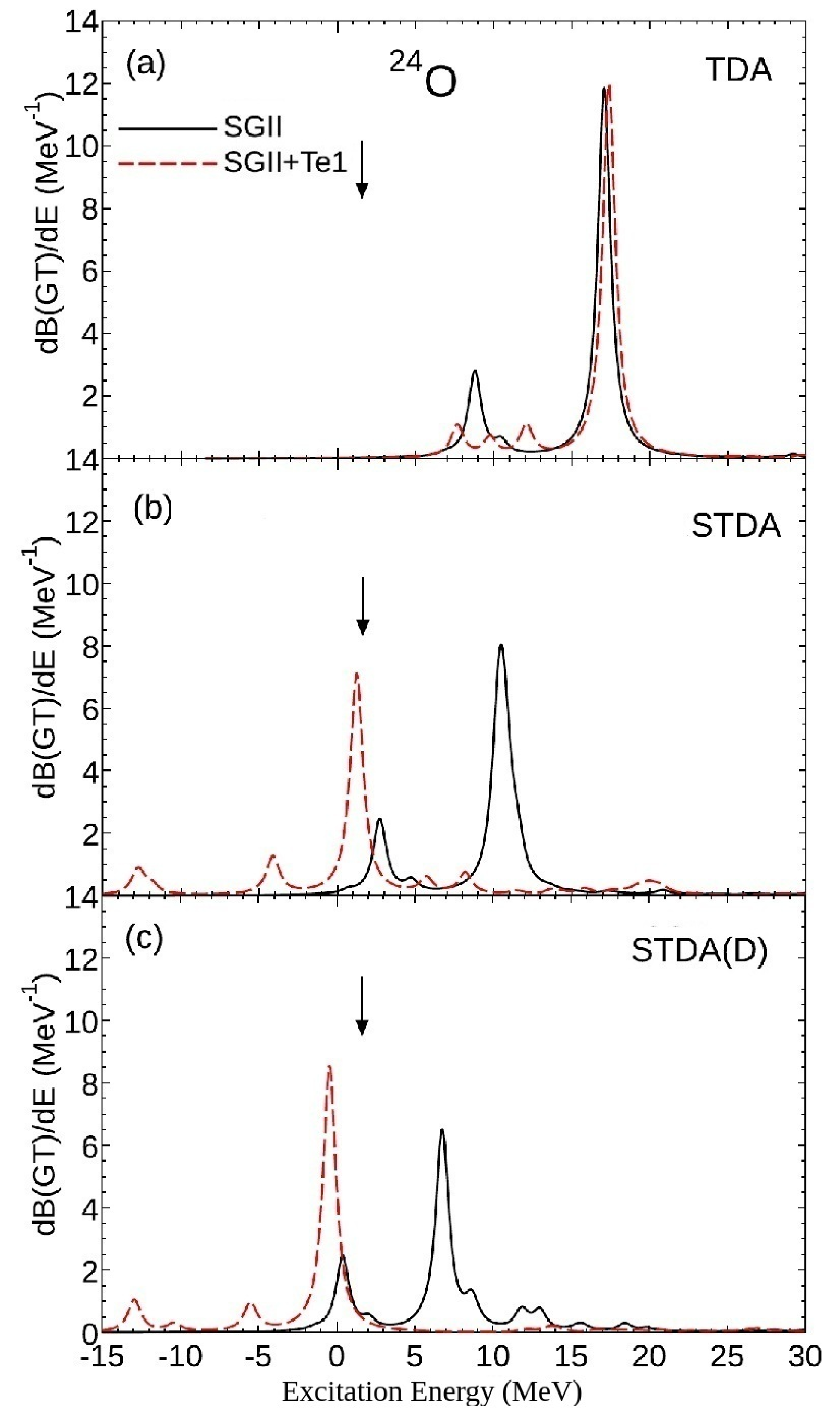}\hfill
\includegraphics[width=.32\linewidth]{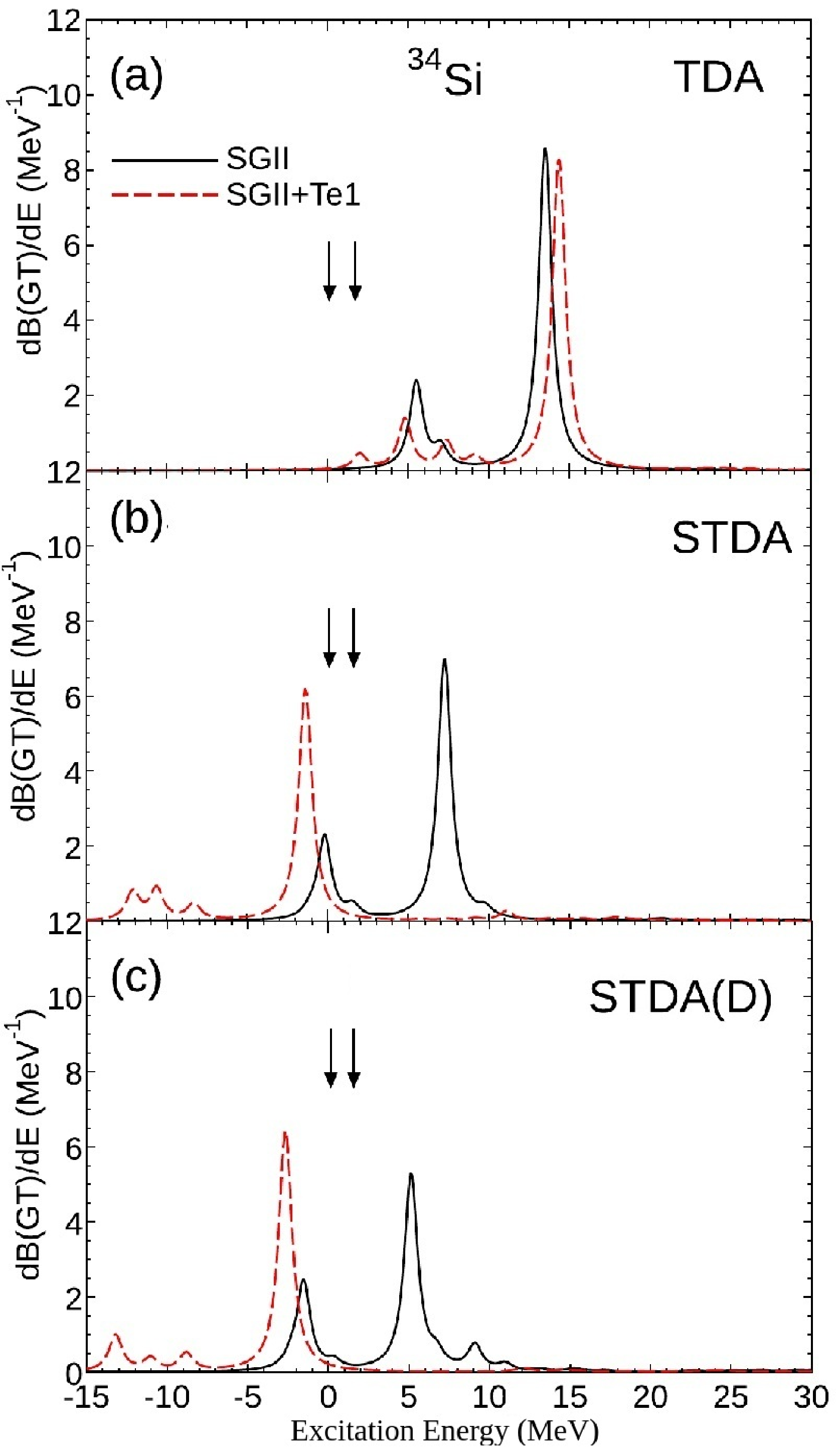}\hfill
\includegraphics[width=.32\linewidth]{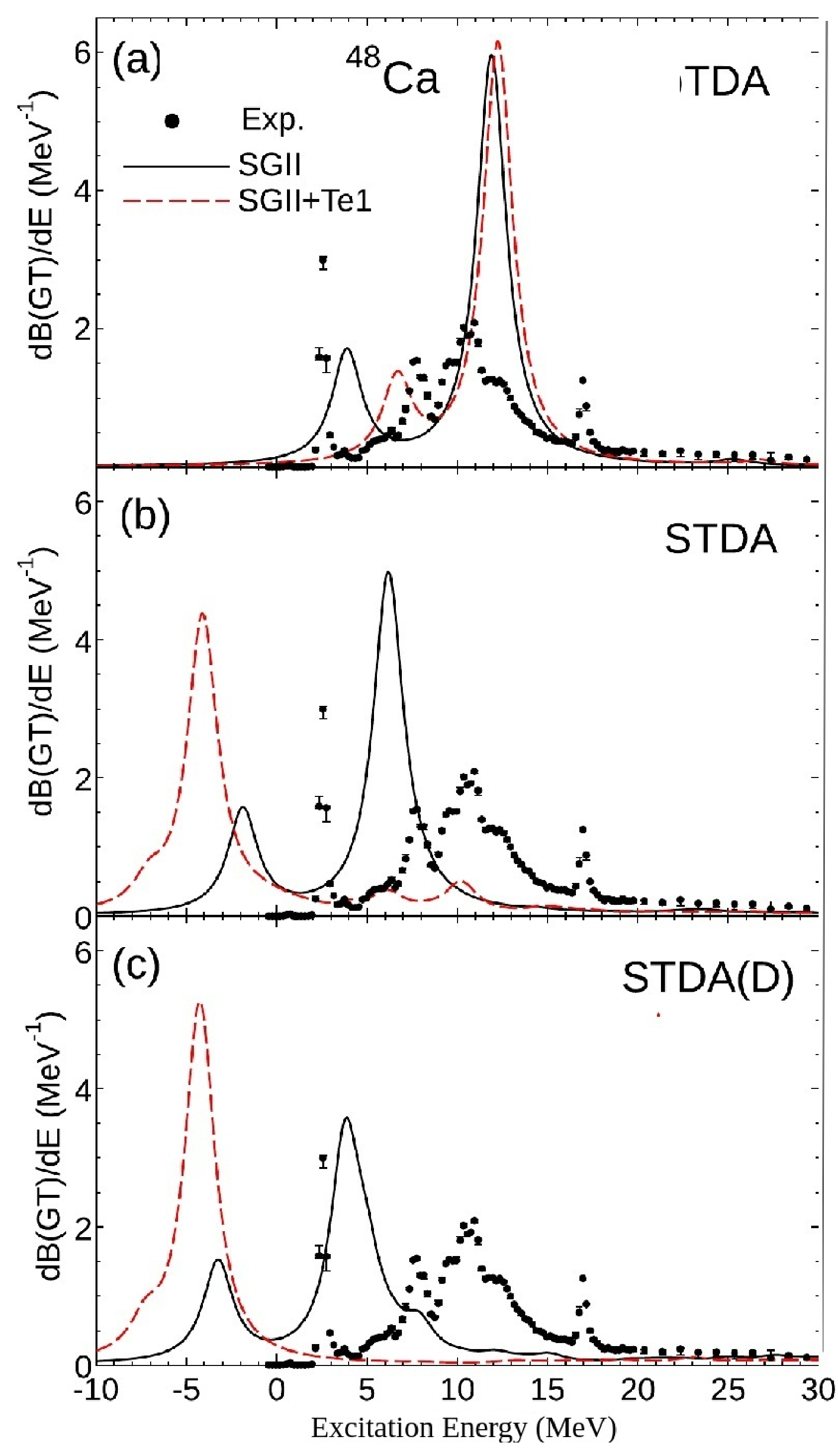}
\caption{GT strength distribution as a function of excitation
energy with respect to its daughter nucleus for $^{24}$O (left), $^{34}$Si (center) and $^{48}$Ca (right). The upper, middle,
and bottom panels are the results for TDA, full STDA, and
STDA in the diagonal approximation, respectively. The solid and dashed lines indicate the results calculated with SGII and SGII+Te1 parameter sets, respectively.
The arrow indicates the experimental peaks \cite{Firestone2007}. Adapted from Ref. \cite{Minato2016}.}
\label{Fig:Minato-3}
\end{figure}
The dependence on the energy cutoff E$_{cut}^{2p-2h}$ was studied, by varying it from 80 to 120 MeV, while keeping $N$ = 13 for $^{24}$O (panel (e)) and $N$ = 12 for $^{48}$Ca (panel (f)). The GT resonances up to 10 MeV exhibit a weak dependence on E$_{cut}^{2p-2h}$ and the high-energy region is even less sensitive. %The following parameters were therefore adopted: $r_{\rm box}$ = 10 fm, E$_{cut}^{2p-2h}$ = 100 MeV, $N$ = 13 for $^{24}$O, and $N$ = 12 for $^{48}$Ca. The same model space as $^{48}$Ca was used for $^{34}$Si.
The GT strength distribution for $^{24}\text{O}$ (left side of Figure \ref{Fig:Minato-3}), $^{34}\text{Si}$ (center), and $^{48}\text{Ca}$ (right side) are studied using the TDA and STDA methods with SGII and SGII+Te1 interactions. The general trend can be summarized as it follows. The GT strength is systematically shifted to lower excitation energy in STDA compared to TDA. For $^{24}\text{O}$, the shift is $\sim6\text{ MeV}$. The total GT transition probability is reduced from $\sim12$ (TDA) to $\sim8$ (STDA) due to $2p-2h$ mixing. For $^{34}\text{Si}$, the shift is $\sim7\text{ MeV}$ (SGII). For $^{48}\text{Ca}$, TDA roughly reproduces the experimental GT GR position, and STDA places it too low. The tensor force (SGII+Te1) further shifts the STDA GT resonances downward, often producing peaks at negative energies (e.g., in $^{24}\text{O}$ and $^{34}\text{Si}$). The TDA fails to show the experimentally observed low-lying $\text{1}^+$ state (e.g., at $1.8\text{ MeV}$ in $^{24}\text{O}$). The STDA, by shifting low-lying TDA resonances produces several resonances near the experimental $\text{1}^+$ state. The low-lying resonance heights and widths are less sensitive to the $2p-2h$ effect than the GT GRs, keeping the $\text{1}p-\text{1}h$ configuration dominant.
 Regarding STDA in the diagonal approximation, one observes that the GT resonance positions are consistently lower than those for full STDA for all nuclei. Beyond this general trend, three observations emerge. First, the difference in GT resonance position between STDA and STDA in the diagonal approximation diminishes as the nuclear mass increases, progressing from the lightest nucleus in this study, $^{24}$O, to the heaviest, $^{48}$Ca. Second, the difference between TDA and STDA(D) is considerably smaller with the SGII+Te1 interaction compared to SGII. This suggests that the diagonal matrix elements of the tensor force component are larger than the off-diagonal elements. Third, the overall GT resonance shape remains largely consistent between STDA and STDA in the diagonal approximation.  Based on these findings, it can be concluded that the diagonal approximation provides  qualitatively reasonable results, in particularly for heavy nuclei or when the tensor force is included.

\begin{figure}
	\includegraphics[width=.5\linewidth]{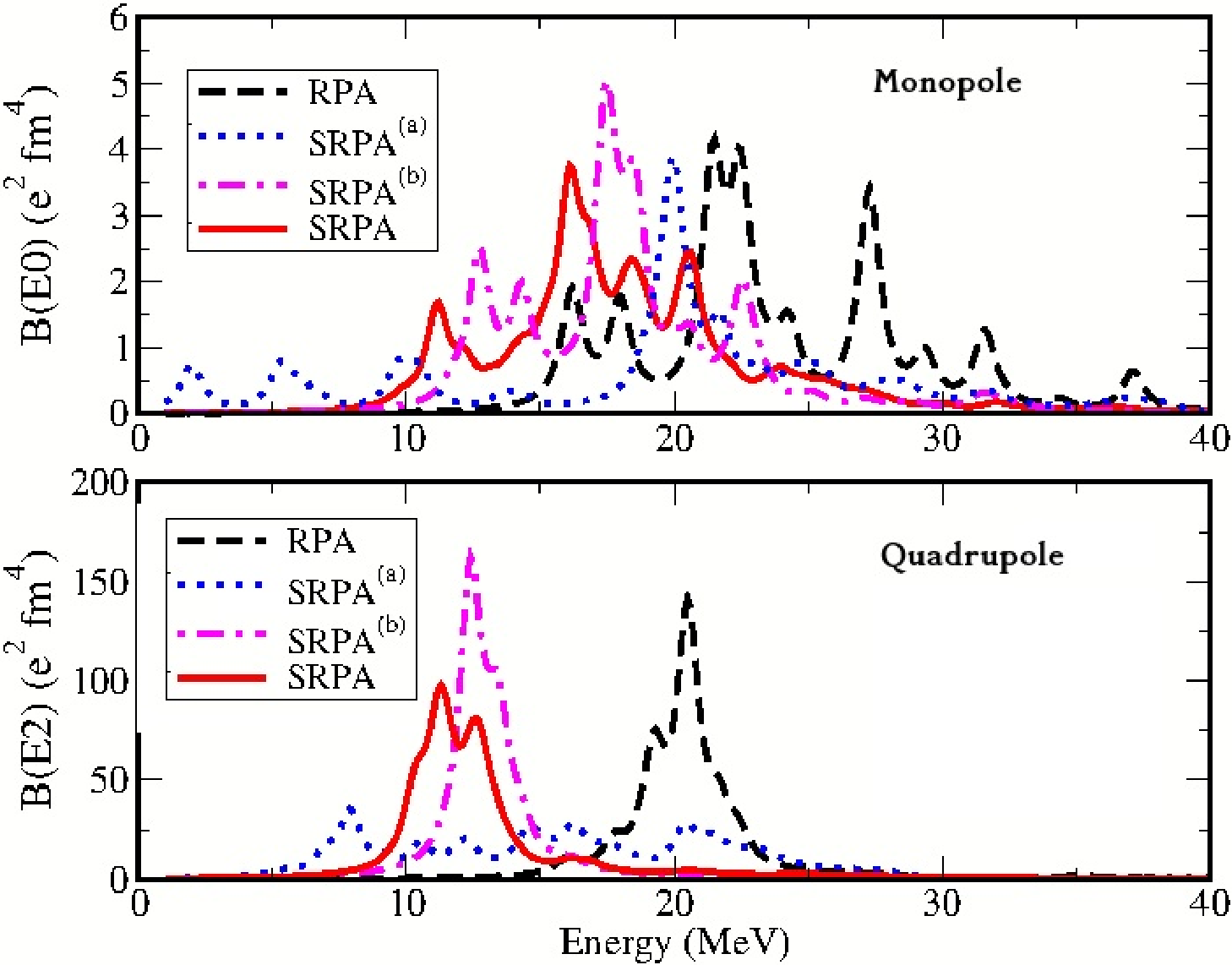}\hfill
	\includegraphics[width=.48\linewidth]{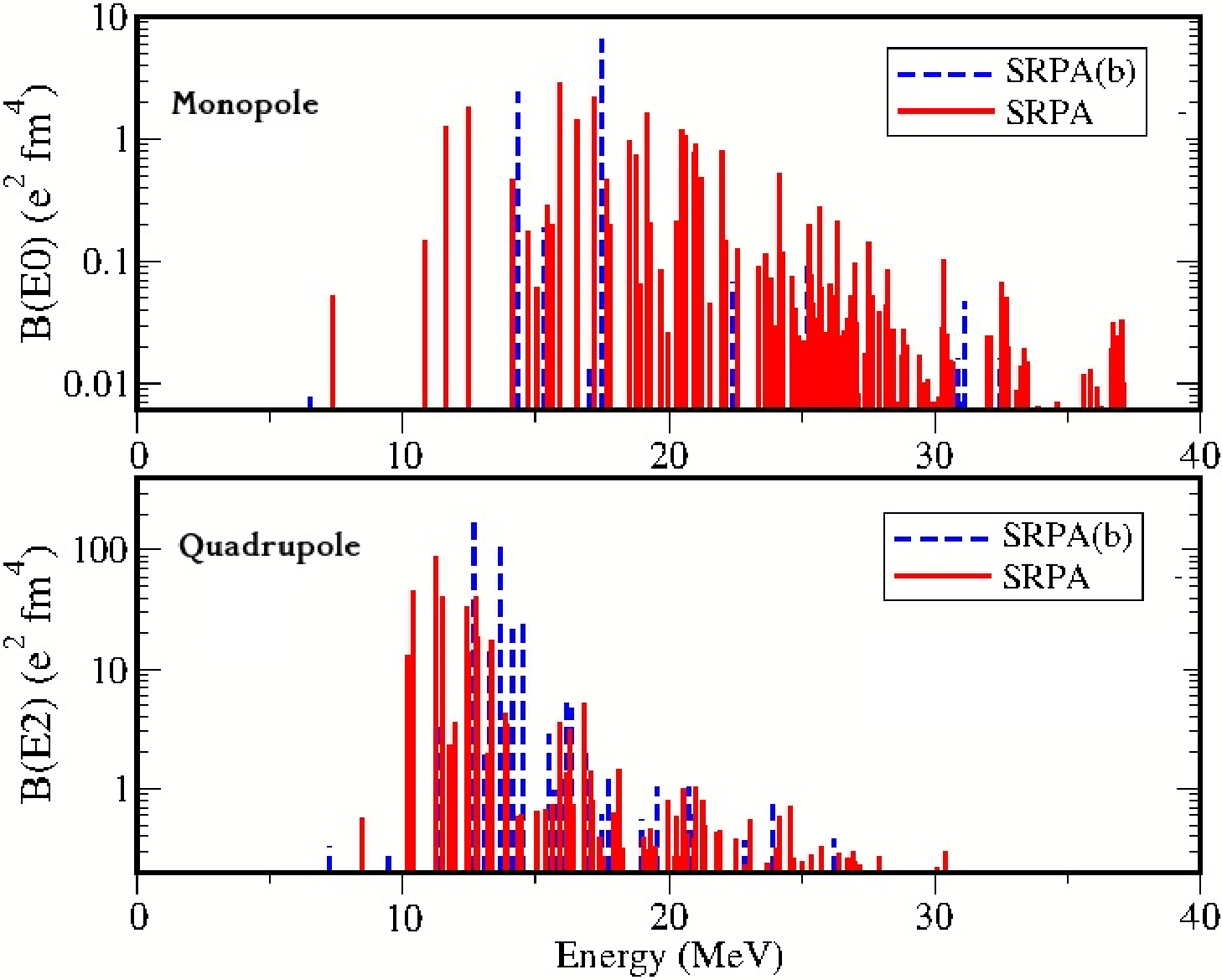}
	\caption{Left side: Folded isoscalar strength 
		distributions obtained for $^{16}$O (monopole (top) and quadrupole (bottom)) within the RPA (black dashed lines), 
		SRPA with all the rearrangement terms as in RPA 
		(SRPA$^{(a)}$, blue dotted lines)
		SRPA without rearrangement terms 
		in beyond-RPA matrix elements (SRPA$^{(b)}$, magenta dot-dashed lines), 
		and full SRPA (red solid lines). Right side: Discrete isoscalar monopole (top) and quadrupole (bottom) strength 
		distributions obtained for $^{16}$O with 
		SRPA without rearrangement terms 
		in beyond-RPA matrix elements (SRPA$^{(b)}$, blue dashed lines), 
		and full SRPA (red solid lines). A logarithmic scale is used in the ordinate. Adapted from Ref. \cite{Gambacurta2011a}. }
	\label{Fig:rearr}
\end{figure}

\subsubsection{Applications with the rearrangement terms}
\label{Sec:ApplicationPart_SRPA_Rear}
The first numerical implementation of the rearrangement terms in SRPA was presented in Ref. \cite{Gambacurta2011a} for the monopole and quadrupole excitations in $^{16}$O. The Skyrme interaction SGII \cite{SGII} was used.
The full SRPA rearrangement terms are compared with the corresponding results discussed in Section \ref{Sec:App_First_SRPA} within two approximated schemes. On the left side of Figure \ref{Fig:rearr}, the isoscalar monopole (upper panel) 
and quadrupole (lower panel) strength 
distributions calculated for the multipole transition operators (\ref{Eq:Op-J02-isos})
%\begin{equation}\label{isos-oper}
%	F_{\lambda}=\sum r_i^2 Y_{\lambda 0}(\hat{r}_i) 
%\end{equation}
are shown. The discrete spectra have been folded with a Lorentzian having a width of 1 MeV for visualization purposes. The RPA profiles (black dashed lines) are plotted together with SRPA profiles obtained by calculating all the rearrangement terms as in RPA (SRPA$^{(a)}$),
SRPA profiles obtained without rearrangement terms 
in beyond-RPA matrix elements (SRPA$^{(b)}$), 
and full SRPA profiles (red solid lines).
The comparison shows that both the 
approximations lead to quite different results with respect to the full SRPA strength distributions. 
In the quadrupole case in particular, one observes that the blue dotted curve (all the rearrangement terms like in RPA) is completely flat and fragmented. 
On the right side of Figure \ref{Fig:rearr}, the discrete spectra are shown in logarithmic scale. One can see that a much more fragmented distribution is found when the full SRPA rearrangement terms (red solid lines) are used with respect to the case where they are neglected (blue dashed-lines). Summarizing, the three distinct SRPA calculation methods yield differing peak structures and fragmentation patterns, showing the importance of a correct evaluation of the rearrangement terms in SRPA calculations, when employing density-dependent forces.

\subsection{SRPA with the Gogny interaction}
\label{Sec:Applications_SRPA_Gogny}
% \subsubsection{Introduction}

In the previous Section, we saw that the SRPA calculations with the Skyrme interaction systematically show a strong shift towards lower energies of the strength distributions with respect to the RPA ones. One possible source of this behavior might be related to the zero-range nature of the Skyrme interaction, which does not provide thus a natural cutoff. In Ref. \cite{Gambacurta2012}, the first application of the SRPA with the Gogny force was presented and discussed.

\begin{wrapfigure}{l}{0.45\textwidth}
	\centering
	%  	\vspace{-3.5mm}
	\includegraphics[width=0.45\textwidth]{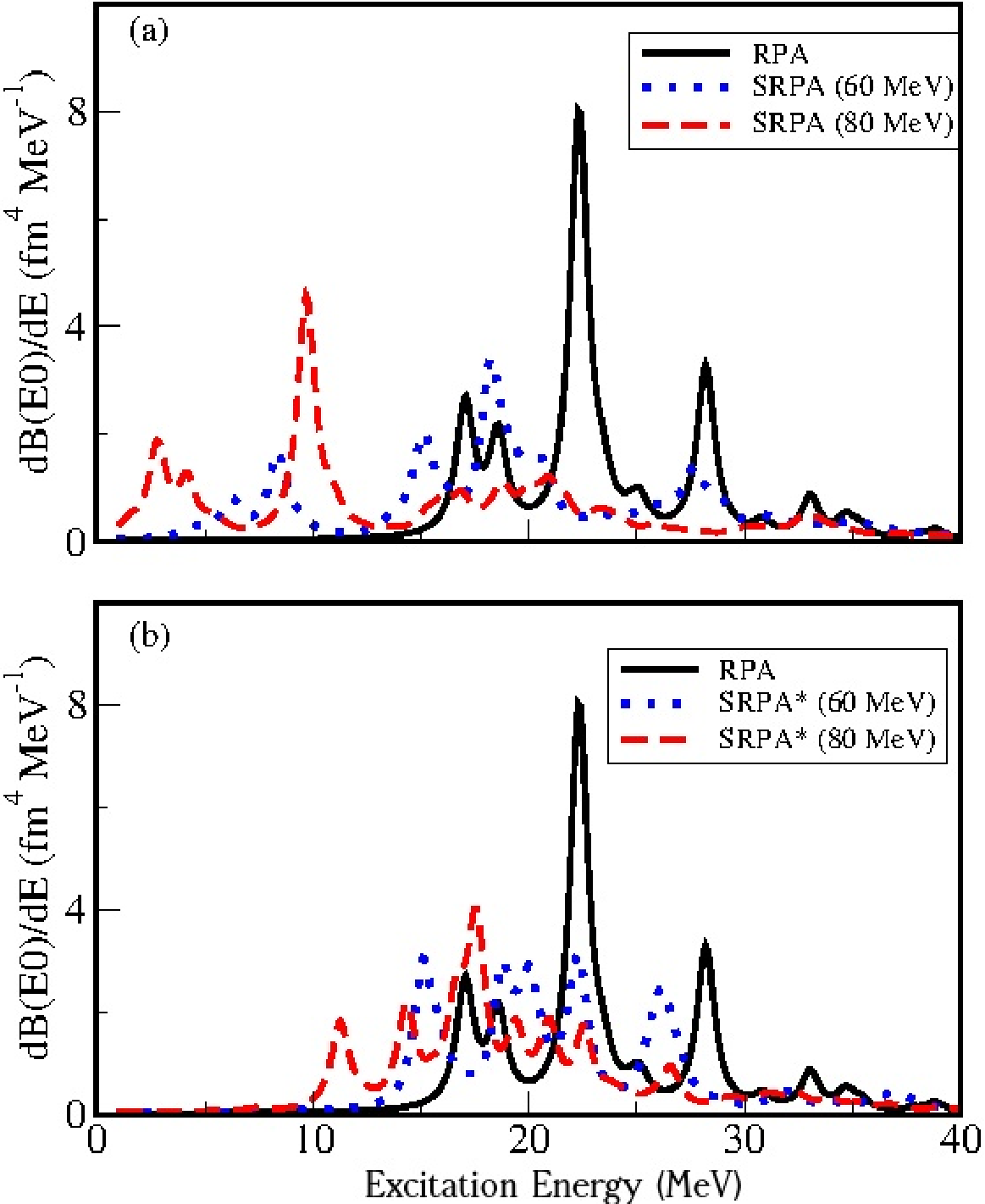}
	\caption{Isoscalar 0$^+$ response in $^{16}$O calculated with the Gogny-RPA (full line)
		and with the SRPA (panel a) and SRPA* (panel b) approach with an energy cutoff on the $2p-2h$
		configurations of 60 (dotted line)
		and 80 (dashed line) MeV.
		Adapted from Ref. \cite{Gambacurta2012}.}
	%  	\vspace{-7mm}
	\label{Fig:Gogny1a}
\end{wrapfigure}
While computationally more intensive than employing a zero-range interaction, the use of a finite-range force offers several key advantages. Firstly, the Gogny force has been designed and calibrated for consistent application in both the HF and pairing channels. In particular, considering that SRPA involves not only the standard RPA-type particle-hole matrix elements but also other terms, such as 4-particle and 4-hole matrix elements, employing a force explicitly tailored to handle these diverse interaction components could be advantageous. A second, and equally important, benefit stems from the finite range of the four central terms of the Gogny force.

\begin{wrapfigure}{l}{0.45\textwidth}
	\centering
	 	\vspace{-0.5cm}
\includegraphics[width=0.4\textwidth]{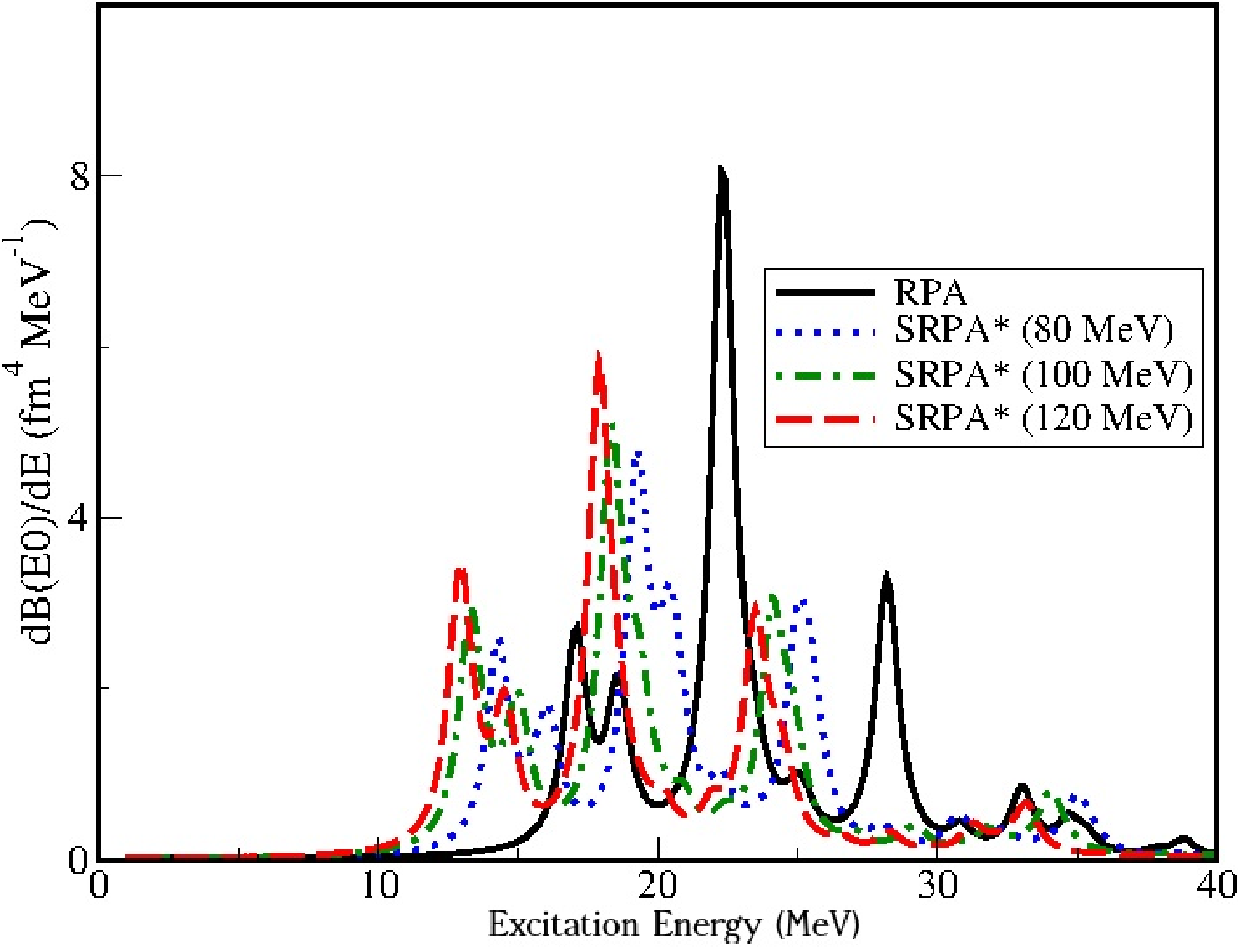}
	%  	\vspace{-1.5mm}
	\caption{Isoscalar monopole response in $^{16}$O
		calculated in the SRPA* case with
		cutoff energies of 80 (dotted line), 100 (dot-dashed line) and
		120 (dashed line) MeV.
		Adapted from Ref. \cite{Gambacurta2012}.}
	%  	\vspace{-7mm}
	\label{Fig:Gogny1b}
\end{wrapfigure}

 This finite range is expected to naturally facilitate convergence with respect to the energy cutoff in the $2p-2h$ configuration space, reducing the influence of high-momentum components and thus mitigating the sensitivity to the cutoff.  While complete convergence is not fully expected due to the presence of zero-range density-dependent and spin-orbit terms, the finite-range central terms significantly contribute to numerical stability.

In the following, the isoscalar monopole response is shown as an illustrative example. The D1S parametrization of the Gogny interaction \cite{D1S} is used. The single-particle and $1p-1h$ spaces were chosen to be sufficiently large to ensure the stability of the EWSR values.

% Specifically, all single-particle states with unperturbed energies below 60 MeV (corresponding to all $1p-1h$ configurations with unperturbed excitation energies up to 100 MeV) were included.

The Coulomb and spin-orbit contributions were omitted from the residual interaction, producing $\simeq$ 5\% violation of the EWSR within the RPA. In constructing the $2p-2h$ space, all configurations with unperturbed energies below a cutoff E$_{cut}$ were considered. The numerical stability of the results with respect to E$_{cut}$ was investigated.  For SRPA calculations with E$_{cut}$ = 60 MeV, a full numerical verification was possible, calculating the full spectrum and obtaining  a 5\% EWSR violation (identical to the RPA case).  Increasing E$_{cut}$ up to 80 MeV, a few imaginary solutions appear but the $m_0$ and $m_1$ values were not strongly affected. However strong differences between the two corresponding spectra  (with E$_{cut}$=60 and 80 MeV, see panel (a) of Fig. \ref{Fig:Gogny1a}) were observed. A more careful analysis pointed out that certain neutron-proton ($\nu\pi$) matrix elements of the interaction, appearing in the beyond-RPA block matrices, are extremely large. Some of these were found to be five to ten times larger than the others. These matrix elements are likely to produce these strong differences and the appearance of imaginary solutions, influencing the stability of the results, particularly the peak structure of the nuclear response. It is worth mentioning that analogous effects associated with large $\nu\pi$ matrix elements have been observed in  applications of the variational multiparticle-multihole configuration method to the low-lying spectroscopy of $^{30}$Si using the Gogny force \cite{Pillet2012}.  These findings corroborate the SRPA results and suggest that these interaction matrix elements deserve careful attention and  tuning within the parameter fitting procedure.

To isolate and analyze these effects, two distinct types of calculations were performed: (a) a full SRPA calculation including all $2p-2h$ configurations, labeled as SRPA; and (b) a restricted SRPA calculation considering only $2p-2h$ configurations composed of pure neutron or pure proton excitations, labeled as SRPA$^*$. The latter does not include
%excludes $2p-2h$ configurations where the two particles and the two holes have different isospin character. This exclusion eliminates
 the $\nu\pi$ matrix elements. The impact of the $\nu\pi$ matrix elements is illustrated in Figure \ref{Fig:Gogny1a}, displaying the isoscalar monopole response calculated using both the full SRPA (a) and the restricted SRPA* (b) schemes, for two different cutoff energies, E$_{cut}$ = 60 and 80 MeV. The corresponding Gogny-RPA results are also included in both panels for comparison.

In the SRPA, the response varies appreciably with the cutoff energy. For E$_{cut}$ = 80 MeV, the main peak is shifted to energies more than 10 MeV lower than the RPA result (panel (a)). In contrast, the SRPA* results (panel (b)) exhibit much greater stability with respect to changes in the cutoff energy. 
\begin{comment}
This stability is also reflected in the centroid energies of the strength distributions. Increasing the cutoff from 60 to 80 MeV shifts the centroid from 20.37 to 15.30 MeV (a 25\% change) in the full SRPA calculations, but only from 23.97 to 22.37 MeV (a 7\% change) in the SRPA* calculations. 	content...
\end{comment}
Furthermore, the nature of the change is different. In the SRPA* case, the change is essentially a shift, whereas the full SRPA strength distribution is significantly altered in shape compared to the RPA result when the $\nu\pi$ matrix elements are included. To further examine the stability of the SRPA* results, calculations with cutoff values of 100 and 120 MeV were performed (Figure \ref{Fig:Gogny1b}). Increasing the cutoff from 80 to 100 MeV shifts the centroid from 22.37 to 21.32 MeV (5\%), and a further increase from 100 to 120 MeV shifts it from 21.32 to 20.49 MeV (4\%). Conversely, the full SRPA results continue to change significantly with increasing cutoff, and for values above 80 MeV, the corresponding equations exhibit imaginary eigenvalues. It is concluded that the expected stability associated with the Gogny interaction is achieved in the SRPA* case, where, by construction, all large $\nu\pi$ matrix elements of the residual interaction in the beyond-RPA blocks are neglected.
\begin{wrapfigure}{l}{0.5\textwidth}
	\centering
	\includegraphics[width=0.7\linewidth]{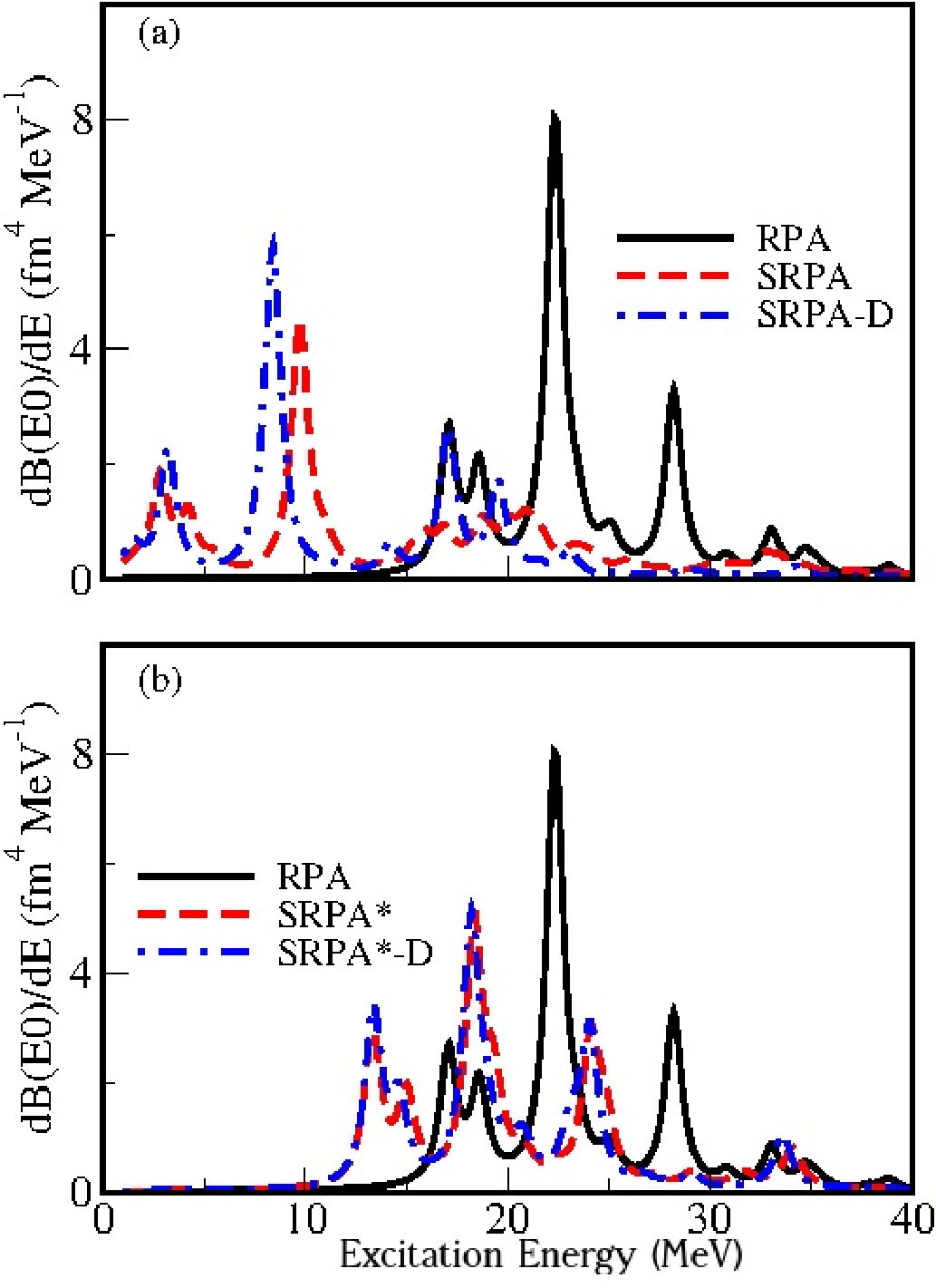}
	\caption{Isoscalar monopole distributions in $^{16}$O
		calculated within the SRPA (panel (a)) and SRPA* (panel (b)) model (dashed lines), compared with the RPA  ones. The comparison with the corresponding spectra obtained in the
		diagonal approximation (dot-dashed lines) is shown. Adapted from Ref. \cite{Gambacurta2012}.}
	\label{Fig:Gogny2}
%	\vspace{-1.9cm}
\end{wrapfigure}
% This suggests that the zero-range density-dependent part of the interaction does not significantly affect the convergence.
%The strong impact of these $\nu\pi$ matrix elements is also evident on the left side of Figure \ref{Fig:Gogny2}, which displays the discrete spectra on a logarithmic scale to emphasize the fragmentation of the response. The SRPA and SRPA* spectra (solid lines) correspond to a cutoff of 80 MeV. For comparison, the corresponding RPA results are also plotted (dashed lines) in both panels. It is observed that not only the energies but also the fragmentation of the peaks is significantly affected in the full SRPA response (panel (a)). While the monopole case is presented here as an example, it has been verified that this behavior (changes in both energies and fragmentation) is also present in the dipole and quadrupole channels within the full SRPA scheme, where all $\nu\pi$ matrix elements of the residual interaction are included.

For the SRPA* calculations, where reasonable convergence with respect to the $2p-2h$ space cutoff is achieved and the anomalous $\nu\pi$ matrix elements are neglected. The experimental centroid is located at 21.13 MeV. The calculated centroid energies are 23.88 MeV in RPA and 20.49 MeV in SRPA*. Comparing the  Gogny-SRPA results with Skyrme ones (Section \ref{Sec:App_First_SRPA}) reveals a similar, though less pronounced, behavior. The energy shift observed in the Gogny case is only comparable to the corresponding Skyrme result when all the large $\nu\pi$ matrix elements are omitted (the SRPA* scheme). Otherwise, their influence is excessively strong, driving the centroid energies to unrealistically low values.

Finally, within the Gogny-SRPA framework, the validity of the diagonal approximation was examined, denoted SRPA-D. Figure \ref{Fig:Gogny2} compares results from the full SRPA (a) and SRPA* (b) models with corresponding SRPA-D results. In both SRPA and SRPA*, neglecting the residual interaction in the $2p-2h$ space does not drastically alter the shifts or shapes of the strength distributions, though the differences are larger in the SRPA case than in the SRPA* case. The relatively small differences between SRPA* and SRPA*-D suggest that the influence of the $\nu\pi$ matrix elements is primarily related to the coupling between $1p-1h$ and $2p-2h$ configurations. In the SRPA* case (panel (b)), the diagonal approximation results are extremely close to the full SRPA* ones, a behavior that persists for larger energy cutoffs. Comparing these results with those obtained using the Skyrme interaction of Section \ref{Sec:Applications_SRPA_Skyrme} reveals that, in the Gogny case, the diagonal approximation yields results much closer to the full calculations, particularly for the monopole response. However, more systematic applications, exploring in particular heavier systems would be needed to confirm this behaviour.

%In summary, Gogny-SRPA calculations reveal that the nuclear response is strongly influenced by certain $\nu\pi$ matrix elements of the residual interaction, especially in channels coupling $1p-1h$ and $2p-2h$ configurations (i.e., those of 3 hole-1 particle type). These matrix elements are absent in HF and standard RPA calculations, and therefore do not contribute to the usual fitting procedures used to determine effective force parameters.

\subsection{Stability issues in the SRPA}\label{Sec:Applications_SRPA_Stability}

The term "self-consistent" in the context of HF plus RPA calculations implies a crucial interplay between the MF and the collective excitation parts of the calculation. It relies on the use of the same underlying interaction governing both the single-particle structure (described by HF) and the collective response (described by RPA). This consistency is essential for a physically meaningful description of nuclear systems. The RPA matrix depends on the so-obtained single-particle energies and on the residual interaction. In a fully self-consistent HF+RPA calculation, this residual interaction must be derived from the same nucleon-nucleon interaction (or functional) used in the HF calculation to generate the MF. This ensures that the description of the collective motion is consistent with the underlying single-particle structure. This scheme is important not only for formal theoretical consistency but also because: (i) it avoids potential double-countings of the nucleon-nucleon interaction, ensuring that the RPA only accounts for the correlations beyond those already included in the MF, (ii) it avoids numerical inconsistencies and potentially unphysical results, as those related, for example, to broken symmetries and mixing with spurious mode components.
Indeed the RPA matrix (\ref{Eq:RPAmat})
%\begin{equation}
%	\left(\begin{array}{cc}
%		\mathcal{A} & \mathcal{B} \\
%		\mathcal{B}^{*} & \mathcal{A}^{*} \\
%	\end{array}\right)
%	\label{Eq:HF-stability}
%\end{equation}
is nothing but the stability matrix of the HF variational problem \cite{THOULESS1960}. In other words, the HF solution guarantees the stability condition of RPA~\cite{RS.80,THOULESS1960}, meaning that the RPA solutions are real and the RPA is free of instabilities. In this case moreover,
the summed energy-weighted strength of a single-particle transition operator in RPA is determined by a specific expectation value
in the HF ground state, Eq. (\ref{Eq:EWSR_RPA}). One of the consequences of this theorem is that symmetries which are broken by the HF ground state are restored by the RPA and corresponding spurious modes have to appear at zero excitation energy. On the contrary, if the RPA has imaginary solutions, the system undergoes a phase transition.

In analogy, one may refer to ''self-consistent'' HF-SRPA, whenever the single-particle energies and residual interaction appearing in the SRPA equations are derived by using the same interaction or functional. The SRPA studies discussed in the previous sections, have shown some general features of this method when applied without approximations, using large model spaces, in a self-consistent manner. In particular, the total strength is preserved, conserving both the $m_0$ and $m_1$ RPA moments, though a strong downward shift in energy of the strength distribution is also observed together with its fragmentation.
This downward shift produced by the coupling with the $2p-2h$ configurations can in some cases be so strong as to push low-energies states (obtained at RPA level) leading to unstable (imaginary) solutions. This issue is particularly problematic for the lowest-lying state associated with a broken symmetry, such as the dipole mode when translational invariance is broken. While self-consistency ensures the RPA stability matrix is positive semi-definite, this condition does not extend to the SRPA matrix (Eq. (\ref{eq_srpa})). Consequently, the SRPA matrix is not necessarily positive semi-definite.
As a consequence, imaginary or complex eigenvalues can be found or
positive (negative) eigenvalues but with negative (positive) norms. 

In Ref. \cite{Papakonstantinou2014}, a detailed numerical analysis focused on this problematic behavior  of the SRPA was performed, illustrating  the potential risks that may occur. The $^{16}$O and $^{48}$Ca are considered as test cases.
\begin{figure}
	\includegraphics[width=.53\linewidth,angle=0]{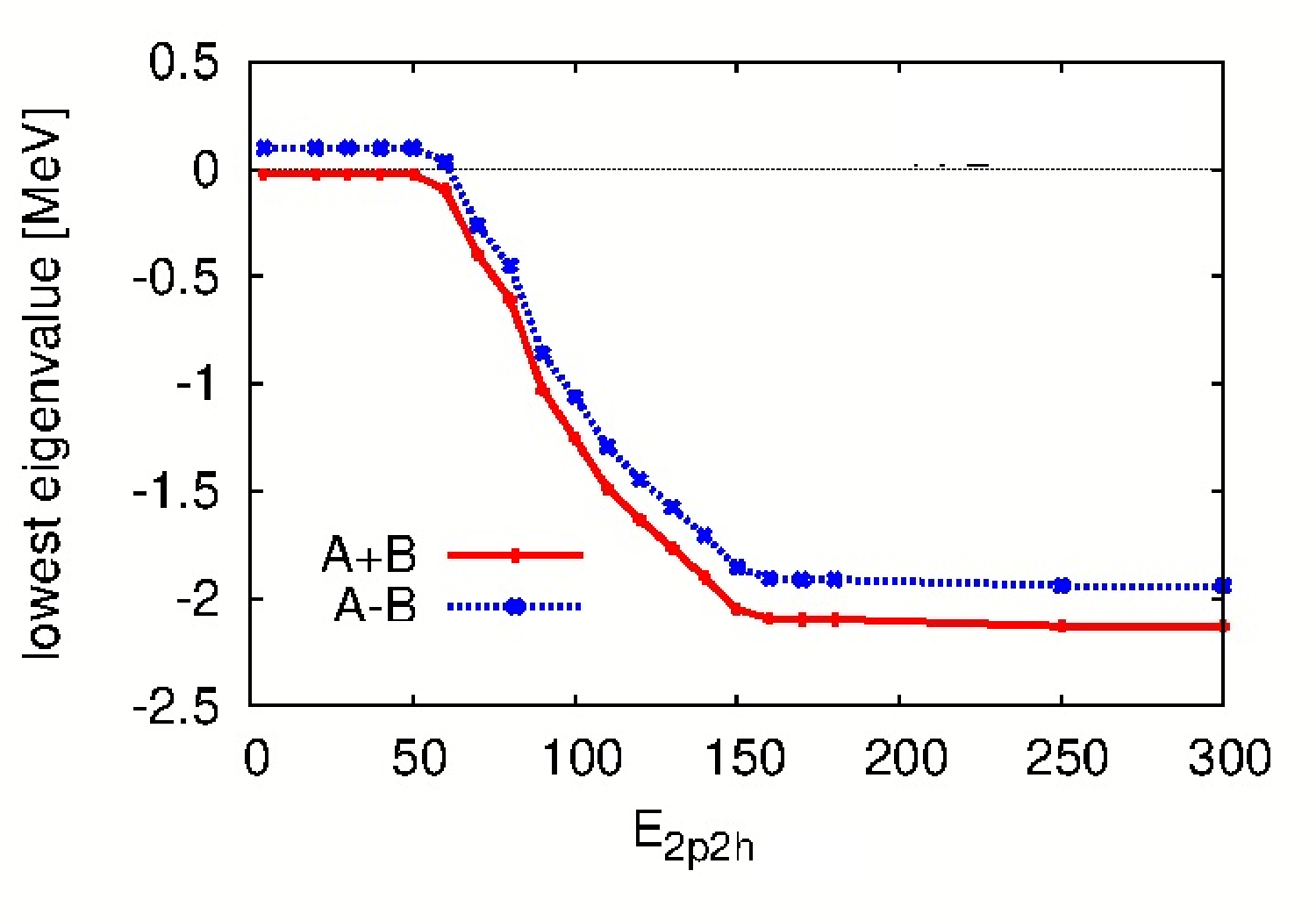}\hfill
	\includegraphics[width=.44\linewidth,angle=0]{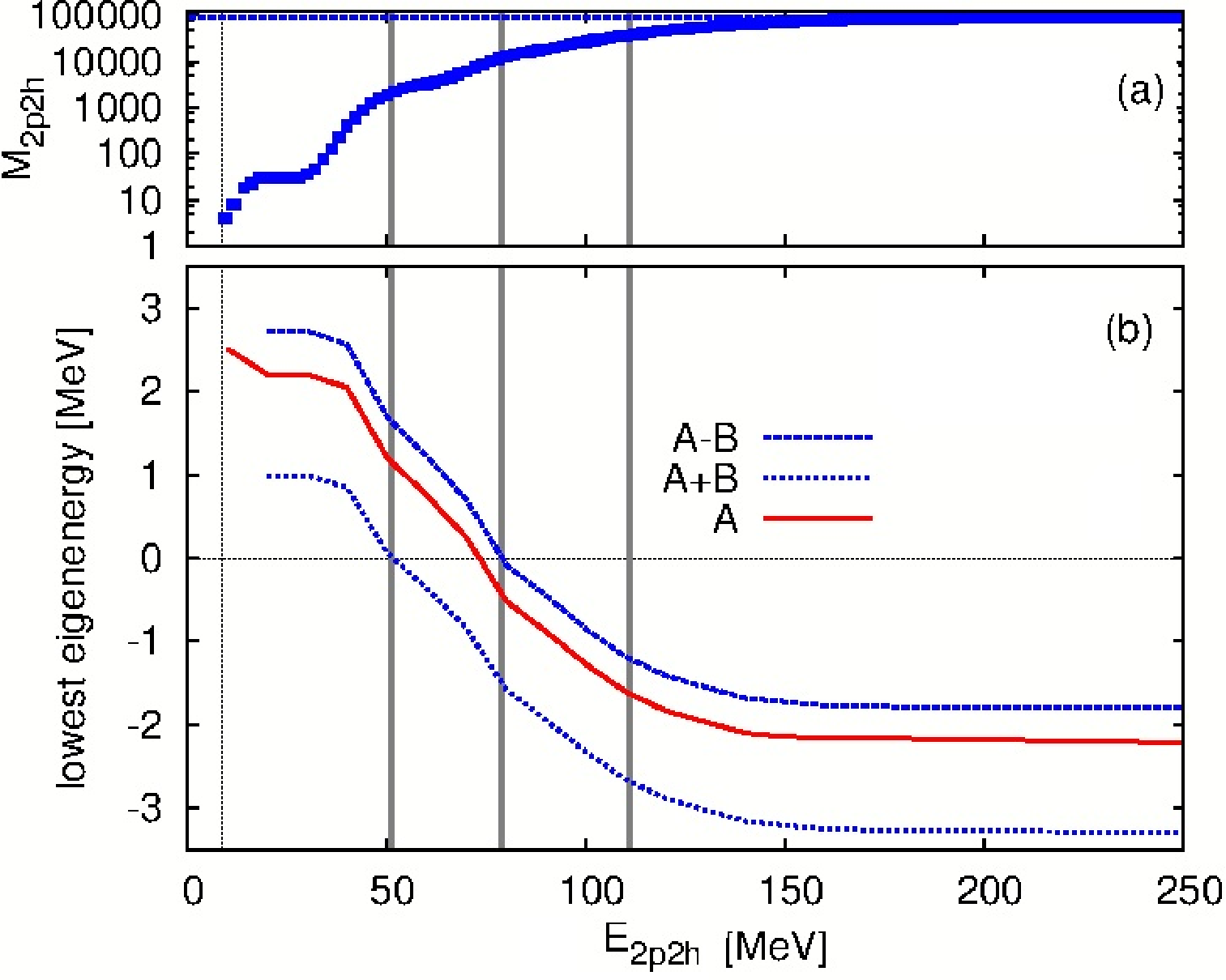}
	\caption{Left side: Lowest eigenvalue of the $A\pm B$ matrices in the $1^-$ channel of $^{16}$O within SRPA a function of the $2p-2h$ energy cutoff. Right side: $2^+$ channel of $^{48}$Ca within SRPA for increasing number of the $2p-2h$ configurations, shown in the upper panel as a function of the cutoff energy $E_{2p-2h}$. In the lower panel (b), the lowest eigenvalue of the $A\pm B$ and $A$ matrices is shown. See the text for more details. Adapted from \cite{Papakonstantinou2014}.}
	\label{Fig:Papa_Thouless_Fig1}
\end{figure}
%
%Starting from the RPA eigenvalues, the number of $2p-2h$ configurations in the SRPA calculations will be gradually increased, up to exhaustion of the available space, in order to study their impact on the low-energy states, on the energy-weighted and non-energy-weighted sums, and on the spurious state.
In the dipole case, the following operators are employed: $O_{E1}^{(1)}\equiv$ (\ref{Eq:Op-J1-em}),
$O_{E1}^{(2)}\equiv$ (\ref{Eq:Op-J1-IV-CM}) and
$O_{\mathrm{sp}}\equiv$ (\ref{Eq:Op-J1-isos}) 
%\begin{equation}
%O_{E1}^{(1)} = e\sum_{\pi} r_{\pi}Y_{10}(\hat{r}_{\pi})
%\end{equation}
%\begin{equation}
%	O_{E1}^{(2)}=\frac{N}{A} e\sum_{\pi }
%	r_{\pi}Y_{10}(\hat{r}_{\pi})
%	-\frac{Z}{A}e \sum_{\nu }r_{\nu}Y_{10}(\hat{r}_{\nu})
%\end{equation}
%
%\begin{equation}
%O_{\mathrm{sp}}=\frac{A}{Z} (O_{E1}^{(1)}-O_{E1}^{(2)})
%\end{equation}
the latter used to study the effect of spurious $1^-$  component. 
In the isoscalar channel,  the second order operator $O_{\mathrm{IS}}^{(1)}\equiv$ (\ref{Eq:Op-J1-isos-r3})
and $O_{\mathrm{IS}}^{(2)}\equiv$ (\ref{Eq:Op-J1-isos-r3-corrected}) will be also employed.
%\begin{equation}
%	O_{\mathrm{IS}}^{(1)} = e\sum_{i=1}^A r_i^3 Y_{10}(\hat{r}_i)
%\end{equation}
%and the intrinsic operator
%\begin{equation}
%	O_{\mathrm{IS}}^{(2)} = e\sum_{i=1}^A (r_i^3 - \eta r_i ) Y_{10}(\hat{r}_i)
%\end{equation}
%	where $\eta = \frac{5}{3}\langle r^2\rangle$~\cite{SGII}.
In the quadrupole case, the isoscalar and isovector operators considered are those defined in Eqs (\ref{Eq:Op-J02-isos}) and (\ref{Eq:Op-J02-isov}).
%\begin{equation}
%		O_{\mathrm{IS}} = e \sum_{i=1}^A r_i^2 Y_{20} (\hat{r}_i)
%\end{equation}
%
%\begin{equation}
%	O_{\mathrm{IV}} = e \sum_{\pi} r_{\pi}^2 Y_{20} (\hat{r}_{\pi}) - 	e \sum_{\nu } r_{\nu}^2 Y_{20} (\hat{r}_{\nu}).
%\end{equation}
For $^{16}$O, the RPA spurious state is obtained at $i49$~keV (imaginary). Gradually increasing the number of $2p-2h$ configurations, one can identify an antinormal state as spurious, being
characterized by a strong transition matrix elements of spurious operators. Its energy gets lower and lower, reaching the value of $-2$~MeV when the $2p-2h$ space is exhausted.
Since in this case $A$ and $B$ are real, the behavior of the spurious state is controlled by the lowest-energy eigenvalues of the $A\pm B$ matrices, that can be seen on the left side of Figure~\ref{Fig:Papa_Thouless_Fig1}. Starting from the RPA, e.g., $E_{2p-2h}=0$, the $A-B$ is positive-definite, while $A+B$ has a negative eigenvalue very close to zero, resulting in a RPA imaginary eigenvalue close to zero. For $E_{2p-2h}$ larger than 60~MeV, $A-B$ also acquires a negative eigenvalue, in this case having thus in SRPA a pair of antinormal states.
\begin{figure}[h]
	\centering
	\includegraphics[width=.34425\linewidth]{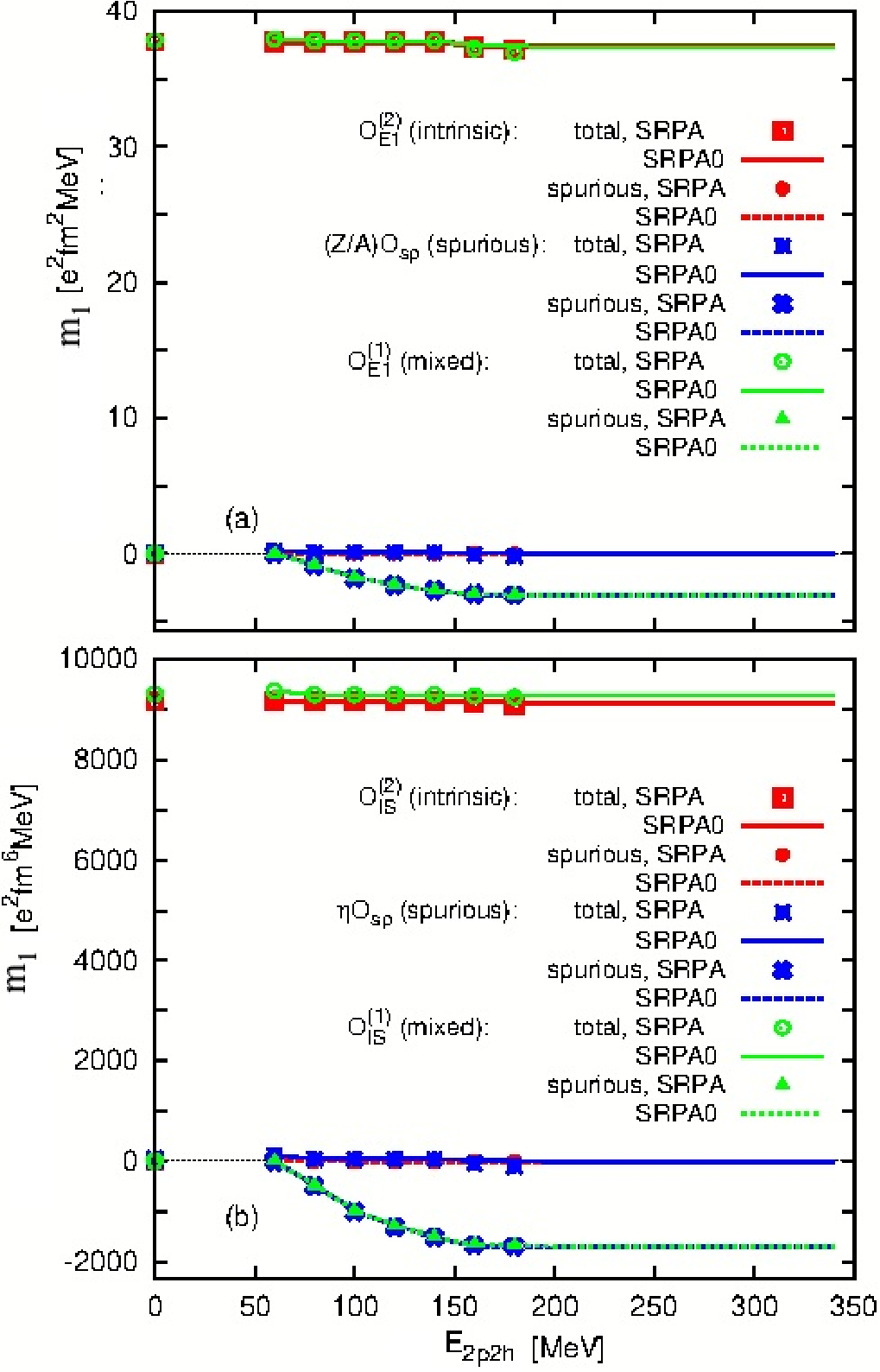}
	%	\hfill
	\includegraphics[width=.333\linewidth]{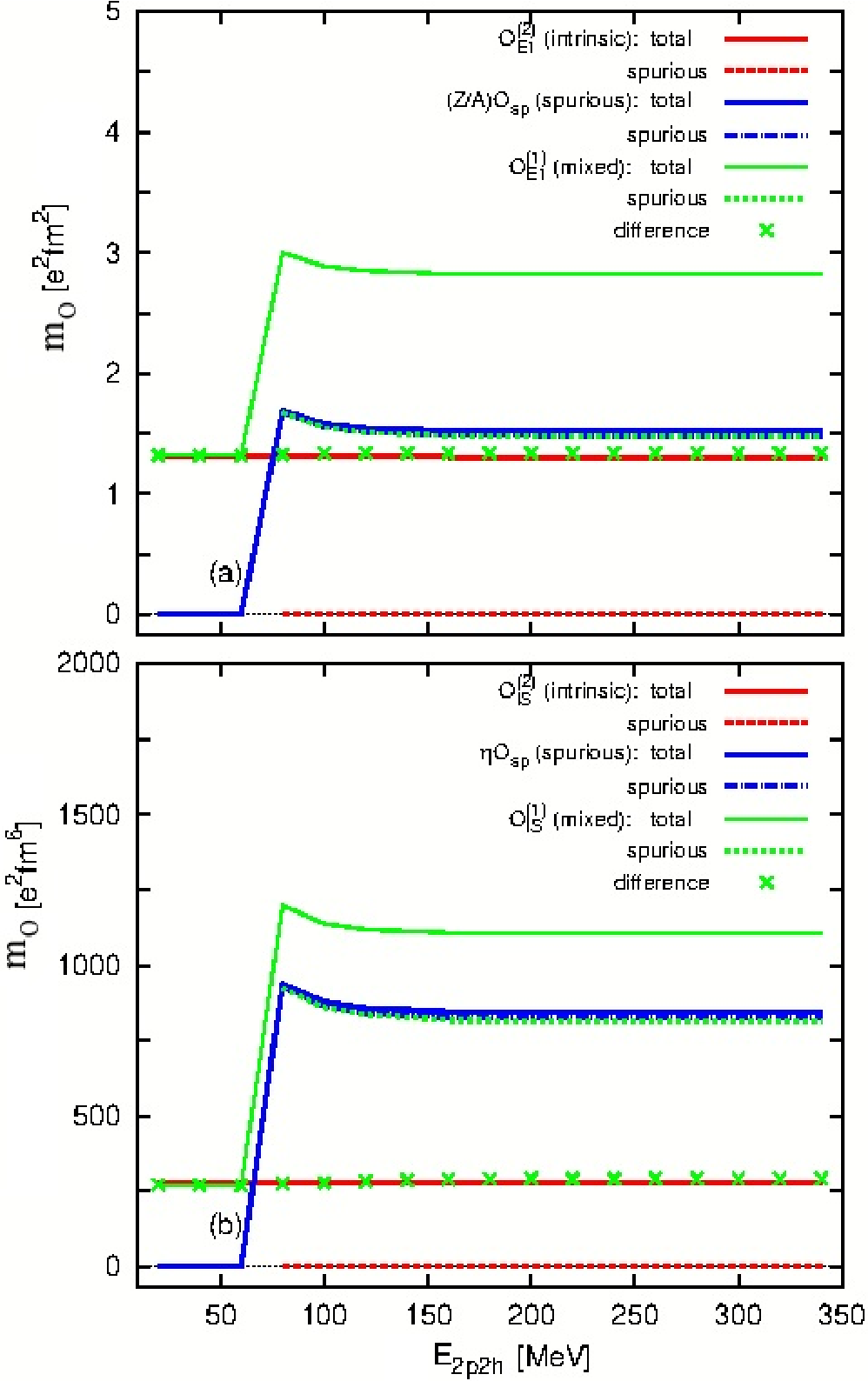}
	\caption{Left side: the energy-weighted sums $m_{1}$ and spurious-state-only contributions for various dipole (isovector in panel (a) and isoscalar in panel (b)) operators in $^{16}$O within SRPA and SRPA in diagonal approximation SRPA(0), as a function of the $2p-2h$ energy cutoff. Right side: The non-energy-weighted sums $m_{0}$ and spurious-state-only contributions for the $1^-$ channel of $^{16}$O obtained in SRPA, for (panel (a) the isovector operators and (panel (b) isoscalar operators, defined in the text.   Adapted from \cite{Papakonstantinou2014}.}
	\label{Fig:Papa_Thouless_Fig2}
%	\vspace{-0.8cm}
\end{figure}
For $^{48}$Ca, we focus on the quadrupole response, which in RPA is characterized by two states: the GR and a low-lying one. In the upper panel (a) of the right side of Figure~\ref{Fig:Papa_Thouless_Fig1}, the number of $2p-2h$ configurations, $E_{2p-2h}$, is shown. In the lower panel (b), the lowest eigenvalues of the $A\pm B$ matrices are plotted. The eigenvalue of the $A$ matrix, corresponding to the TDA or STDA limit, is also shown. All eigenvalues are positive at $E_{2p-2h}=0$, then decrease until becoming negative at 51 MeV for $A+B$ and 78 MeV for $A-B$. The first two vertical gray lines in the figure mark these sign changes. One can thus expect a pair of imaginary SRPA solutions within this $E_{2p-2h}$ interval (51–78 MeV), also reflected in a violation of the EWSR. Beyond 78 MeV, where both $A\pm B$ matrices give one negative eigenvalue, a pair of antinormal SRPA solutions is expected. Beyond $E_{2p-2h}\approx 111$ MeV (the third gray line), a quartet of complex conjugate solutions appears. More precisely, the first positive-norm SRPA solution and the adjoint of the second positive-norm (and positive-energy) SRPA solution are degenerate.

On the left side of Figure~\ref{Fig:Papa_Thouless_Fig2}, the energy-weighted sums and the spurious state, are shown as a as a function of the $2p-2h$ energy cutoff,
for the isovector and isoscalar case in panel (a) and (b), respectively. In the upper panel (a), the sums are shown for the three isovector operators
$O_{E1}^{(2)}$ (intrinsic), $O_{sp}$ (spurious), and $O_{E1}^{(1)}$ (mixed).
The SRPA and SRPA0 (diagonal approximation) values are shown, showing no noticeable difference, as well as the contribution of the spurious state only.
The total energy-weighted sums for all operators remain consistent between SRPA(0) and RPA, irrespective of $E_{2p-2h}$. While the spurious state makes no contribution to the intrinsic operator sum, its contribution to the spurious operator sum nearly cancels out with the contribution from the physical states (not shown), resulting in a total sum that is essentially zero across all values of $E_{2p-2h}$. Furthermore, the total sum for the mixed operator closely approximates that of the intrinsic operator, despite significant variations in the spurious state's contribution. For the isoscalar case (panel (b)), the conclusions mirror those drawn for the isovector operators. Specifically, all energy-weighted sums remain essentially identical in SRPA/SRPA(0) to the RPA value, independently of $E_{2p-2h}$. The spurious state's contribution to the intrinsic operator sum is negligible, whereas it makes a negative contribution to the mixed operator sum. Even at the RPA level, slight differences exist between the sums for the approximately intrinsic and mixed operators.
The results presented above confirm that the presence of SRPA spurious states at finite energies does not violate Thouless' theorem. However, SRPA solutions can contain finite spurious admixtures within the physical spectrum.
\begin{comment}
Specifically, the energy-weighted sum for the spurious operator includes contributions from physical solutions that collectively match the amplitude of the finite-energy spurious state's contribution, but with opposite sign, resulting in a zero total sum. Similarly, t
\end{comment}
The contribution of physical states to the energy-weighted sum for a mixed operator differs from that of the corresponding intrinsic operator. Therefore, intrinsic operators is essential in SRPA.

Let us now consider the non-energy-weighted sum $m_0$, which is the same in RPA and SRPA (see Section \ref{Sec:FormalPart_SRPA_Moments}). The right panels of Figure~\ref{Fig:Papa_Thouless_Fig2} shows the effect of the spurious state's appearance on $m_0$. The SRPA and SRPA0 (diagonal approximation) results are practically identical. A line is drawn between the results for $E_{2p-2h}=60$ and 80 MeV simply to guide the eye. The $m_0$ exhibits a discontinuity related to the appearance of an imaginary solution for low cutoff values, as well as in RPA. In both the isoscalar and isovector cases, it is evident that when the intrinsic operator is used, $m_0$ is unaffected by the spurious state. However, in the case of the mixed operator, it receives significant contributions from both the spurious state and the physical states. Moreover, the $m_0$ for the spurious operator derives mostly from the spurious state, although admixtures within the physical spectrum are introduced. %This analysis also confirms the necessity of using intrinsic operators.

\begin{figure}
	\centering
	\includegraphics[width=.62\linewidth,angle=0]{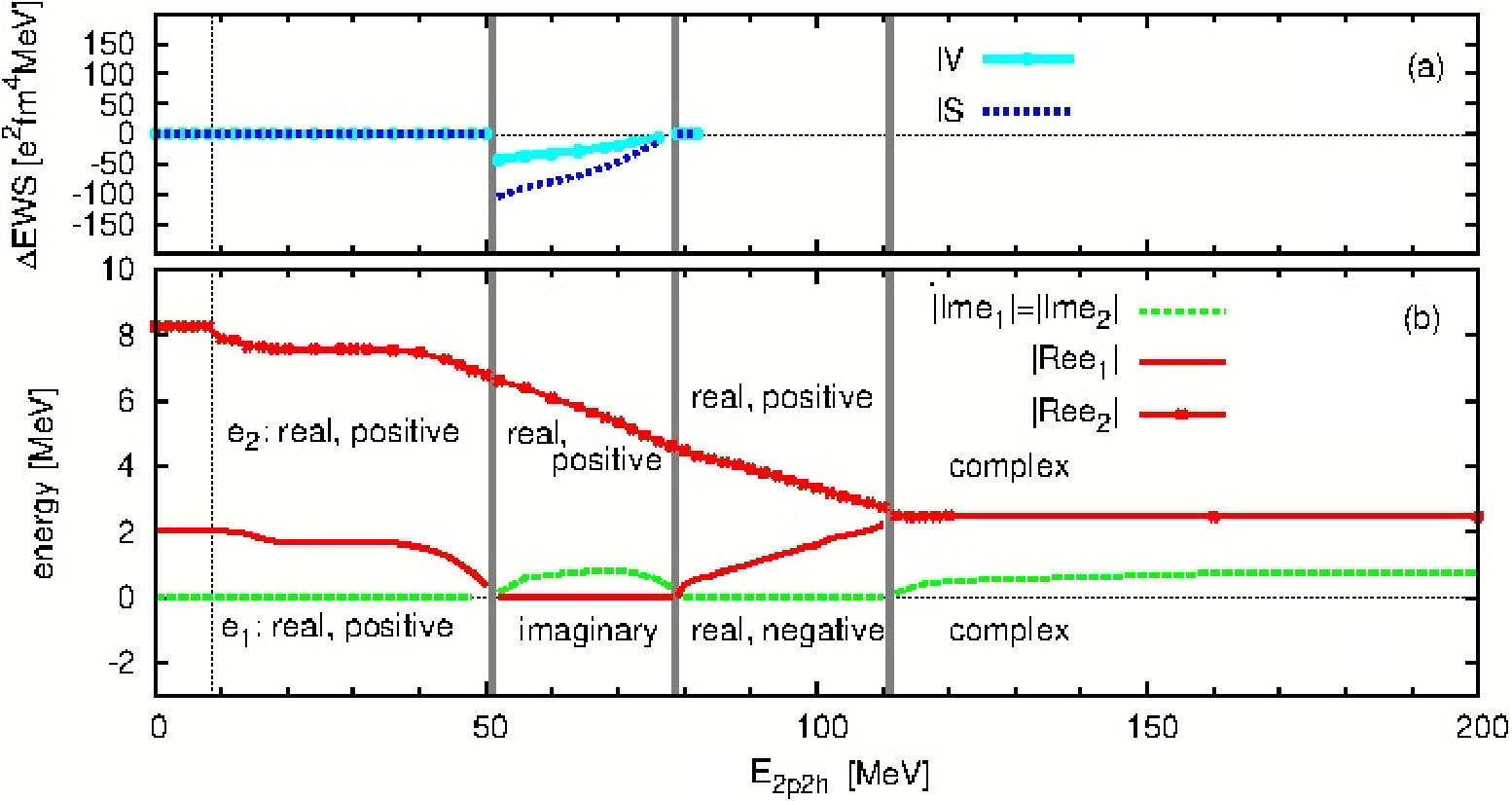}
	\caption{(a) Deviations from the RPA energy-weighted sums and (b) lowest SRPA eigenvalues as a function of the $2p-2h$ energy cutoff, for the $2^+$ channel of $^{48}$Ca within SRPA.
		See the text for more details. Figure from \cite{Papakonstantinou2014}.}
	\label{Fig:Papa_Thouless_Fig3}
\end{figure}
 Finally, in Figure ~\ref{Fig:Papa_Thouless_Fig3}, one can see the deviations of the SRPA energy-weighted sums from the RPA ones for the isoscalar and isovector quadrupole operators (panel (a)) and the lowest (positive-norm) SRPA eigenvalues, denoted as $e_1$ and $e_2$. When $e_1$ and $e_2$ are real the energy-weighted sums are the same as in RPA. When instead the energies become imaginary, and thus their corresponding strength do not contribute to the sum, deviations are seen. Note that for $E_{2p-2h}>85$~MeV the model space is too large, therefore the sum rules were not calculated. However, one can study the SRPA instability. When both $A\pm B$ are positive definite (as seen on the right side of Fig.~\ref{Fig:Papa_Thouless_Fig1}), both $e_{1,2}$ are real and positive, resulting in normal solutions. As $A+B$ acquires a negative eigenvalue, $e_1$ becomes imaginary. In the region where both $A\pm B$ have a negative eigenvalue, $e_1$ corresponds to an antinormal solution. Beyond $E_{2p-2h}\approx 111$ MeV, the two solutions and their adjoints merge into a quartet with complex conjugate eigenvalues, $\pm e_{1,2}$, of equal magnitude.

These results show that the SRPA can violate the stability condition producing numerical instabilities. In particular, spurious strength can appear at finite energy even when calculations are performed self-consistently. These spurious admixtures are not expected to be important in the GR region but can be significant in the low-energy spectrum. Therefore, the use of intrinsic operators is even more important, and the significance of these admixtures should be carefully assessed.  It is expected that the stronger the matrix elements of the residual interaction (especially those coupling $1p-1h$ and $2p-2h$ configurations), the more serious these issues become. In the case of attractive terms, the progressive decrease of the lowest energies likely produces imaginary solutions. Similarly, if repulsive matrix elements are sufficiently strong, they can also lead to imaginary solutions, as the SRPA matrix is no longer positive semi-definite producing a violation of the stability (Thouless) condition. One possible way to mitigate these problems, using a correlated ground state and including higher-order correlations to satisfy the stability condition, has been discussed \cite{Papakonstantinou2009,Tohyama2010, Gambacurta2006, Gambacurta2010b}. However, as we will see in the next section, the subtraction procedure within the EDF approach is a powerful way to solve or drastically mitigate these issues.

%\newpage
	\newpage
 \section{Applications of the SSRPA for charge-conserving excitations}\label{Sec:Applications_SSRPA_CC}
 \subsection{First applications of the SSRPA}\label{Sec:Applications_SSRPA_O16}
 The subtraction procedure was implemented for the first time in the SRPA case in Ref. \cite{Gambacurta2015} and applied to the study of the nuclear response in $^{16}$O. These calculations included all $1p-1h$ configurations below 100 MeV, such as to satisfy the EWSR within less than 1\% while the $2p-2h$ configuration spaces were limited to 70 MeV and 50 MeV for the monopole and quadrupole cases, respectively. These energy cutoffs resulted in manageable $2p-2h$ configuration numbers, enabling the full inversion of the $A_{22\prime}$ matrix required for the subtraction procedure. Two subtraction schemes were presented: SSRPA$_F$, representing the full subtraction  and SSRPA$_D$, denoting the subtracted SRPA with the diagonal approximation for the correction terms. 

\begin{figure}[h]
	\includegraphics[width=.45\linewidth]{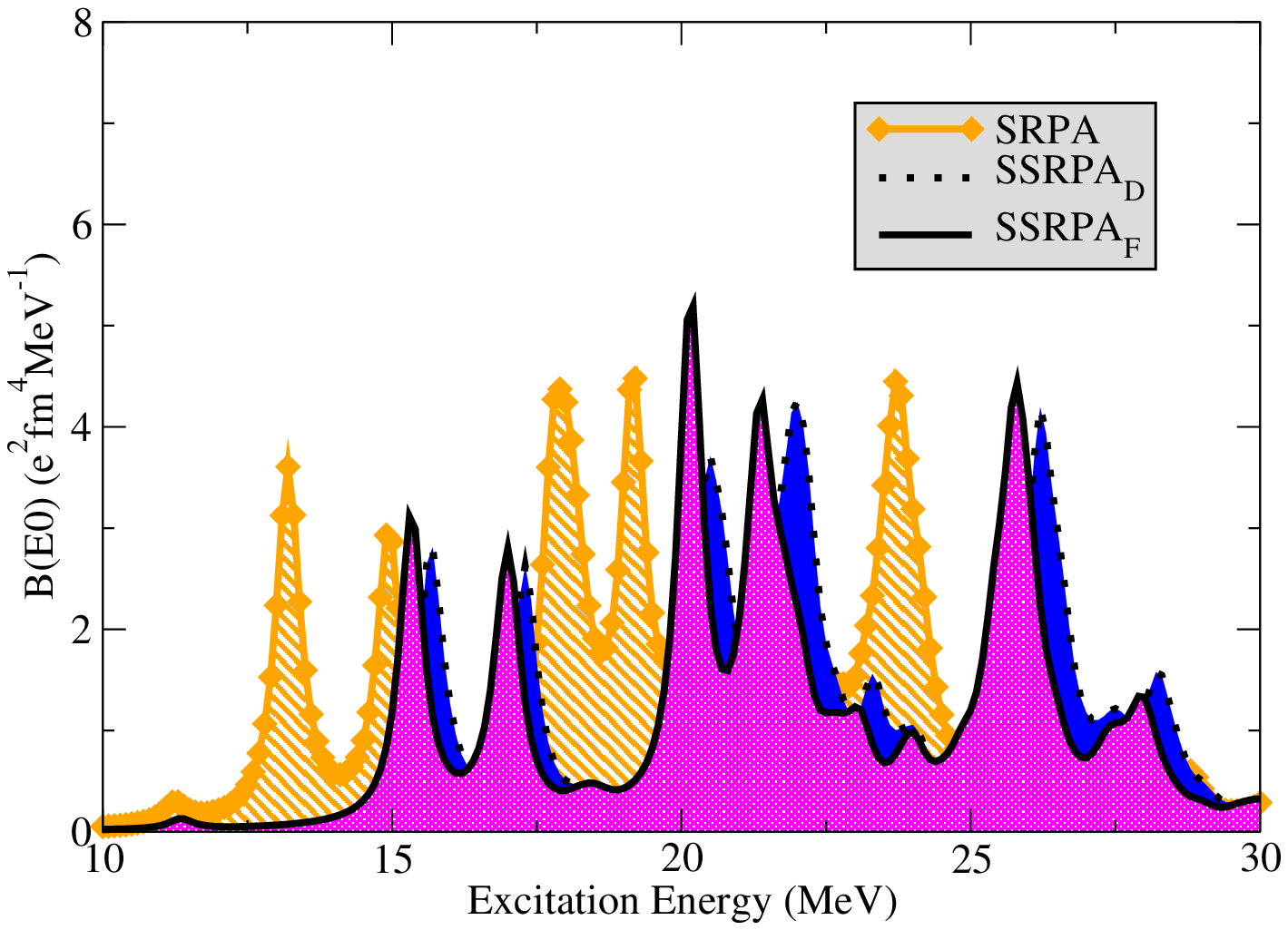}\hfill
	\includegraphics[width=.5\linewidth]{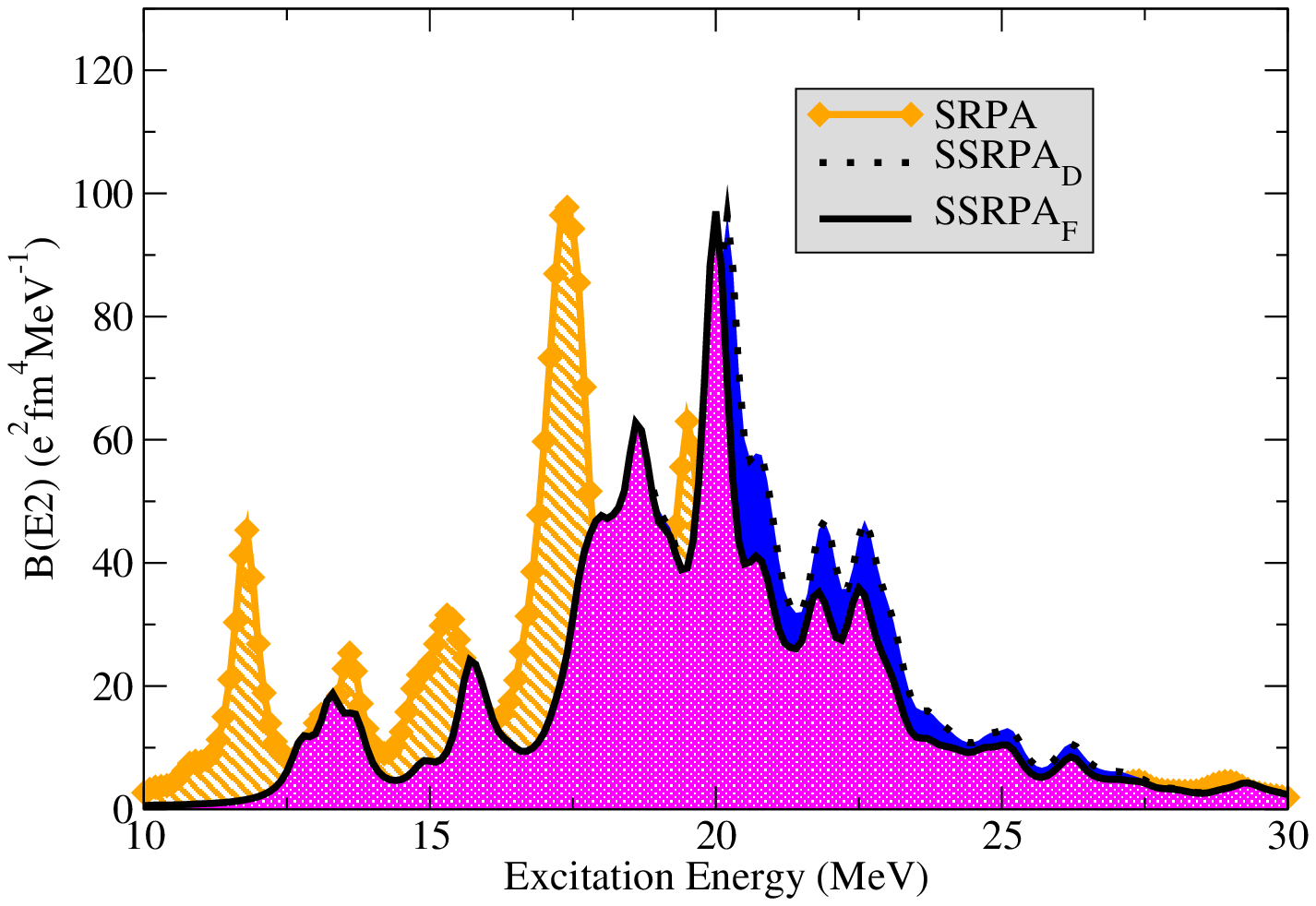}
	\caption{Isoscalar monopole (left side ) and quadrupole (right side) for the nucleus $^{16}$O,
		calculated in the SRPA without subtraction (orange diamonds and orange area),
		in the SSRPA$_F$ (black solid line and magenta area) and in
		the SRPA$_D$ (black dotted line and blue area), with a cutoff in
		the correction terms at 70 (50) MeV, for the monopole (quadrupole) case. Adapted from Ref. \cite{Gambacurta2015}.}
	\label{Fig:Sub-Fig3}
\end{figure}

 Figure \ref{Fig:Sub-Fig3}  compares the full (SSRPA$_F$) and diagonal (SSRPA$_D$) subtraction results for the monopole and quadrupole corrections. A close similarity between SSRPA$_F$ and SSRPA$_D$ is observed, with SSRPA$_D$ exhibiting a slight shift towards higher excitation energies.
\begin{comment}
	To elucidate the additional energy shift observed in SSRPA$_D$, Figure \ref{Fig:Sub-Fig4} presents the diagonal components of the correction for the monopole (left) and quadrupole (right) channels. These are calculated for each $1p-1h$ configuration using both SSRPA$_F$ and SSRPA$_D$ with the largest cutoff applied for each multipolarity. The subtraction-induced correction modifies the diagonal elements of the RPA $A$ matrix, thereby shifting the unperturbed $1p-1h$ excitation energies. The figure shows that the diagonal correction term is consistently larger in SSRPA$_D$ compared to SSRPA$_F$, explaining the additional spectral shift observed in the former. While systematic, this difference remains small. This finding suggests that the dominant effect of the correction arises from its diagonal component.
\end{comment}

\begin{figure}
\includegraphics[width=.48\linewidth]{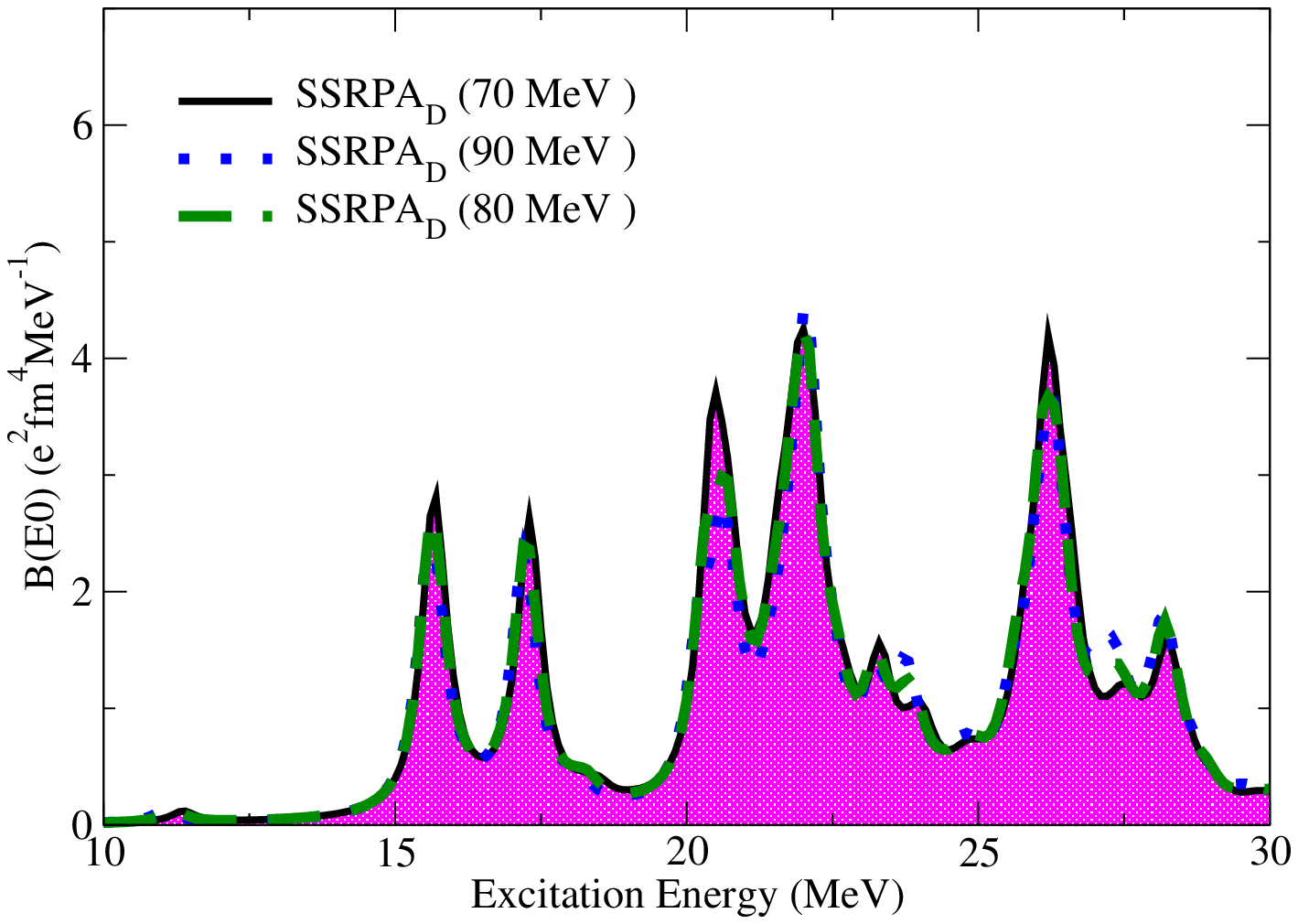}\hfill
\includegraphics[width=.45\linewidth]{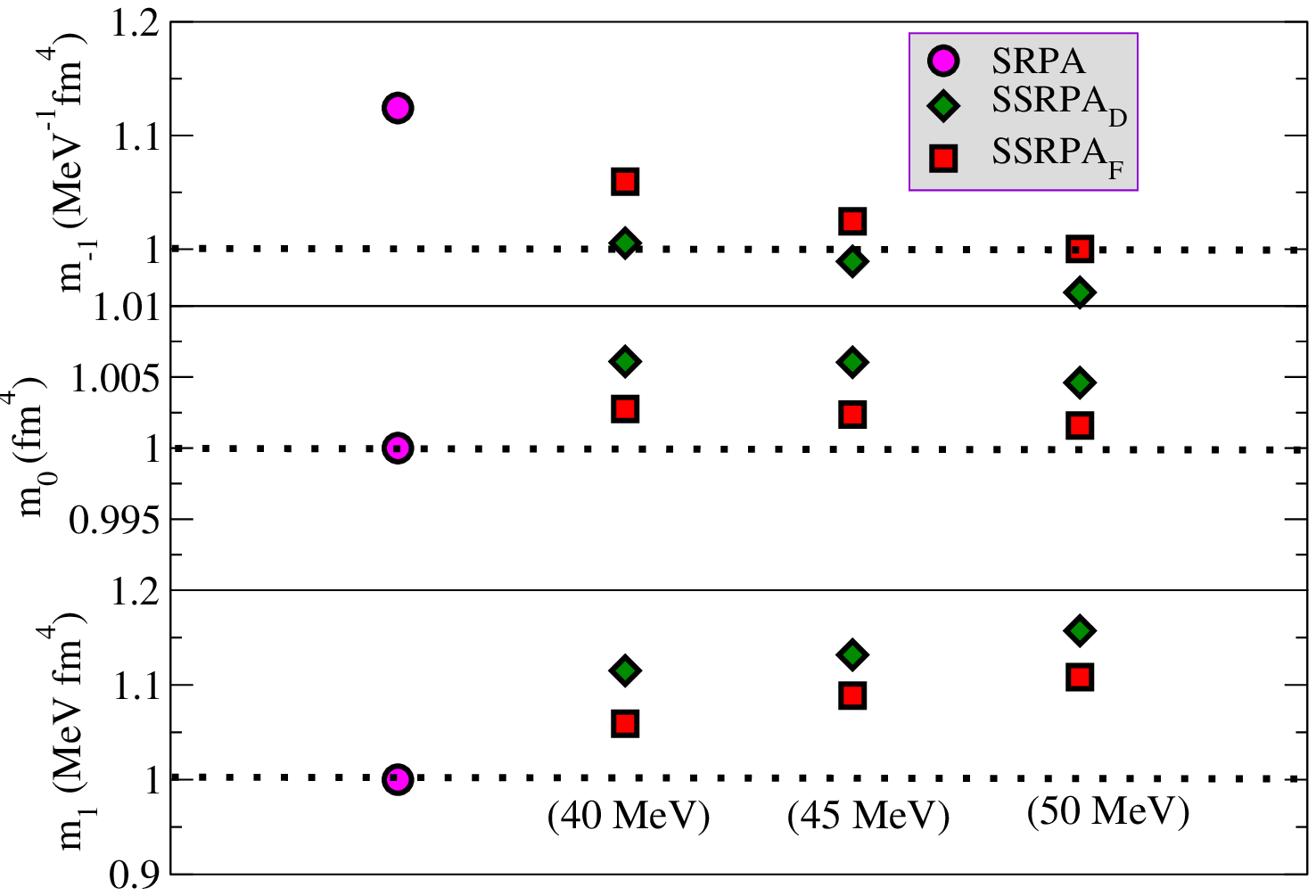}
\caption{Left side: Isoscalar monopole response for $^{16}$O in the diagonal
approximation with cutoff for the correction terms at 70 (black line and
magenta area), 80 (green dashed line), and 90 (blue dotted line) MeV.
Right side: Ratios of the moments $m_{-1}$, $m_0$, and $m_1$ of the
quadrupole strength distribution in the SRPA (purple circles), the
SSRPA$_F$ (red squares), and the SSRPA$_D$ (green diamonds)
to those in the RPA for increasingly high cutoffs in the correction terms, at 40, 45, and 50 MeV. Adapted from Ref. \cite{Gambacurta2015}.
}
\label{Fig:Sub-Fig5}
\end{figure}

The stability of the results is the studied within the diagonal approximation, focusing on the $0^+$ channel, increasing the $2p-2h$ energy cutoff up to 90 MeV. Figure \ref{Fig:Sub-Fig5} displays the corresponding results.  The three strength functions exhibit a weak dependence on the energy cutoff, a trend mirrored in the lowest $0^+$ state. Its energy at 70, 80, and 90 MeV cutoffs is 6.26 MeV, 6.13 MeV, and 5.96 MeV, respectively, representing a difference of only 5\% between the highest and lowest values. This demonstrates that the subtraction procedure not only corrects the SRPA energy shifts for GRs, but also yields significantly more robust (cutoff-insensitive) predictions for both GRs and low-lying states.

\begin{figure}
\includegraphics[width=.48\linewidth]{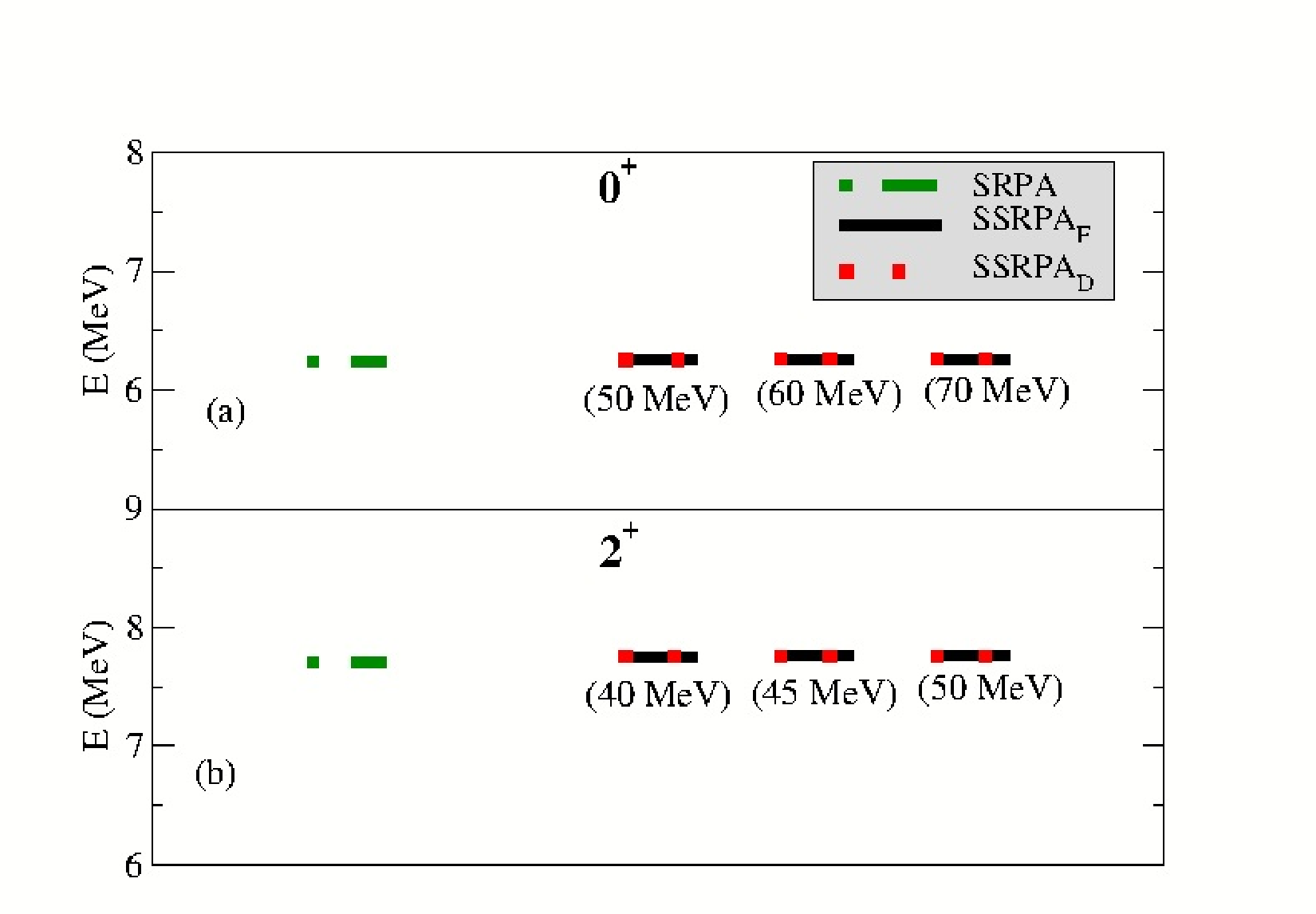}\hfill
\includegraphics[width=.47\linewidth]{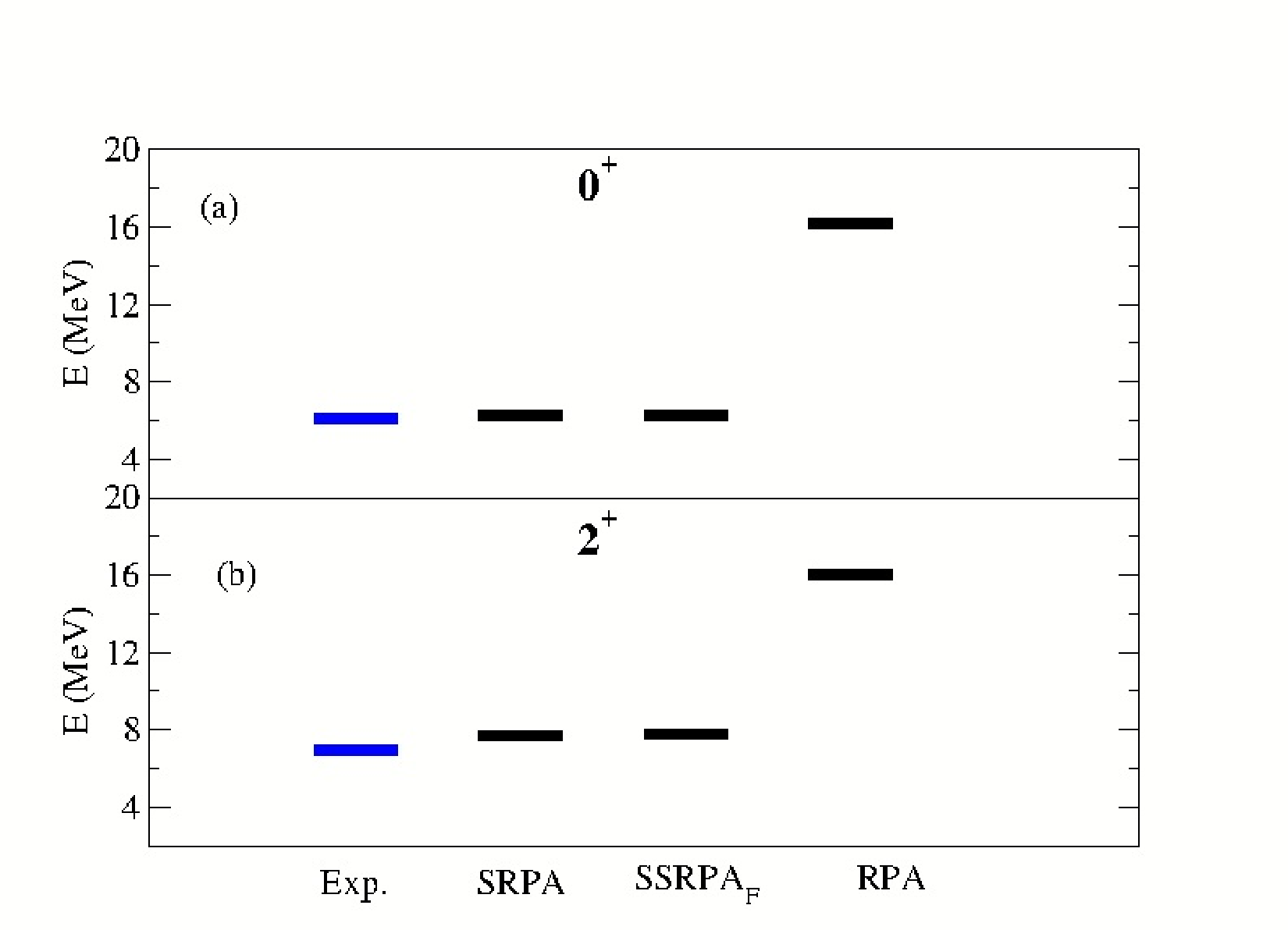}
\caption{Left side: First 0$^+$ (a) and 2$^+$ (b) states for $^{16}$O calculated with
the standard SRPA, the SSRPA$_F$, and the
SSRPA$_D$, with different cutoffs in the correction terms (in parentheses).
Right side: Comparison of values from the standard SRPA, the SSRPA$_F$, the RPA, and experiment for the energy of the first low--lying 0$^+$ (a) and 2$^+$ (b)
states. Adapted from Ref. \cite{Gambacurta2015}.
}
\label{Fig:Sub-Fig6}
\end{figure}

The moments $m_{-1}$, $m_0$, and $m_1$ of the strength distribution for the isoscalar quadrupole case offer further insight. Figure \ref{Fig:Sub-Fig5} (right) shows the ratios of these moments, calculated using SRPA, SSRPA$_F$, and SSRPA$_D$, to their RPA counterparts. As expected, $m_0$ and $m_1$ are identical in RPA and SRPA. However, the subtracted-SRPA moments differ. The upward shift induced by subtraction needs a larger $m_1$ and a smaller $m_{-1}$ compared to the unsubtracted case, a trend confirmed by the figure. Furthermore, as expected, the full subtraction (at maximum cutoff) yields values closer to the RPA results than the diagonal approximation. The inverse moment $m_{-1}$ in SSRPA$_F$ with maximum cutoff is identical to the RPA value, as required. This equality holds only under conditions of full coherence, meaning full matrix inversion and identical $2p-2h$ spaces in both the matrices and the correction term.
\begin{figure}
	\includegraphics[width=.46\linewidth]{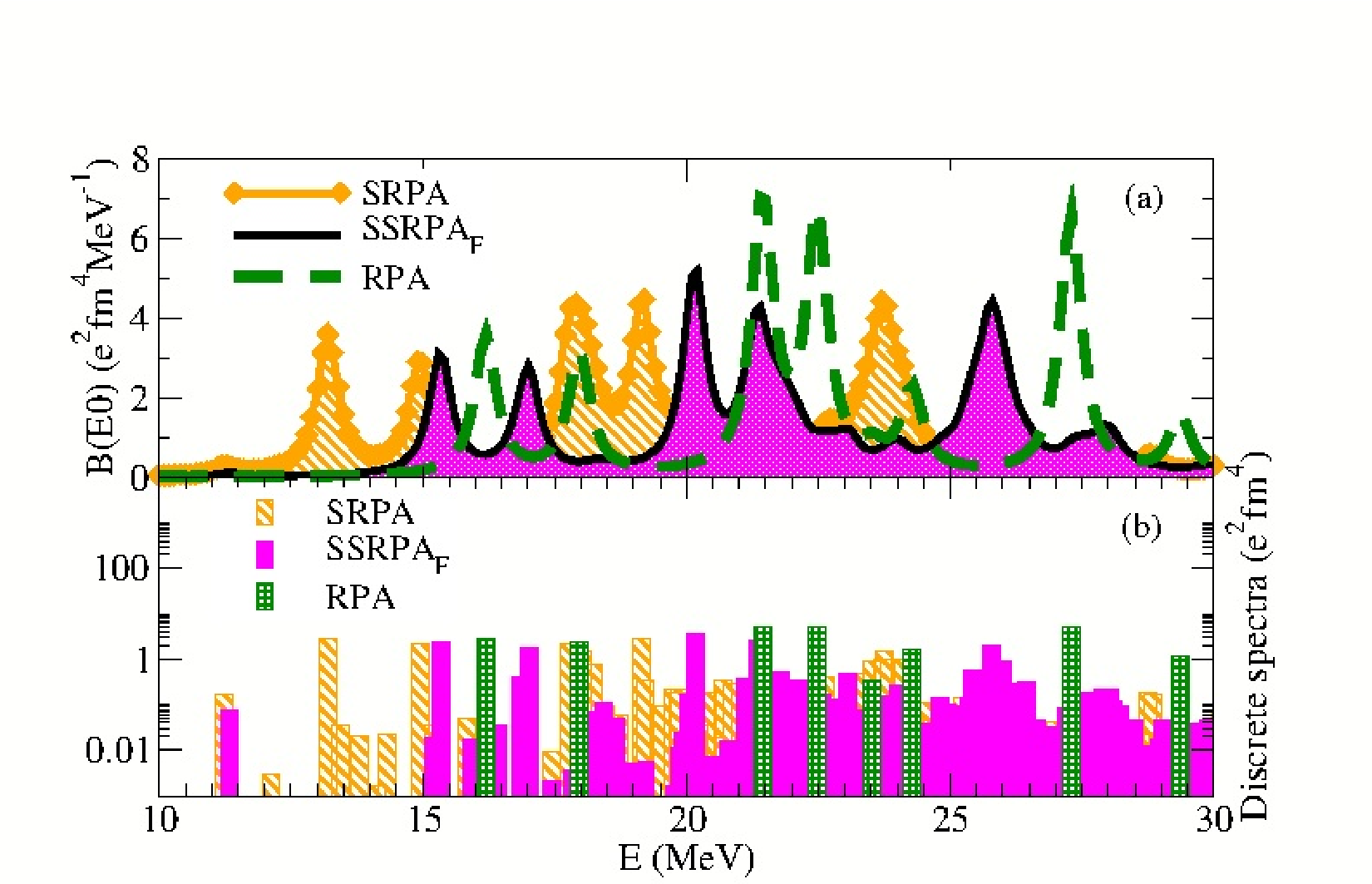}\hfill
	\includegraphics[width=.5\linewidth]{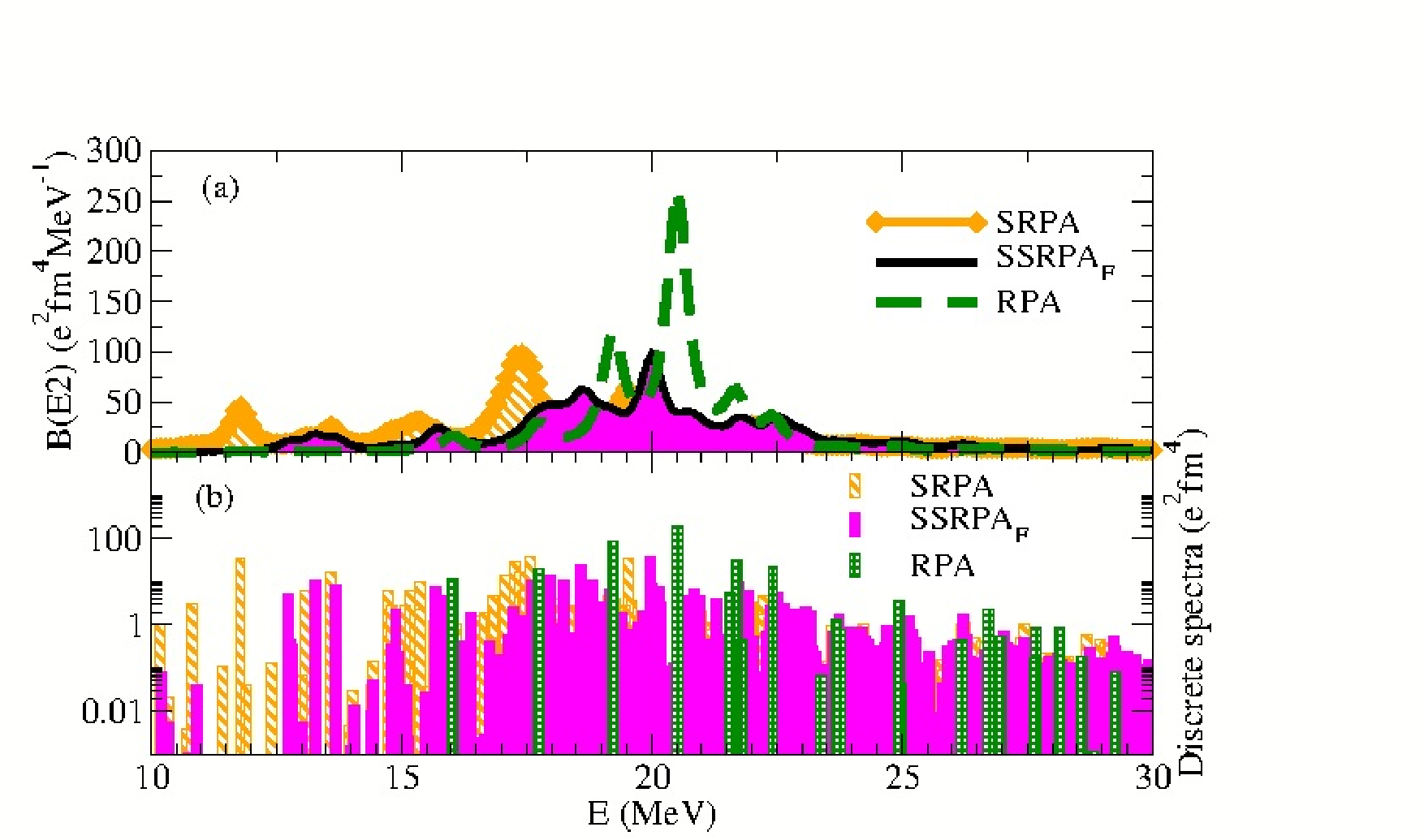}
	\caption{Isoscalar monopole (left side ) and quadrupole (right side ) for the nucleus $^{16}$O in the standard SRPA
		(orange diamonds and orange area), RPA (blue dashed line), and
		the SSRPA$_F$ (black solid line and magenta area); (b): Discrete
		spectra (binned strength) obtained with the SRPA (orange dashed
		bars), the RPA (blue dotted bars) and the
		SSRPA$_F$ (magenta solid bars). Adapted from Ref. \cite{Gambacurta2015}. }
	\label{Fig:Sub-Fig7}
\end{figure}
Let's now discuss the low-lying $0^+$ and $2^+$ excited states for the $^{16}$O nucleus. These states, predominantly multi-particle multi-hole in nature \cite{Zuker1968}, can not be described within the standard RPA. Therefore, investigating the influence of $2p-2h$ configurations on their description, as well as the effects of the subtraction procedure on their energies, is of considerable interest. Figure \ref{Fig:Sub-Fig6} (left side) displays the energies of the lowest $0^+$ and $2^+$ states obtained using SRPA, SSRPA$_F$, and SSRPA$_D$ with varying $2p-2h$ correction term cutoffs. The impact of the correction differs from that observed for GRs. The subtraction method does not significantly alter these low-lying states, as they are primarily $2p-2h$ character. Consequently, these states remain largely unaffected by the subtraction procedure, which operates exclusively within the $1p-1h$ sector of the SRPA matrix.

On the right side of Figure \ref{Fig:Sub-Fig6}, we compares the SRPA, SSRPA$_F$, and RPA energies with the experimental ones
from experiment \cite{Ajzenberg1982} for the first 0$^+$ and 2$^+$ states. One can see that RPA energies strongly overestimate the experimental data while the SRPA results (with or without subtraction) are in good agreement with experiment. However, the corresponding transition probabilities are much smaller than the experimental one, as it will be discussed in the following Section.
%As just mentioned, the subtraction procedure does not affect the energies of these states due to their dominant $2p-2h$ nature. 

Figure \ref{Fig:Sub-Fig7} (top panels) compares the monopole (left side) and quadrupole (right) strength distributions obtained from standard SRPA, SSRPA$_F$, and RPA calculations. The substantial shift observed in standard SRPA relative to RPA is significantly strongly reduced by the subtraction procedure. The lower panels of these figures present the corresponding discrete (binned) strengths. Both SRPA models (with and without subtraction) naturally describe the fragmentation and width of the excitation through their inclusion of a dense spectrum of discrete $2p-2h$ configurations. While the subtraction procedure does not alter this fragmentation, it does induce an energy shift towards higher excitation energies.

These results show that the subtraction method yields stable SSRPA solutions. As a matter of fact, the subtraction method acts like a regularization procedure.  The subtracted SRPA provides robust, stable predictions with weak dependence on the $2p-2h$ configuration cutoff. By eliminating double counting, this method significantly reduces the large, anomalous downward energy shift that standard SRPA calculations systematically exhibit compared to RPA.

\begin{wrapfigure}{l}{0.5\textwidth}
	\centering
%  	\vspace{-3.5cmm}
	\includegraphics[width=0.45\textwidth]{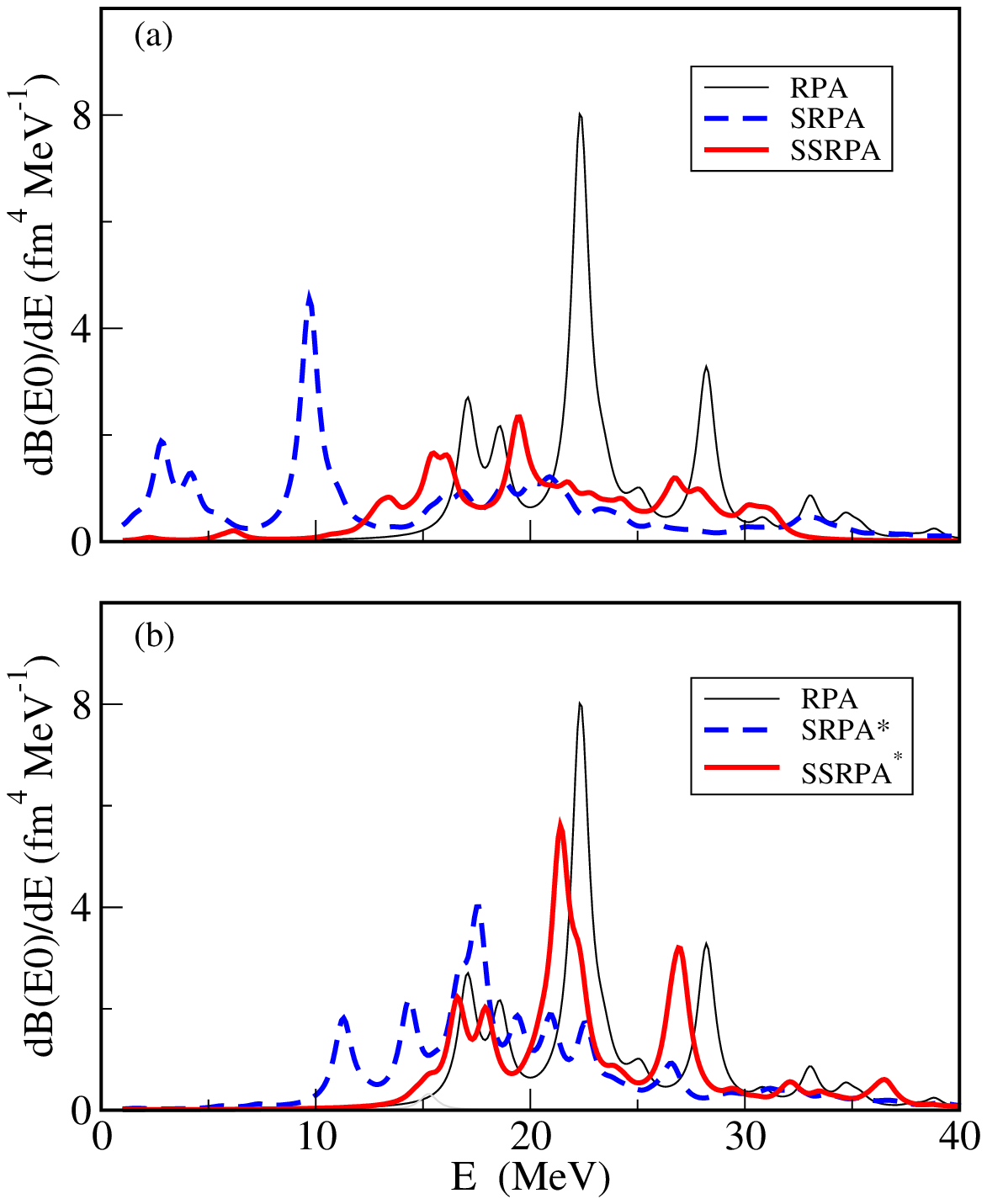}
	\vspace{-1.5mm}
	\caption{Panel (a): comparison between the
Gogny-RPA, SRPA and SSRPA results for the monopole response in $^{16}$O. Panel (b): same as panel (a) but for the SRPA and SSRPA cases. Adapted from Ref. \cite{Gambacurta2016}. }
	%	 	\vspace{-7mm}
	\label{Fig:SSRPA_Gogny}
\end{wrapfigure}

The subtraction procedure was also applied in the Gogny case in Ref. \cite{Gambacurta2016}. The comparison between Gogny-based subtraction results, SRPA, and RPA is presented in Figure \ref{Fig:SSRPA_Gogny} for the monopole case in $^{16}$O. As discussed in Section \ref{Sec:Applications_SRPA_Gogny}, the SRPA* results refer to calculations where the $\nu\pi$ matrix elements of the residual interaction have been omitted.  The subtraction method shifts the strength distributions upward, similar to the Skyrme case. Panel (a) highlights that the full SSRPA produces a stronger shift, mitigating the large downward energy displacement caused by the $\nu\pi$ matrix elements. However, the SSRPA and SSRPA* results still differ notably: the former accounts for 65\% of the EWSR (with a centroid at 20.82 MeV) below 40 MeV, while the latter reaches 91\% (at 23.82 MeV). In both cases, the remaining strength is located above 40 MeV, characterized by strong fragmentation over $2p-2h$ states.

The subtraction method is conceptually tied to the EDF framework, where double counting between static and dynamical correlations is unavoidable. Therefore, for SRPA based on EDFs (Skyrme, Gogny, relativistic), subtraction is required to obtain reliable  and stable results. For calculations based on realistic Hamiltonians, subtraction is not formally required because, in principle, there is no double counting. However, at present, no effective interactions capable of reasonably reproducing nuclear structure properties are entirely free from fitting procedures. Ideally, in the case of realistic Hamiltonians, such fitting is restricted to nucleon-nucleon scattering data and properties of very light systems (Helium-4 in the UCOM case). Therefore, any potential double counting should be negligible compared to that found in phenomenological forces. On the other hand, the subtraction procedure could be viewed as a regularization technique that might alleviate common issues (such as slow convergence, strong sensitivity to cutoffs, and instabilities related to the violation of the Thouless theorem) also encountered in SRPA calculations based on realistic forces. However, as there are no studies in the literature exploring SSRPA with realistic forces, this possibility here is only mentioned. In the following Sections, we will present and discuss different applications of EDF-based SSRPA studies.

% \newpage
\subsection{The role of the tensor force in SSRPA calculations}\label{Sec:Applications_SSRPA_Tensor}
\begin{figure}
	\includegraphics[width=.5\linewidth]{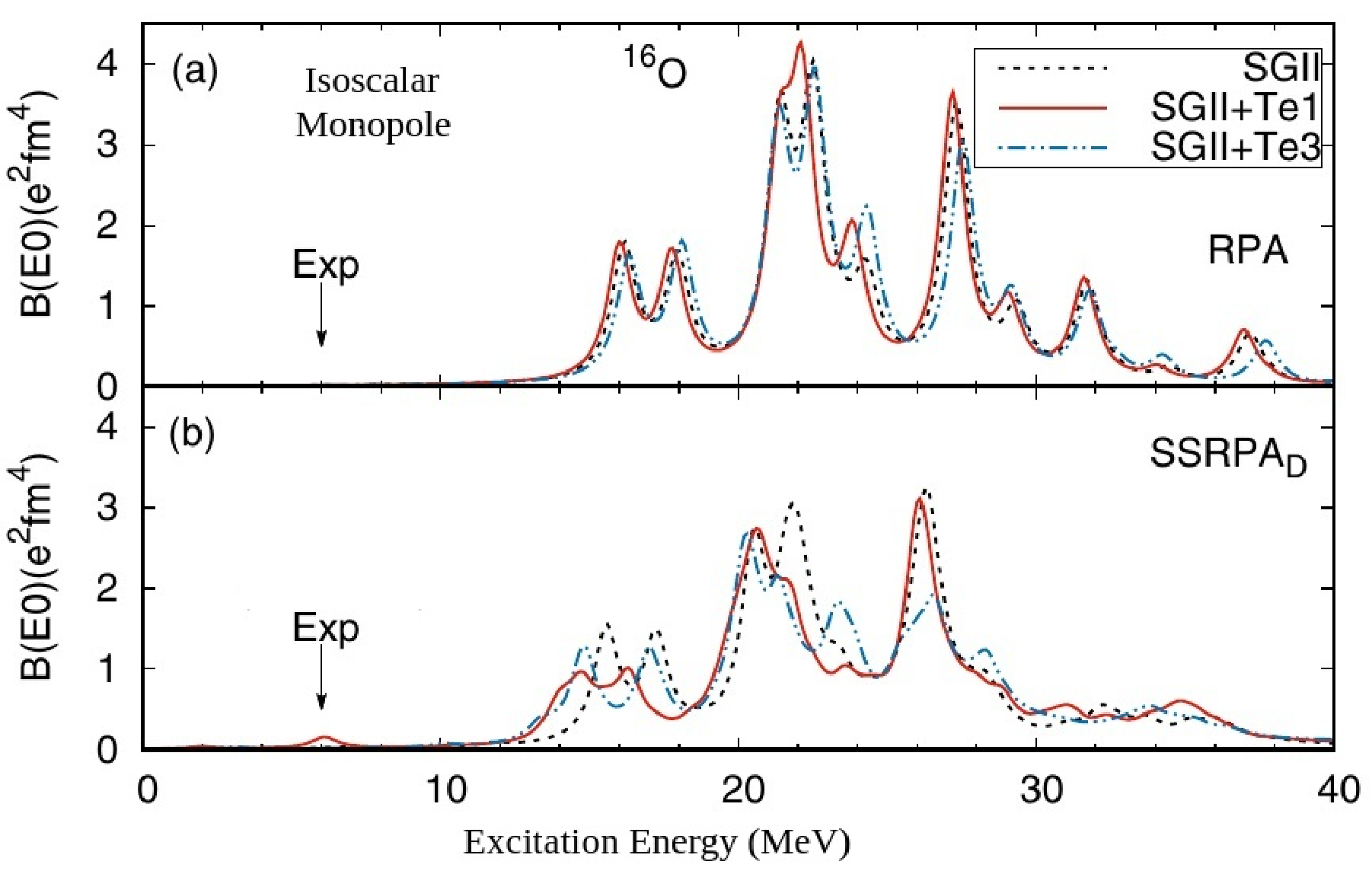}\hfill
	\includegraphics[width=.5\linewidth]{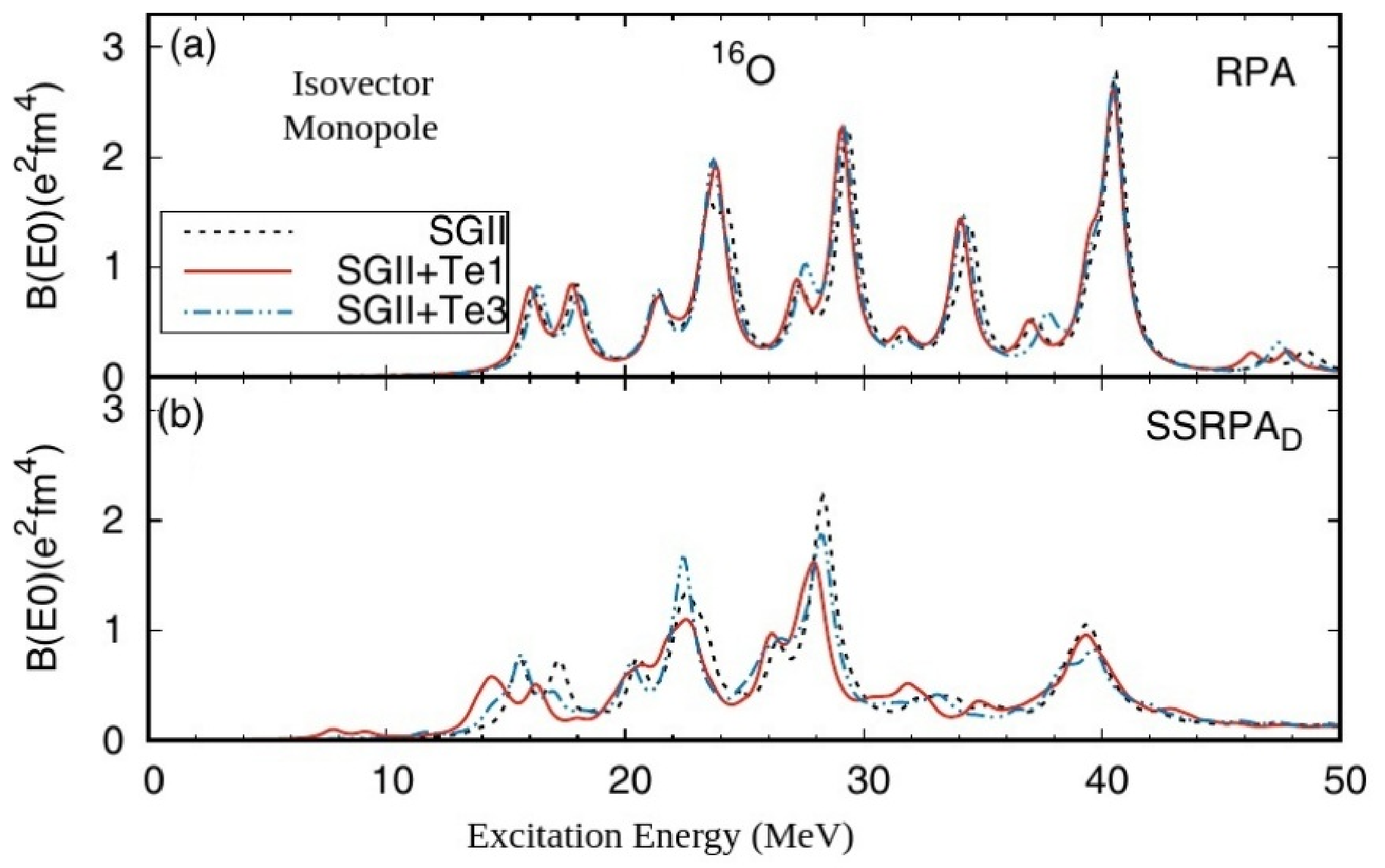}
	\caption{Isoscalar (left side) and isovector (right side) monopole response in $^{16}$O obtained in RPA (panel (a)) and SSRPA (panel (b)) with the interactions SGII (black dashed-dotted lines), SGII + Te1 (red solid lines), and SGII + Te3 (blue dashed-dotted lines). The arrow indicates the experimental first excited state \cite{Tilley1993}. Adapted from Ref. \cite{Sagawa2021}.}
	\label{Fig:Tensor_1}
\end{figure}

The tensor force is an important component of the nucleon-nucleon interaction. Its role is crucial in explaining the evolution of magic numbers in neutron-rich nuclei both concerning the ground state properties and collective excitations \cite{Otsuka2005,Otsuka2006,Sorlin2008,Colo2007,Colo2008,Brink2007,Wang2020,Hu2020,DeDonno2016,Deloncle2017}. Usually, the tensor interaction's effect on the ground state of spin-saturated nuclei is minimal. However, it significantly affects spin-dependent excitations within the RPA model, while having a relatively minor impact on spin-independent normal parity excitations. The Skyrme tensor force was introduced within the SSRPA framework for the first time in Ref. \cite{Sagawa2021}. The authors specifically investigate the tensor force's effect on normal-parity collective excited states in $^{16}$O and $^{40}$Ca, incorporating the $2p-2h$ configuration space.  The study was performed by using the subtraction procedure in the diagonal approximation, that was confirmed to give results close to the full one, as discussed in the previous Section. Many tensor parameter sets are available in the literature, see for example \cite{Lesinski2007}, where changing the signs and strengths of the coupling constants, the influence of the tensor effect may change significantly. In Ref. \cite{Sagawa2021}, the SGII + Te1 and Te3 parametrizations \cite{Bai2011} were employed in order to study the effect of the tensor with respect to the original, tensor-free, SGII interaction \cite{SGII}.
More precisely,
 the two parameterizations SGII + Te1 and SGII + Te3 are characterized by different
 strengths of the triplet-even (T) and triplet-odd tensor (U) terms equal to  $(T,U)=(500,-350)$ anf $(650,200)$ MeV fm$^5$, respectively.
The two parameter sets can be considered extreme cases, as the strength of the triplet-odd tensor term has opposite signs.
\begin{figure}
	\includegraphics[width=.5\linewidth]{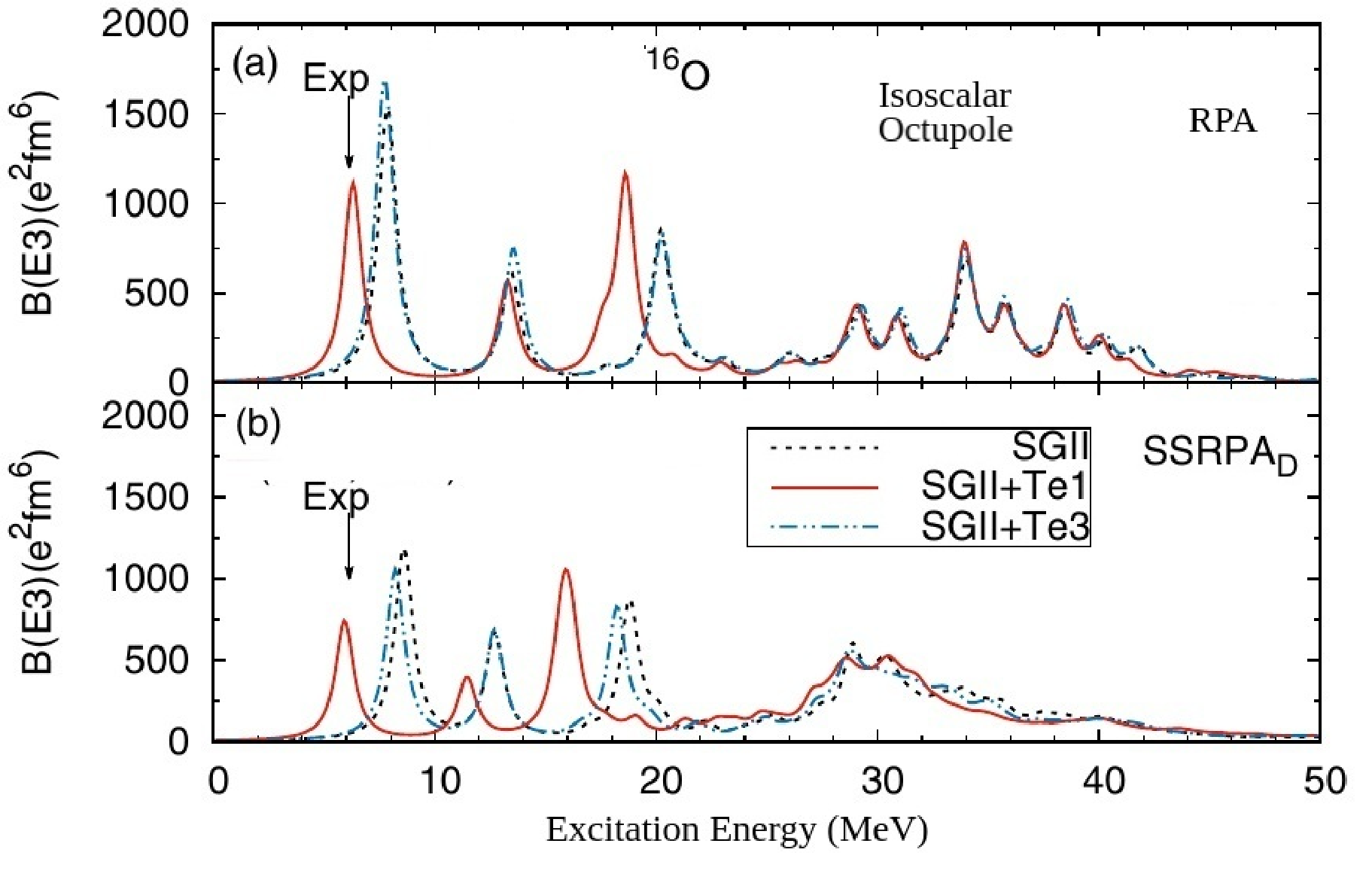}\hfill
	\includegraphics[width=.5\linewidth]{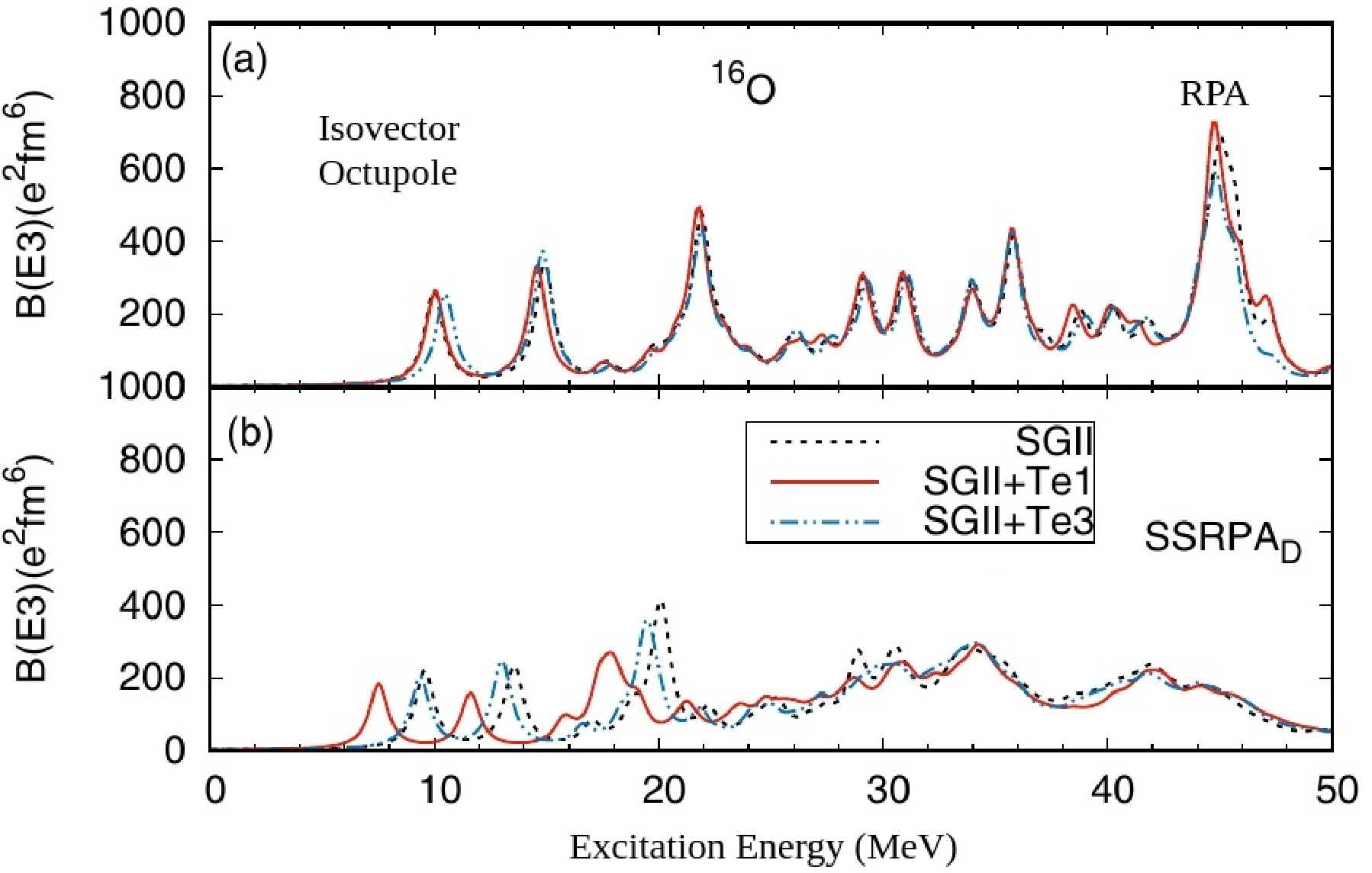}
	\caption{As in Figure \ref{Fig:Tensor_1} but for the octupole response in $^{16}$O, data from \cite{Kibedi2002}. Adapted from Ref. \cite{Sagawa2021}.}
	\label{Fig:Tensor_3}
\end{figure}

The left (right) side of Figure \ref{Fig:Tensor_1} illustrates the isoscalar (isovector) RPA (panels (a)) and SSRPA (panels (b)) monopole strength distributions for $^{16}$O, with and without the inclusion of the tensor force. The results obtained from different interactions are distinguished by: SGII (black dashed-dotted lines), SGII + Te1 (red solid lines), and SGII + Te3 (blue dashed-dotted lines). In the RPA calculations, the impact of the tensor interaction is very small. Specifically, SGII + Te3 yields results nearly identical to SGII, while SGII + Te1 induces a slight downward shift in strength, within a few tens of keV. In the SSRPA calculations, the tensor force's influence on the strength distribution above 20 MeV is observable but not very strong.
% With SGII + Te3, the strength is distributed across five smaller peaks, whereas with SGII + Te1, the two peaks around 22 MeV merge into a single peak at approximately 20.5 MeV. However, in the energy region below 20 MeV, the tensor interaction's effect becomes more pronounced. With SGII + Te3, the two peaks near 18 MeV shift downward by about 1 MeV. This shift is further amplified with SGII + Te1. 
 In the RPA calculations, no states are found below 10 MeV. Conversely, in the SSRPA calculations without the tensor force, and with SGII + Te3, the first 0+ excited state emerges below 10 MeV, exhibiting improved agreement with the experimental energy. Nevertheless, the strength in these calculations remains weak, approximately 0.01 $e^2fm^4$. In the SSRPA calculations using SGII + Te1, a stronger transition strength is observed at 6.1 MeV, with a strength of 0.23 $e^2fm^4$. This contrasts with the experimentally measured first excited 0$^+$ state at 6.05 MeV, which has a significantly larger strength of 3.66 $e^2fm^4$. In the isovector channel, the tensor force has even a smaller effect.
%  does not produce a significant effect on the RPA strength distribution. In SSRPA models using SGII and SGII + Te3, some states are present, but no discernible strength is distributed below 10 MeV. However, with SGII + Te1, visible strength is observed in this lower energy region.

\begin{comment}
	The left (right) side of Figure \ref{Fig:Tensor_2} shows the isoscalar (isovector) quadrupole response in $^{16}$O. In the isoscalar case, the RPA model, regardless of tensor force inclusion, shows no states at low energy. Furthermore, the tensor force has a negligible impact on the strength distributions within the RPA framework, as i the monopole case. In SSRPA calculations, using SGII or SGII + Te3, states emerge below 10 MeV, though with very weak strength, approximately 0.08 $e^2fm^4$. Conversely, SSRPA calculations with SGII + Te1 reveal a downward shift of the main peaks by about 2 MeV, and a state with a strength of approximately 0.55 $e^2fm^4$ is observed below 10 MeV. For the isovector case, in the RPA calculations, the tensor force's effect remains negligible. However, the influence of different tensor forces becomes apparent in the SSRPA results.
\end{comment}

The left (right) side of Figure \ref{Fig:Tensor_3} displays the isoscalar (isovector) octupole response in $^{16}$O, calculated using RPA (panels (a)) and SSRPA (panels (b)).
 In the isoscalar case, the tensor effect on 3$^-$ states differs somewhat from the 0$^+$ case. In RPA calculations, SGII and SGII + Te3 produce nearly identical results, whereas SGII + Te1 induces a substantial effect below 25 MeV. Specifically, the lowest peak at around 8 MeV shifts downward to approximately 6 MeV, with a B(E3) value of 1441 $e^2fm^6$, closely matching the experimental value of 1300 $e^2fm^6$ at 6.13 MeV. In SSRPA calculations, SGII and SGII + Te3 produce very similar results. With SGII + Te1, the tensor force's influence is stronger, shifting the peak at approximately 18 MeV to about 16 MeV. Additionally, the low-energy peak shifts to about 5.9 MeV with a strength of 1157 $e^2fm^6$, also aligning well with experimental data. For the isovector case, in the RPA calculations, the tensor force has almost no effect on the strength distributions. In SSRPA calculations, the tensor force in SGII + Te1 visibly shifts the strength downward below 20 MeV, while SGII + Te3 yields strength distributions nearly identical to those obtained without the tensor force.
\begin{figure}
	\includegraphics[width=.5\linewidth]{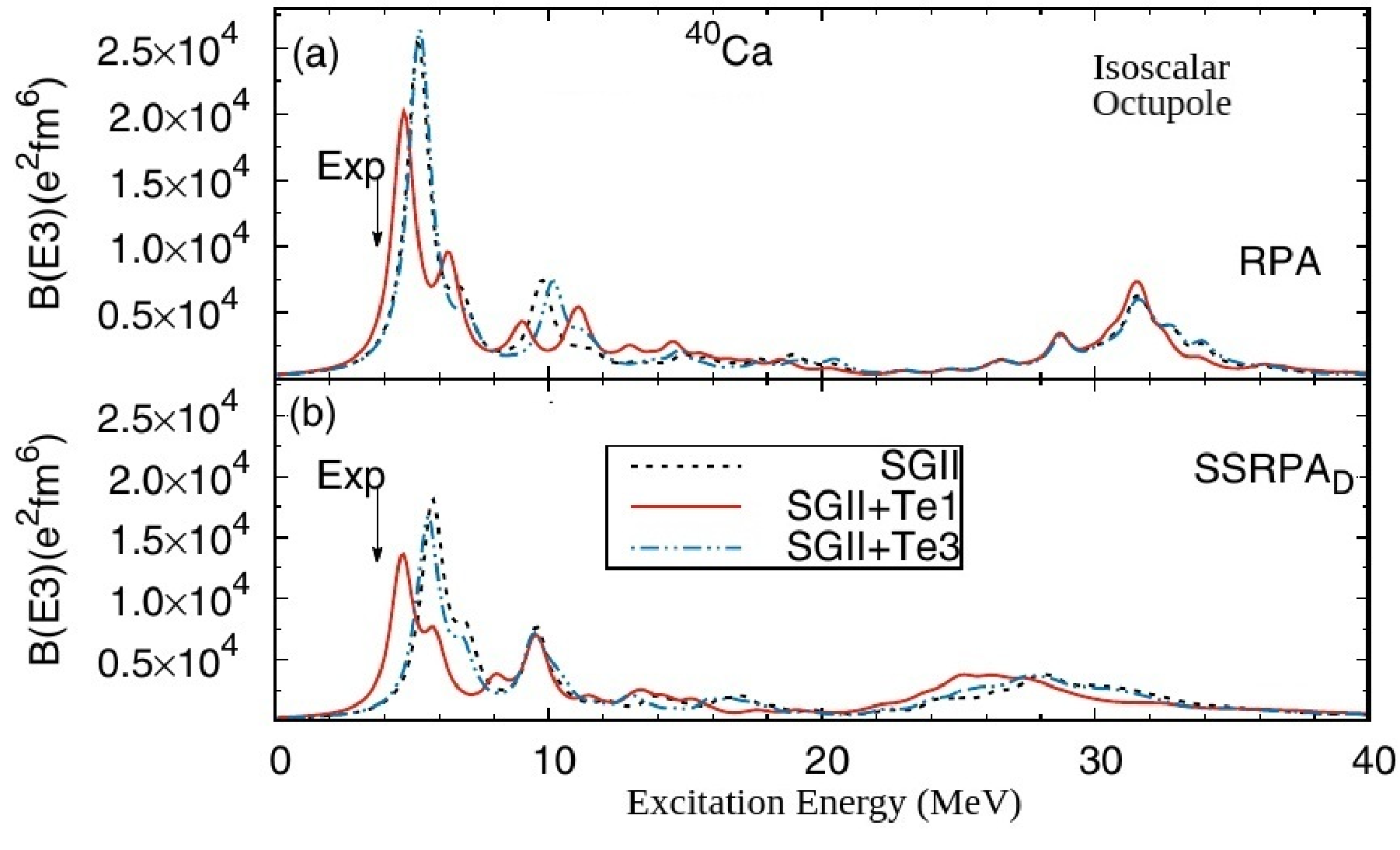}\hfill
	\includegraphics[width=0.5\linewidth]{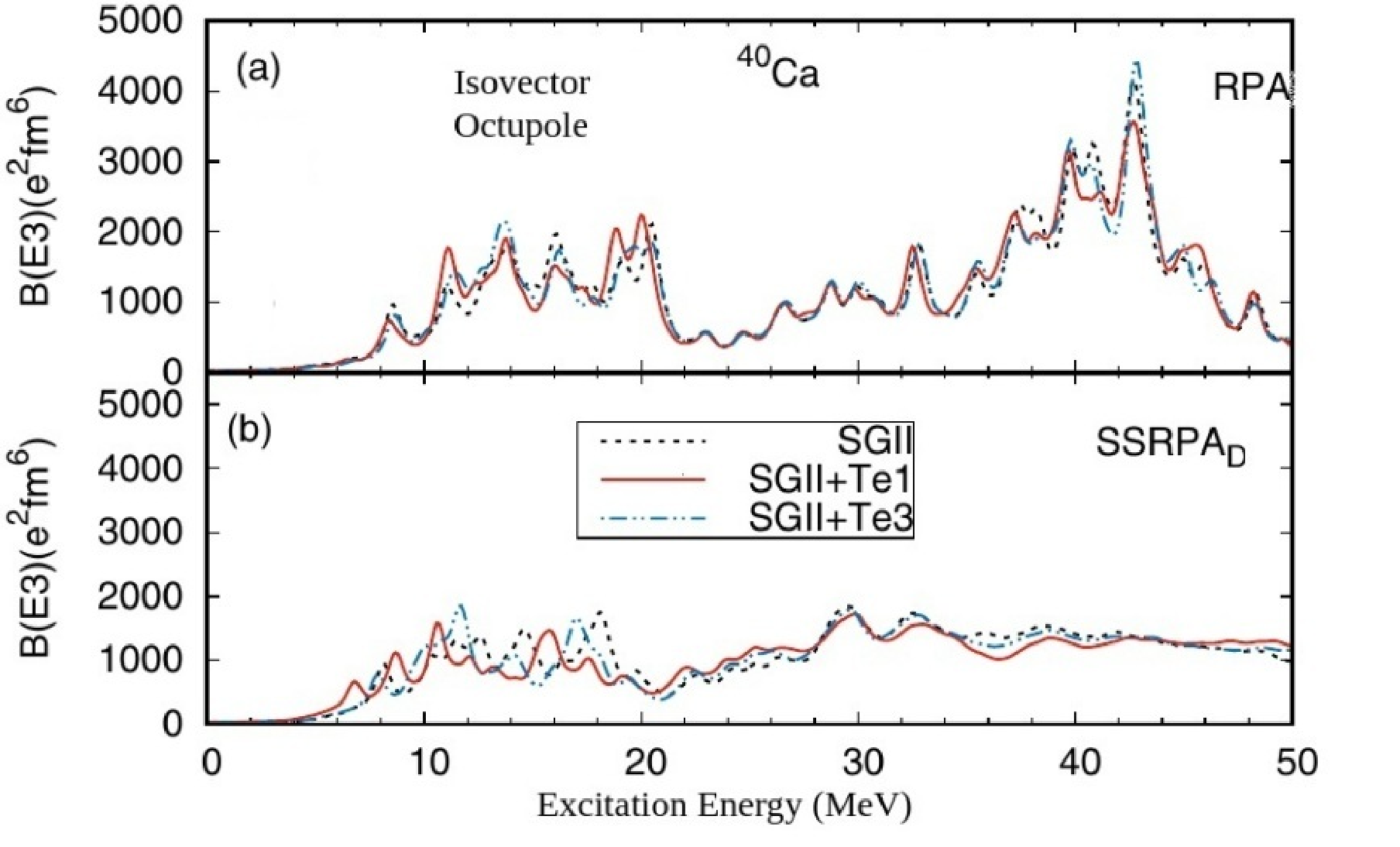}
	\caption{As in Figure \ref{Fig:Tensor_1} but for octupole response in $^{40}$Ca, the energy of the first excited state is taken from \cite{Ulrickson1977}. Adapted from Ref. \cite{Sagawa2021}.}
	\label{Fig:Tensor_6}
\end{figure}
Similar results are obtained for $^{40}$Ca, here we recall only the results in the octupole case.
\label{Sec:Applications_SSRPA_Monopole}
\begin{figure}
	\includegraphics[width=.45\linewidth]{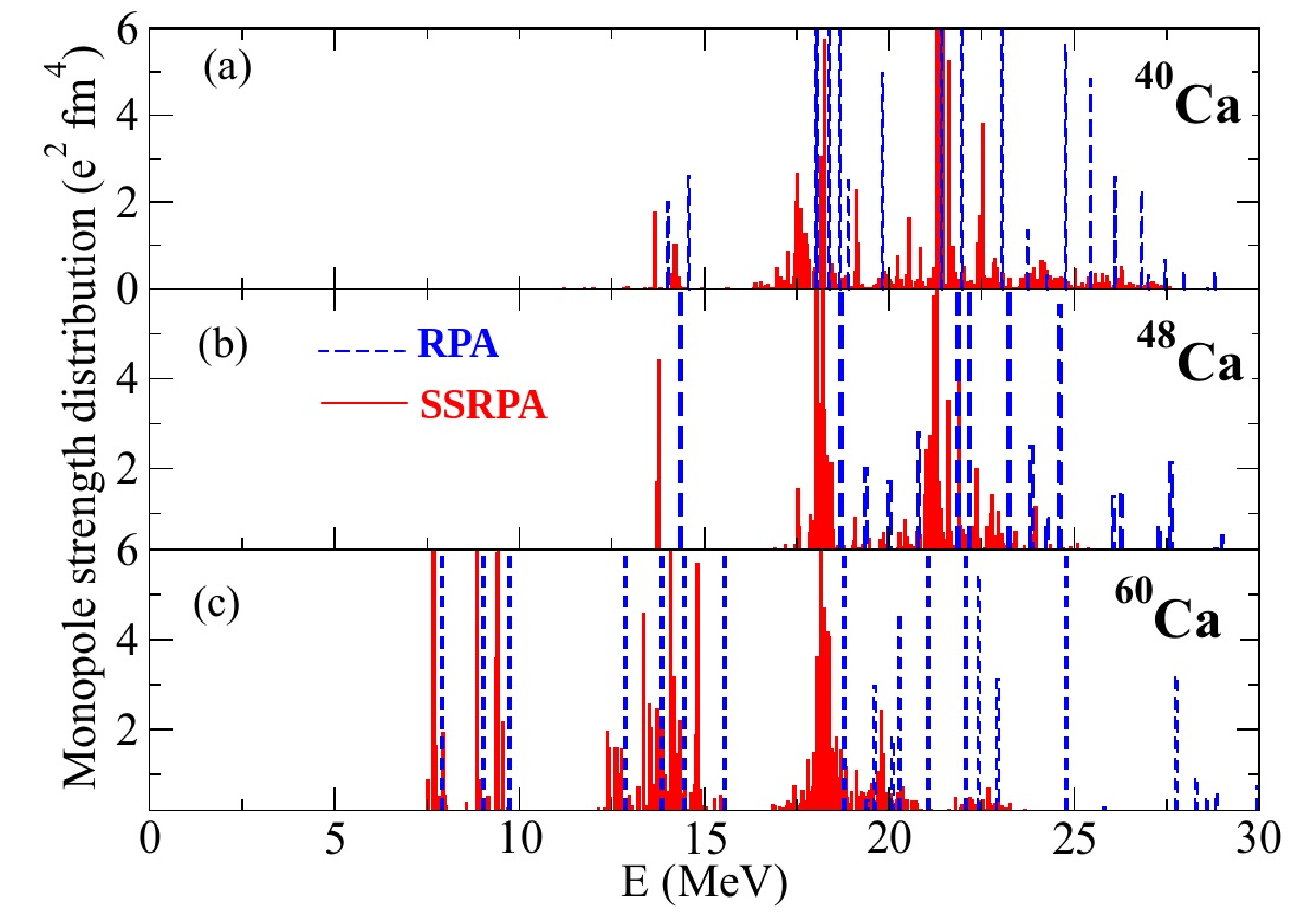}\hfill
	\includegraphics[width=.5\linewidth]{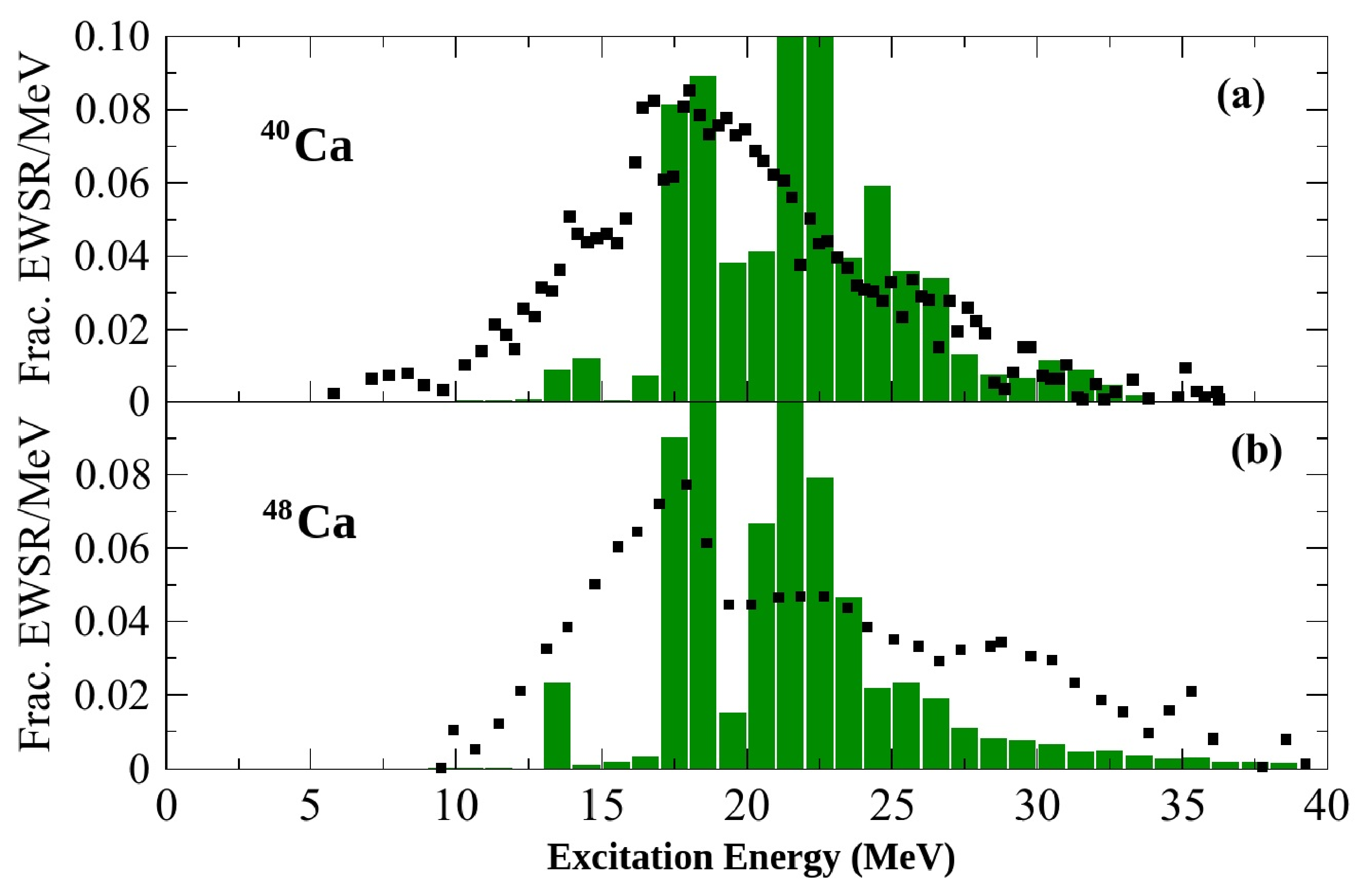}
	\caption{Left side: Monopole strength distribution computed with RPA (dashed blue bars) and SSRPA (full red bars) for $^{40}$Ca (panel (a)), $^{48}$Ca (panel (b)) and $^{60}$Ca (panel (c)). Right side: Experimental data \cite{Lui2011} (black points) compared with the SSRPA predictions, green bars, for $^{40}$Ca (panel (a)) and $^{48}$Ca (panel (b)). Adapted from Ref. \cite{Gambacurta2019}.}
	\label{Fig:Monopole_fig1}
\end{figure}
  Figure \ref{Fig:Tensor_6} show the isoscalar (isovector) octupole response in $^{40}$Ca, obtained by RPA (panel (a)) and SSRPA (panel (b)) calculations with different tensor forces or without tensor interaction. In RPA calculations, SGII + Te3 produces nearly the same strength distribution as SGII. With SGII + Te1, the low-lying states shift slightly downward to about 4.7 MeV, with a strength of 30.5 $\times$ 10$^3$ $e^2fm^6$, which compares well with the experimental data at 3.74 MeV with a strength of 18.4 $\times$ 10$^3$ $e^2fm^6$. In SSRPA calculations, SGII + Te3 again has a negligible effect. With SGII and SGII + Te1, the lowest energy state appears at approximately 5.6 and 4.6 MeV, respectively, with strengths of 27 $\times$ 10$^3$ and 19.9 $\times$ 10$^3$ $e^2fm^6$. The tensor effect with SGII + Te1 improves the results compared to SGII and aligns well with experimental data. The isovector octupole strength distributions in $^{40}$Ca are shown on the right side. The strengths are distributed over a broad energy range above 8 MeV. The tensor force effect is not visible in both RPA and SSRPA models above 20 MeV. Below 20 MeV, SGII + Te1 slightly expands the strength distributions to a wider energy range, and some strengths shift downward around 6 MeV.
  The results indicate that  the SGII + Te1 paratremization describes the experimental data better than SGII + Te3. In the latter, the triplet-even and triplet-odd tensor forces cancel each other out in the $0^+$ and $3^+$ states, vanishing their cumulative effect. On the contrary, these terms act constructively  in the SGII + Te1 case, leading to improved results.

\begin{comment}
The study demonstrates that the tensor force within the SGII + Te1 interaction can significantly impact the coupling between $1p-1h$ and $2p-2h$ model spaces. For isoscalar monopole transitions in $^{16}$O and $^{40}$Ca, SSRPA calculations show that the tensor force plays an important role in enhancing the strength of low-lying states below 10 MeV. Regarding the isoscalar quadrupole state, the tensor force produces negligible effects in RPA calculations. However, in SSRPA calculations, it shifts the main peak by approximately 1 to 2 MeV and also increases the strength below 10 MeV, though slightly. For the octupole case, the tensor force visibly shifts the lowest peak downward by about 2 to 3 MeV in $^{16}$O and 1 to 1.5 MeV in $^{40}$Ca. This results in substantial improvements in predicting the lowest 3$^-$ state, both in excitation energies and transition strengths. This effect is even evident at the RPA level. SSRPA calculations clearly show different effects depending on the tensor interaction employed, e.g., SGII + Te1 and SGII + Te3, where the triplet-odd tensor term varies from -350 to 200 MeV fm$^5$. The effects of the triplet-even and triplet-odd tensor forces within SGII + Te3 largely cancel each other, resulting in no significant overall impact. Conversely, in the SGII + Te1 case, the two tensor terms reinforce each other, leading to significant improvements in the numerical results.
\end{comment}

\subsection{Monopole response and soft modes}

%The ISGMR is a collective compression mode, which besides its intrinsic interest related to the properties of atomic nuclei, it is also strongly connected with the nuclear EoS, more precisely to its incompressibility. The nuclear incompressibility is then also a key quantity influencing phenomena ranging from heavy-ion collisions to neutron star structure.
%  Low-energy collective modes, such as the pygmy dipole resonance (PDR) \cite{Bracco2019,Savran2013} and pygmy quadrupole resonance \cite{Pellegri2015, Spieker2016}, arise from the oscillation of a neutron skin against proton-neutron saturated core. These low-energy resonances offer insights into the neutron skin thickness and the symmetry energy component of the EOS.
 In neutron-rich nuclei, low-energy collective modes, such as the pygmy dipole resonance (PDR) \cite{Bracco2019,Savran2013} and pygmy quadrupole resonance \cite{Pellegri2015, Spieker2016},
are usually described as the oscillation of a neutron skin against proton-neutron saturated core. In both cases, the microscopic nature and properties of these states
are still to be fully understood. This is even more debated in the quadrupole case, especially concerning their collective nature \cite{Yuksel2018,Lanza2019}.
However, these low-energy resonances are of great interest also because they might offer insights into the neutron skin thickness and the symmetry energy component of the EOS.
 On the other hand, low-lying monopole excitations (also referred as soft modes) in neutron-rich nuclei have received considerably less attention than their counterparts in the dipole and quadrupole channels. Theoretical predictions have indicated the existence of these excitations in neutron-rich isotopes of calcium, nickel, lead, and tin  \cite{Capelli2009,Khan2011,Khan2013,Hamamoto2014}. Despite several experimental attempts to detect these modes in exotic nuclei, a definitive experimental signature remains elusive. Recent experimental efforts have focused on nickel isotopes, employing active targets as detectors. Specifically, studies have been conducted on $^{56}$Ni using deuteron probes \cite{Monrozeau2008} and on $^{68}$Ni utilizing alpha and deuteron scattering \cite{Vandebrouck2015}. $^{68}$Ni, because of its strong neutron excess, might be a favorable candidate for exhibiting a soft breathing mode. However, limitations in the experimental setup of the study described in Ref. \cite{Vandebrouck2015} hindered the unambiguous observation of soft monopole excitations. The majority of existing theoretical calculations rely on the MF approximation. Therefore, an investigation incorporating BMF effects would provide valuable insights into these excitation modes, potentially yielding a more comprehensive and general analysis of their characteristics.
 In Ref. \cite{Gambacurta2019}, the SSRPA was employed to investigate the properties of soft monopole modes. %Numerical details can be found in Ref. \cite{Gambacurta2019}.
 The evolution of the low-lying monopole strength was first studied in Ca isotopes, by increasing the neutron number from $N=20$ in $^{40}$Ca to $N=28$ and $N=40$ in $^{48}$Ca and $^{60}$Ca. After that, three $N=20$ neutron-rich isotones $^{36}$S and $^{34}$Si were also studied, and finally, the $^{68}$Ni case, where the isospin asymmetry is comparable to that of $^{48}$Ca and $^{34}$Si, but the number of nucleons is higher, was investigated.

On the left side of Figure \ref{Fig:Monopole_fig1}, the SSRPA results for the Ca isotopes are shown. In panel (a), the $^{40}$Ca case is shown as a starting point, being a nucleus without neutron excess. In addition to the easily identifiable  ISGMR region, the figure reveals strength around 14 MeV. For comparative purposes, the RPA strength distribution is also shown. Both models predict a low-lying excitation, with SSRPA exhibiting a shift to lower energies and increased fragmentation due to $2p-2h$ configuration coupling. In panel (b), the $^{48}$Ca is then shown, having a neutron excess $\delta=(N-Z)/A=0.17$. Consistent with $^{40}$Ca, SSRPA induces a shift towards lower energies and greater fragmentation compared to RPA. Given the availability of experimental ISGMR strength distributions for both $^{40}$Ca and $^{48}$Ca \cite{Lui2011}, the right side of Figure \ref{Fig:Monopole_fig1} provides a comparison between SSRPA predictions and experimental results. The EWSR/MeV fractions are reported in panel (a) and (b) for $^{40}$Ca and $^{48}$Ca, respectively. %The SSRPA results demonstrate a significant improvement in strength fragmentation description compared to RPA, with RPA discrete spectra shown on the left side. %Figure 10 of Ref. \cite{Lui2011}, compares experimental distributions for $^{40}$Ca and $^{48}$Ca with folded RPA distributions using various Skyrme parametrizations. That figure shows that the RPA model systematically predicts strength distributions shifted to higher energies than experimental data.
 The RPA (SSRPA) centroid energies using the SGII parametrization are 21.3 (20.7) MeV for $^{40}$Ca and 20.7 (20.4) MeV for $^{48}$Ca, respectively. Centroids are calculated using $\sqrt{m_1/m_{-1}}$, where $m_1$ and $m_{-1}$ are the energy-weighted and the inverse energy-weighted moments. Experimental centroid values are 18.3 MeV for $^{40}$Ca and 19.0 MeV for $^{48}$Ca. These results indicate that SSRPA predictions better reproduce the experimental values than RPA. Finally, in panel (c) of the left side of Figure \ref{Fig:Monopole_fig1}, the extreme case of $^{60}$Ca, recently discovered \cite{Tarasov2018}, is shown. Also in this case,
going from RPA to SSRPA, the strength distributions show a similar effects as observed for $^{40}$Ca and $^{48}$Ca. The strength distribution is localized in three main energy regions: 5-11 MeV, 11-16 MeV, and above 16 MeV (ISGMR). The 5-11 MeV region is not predicted for the less neutron-rich $^{48}$Ca.
\begin{figure}
	\centering
	\includegraphics[width=.4\linewidth]{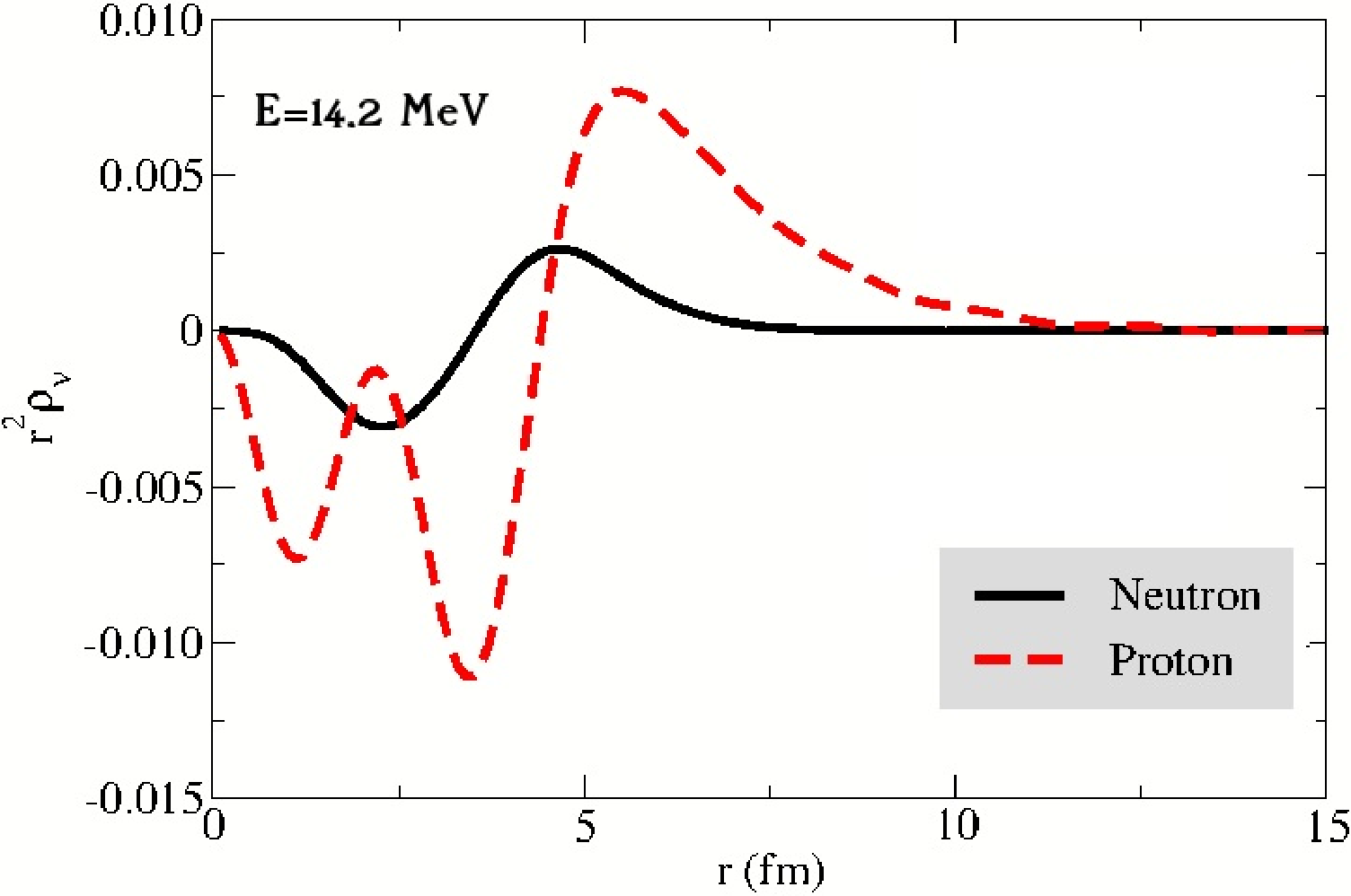}
	\includegraphics[width=.4\linewidth]{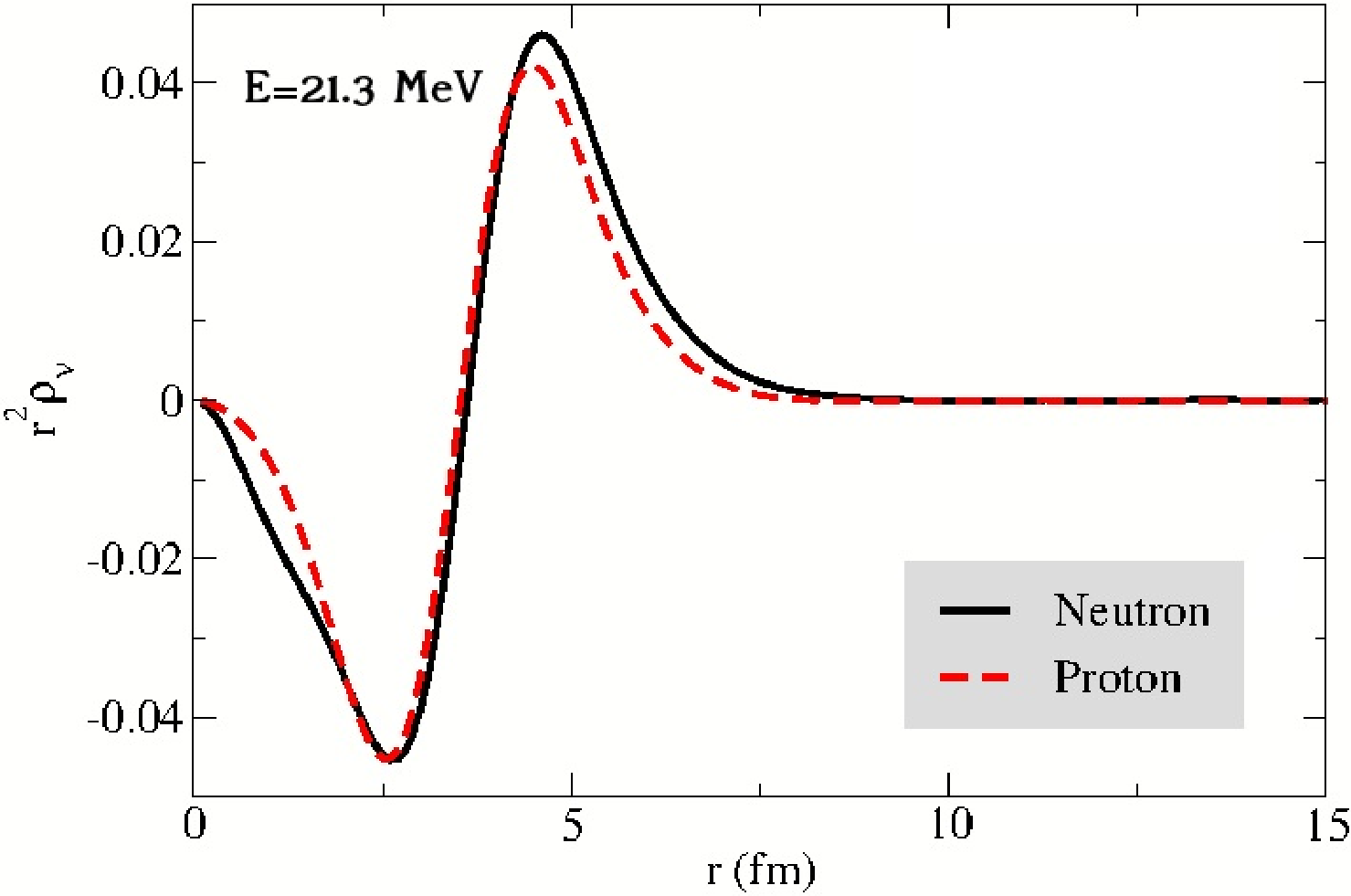}
	\caption{Neutron and proton transition densities multiplied by $r^2$ (in units of fm$^{-1}$) for 
		$^{40}$Ca associated with the SSRPA energy peak located at 14.2 (left side:) MeV and at 21.3 MeV (right side). Adapted from Ref. \cite{Gambacurta2019}.}
	\label{Fig:Monopole_fig2}
\end{figure}
\begin{figure}[h]
	\centering
	\includegraphics[width=.4\linewidth]{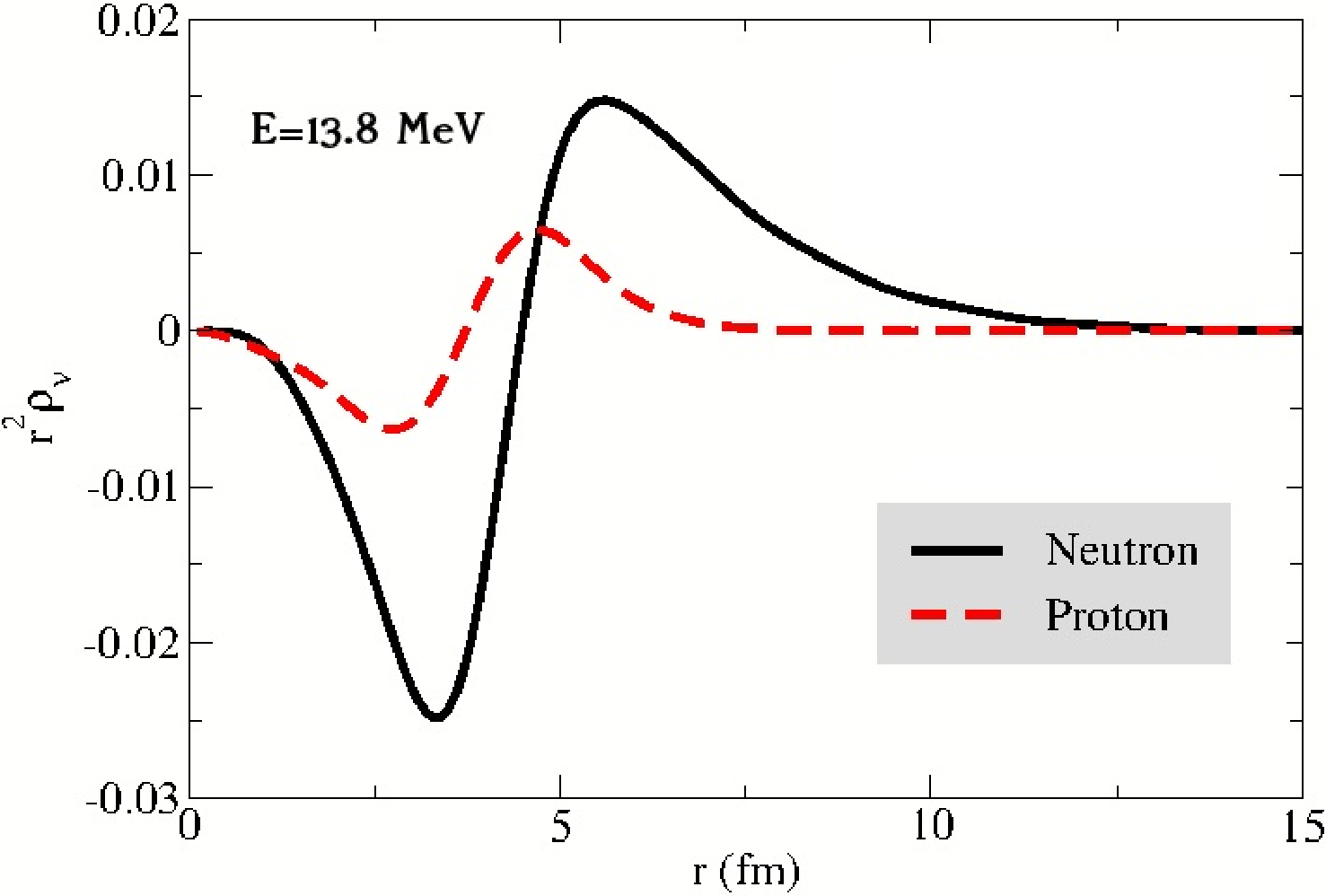}
	\includegraphics[width=.4\linewidth]{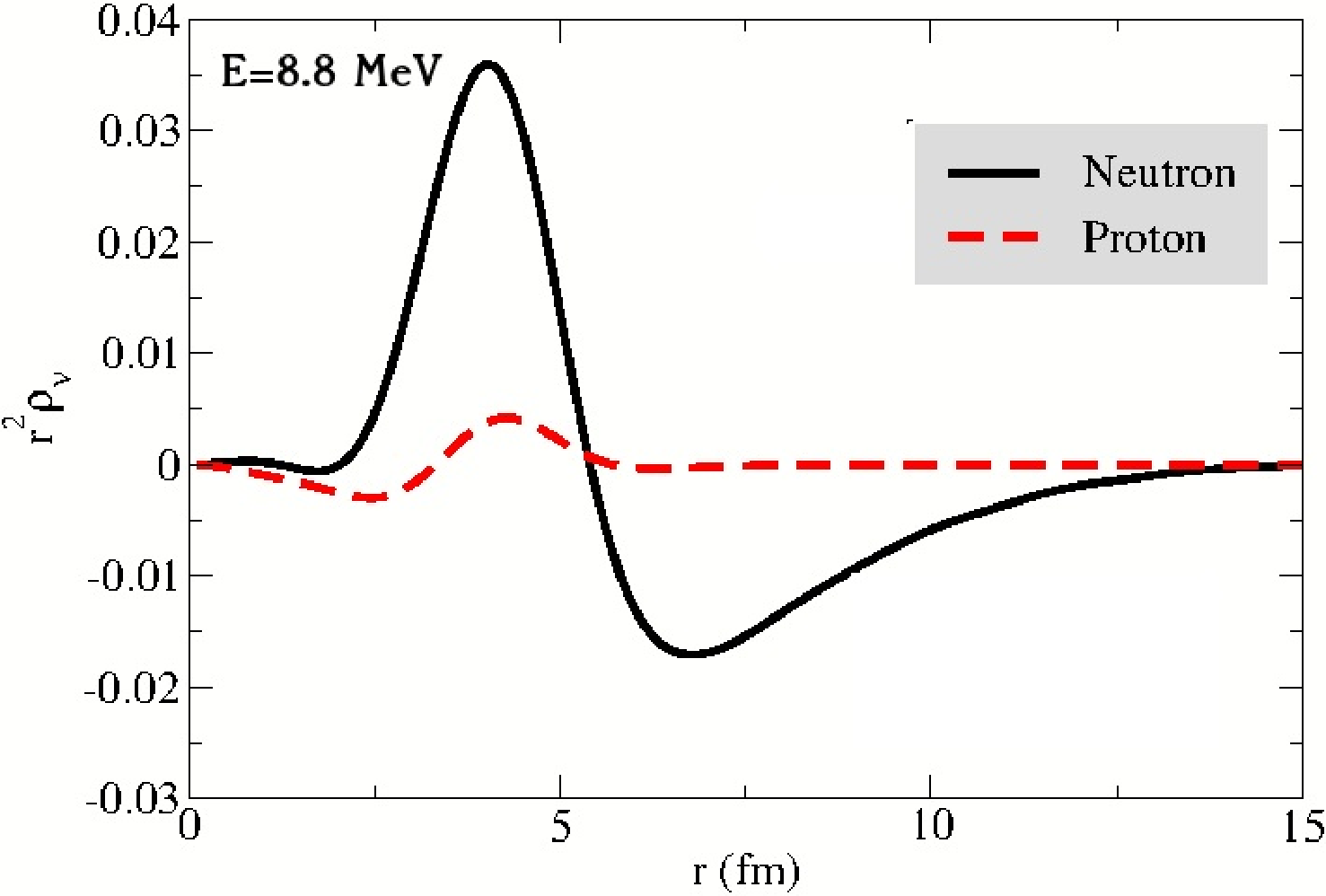}
	\caption{Neutron and proton transition densities multiplied by $r^2$ (in units of fm$^{-1}$) associated with the SSRPA energy peak located at 13.8 for
		$^{48}$Ca (left side:) MeV and at 8.8 MeV for 
		$^{60}$Ca (right side). Adapted from Ref. \cite{Gambacurta2019}.}
	\label{Fig:Monopole_fig3}
\end{figure}
A closer insight on the nature of the states can be drawn analyzing their transition densities.
In the case of $^{40}$Ca, in Figure \ref{Fig:Monopole_fig2}, one can see the SSRPA neutron and proton transition densities multiplied by $r^2$ and associated with the energy peak located at 14.2 MeV (left side) and at 21.3 MeV (right side). The latter, corresponds to the ISGMR and it shows the expected profile where neutrons and protons oscillate in phase. For the state at lower energy, one can see a different behavior, having a dominant proton contribution. Moreover, the state is characterized by strong $1p-1h$ and $2p-2h$ mixing, and the $1p-1h$ component is mostly given entirely by the proton $1p-1h$ configuration $[\pi 3s_{1/2}, \pi 2s_{1/2}]^{J=0}$. The RPA prediction is quite similar, the energy being slightly higher, having of course only $1p-1h$ nature. In the SSRPA case, the mixing with higher-order configurations leads to a strong contribution coming from the $2p-2h$ sector ($\approx 60\%$), with a strong spreading over several $2p-2h$ configurations. The percentage of the EWSR computed in SSRPA up to 15 MeV is 2.13\%.

%The state at 13.8 MeV for $^{48}$Ca. exhibits a dominant composition of the $1p-1h$ configuration $[\nu 2f_{7/2},\nu 1f_{7/2}]^{J=0}$ and the $2p-2h$ configuration $[ \nu 2f_{7/2}, \nu 4f_{7/2}]^{J_P}[ \nu 1f_{7/2}, \nu1f_{7/2}]^{J_H}]^{J=0}$, where $J_P=J_H=$ 0, 2, 4. The $J_P=J_H=4$ component is the most significant. Other $2p-2h$ configurations contribute negligibly. The left side of Figure \ref{Fig:Monopole_fig3} presents the SSRPA neutron and proton transition densities for this peak, illustrating a substantial evolution in the excitation mode's physical nature compared to $^{40}$Ca. This excitation is now predominantly neutron-driven. The percentage of the EWSR computed up to 15 MeV is 2.5\%, slightly exceeding that of $^{40}$Ca. The transition probability of this state is primarily determined by the most significant $1p-1h$ contribution, though several other $1p-1h$ configurations, primarily neutron configurations, contribute to the total $B(E0)$. Consequently, this state exhibits a slightly higher degree of collectivity, albeit still weakly collective, compared to the proton state in $^{40}$Ca. As expected, increasing the neutron excess from $N=20$ to $N=28$ results in the low-lying excited mode acquiring a dominant neutron-driven character, although a small proton contribution remains. As demonstrated by the transition densities, the dominant neutron contribution is not confined to the nuclear surface but extends throughout the nuclear volume.

The left side of Figure \ref{Fig:Monopole_fig3} shows the SSRPA transition densities  of the $\text{0}^+$ state in $^{48}\text{Ca}$ located at $13.8\text{ MeV}$. It is primarily a neutron-driven excitation, a significant shift from the proton-driven mode in $^{40}\text{Ca}$ due to the increased neutron excess. This state is mainly composed of the $\text{1}p-\text{1}h$ configuration $[\nu 2f_{7/2},\nu 1f_{7/2}]^{J=0}$ and a specific $\text{2}p-\text{2}h$ configuration with $J_P=J_H=4$ being the most significant component. Though weakly collective, it exhibits a slightly higher degree of collectivity than the $^{40}\text{Ca}$ state, with its  transition probability dominated by the main $\text{1}p-\text{1}h$ neutron configurations. The calculated strength up to $15\text{ MeV}$ is $2.5\%$ of the EWSR, slightly exceeding that of $^{40}\text{Ca}$. The dominant neutron contribution extends throughout the nuclear volume, not just the surface.

In $^{60}\text{Ca}$, the low-energy spectrum is strongly populated, with a prominent peak at $8.8\text{ MeV}$ (Figure \ref{Fig:Monopole_fig3} right). This state is $\sim 70\%$ the $\text{1}p-\text{1}h$ configuration $[\nu 2f_{5/2},\nu 1f_{5/2}]^{J=0}$, while the remaining $30\%$ is attributed to highly fragmented $\text{2}p-\text{2}h$ neutron configurations. It is strongly neutron-driven, similar to but more pronounced than in $^{48}\text{Ca}$, and contributes $5.13\%$ to the EWSR up to $11\text{ MeV}$, showing enhanced collectivity over the low-lying states in $^{48}\text{Ca}$. The region from $11\text{ to }16\text{ MeV}$ contains the most collective state at $14.1\text{ MeV}$, which is predominantly ($86\%$) $\text{2}p-\text{2}h$ in nature, with highly fragmented components and two significant $\text{1}p-\text{1}h$ configurations: $[\nu 4p_{3/2}, \nu 2p_{3/2}]^{J=0}$ and $[ \nu 3f_{5/2}, \nu 1f_{5/2}]^{J=0}$. The total EWSR up to $16\text{ MeV}$ is $26.81\%$, indicating that states in this higher energy window are considerably more collective, contributing over $20\%$ of the total strength. 
%Overall, the low-lying excitation modes are consistent between $^{48}\text{Ca}$ and $^{60}\text{Ca}$ (neutron excitation extending throughout the volume), but collectivity is significantly increased in $^{60}\text{Ca}$. Comparing states around $14\text{ MeV}$, the excitation has evolved from a predominantly single-particle, low-collectivity mode in $^{48}\text{Ca}$ to a dominant $\text{2}p-\text{2}h$ nature in $^{60}\text{Ca}$, leading to substantial enhancement of collectivity due to greater configuration mixing.

\begin{figure}
	\includegraphics[width=.47\linewidth]{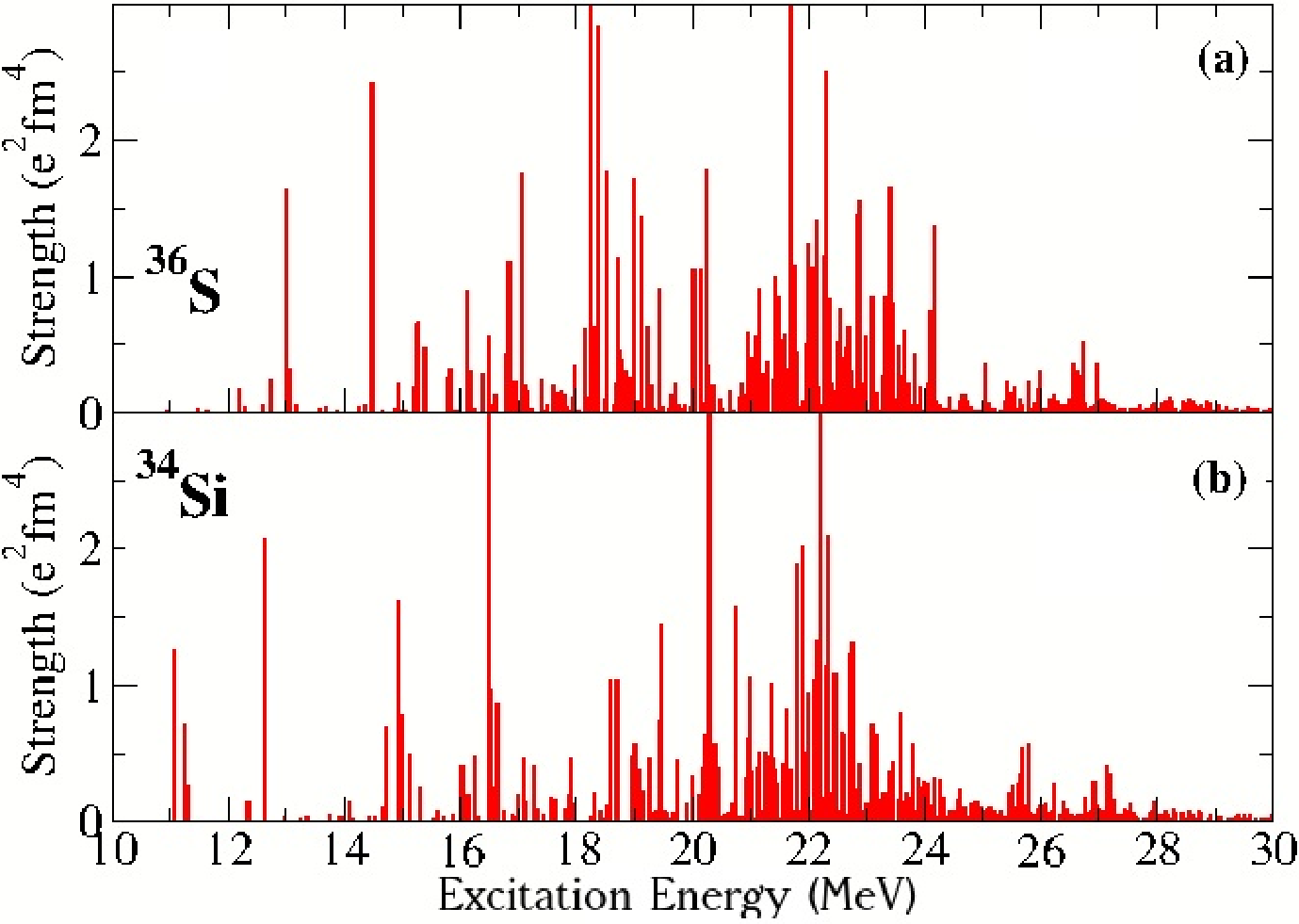}\hfill
	\includegraphics[width=.5\linewidth]{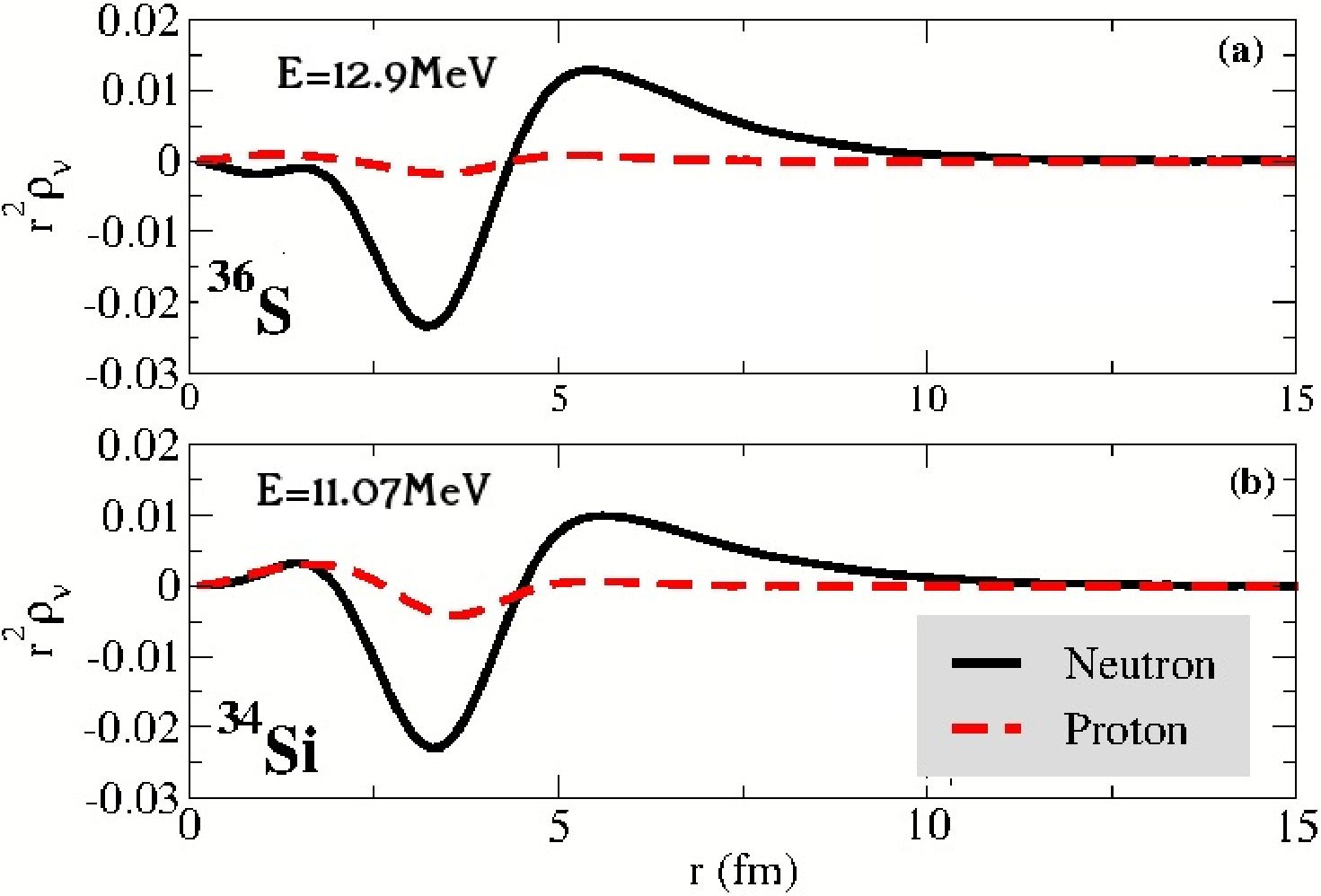}
	\caption{Left side: Monopole isoscalar strength distributions calculated for the nuclei 
		$^{36}$S in panel	(a) and $^{34}$Si in panel	(b). Right side: Neutron and proton transition densities multiplied by $r^2$ (in units of fm$^{-1}$) for $^{36}$S in panel (a) and $^{34}$Si in panel (b) associated with the energy peak located at 12.99 (11.07) MeV for $^{36}$S (34S). Adapted from Ref. \cite{Gambacurta2019}.}
	\label{Fig:Monopole_fig4}
\end{figure}
%
%Following the analysis of low-lying monopole mode evolution in Ca isotopes from $N=20$ to $N=40$, o
One can perform a similar analysis for $N=20$ isotones, specifically $^{40}$Ca, $^{36}$S, and $^{34}$Si. The left side of Figure \ref{Fig:Monopole_fig4} displays the isoscalar monopole strength distributions for $^{36}$S and $^{34}$Si. Both nuclei exhibit strength above 10 MeV. $^{36}$S shows a first peak around 13 MeV, while $^{34}$Si first peak is approximately 2 MeV lower. Analysis of these excitations reveals a significant mixing between $1p-1h$ and $2p-2h$ excitations, with the $1p-1h$ contribution in both nuclei driven by the neutron single-particle configuration $[\nu 2d_{3/2}, \nu 1d_{3/2}]^{J=0}$. The $\sim$ 2 MeV difference in the first peak excitation energies is primarily attributed to the energy difference of this dominant neutron single-particle configuration. 
For $^{34}$Si ($^{36}$S), the $1p-1h$ contribution to the peak at 11.07 (12.99) MeV is 54 (52) \%. The remaining contribution arises from the mixing of various $2p-2h$ configurations. The right side of Figure \ref{Fig:Monopole_fig4} show the transition densities of the two low-lying collective states, both showing similar  neutron-driven profiles. %For $^{34}$Si, the $1p-1h$ configuration $[\nu 2d_{3/2}, \nu 1d_{3/2}]^{J=0}$ constitutes 53 \% of the total excitation composition, and the rest is spread over several $2p-2h$ configurations collectively contributing to 31 \%. For $^{36}$S, the $1p-1h$ configuration $[\nu 2d_{3/2}, \nu 1d_{3/2}]^{J=0}$ contributes 51 \% of the total composition, and the $2p-2h$ configurations together contribute 29 \%.

\begin{figure}
	\centering
	\includegraphics[width=0.9\linewidth]{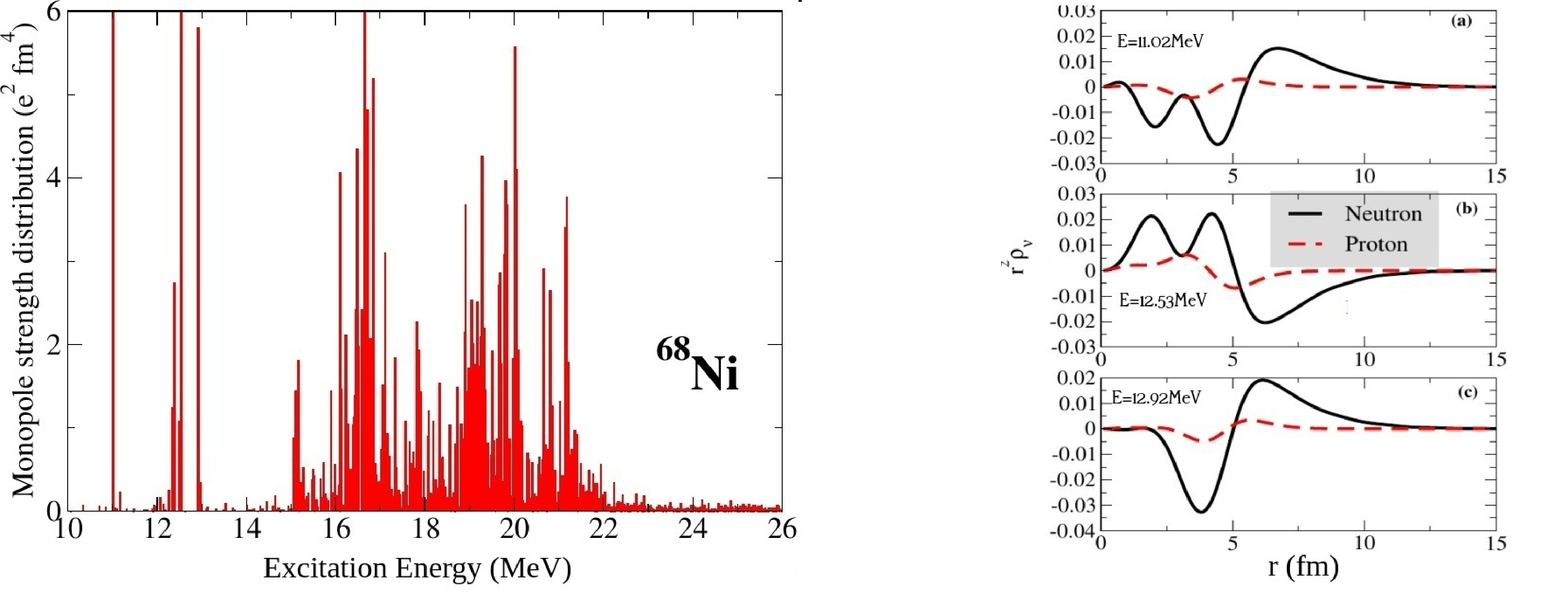}\hfill
	\caption{Left side: Monopole isoscalar strength distribution for $^{68}$Ni. 
		Right side: Neutron and proton transition densities multiplied by $r^2$ (in units of fm$^{-1}$) associated with the peaks located at 11.2, 12.53, and 12.92 MeV in the monopole spectrum of 
		$^{68}$Ni. Adapted from Ref. \cite{Gambacurta2019}.}
		\label{Fig:Monopole_fig5}
\end{figure}
Finally, the heavier neutron-rich nucleus $^{68}$Ni is considered. By comparing it with a lighter nucleus, such as $^{34}$Si (possessing a similar isospin asymmetry), one can investigate whether the SSRPA model predicts alterations in the nature and collectivity of the soft mode for a heavier exotic system located in a distinct region of the nuclear chart.
\begin{wrapfigure}{l}{0.5\textwidth}
	\centering
	\vspace{-3.5mm}
	\includegraphics[width=0.45\textwidth]{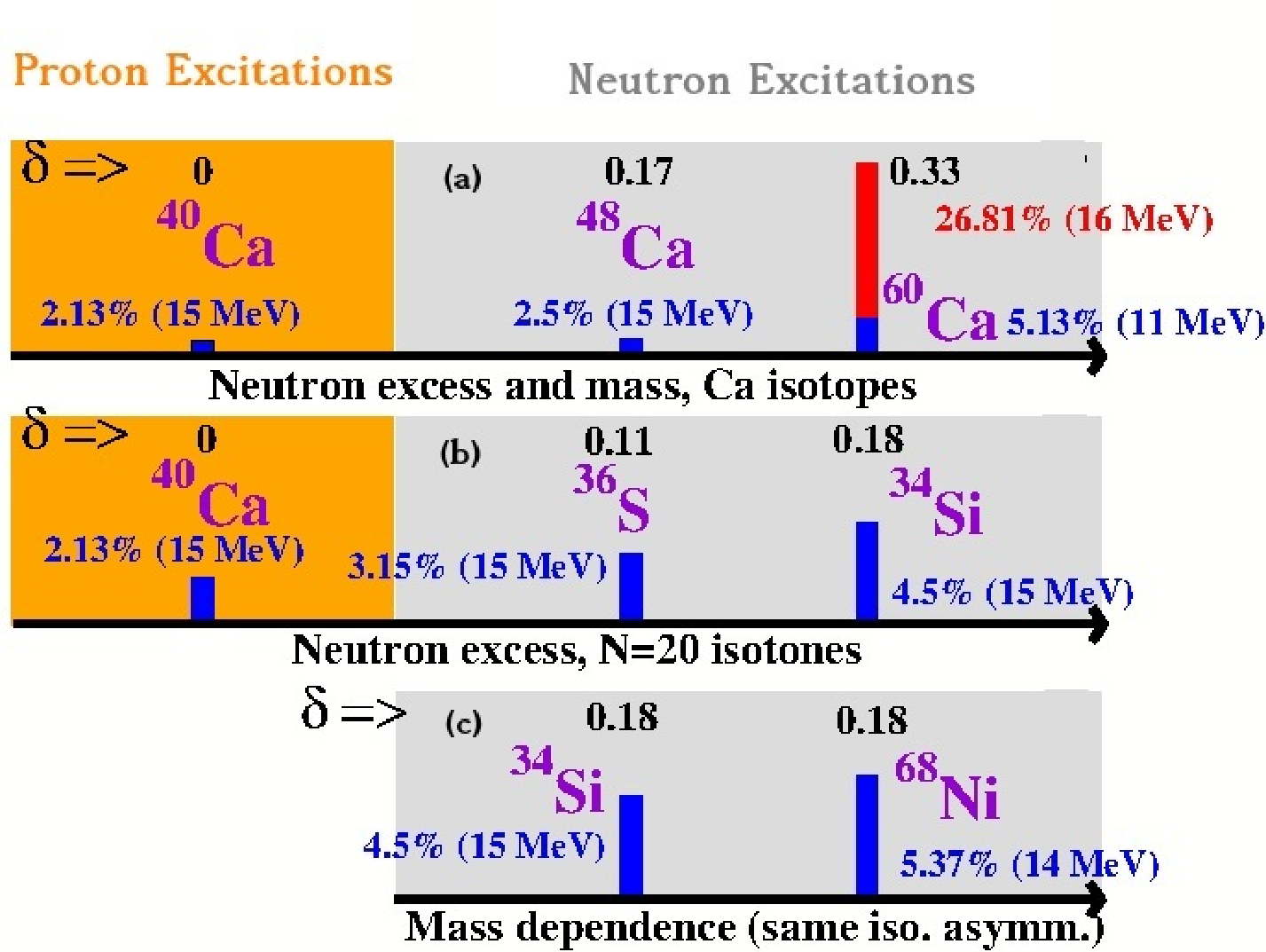}
	\vspace{-1.5mm}
	\caption{Percentages of the EWSR (up to the energy in parentheses) for: (a) Ca isotopes, (b) $N=20$ isotones, (c) Evolution as a function of the mass for two nuclei with identical isospin asymmetry, $^{34}$Si and $^{68}$Ni. The isospin asymmetry $\delta$ corresponding to each nucleus is provided. Adapted from Ref. \cite{Gambacurta2019}.}
	%	 	\vspace{-7mm}
	\label{Fig:Monopole_fig6}
\end{wrapfigure}
 The comparison between $^{34}$Si and $^{68}$Ni is particularly significant because these two isotopes exhibit comparable shell structures. Both possess double shell closures, with a spin-orbit type for protons and a harmonic-oscillator type for neutrons. %Consequently, any potential modifications to the soft mode's nature and collectivity are expected to be minimally influenced by their respective shell structures and primarily driven by the mass difference between the two systems. s
The response function for $^{68}$Ni is shown on the left side of Figure \ref{Fig:Monopole_fig5}. The presence of multiple peaks below 14 MeV is observed. The percentage of the EWSR computed below 14 MeV is 5.37\%. Compared to the lighter system $^{34}$Si, this percentage is higher, indicating an enhancement of collectivity. However, this collectivity enhancement is significantly less pronounced than that observed when transitioning from $^{48}$Ca to $^{60}$Ca. This suggests that collectivity is more strongly dependent on neutron excess than on system mass. The neutron and proton transition densities corresponding to the three peaks located at 11.02, 12.53, and 12.92 MeV are plotted on the right side of Figure \ref{Fig:Monopole_fig5}. For the three peaks there is a dominant neutron $1p-1h$ contribution given by the configurations $[\nu 3p_{1/2},\nu 2p_{1/2}]^{J=0}$, $[\nu 3p_{3/2},\nu 2p_{3/2}]^{J=0}$, and $[\nu 2f_{5/2},\nu 1f_{5/2}]^{J=0}$, respectively, whereas the $2p-2h$ contribution corresponds to 0.18\%, 
0.49 \%, and 0.29 \% of the EWSR for the three cases, respectively. We conclude that the mixing with $2p-2h$ configurations is very weak in this nucleus for the soft monopole excitations under study.

Figure \ref{Fig:Monopole_fig6} summarize the evolution of the low--energy contribution to the EWSR for Ca isotopes as a function of the mass and the neutron excess (a), for $N=20$ isotones as a function of the neutron excess (b), and for two nuclei with the same isospin asymmetry but different masses (c). 
The study confirms the prediction already given at RPA level of soft monopole modes related to the neutron excess. However, compared to the dipole and quadrupole case, some differences are found. The transition density profiles clearly show that these state are neutron-driven, the dominant neutron contribution extends over the entire volume of the system and it is not localized only at the surface of the nucleus. Moreover, the wave-function in terms of $1p-1h$ and $2p-2h$ components shows that the low-lying states are typically composed by single $1p-1h$ configurations, indicating the absence of a genuine  collectivity, compared to the dipole and quadrupole case.

\subsection{SSRPA studies in the dipole channel }
\label{Sec:Applications_SSRPA_Dipole}
\subsubsection{The $^{40,48}$Ca case}
\label{Sec:Applications_SSRPA_Dipole_Ca}
\begin{figure}
	\includegraphics[width=.435\linewidth]{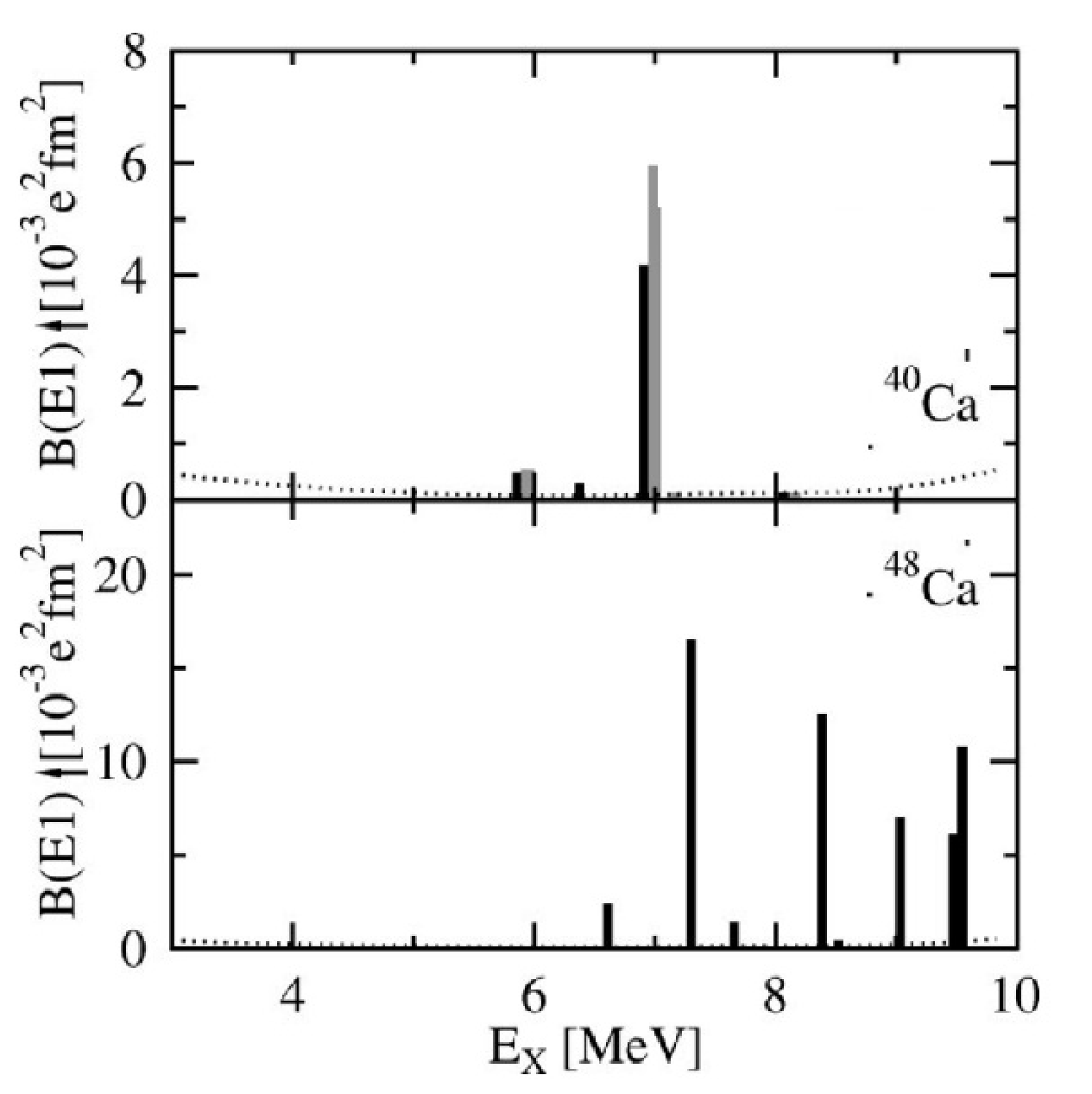}\hfill
	\includegraphics[width=0.545\linewidth]{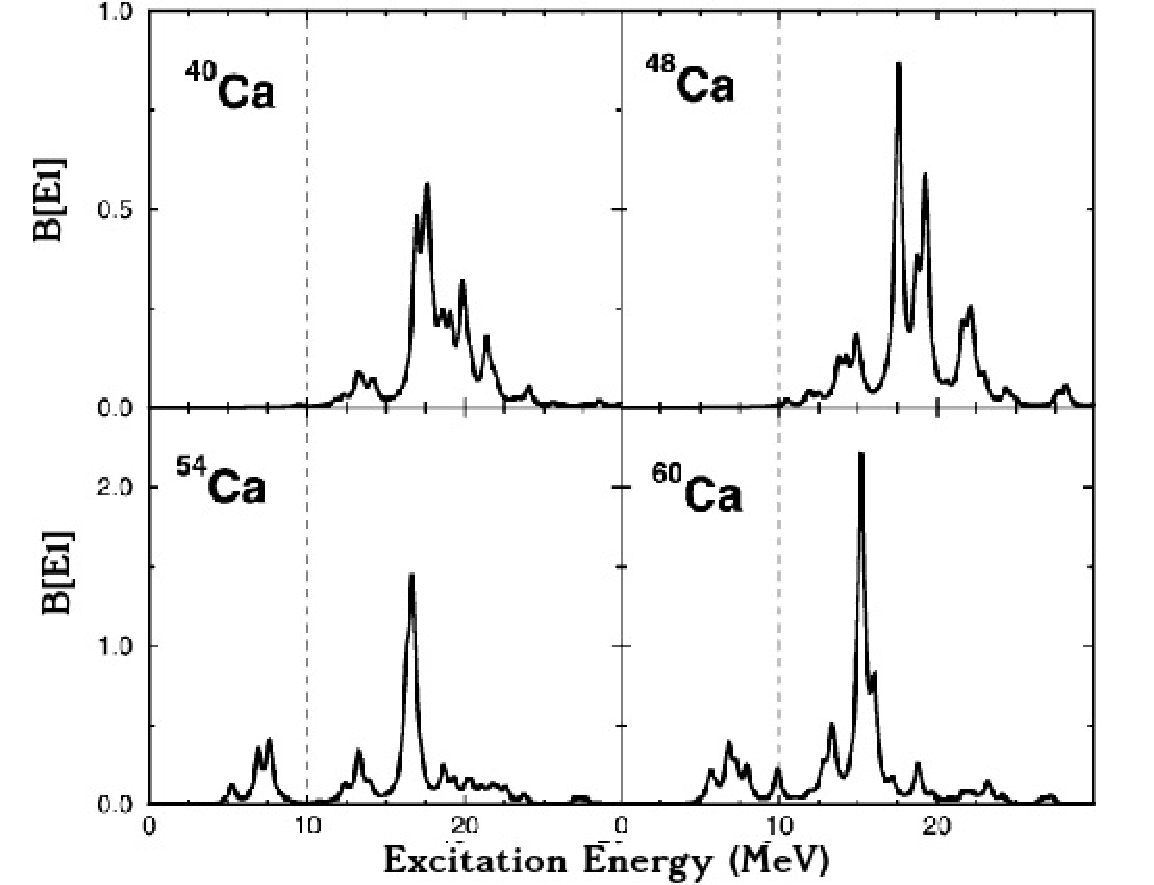}
	\caption{Left side: Electric dipole strength distributions for $^{40}$Ca (upper panel) and $^{48}$Ca (lower panel). The experimental data from \cite{Hartmann2002} (black lines) are compared with those ENSDF data base (gray lines). Adapted from Ref. \cite{Hartmann2002}. Right side: Relativistic RPA isovector dipole strength distributions in Ca isotopes. Adapted from Ref. \cite{Vretenar2001}.
	}
	\label{Fig:SRPA_Ca40_48}
\end{figure}

 The electric  dipole  and quadrupole  strength distributions in  $^{40}$Ca and $^{48}$Ca was measured using the photon scattering method \cite{Hartmann2002}. The study revealed a comparable summed B(E2) strength below 10 MeV for both isotopes. However, a significant difference was observed in the B(E1) strength. %The experimental setup and particle emission thresholds of the nuclei limited the measurements to an energy range up to 10 MeV. 
 Specifically, as shown on the left side of Figure \ref{Fig:SRPA_Ca40_48}, a low-energy strength development between 5 and 10 MeV is observed in $^{48}$Ca, but not in $^{40}$Ca. 

Theoretically, both relativistic and non-relativistic (Q)RPA models are not able to reproduce the low-lying response in $^{40,48}$Ca. These models either overestimate excitation energies or fail to capture the observed fragmentation of spectral peaks. For example, relativistic RPA calculations of the Zagreb's group predict no strength below 10 MeV \cite{Vretenar2001} in $^{40,48}$Ca.

\begin{wrapfigure}{l}{0.5\textwidth}
	\centering
		\includegraphics[width=0.9\linewidth]{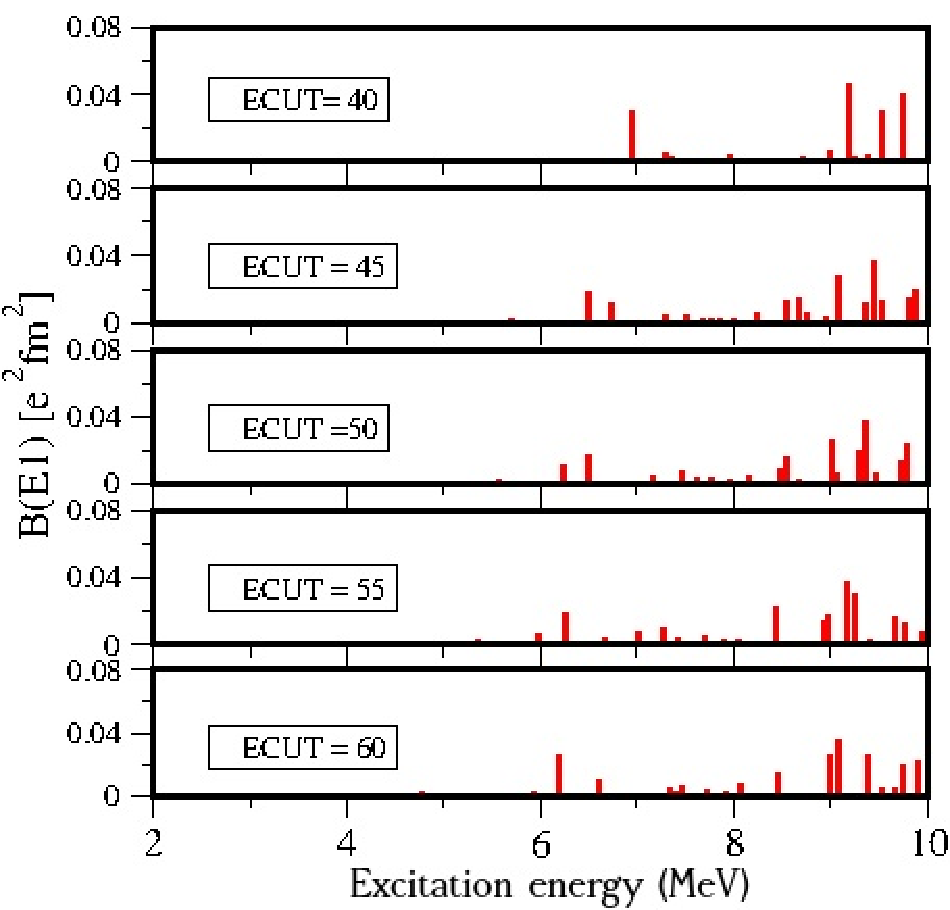}
	\caption{SRPA dipole strength distribution for the $^{48}$Ca for increasing values of the energy cutoff (in MeV units) on the $2p-2h$ configurations. Adapted from Ref. \cite{Gambacurta2011}.}
	\label{Fig:Dipole_Ca40_48_Ecut}
\end{wrapfigure}

 However, as shown on the right side of Figure \ref{Fig:SRPA_Ca40_48}, further increasing the neutron excess results in the appearance of low-lying strength in the $^{54,60}$Ca isotopes. This development of strength, correlated with neutron excess, suggests that the low-lying states in $^{48}$Ca might also be PDR states, such as oscillations of the neutron skin against the neutron-proton saturated core. 
 Also non-relativistic Skyrme-RPA calculations show a similar behavior, in particular no strength below 10 MeV is found in $^{48}$Ca. It might be thus interesting to study whether the inclusion of the $2p-2h$ configurations can affect the low-energy part of the spectrum. Before considering the SSRPA case, useful and instructive information can be drawn by considering the SRPA results. Figure \ref{Fig:Dipole_Ca40_48_Ecut} illustrates the SRPA dipole strength for the transition operator ((\ref{Eq:Op-J1-IV-CM}))
%\begin{equation}\label{trans-operator}
%	F_{10}=e\frac{N}{A} \sum_{i=1}^{Z} r_i Y_{10}(\Omega_h)-
%e\frac{Z}{A}\sum_{i=1}^{N} r_i Y_{10}(\Omega_h)
%\end{equation}
% for $^{40}$Ca (left side) and $^{48}$Ca (right side) as a function of the energy cutoffS for the 2p-2h configurations included in the calculations. The increasing of the energy cutoff leads to the development of low-lying strength, with this effect being more pronounced in $^{48}$Ca.
 for $^{48}$Ca  as a function of the energy cutoff for the $2p-2h$ configurations. The increasing of the energy cutoff leads to the development of low-lying strength.

\begin{wrapfigure}{l}{0.5\textwidth}
	\centering
	%\begin{figure}
	\centering
	\includegraphics[width=0.9\linewidth]{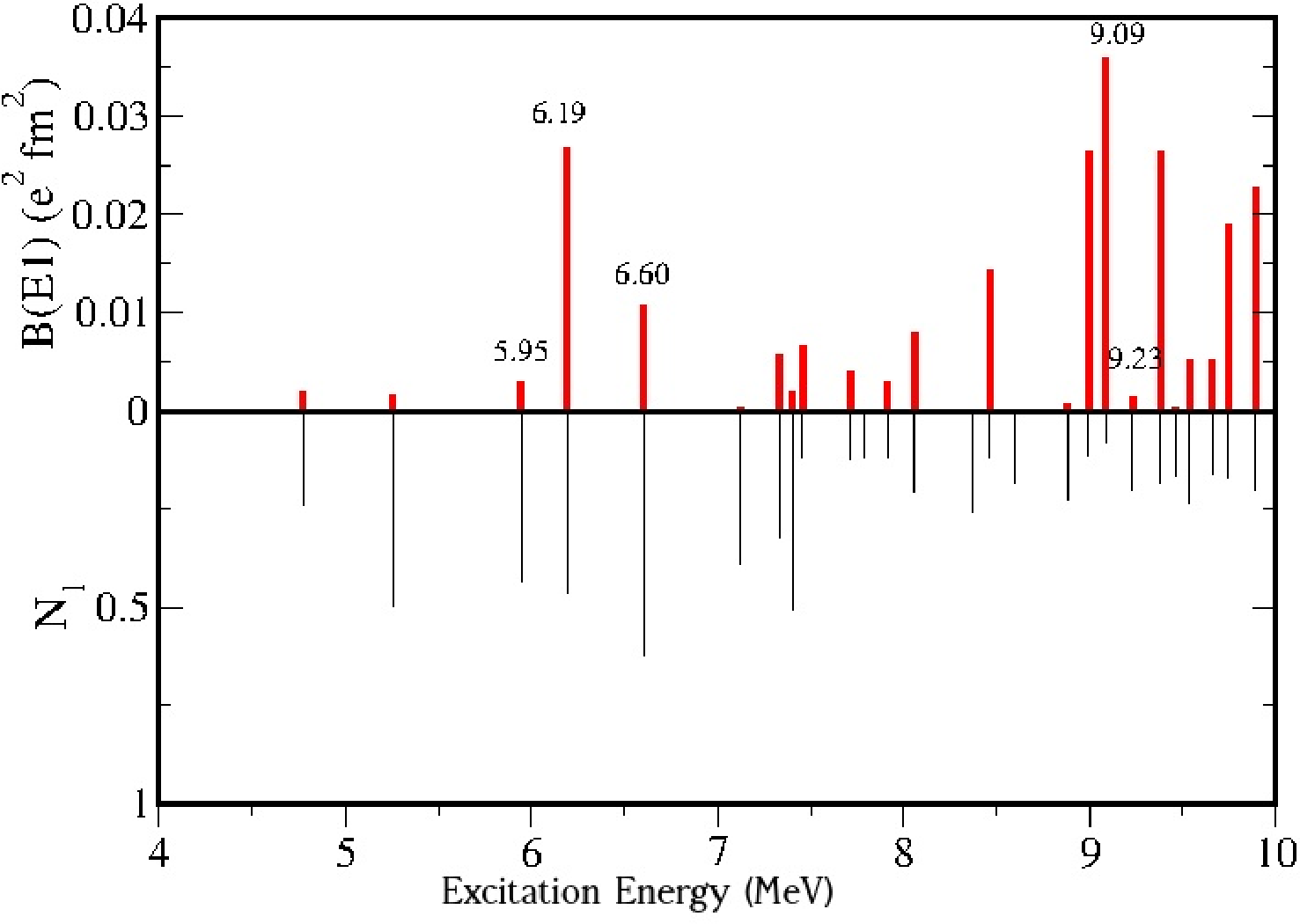}
	\caption{For each B(E1) state the $N_1$ contribution to the norm of the state defined in Eq. \ref{Eq:norm_srpa} (lower panel) is shown for $^{48}$Ca. In the upper panel, the energy in MeV units of some states is reported, for which the corresponding transition densities are shown in Figure \ref{Fig:Ca48_TDR}.  Adapted from Ref. \cite{Gambacurta2011}.}
	\label{Fig:Ca48_N1}
	%\end{figure}
\end{wrapfigure}

The composition of excitation modes, specifically the interplay between $1p-1h$ and $2p-2h$ configurations, provides a  valuable insight.  Considering the SRPA normalization condition,
\begin{displaymath}
	\sum_{ph}(\mid X_{ph}^{\nu}\mid^2-\mid Y_{ph}^{\nu}\mid^2)+
	\sum_{p<p',h<h'}(\mid X_{php'h'}^{\nu}\mid^2-\mid Y_{php'h'}^{\nu}\mid^2)
	\end{displaymath}
	\begin{equation}
	= N_1+N_2=1,
	\label{Eq:norm_srpa}
\end{equation}
where $N_1$ ($N_2$) the amount of $1p-1h$ ($2p-2h$) content in the wave-function. In Figure \ref{Fig:Ca48_N1} for each state, the corresponding $N_1$ value is plotted. One can see  that all excitations exhibit a mixing of $1p-1h$ and $2p-2h$ configurations. Excitations with a higher $1p-1h$ content (approximately 50\%) can be interpreted as RPA excitations, originally located at higher energies (around 11 MeV), which have been shifted to lower energies through coupling with $2p-2h$ configurations. Furthermore, the SRPA spectrum displays several states characterized by a dominant $2p-2h$ nature and a relatively large B(E1) value, despite their low $1p-1h$ content, as evidenced in the energy region around 9 MeV.

%\begin{wraptable}{r}{0.48\textwidth}
%	\begin{tabular}{|c|c|c|c|}
%		\hline
%		&&$^{48}$Ca&$^{40}$Ca\\
%		\hline
%		$\sum B(E1)$&SRPA&230 &23\\
%		(10$^{-3}$ e$^{2}$ fm$^{2}$)& Exp& 68.7 $\pm$ 7.5 & 5.1 $\pm$ 0.8 \\
%		\hline
%		
%		$\sum_i E_i B_i(E1)$&SRPA&1964&211\\
%		10$^{-3}$ e$^{2}$ fm$^{2}$ MeV &Exp &570 $\pm$ 62 &35 $\pm$ 5 \\ 
%		\hline
%		$E_{centroid}$ &SRPA &8.54 &9.17\\
%		MeV&Exp&8.40&6.80 \\
%		\hline	
%	\end{tabular}
%	\caption{
%		Total $B(E1)$ and EWSRs integrated up to 10 MeV and corresponding centroid energies obtained in SRPA compared with the experimental values \cite{Hartmann2002} for the $^{40,48}$Ca isotopes. From Ref. \cite{Gambacurta2011}.} 
%	\label{Tab:Ca48-SRPA-Exp}
%	\vspace{1cm}
%\end{wraptable}

\begin{figure}
	\centering
	\includegraphics[width=0.45\linewidth]{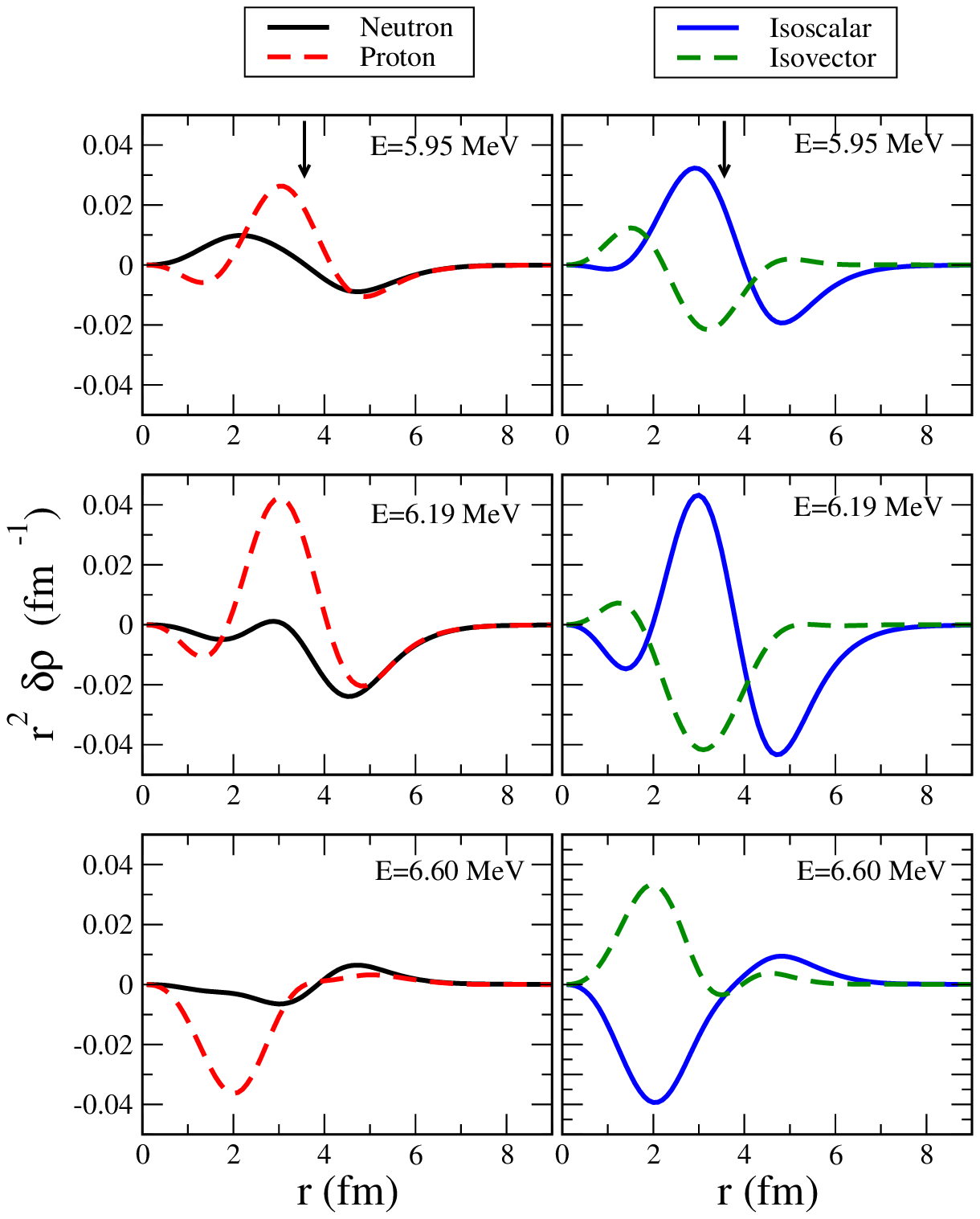}\hfil
	\includegraphics[width=0.45\linewidth]{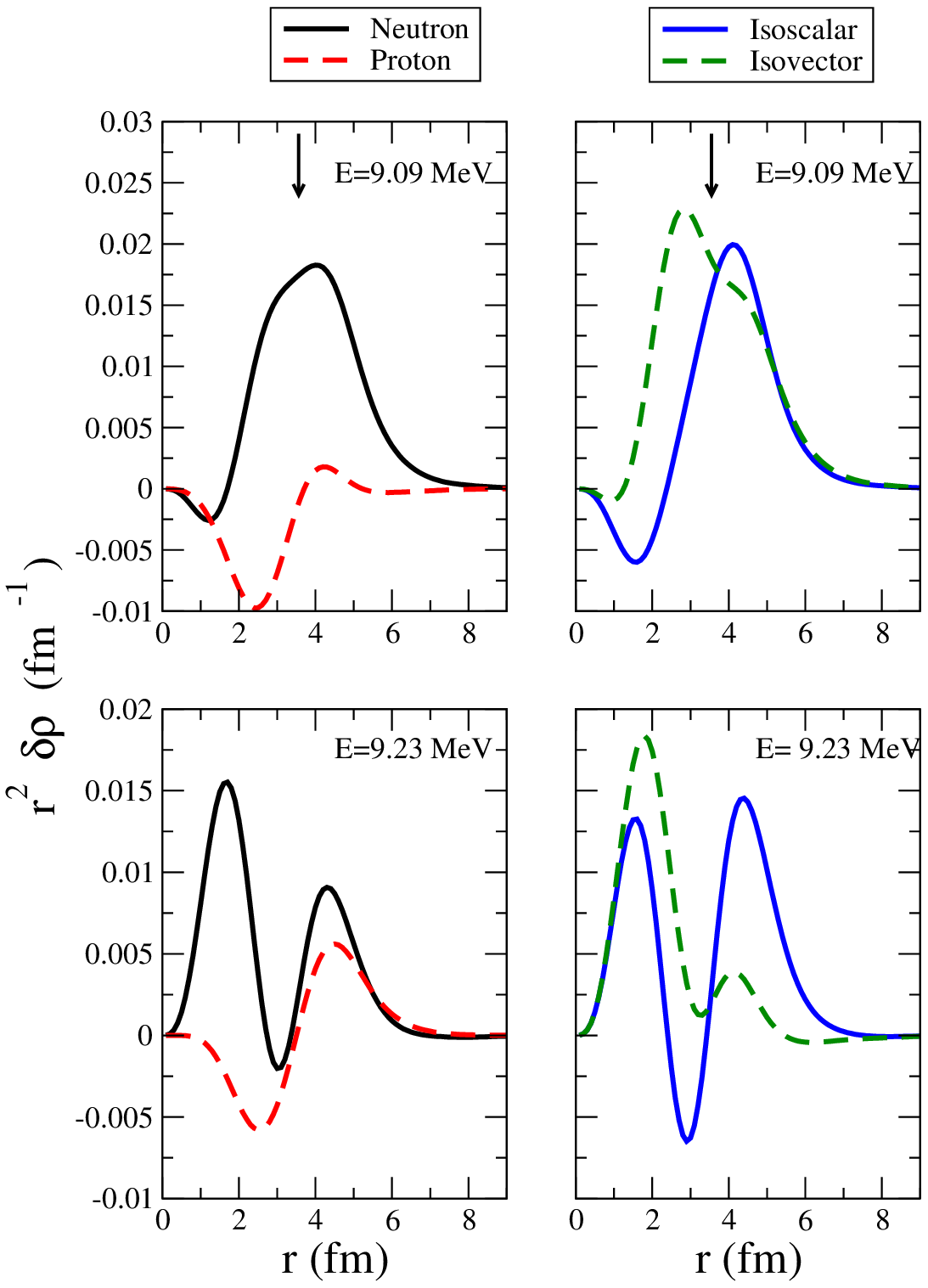}
	\caption{Transition densities for $^{48}$Ca associated to the different SRPA peaks (see Figure \ref{Fig:Ca48_N1}). Adapted from Ref. \cite{Gambacurta2011}.}
	\label{Fig:Ca48_TDR}
\end{figure}

The nature of the low-energy peaks can be further analyzed by looking at the associated transition densities. In Figure \ref{Fig:Ca48_TDR}, the transition densities for several sates identified by their excitation energy (see Figure \ref{Fig:Ca48_N1}) are shown. For each state,
the neutron, proton, isoscalar and isovector transitions densities are plotted. Analysis of transition densities reveals that the low-lying states in $^{48}$Ca do not consistently exhibit a clear neutron skin oscillation against the core. Instead, protons and neutrons often oscillate in phase, particularly in the nuclear surface. States at 5.95 MeV and 6.60 MeV show strong isoscalar and isovector transition density oscillations, leading to near-zero B(E1) values due to cancellations. Similar cancellations occur at 9.23 MeV. However, the 9.09 MeV state, the most collective, displays a dominant neutron transition density with out-of-phase proton oscillations in the interior, resulting in strong isoscalar-isovector mixing, and a different transition density profile.

%
%
%
%
%\begin {table}
%\begin{center}
%	\begin{tabular}{|c|c|c|c|}
%		\hline
%		&&$^{48}$Ca&$^{40}$Ca\\
%		\hline
%		$\sum B(E1)$&SRPA&230 &23\\
%		(10$^{-3}$ e$^{2}$ fm$^{2}$)& Exp& 68.7 $\pm$ 7.5 & 5.1 $\pm$ 0.8 \\
%		\hline
%		
%		$\sum_i E_i B_i(E1)$&SRPA&1964&211\\
%		10$^{-3}$ e$^{2}$ fm$^{2}$ MeV &Exp &570 $\pm$ 62 &35 $\pm$ 5 \\ 
%		\hline
%		$E_{centroid}$ &SRPA &8.54 &9.17\\
%		MeV&Exp&8.40&6.80 \\
%		\hline	
%		
%	\end{tabular}
%	\caption{
%		Total $B(E1)$ and EWSRs integrated up to 10 MeV and corresponding centroid energies obtained in SRPA compared with the experimental values \cite{Hartmann2002} for the $^{40,48}$Ca isotopes. From Ref. \cite{Gambacurta2011}.} 
%		\label{Tab:Ca48-SRPA-Exp}
%\end{center}
%\end{table}
%Table \ref{Tab:Ca48-SRPA-Exp} presents a comparison of the total B(E1) strength, the EWSR integrated up to 10 MeV, and the corresponding centroid energies for $^{40}$Ca and $^{48}$Ca, comparing SRPA calculations with experimental data \cite{Hartmann2002}. The SRPA-calculated total B(E1) strength significantly overestimates the experimental value, by nearly a factor of four. Similar discrepancies are observed in the EWSRs, while, the SRPA-calculated centroid energy aligns closely with the experimental measurement. 

A more reliable description can be obtained  by using the SSRPA model, eliminating double-counting issues, as shown in Ref. \cite{Gambacurta2018}. The latter presents the first application of a fully self-consistent scheme, incorporating all interaction terms consistently with the MF description. In particular, the spin-orbit and Coulomb terms are included, differently from the previously discussed results. Two Skyrme parametrizations, SGII \cite{SGII} and SLy4 \cite{SLY4}, are employed. 
%Due to the high number of $2p-2h$ configurations, a diagonal approximation was implemented in the subtraction procedure. 

%This approximation is not in the diagonalization matrix, was validated in Reference [18], demonstrating results closely aligned with the exact subtraction procedure. Conversely, employing the diagonal approximation in the diagonalization matrix yields significantly different results, highlighting the crucial role of residual interactions among 2p-2h configurations in accurately describing the fragmentation of the low-lying response and the centroid energy within the Giant Dipole Resonance (GDR) region.

\begin{figure}
	\includegraphics[width=.45\linewidth]{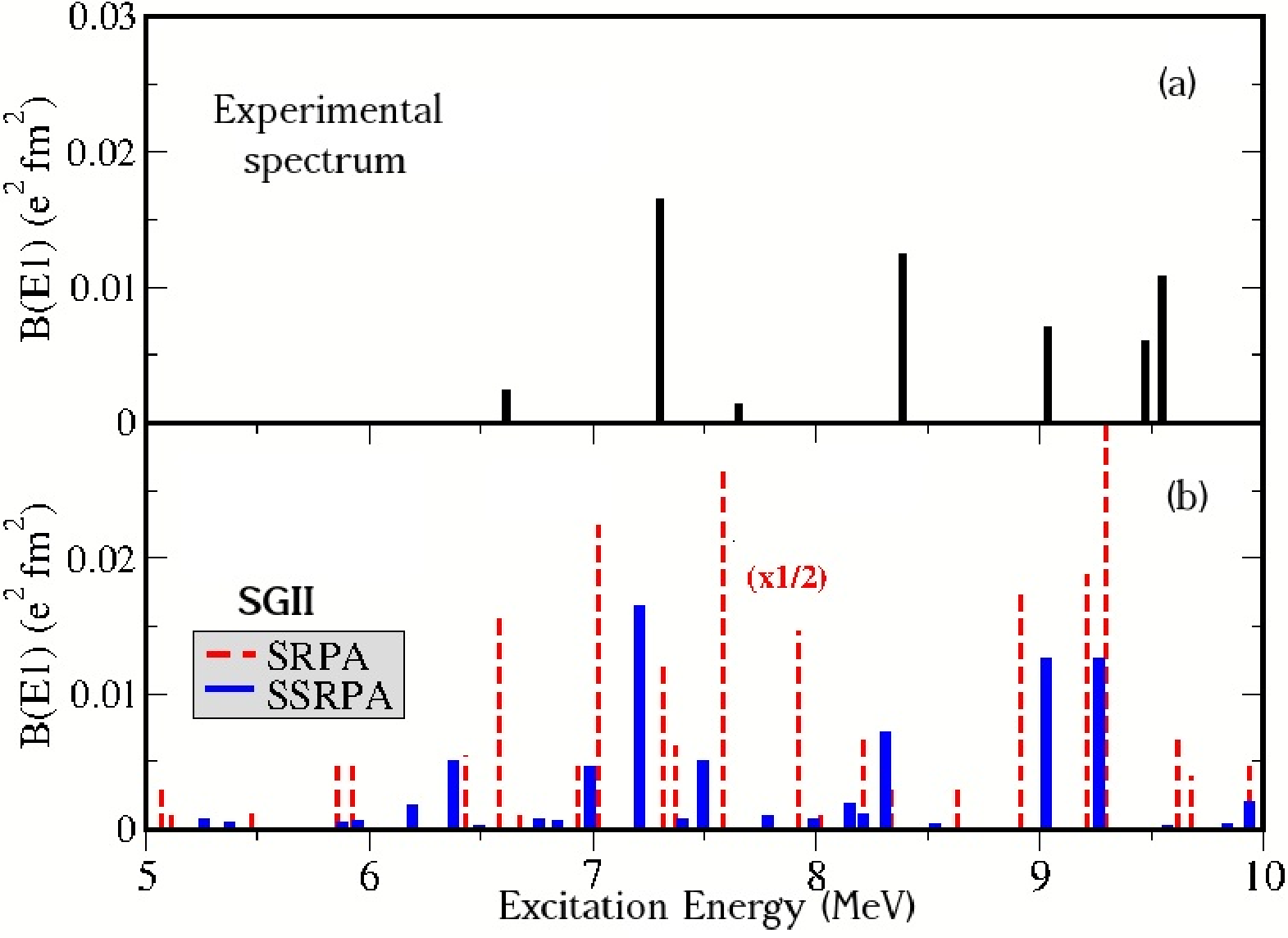}\hfill
	\includegraphics[width=.45\linewidth]{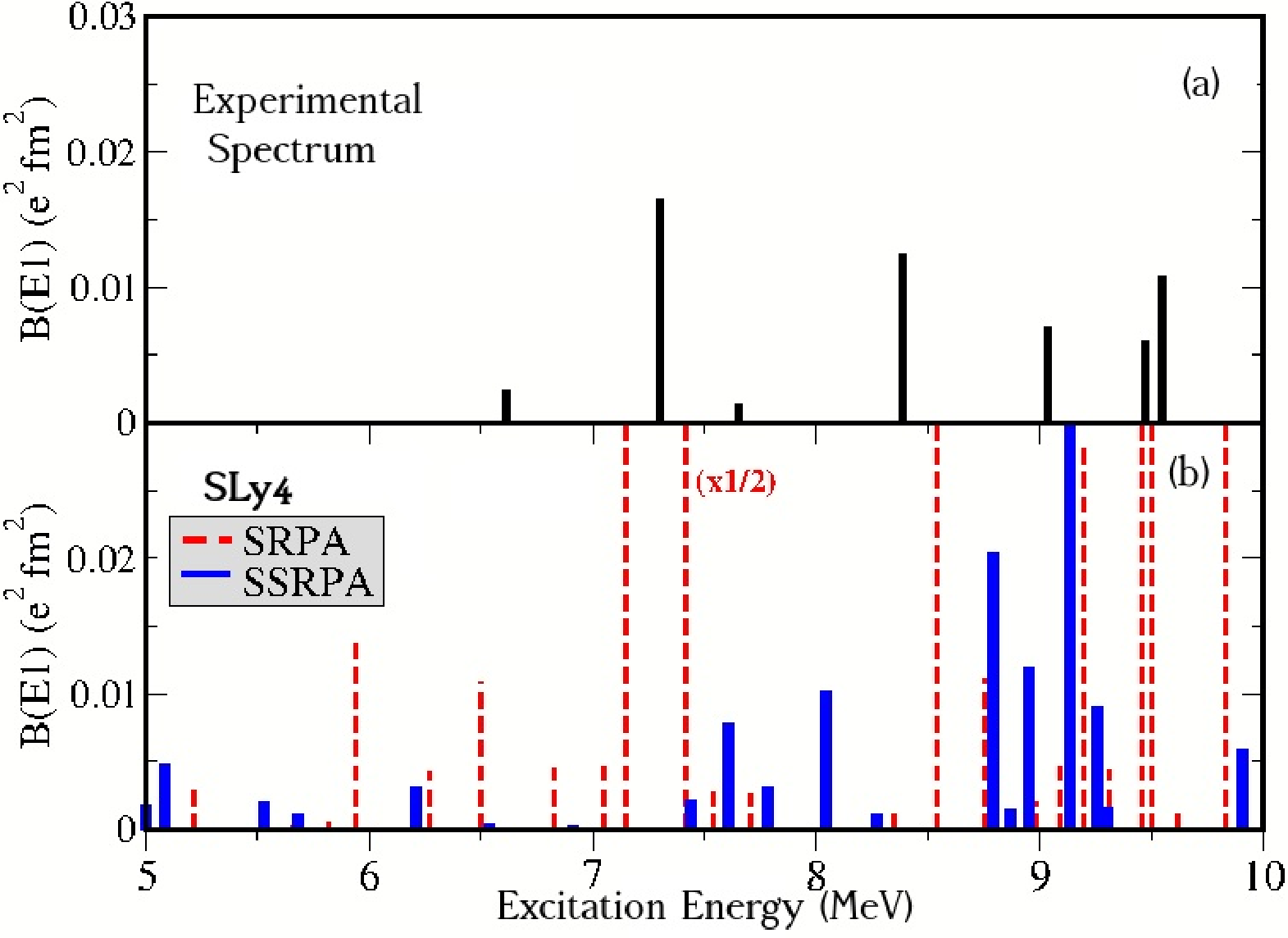}
	\caption{Experimental B(E1) values (panels (a)) from Ref. \cite{Hartmann2002} compared with the theoretical ones (panels (b)) for $^{48}$Ca obtained within the standard SRPA (dashed red bars; the values have been divided by 2) and with the SSRPA (blue thick bars), employing the Skyrme parametrization SGII (left side) and SLy4 (right side). Adapted from Ref. \cite{Gambacurta2018}.}
	\label{Fig:Ca48_Exp_SSRPA}
\end{figure}

On the left (right) side of Figure \ref{Fig:Ca48_Exp_SSRPA}, the experimental transition probabilities are compared with the theoretical ones, for the SGII (SLy4) force. One can see that the standard SRPA model significantly overestimates the experimental total strength. The SSRPA model, however, demonstrates a very good agreement with the experimental data. The model accurately reproduces the fragmentation of states and the energy positions of the main peaks. The total strength integrated between 5 and 10 MeV is summarized in the first row of Table \ref{Tab:Ca48-SSRPA-Exp}. The table presents the experimental result (first column), the SRPA values (second column), and the SSRPA values (third column) for the SGII parametrization. The SRPA value is eight times larger than the experimental result, whereas the SSRPA summed strength is very close to the measured value. A similar trend is observed for the corresponding EWSR, shown in the second row. The corresponding centroid energies are reported in the third row.
Also in the case of the SLy4 parametrization, a similar behavior is found, e.g., a clear improvement with the subtraction procedure, which significantly reduces the strength. However, the SLy4 results are less satisfactory compared to the SGII parametrization in terms of agreement with experimental data. The fragmentation of strength is less accurately reproduced, and the summed strength between 5 and 10 MeV is less consistent with the experimental value compared to the SGII case. Nevertheless, the subtraction procedure yields a clear improvement. The summed strength between 5 and 10 MeV without subtraction is 15 times larger than the experimental value (fourth column of Table \ref{Tab:Ca48-SSRPA-Exp}) and is reduced to twice the experimental measurement by the subtraction procedure (fifth column of Table \ref{Tab:Ca48-SSRPA-Exp}).

\begin{figure}
	\includegraphics[width=.45\linewidth]{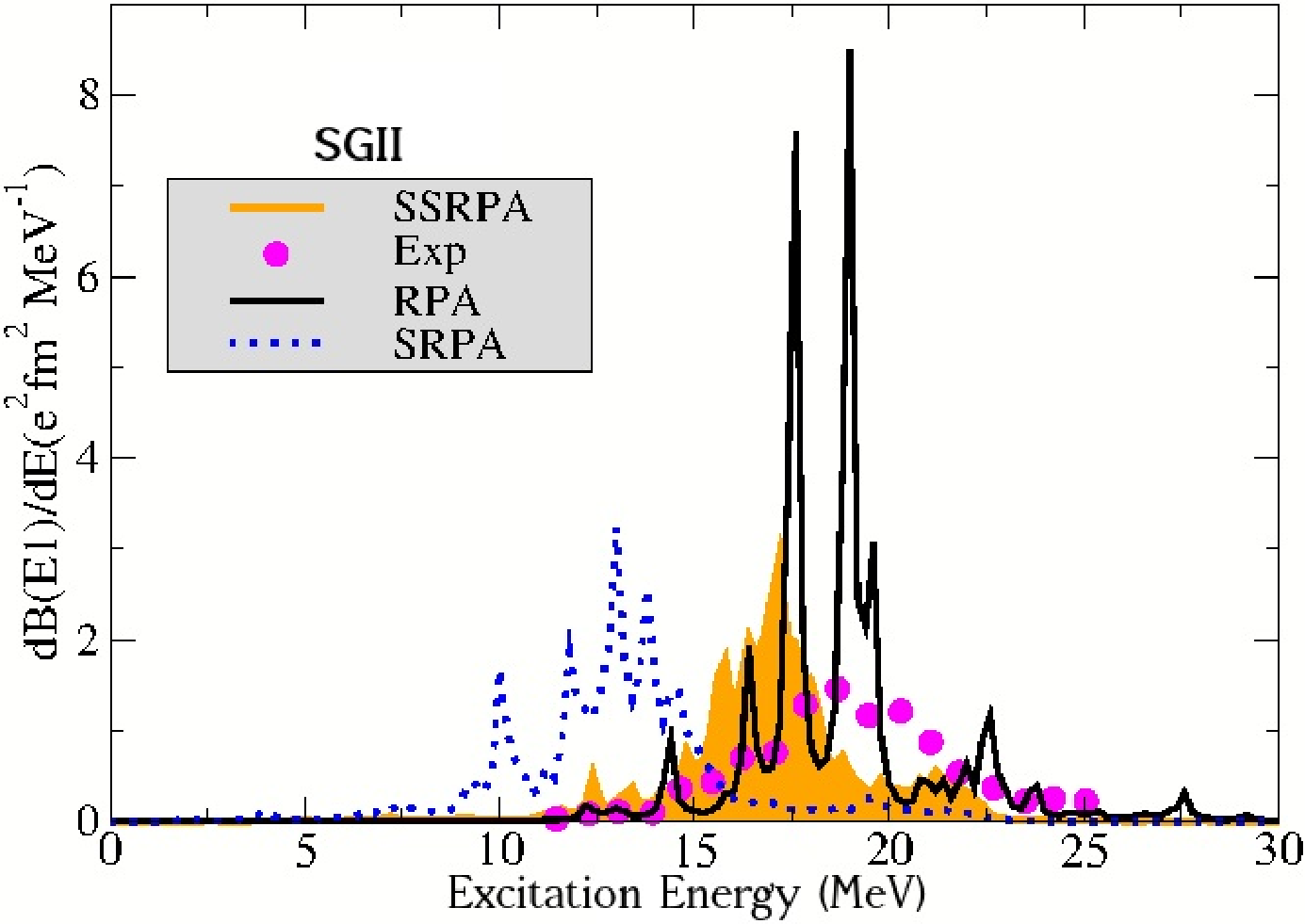}\hfill
	\includegraphics[width=.45\linewidth]{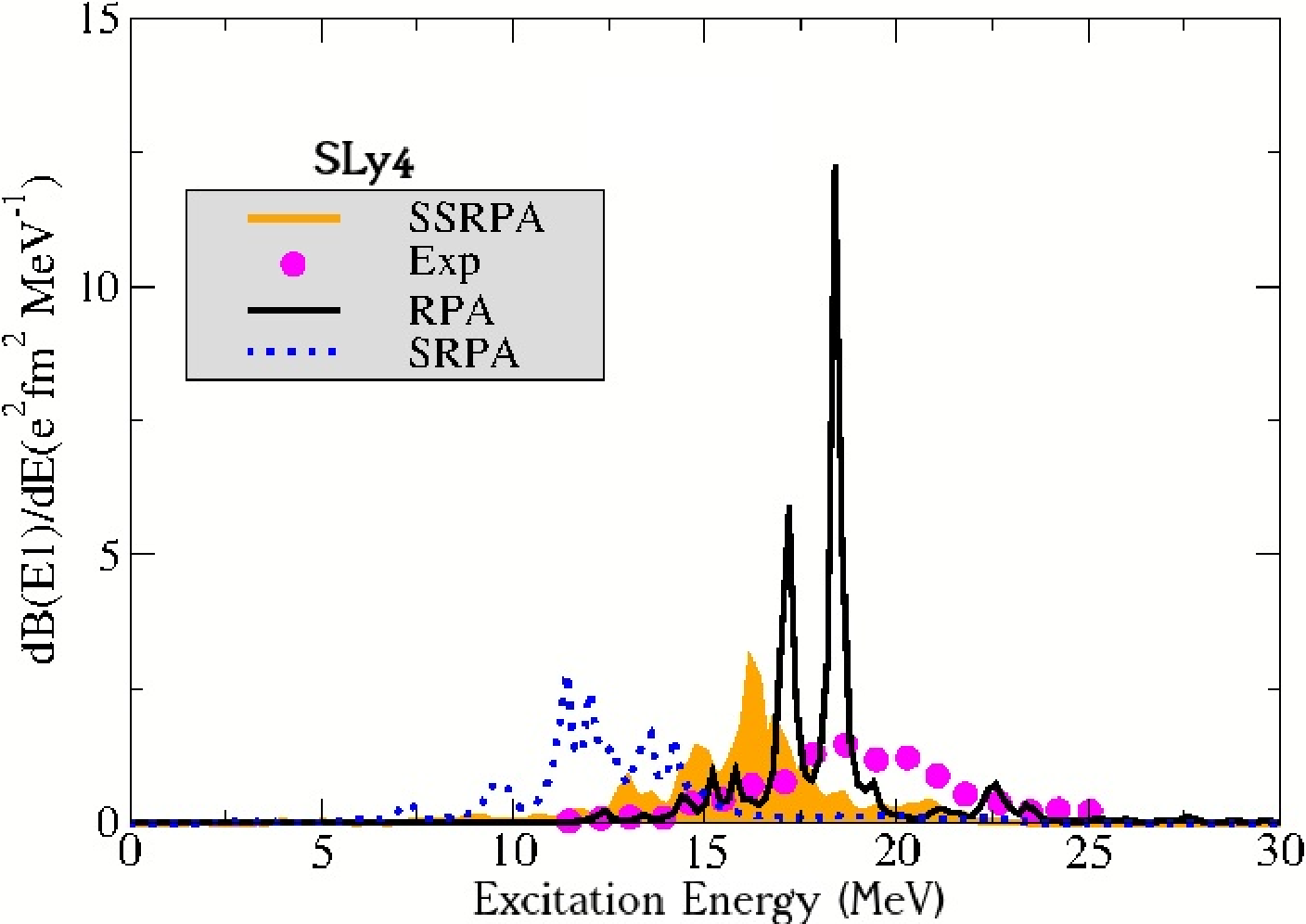}
	\caption{ Dipole strength distributions for $^{48}$Ca obtained in RPA (solid black line), SRPA (blue dotted line), and SSRPA (orange line and area), compared with the experimental distributions (magenta circles) of Ref. \cite{Birkhan2017}. Left side: SGII interaction. Right side: SLy4 interaction. Adapted from Ref. \cite{Gambacurta2018}.}
	\label{Fig:Ca48-SSRPA-GDR}
\end{figure}

\begin{wraptable}{r}{0.47\textwidth}
	%\begin {table} 
	%\begin{center}
	\begin{tabular}{cccccc}
		% \hline
		% \hline
		\hline
		\hline
		& Exp & SRPA & SSRPA & SRPA & SSRPA\\
		& & SGII & SGII & SLy4 & SLy4 \\
		\hline
		$m_0$ & 0.068 & 0.563 & 0.078 & 1.012& 0.126\\
		& $\pm$ 0.008 & & & & \\
		\hline
		$m_1$ & 0.570 & 4.618& 0.621 & 8.795& 1.062\\
		& $\pm$ 0.062 & & & & \\
		$E_c$ &8.38&	8.20&7.96&8.69&8.43\\
		\hline
		\hline
	\end{tabular}
	%\end{center}
	\caption{ Experimental and theoretical $m_0=\sum B(E1)$ in ( e$^{2}$ fm$^{2}$) and $m_1=\sum_i E_i B_i(E1)$ in (MeV e$^{2}$ fm$^{2}$) 
		summed between 5 and 10 MeV. The corresponding centroids are shown in the third row in MeV units. Adapted from Ref. \cite{Gambacurta2018}.}
	\label{Tab:Ca48-SSRPA-Exp}
	%\end {table} 
\end{wraptable}
Let's now consider the IVGDR region, the experimental data \cite{Birkhan2017} spanning 15 to 25 MeV. Figure \ref{Fig:Ca48-SSRPA-GDR} presents the strength distributions calculated with the SGII (left side) and SLy4 (right side) parametrization, comparing RPA, SRPA, and SSRPA results with experimental data. The theoretical distributions are folded with a Lorentzian function of 0.25 MeV width. Both SRPA and SSRPA exhibit more pronounced spreading, which, in these instances, reflects a physical width resulting from the dense distribution of $2p-2h$ configurations. In particular, the theoretical width demonstrates satisfactory agreement with the experimental distribution. The experimental centroid energy ($E_C$) and width ($\Gamma$) values are 18.9 $\pm$ 0.2 MeV and 3.9 $\pm$ 0.4 MeV, respectively \cite{Birkhan2017}. The theoretical centroid energies and widths are computed using the following expressions:
\begin{equation}
	\nonumber
	E_C = \frac{m_1}{m_0}, \;\;\; \Gamma_C=\sqrt{m_2 / m_0 -(m_1/m_0)^2}.
\end{equation}

 The RPA calculations with the SGII interaction, which accurately reproduce the centroid of the strength distribution ($E_C$ = 18.6 MeV), fail to produce a broad distribution, instead showing the strength divided between two main discrete states. The SRPA model naturally generates physical fragmentation ($\Gamma$ = 2.4 MeV), but significantly shifts the spectrum to lower energies ($E_C$ = 13.5 MeV) compared to RPA and experimental data. The SSRPA model effectively corrects this shift ($E_C$ = 17.4 MeV), although a slight underestimation of approximately 1.5 MeV with respect to the experimental centroid remains. The width remains similar to the SRPA case ($\Gamma$ = 2.5 MeV). The discrepancy in width, approximately 1 MeV, could arise from coupling to the continuum (escape width), which is not considered, or from coupling to more complex configurations.
\begin{figure}
	\includegraphics[width=.45\linewidth]{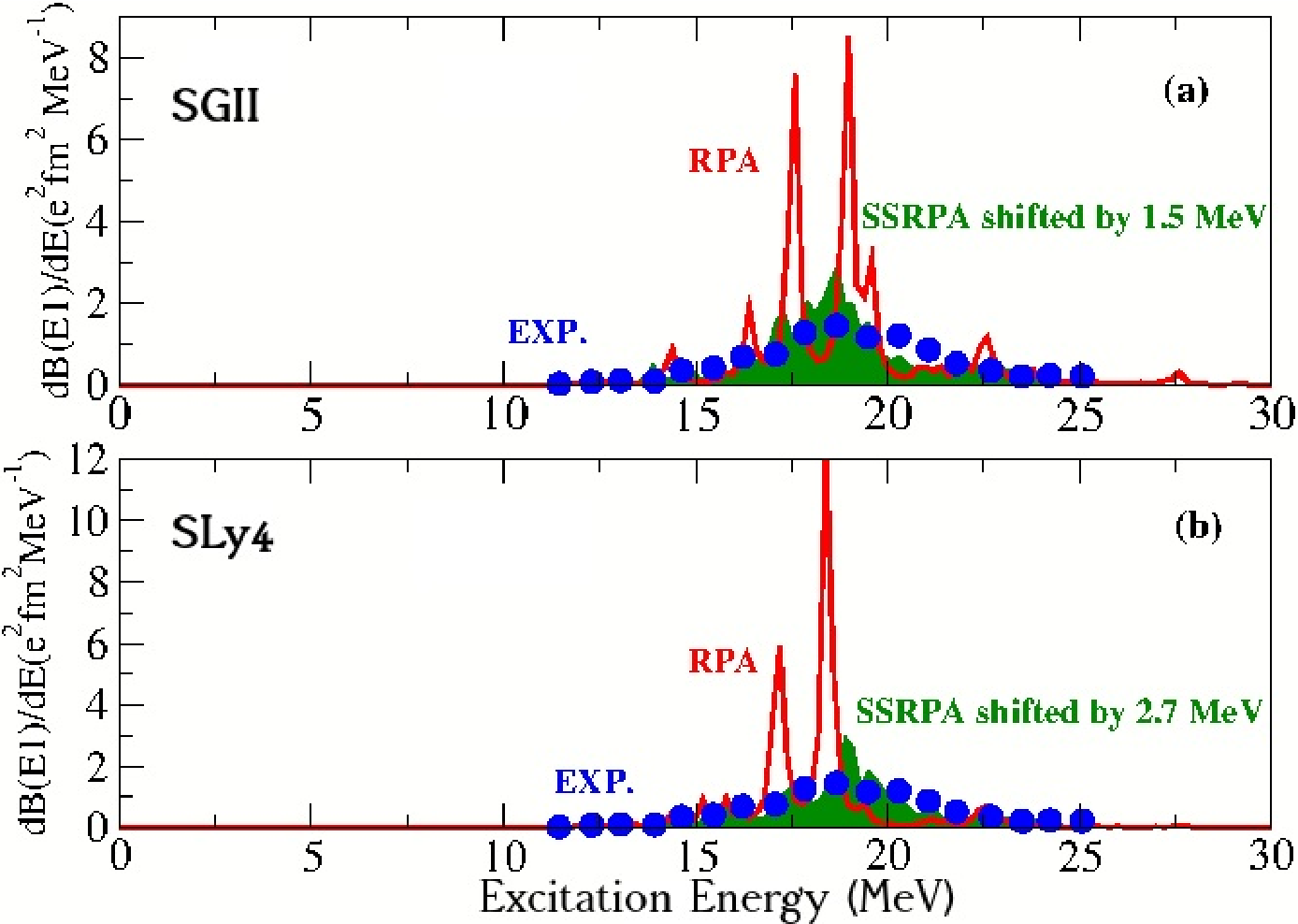}\hfill
	\includegraphics[width=.45\linewidth]{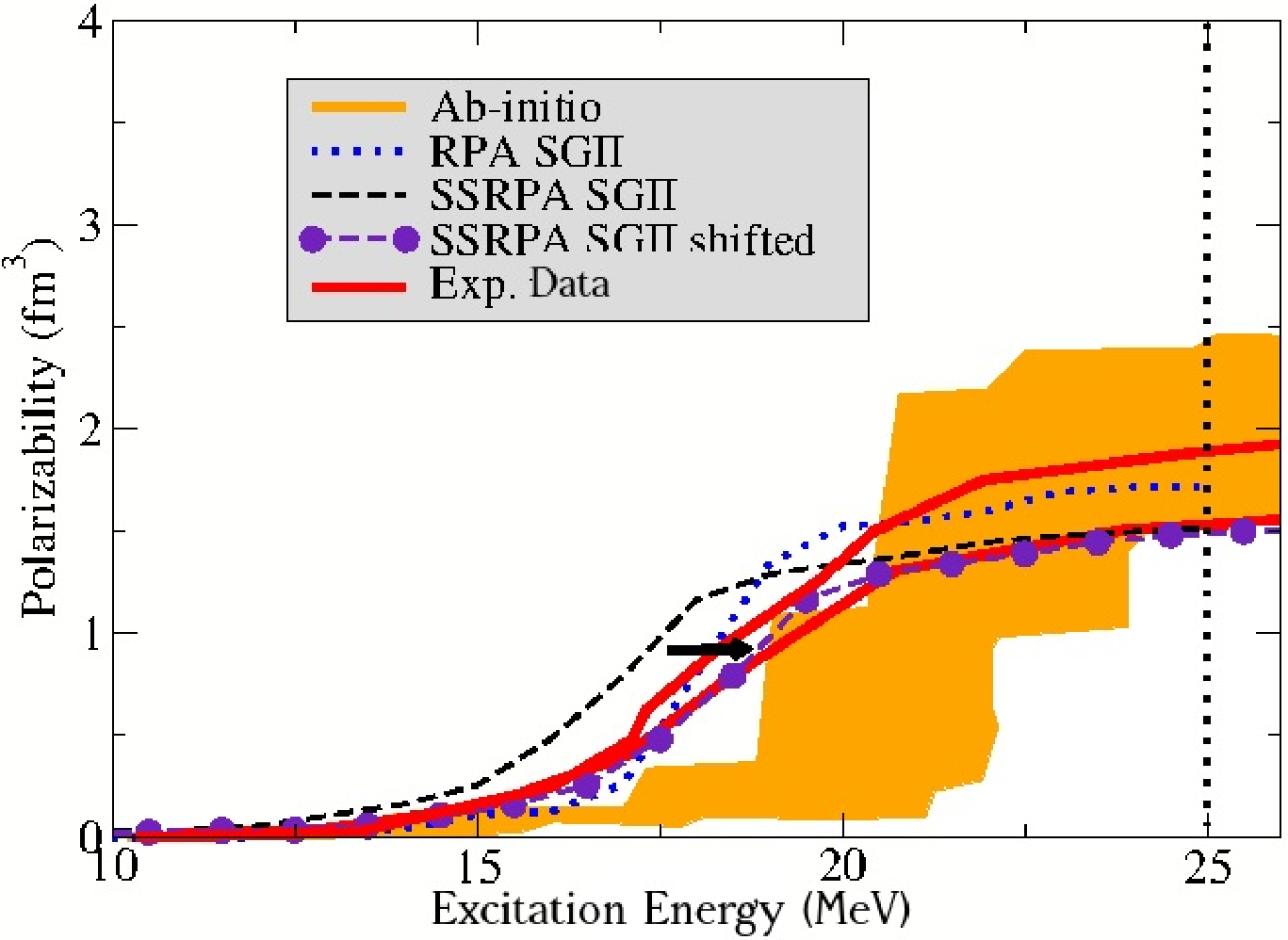}
	\caption{Left side: SSRPA results for $^{48}$Ca shifted by 1.5 MeV (green area) compared with the RPA strength (red line) and with the experimental values (blue circles), obtained with the parametrization SGII in panel (a). In panel (b), same results are shown but for the SLy4 case, the shift being in this case of 2.7 MeV. Right side:
		Electric dipole polarizability as a function of the excitation energy. The area between the two solid red lines correspond to the experimental data, while the
		orange area to {\it ab--initio} results. Results obtained with the RPA (blue dotted line) and the SSRPA (black dashed line) models are displayed for the SGII case. The violet circles show the SSRPA results shifted upwards by 1.5 MeV. Adapted from Ref. \cite{Gambacurta2018}.	}
	\label{Fig:Ca48-SSRPA-POLA}
\end{figure}
The right side panel of Figure \ref{Fig:Ca48-SSRPA-GDR} show the results for the SLy4 parametrization. Similar observations are made. The theoretical centroid energies are 18.0, 13.4, and 16.2 MeV for RPA, SRPA, and SSRPA, respectively. The widths are 2.5 and 2.1 MeV for SRPA and SSRPA, respectively. For the IVGDR region, the SLy4 parametrization yields less satisfactory results than SGII. Specifically, the SSRPA centroid energy is more than 2.5 MeV lower than the experimental centroid.

The left side of Figure \ref{Fig:Ca48-SSRPA-POLA} displays the SSRPA strength distributions obtained with SGII (a) and SLy4 (b), shifted upwards by 1.5 MeV (a) and 2.7 MeV (b), respectively, alongside the RPA strength distributions and experimental data. As mentioned, the SSRPA calculations were performed by employing the diagonal approximation in the subtraction procedure, so these shift might be partially reduced when the full subtraction is applied.

The electric dipole polarizability $\alpha_D$ can be calculated by means of the following relation
\begin{equation}
	\alpha_D = \frac{8 \pi}{9} \int \frac{B(E1,E_x)}{E_x} dE_x.
	\label{polari}
\end{equation}
The photoabsorption cross section was measured between 10 and 25 MeV \cite{Birkhan2017}, with a corresponding contribution to the electric dipole polarizability of 1.73 $\pm$ 0.18 fm$^3$. 
%Experimentally, the contribution below 10 MeV is negligible. Using the SGII and SLy4 parametrizations, the study finds the following values below 10 MeV: RPA yields $\alpha_D$ = 6 $\times$ 10$^{-4}$ and 3 $\times$ 10$^{-3}$ fm$^3$, while SSRPA produces 10$^{-3}$ and 5 $\times$ 10$^{-2}$ fm$^3$, respectively. 

The right side of Figure \ref{Fig:Ca48-SSRPA-POLA} presents the calculated $\alpha_D$ values as a function of the integration upper limit, extending to 25 MeV. Only the SGII results are shown. The calculated curve is compared with experimental and theoretical results extracted from Figure 4(b) of Ref. \cite{Birkhan2017}. Integrating up to 25 MeV is insufficient to achieve saturation of the electric dipole polarizability. To obtain converged experimental values, the $^{48}$Ca data  were  combined (up to 25 MeV) with those from $^{40}$Ca \cite{Ahrens1975} and correcting for the needed shift as explained in Ref. \cite{Hashimoto2015}. 
%They shifted the $^{40}$Ca photoabsorption data from Reference \cite{arhens} based on the centroid energy difference predicted by the formula $E_C=31.2 A^{-1/3}+20.6 A^{-1/6}$ \cite{mass}. 
The SRPA-based calculations are limited by the computational demands. Therefore, the integration is performed only up to 25 MeV, and the results are compared to the corresponding 25 MeV data from Reference \cite{Birkhan2017} (area between the red solid lines up to the vertical dotted line). The theoretical \textit{ab initio} results  based on coupled-cluster calculations \cite{Miorelli2016,Bacca2014}, are also plotted. These results are represented by an orange band, corresponding to various chiral Hamiltonians with interactions that yield reasonable saturation properties for symmetric nuclear matter. A comparison of the \textit{ab initio} results with the experimental band (area between the two red lines) reveals that the theoretical strength distributions, centered around 20 MeV, are significantly less spread than the experimental ones. The theoretical band shows a rapid increase in dipole polarizability around 20 MeV, whereas the experimental $\alpha_D$ increases more gradually in the IVGDR region. On the contrary, the SSRPA results show a much smoother behaviour following better the experimental one.  %Furthermore, the \textit{ab initio} calculations overestimate the centroid energy, placing it above approximately 20 MeV.

%The SGII-SSRPA curve in this study closely follows the experimental profile in the GDR region, accurately reproducing the spreading, as indicated by the slope. The underestimation of the centroid is evident as a slight overall shift of the curve relative to the experimental band. Shifting the curve upwards by 1.5 MeV (violet circles and dashed line) further illustrates the precise reproduction of the slope. The RPA results, like the \textit{ab initio} calculations, show a steep increase in polarizability within a narrow energy range. The Skyrme RPA polarizability at 25 MeV aligns well with the experimental band, suggesting good agreement. It is important to note that, upon integrating to higher energies where polarizability converges, the RPA and SSRPA polarizabilities must converge to the same value, by construction. The subtraction procedure ensures identical moments $m_{-1}$ in both RPA and SSRPA. SSRPA convergence is expected to be slower. A small difference between the two calculations is anticipated due to the diagonal approximation used in the subtraction procedure's corrective term.
%\newpage
\subsubsection{The $^{68}$Ni case}
\label{Sec:Applications_SSRPA_Dipole_Ni68}
The SSRPA approach was also applied to the study of the dipole response in $^{68}$Ni in Ref. \cite{Grasso2020}, with particular attention on the low-energy part of the spectrum.
From the experimental point of view, the first measurement of the low-lying dipole strength in the unstable nucleus $^{68}$Ni was performed, using virtual photon scattering at 600 MeV/nucleon (relativistic Coulomb excitations) at GSI \cite{Wieland2009}. Their findings indicated a strength concentration around 11 MeV, contributing 5\% to the EWSR . Subsequently,  relativistic Coulomb excitation experiments were also performed at GSI \cite{Rossi2013},  to determine the electric dipole polarizability of the same nucleus. This study yielded a slightly different centroid location of 9.55 MeV and a 2.8\% contribution to the EWSR. The discrepancy in centroid values was attributed to a potential energy-dependent branching ratio. The lower-energy portion of the PDR is expected to exhibit a mixed isoscalar/isovector nature, usually referred as isospin splitting \cite{Savran2013,Bracco2019}, excited by both isoscalar and isovector probes, while the higher-energy part, approaching IVGDR tail, is predominantly an isovector excitation. In Ref. \cite{Martorana2018}, the first measurement of $^{68}$Ni using an isoscalar probe (isoscalar $^{12}$C target) was performed at INFN-LNS , reporting a centroid around 10 MeV and a 9\% contribution to the EWSR.
\begin{figure}
	\includegraphics[width=.44\linewidth]{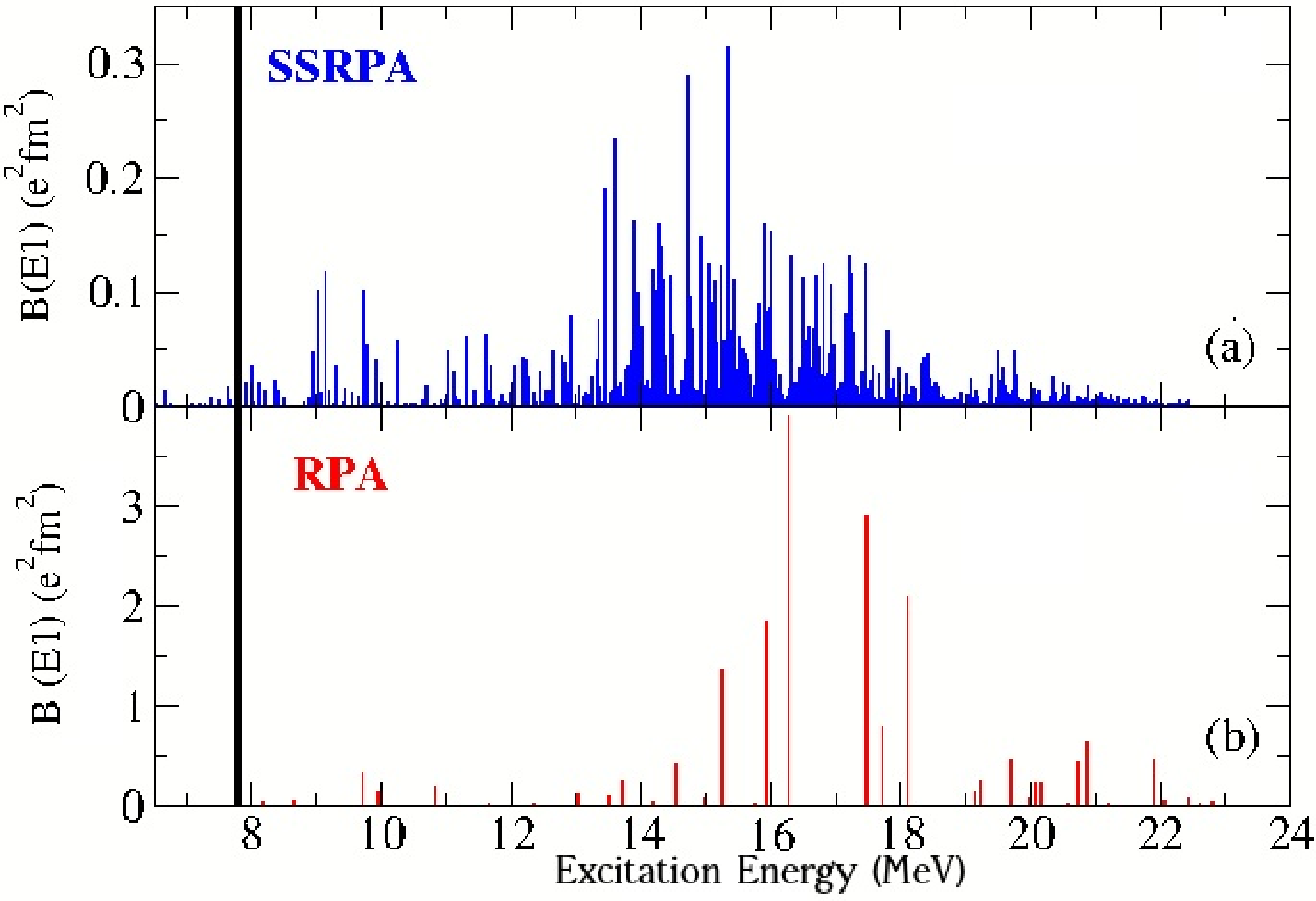}\hfill
	\includegraphics[width=.46\linewidth]{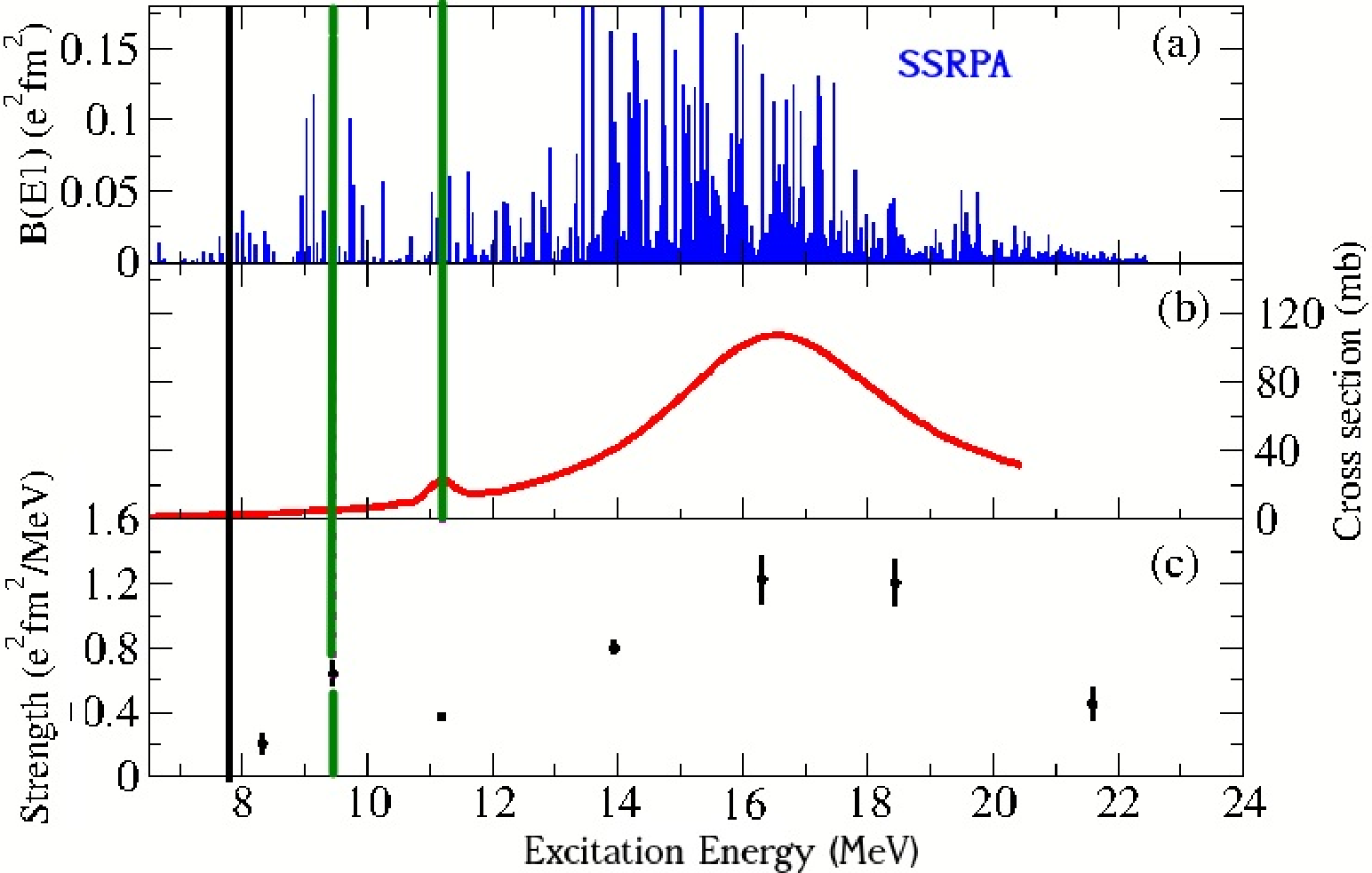}
	\caption{Left side: SSRPA dipole strength distribution obtained for $^{68}$Ni with the parametrization SGII in SSRPA (panel (a)) and RPA (panel (b)). The vertical black line represents the neutron threshold.
		Right side: (panel (a)) SSRPA dipole strength distribution obtained for $^{68}$Ni with the parametrization SGII; (panel (b)) photoabsorption cross section extracted from \cite{Wieland2009}; (c) $E1$ strength distribution from \cite{Rossi2013}. The vertical black line represents the neutron threshold. The vertical green lines correspond to the energy values of the two experimental low--energy centroids measured in Ref. \cite{Wieland2009} (panel (b)) and in Ref. \cite{Rossi2013} (panel (c)). Adapted from Ref. \cite{Gambacurta2020}.}
	\label{Fig:Ni68_Dipole_Full}
\end{figure}

Given these three slightly varying experimental centroids, located at approximately 11, 9.55, and 10 MeV, it is interesting to study the impact of the BMF correlations introduced in the SSRPA. On the left side of Figure \ref{Fig:Ni68_Dipole_Full}, the SSRPA (panel(a)) and RPA (panel(b)) strength distributions for the isovector dipole transition operator (\ref{Eq:Op-J1-IV-CM})
%\begin{equation}
%	\label{op}
%	T_{1M}= \frac{Z}{A}\sum_{n=1,N} r_n Y_{1M} (\hat{r}_n)-\frac{N}{A}\sum_{p=1,Z} r_p Y_{1M}(\hat{r}_p)
%\end{equation}
are plotted for the SGII interaction \cite{SGII}. The procedure described in Ref. \cite{RocaMaza2012} was applied in order to project out possible admixtures with spurious components.

 %Since the calculations are fully self--consistent, it was found that the transition probability and the EWSR percentage are slightly affected by (by less than 0.01 \%).  In the figure, the vertical line located at 7.792 MeV indicates the neutron threshold.
\begin{wrapfigure}{r}{0.45\textwidth}
\vspace{-13mm}
	\centering
% 	\vspace{-5mm}
	\includegraphics[width=0.4\textwidth]{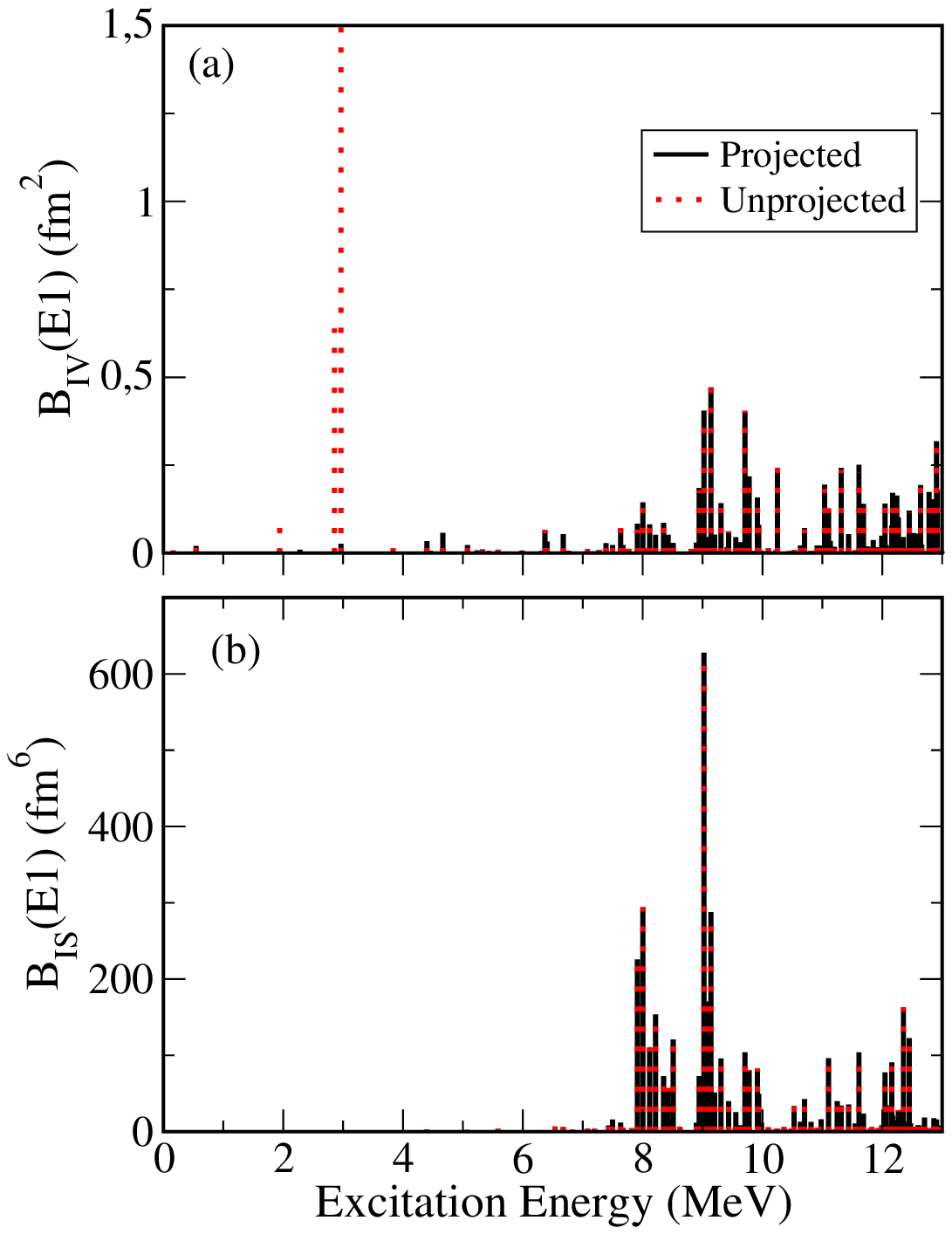}
	 	\vspace{-5.5mm}
	\caption{Isovector (a) and isoscalar (b) low--lying strength
		distributions in $^{68}$Ni, corresponding respectively
		to the operators ((\ref{Eq:Op-J1-isov})) and ((\ref{Eq:Op-J1-isos-r3-corrected})). Full lines represent the
		projected results whereas the dotted lines are obtained without correcting for spurious
		components. Adapted from Ref. \cite{Gambacurta2020}.}
	% 	\vspace{-7mm}
	\label{Fig:Ni68_Dipole_Spurious}
\end{wrapfigure}

The energy region of the spectrum above approximately 12 MeV corresponds to the IVGDR mode. One can see that the SSRPA spectrum exhibits a significantly higher density of states compared to the RPA spectrum, introducing the physical fragmentation and width of the resonance. In the region below 12 MeV the SSRPA spectrum displays a greater concentration of strength, which, as will be discussed, results in a higher percentage of the EWSR calculated up to a low-energy cutoff that separates the PDR from the IVGDR.  Furthermore, the SSRPA low-energy distribution reveals peaks clustered around 9 and 10 MeV, along with several peaks just above 11 MeV. This suggests that the SSRPA model predicts a significant amount of strength in the energy regions where the three experimental centroids are located. In contrast, the RPA spectrum's low-energy region contains far fewer peaks, with the highest located between 9.5 and 10 MeV. 
Another isolated peak resides below 11 MeV, and virtually no strength is observed above 11 MeV. The coupling with the $2p-2h$ configurations strongly enhances the strength fragmentation, leading to improved coverage of the experimental centroid region in SSRPA.
The figure illustrates a separation at approximately 12 MeV between the low-energy strength and the strength associated with the tail of the IVGDR. The percentage of the EWSR up to 12 MeV is  is 3.75\% for SSRPA, compared to 2.35 \% for RPA. %For consistency, all subsequent EWSR percentage calculations are calculated up to 12 MeV.

\begin{wrapfigure}{r}{0.5\textwidth}
	\centering
	% 	\vspace{-3.5mm}
	\includegraphics[width=0.45\textwidth]{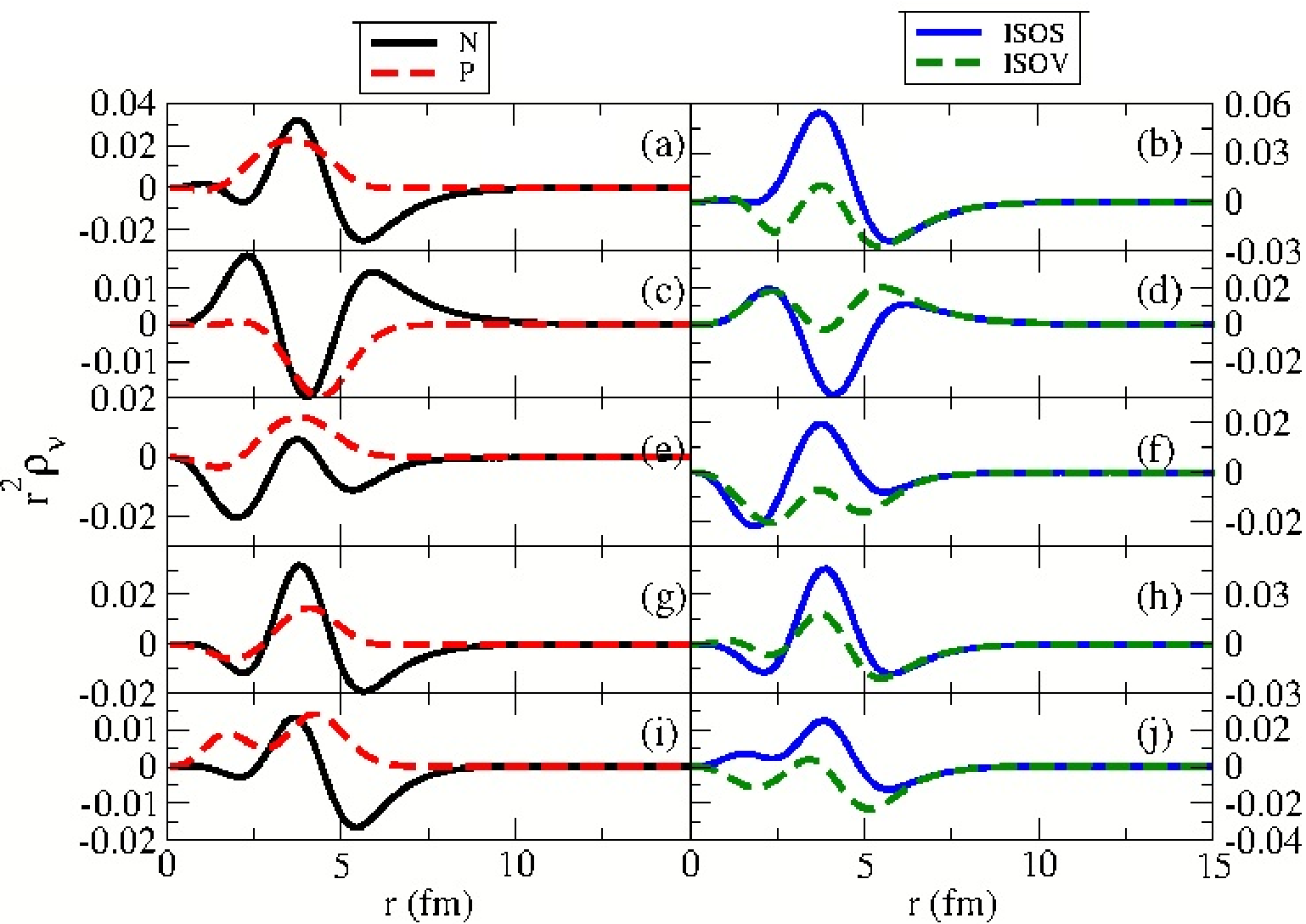}
	% 	\vspace{-1.5mm}
	\caption{ Transition densities multiplied by $r^2$ for the peaks located at 9.14 ((a) and (b)), 9.70 ((c) and (d)), 10.25 ((e) and (f)), 11.10 ((g) and (h)), and 11.31 ((i) and (j)) MeV. Neutron and proton transition densities: panels (a), (c), (e), (g), and (i), and isoscalar and isovector transition densities: panels (b), (d), (f), (h), and (j). The used Skyrme parametrization is SGII. Adapted from Ref. \cite{Gambacurta2020}.}
	% 	\vspace{-7mm}
	\label{Fig:Ni68_Dipole_TDR}
\end{wrapfigure}
 The experimental low-energy contributions to the EWSR were found to be 5, 2.6, and 9\% in the three experiments referenced. The right side of Figure \ref{Fig:Ni68_Dipole_Full} presents a comparison of SSRPA results with experimental data. This comparison focuses on the energy peak locations of the measured and predicted excitation spectra, noting that the vertical axes represent different quantities across the three panels. Panel (a) displays the SSRPA B(E1) distribution. Panels (b) and (c) show the photoabsorption cross section \cite{Wieland2009} and E1 strength distribution \cite{Rossi2013}. We recall that the low-energy centroid from Ref. \cite{Martorana2018} is located at 10 MeV, while the centroids from Refs \cite{Wieland2009} and \cite{Rossi2013} are situated at approximately 11 and 9.55 MeV, respectively, as shown on the right side of Figure \ref{Fig:Ni68_Dipole_Full}.

In order to have a more complete and quantitative insight on the isospin nature of the low--lying excitations,
we plot in Figure \ref{Fig:Ni68_Dipole_Spurious} the transition probabilities associated with the dipole isovector   (\ref{Eq:Op-J1-isov})
%\begin{equation}
%	\label{op_iv}
%	T_{1M}^{IV}= \sum_{n=1,N} r_n Y_{1M} (\hat{r}_n)-\sum_{p=1,Z} r_p Y_{1M}(\hat{r}_p)
%\end{equation}
and the isoscalar operator (\ref{Eq:Op-J1-isos-r3-corrected}).
%
%\begin{equation}
%	\label{op_is}
%	T_{1M}^{IS}= \sum_{i=1,A} (r_i^3 - \frac{5}{3} \langle r^2 \rangle r_i )Y_{1M} (\hat{r}_i)
%\end{equation}
%dipole operators.
 Both unprojected and projected (that is, with spurious--component corrections performed as in Ref. \cite{RocaMaza2012}) results are shown. The figure also clearly demonstrates that the projection of spurious components predominantly affects the lowest states, located around 3 MeV, which correspond to the spurious mode. The comparison shows also a very strong isospin mixing that might explain that different states may be excited with different probes. More specifically, the isovector distribution the strength values are more or less comparable among themselves in the whole 
energy window, from $\sim$ 7 to $\sim$ 13 MeV, in the isoscalar distribution one observes that the strength is more important in the lower--energy part (below $\sim$ 10 MeV) than in the higher--energy part of the spectrum. This indicates the presence of an isoscalar/isovector splitting.

Figure \ref{Fig:Ni68_Dipole_TDR} displays transition densities for selected peaks within the low-lying spectrum, whose energies are reported in the caption of the figure. 
%Specifically, it presents densities for two peaks near 9.55 MeV (9.14 MeV, panels (a) and (b); 9.70 MeV, panels (c) and (d)), one peak around 10 MeV (10.25 MeV, panels (e) and (f)), and two peaks near 11 MeV (11.10 MeV, panels (g) and (h); 11.31 MeV, panels (i) and (j)). Neutron/proton transition densities are shown in panels (a), (c), (e), (g), and (i), while isoscalar/isovector transition densities are shown in panels (b), (d), (f), (h), and (j).
 Considering the neutron and proton transition densities (left panels) one can notice a consistent dominant neutron contribution at the nuclear surface, a characteristic feature of a pygmy excitation according to its traditional interpretation. The analysis of the isoscalar and isovector transition densities (right panels) confirm the isosping mixing of these states. %In essence, there is no discernible isospin splitting. Such splitting would manifest as a mixed isoscalar/isovector nature in the lower energy region and a predominantly isovector nature in the higher energy portion of the low-lying spectrum.

\subsubsection{Magnetic dipole excitations}
\label{Sec:Applications_SSRPA_Magnetic}

The magnetic dipole ($M1$) excitations in atomic nuclei have been the subject of extensive experimental and theoretical investigations for several decades. The low-energy part ($\sim$ 3MeV) of the magnetic dipole ($M1$) response is typically interpreted as the orbital scissors mode and it appears mostly in deformed nuclei. At higher energy, the isovector spin-flip $M1$ excitation appears involving like particles is analogous to the charge-exchange GT transition. It is predominantly induced by $1p-1h$ transitions between spin-orbit partner states in the vicinity of the Fermi surface. The $M1$ excitation is a sensitive probe of the spin-isospin channel of the nuclear effective interaction. It exhibits a characteristic quenching phenomenon, a common issue in spin-isospin responses such as $\beta$ decay and GT strength. As it will be discussed in Section \ref{Sec:Applications_SSRPA_GT} and shown in Refs \cite{Gambacurta2020,Gambacurta2022,Sagawa2022,Sagawa2023,Gambacurta2025}, 
in the GT case, the inclusion of $2p-2h$ configurations within the SSRPA strongly reduces the quenching problem. In Ref. \cite{Yang2024}, the SSRPA was then applied also for the study of the magnetic dipole excitations for $^{48}$Ca, $^{90}$Zr, and $^{132}$Sn, associated to the transition operator (\ref{Eq:Op-J1-magnetic}).
%\begin{equation}
%	F=\frac{3}{4\pi}\sum_i^A\big[g_i^l \mathbf{l_i}+ g_i^s \mathbf{s_i}\big] \mu_N
%	\label{Eq:M1op}
%\end{equation}
%where $\mathbf{l_i}$ and $\mathbf{s_i}$ are the orbital and spin angular momenta, respectively, 
%$g_l$ and $g_s$ the corresponding gyromagnetic factors of nucleons and $\mu_N$ is the Bohr magneton of a nucleon \cite{Fujita2011}.

\begin{figure}
	\centering
	\includegraphics[width=.5\linewidth]{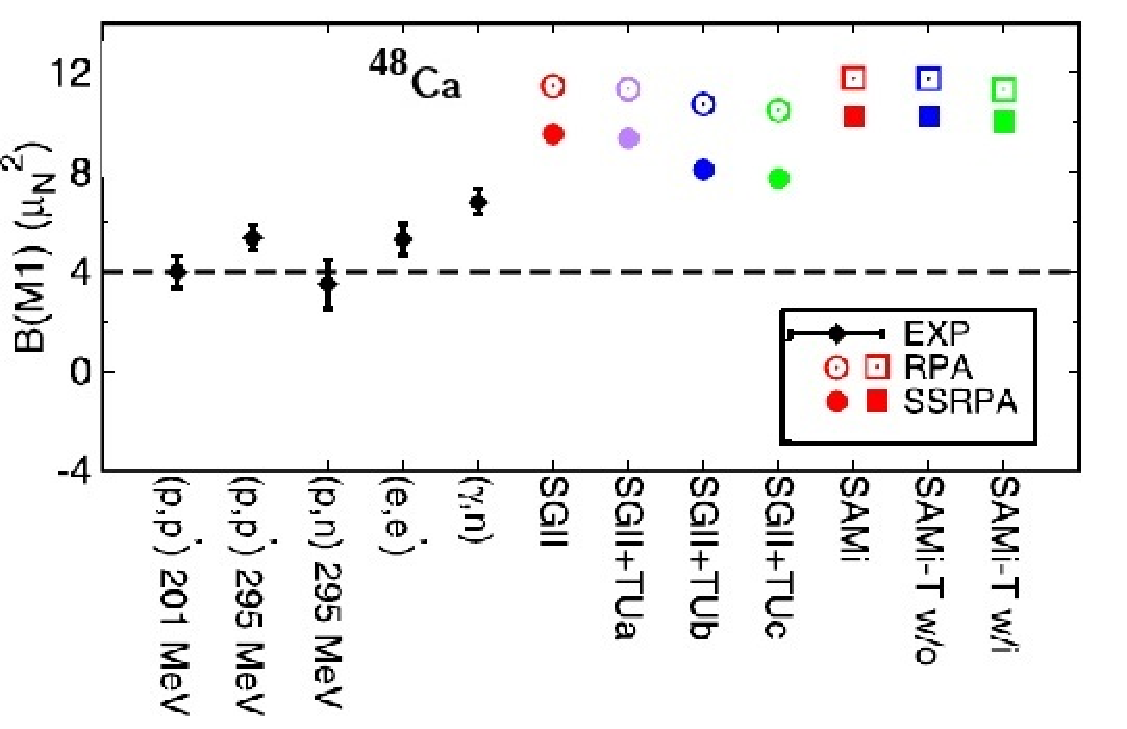}
	\caption{The cumulative sums of $B(M1)$ strength integrated up to 15 MeV in $^{48}$Ca obtained in different measurements compared with the RPA (empty symbols) and SSRPA (filled
		symbols) with different interactions. The experimental data \cite{Steffen1980,Steffen1983,Tompkins2011,Birkhan2016,Mathy2017} are shown
		by the black squares with error bars. Adapted figure from Ref. \cite{Yang2024}.}
	\label{Fig:Magnetic_1}
\end{figure}

%In Figure \ref{Fig:Magnetic_0} the convergence of the SSRPA strength distributions with respect to the energy cutoff on the $2p-2h$ configurations for $^{48}$Ca $^{90}$Zr and $^{132}$Sn is shown for the SGII\cite{SGII} force. 
%Similar checks were performed for other forces employed in the following, namely the SAMi \cite{SAMI} and SAMi-T \cite{SAMIT} in order to study the effect of the tensor on the response. 
In Figure \ref{Fig:Magnetic_1} the total $B(M1)$ strengths up to 15 MeV for $^{48}$Ca measured in several experiments are compared with those calculated in RPA and SSRPA using different forces. Different parameter set
tensor terms T and U for the SGII force are used, (see Ref. \cite{Yang2024} for more detail), together with the SAMi \cite{SAMI} and SAMi-T \cite{SAMIT}.  More precisely the SGII+TUa (T = 500, U = +150), the  SGII+TUb (T = 500, U = -280) , the SGII+TUc (T = 500, U = -350), and The SAMi-T (T = 415.45, U = -95.53) MeV fm$^5$ have been used.
%Keeping the triplet-even term T = 500 MeV $fm^5$ , the triplet-odd term is varied from a positive value U=150, to -280 and -350 MeV $fm^5$ for sets TUa, TUb, and TUc, respectively. For the SAMi-T, the parameters are T=415.45 and Y=-95.53 MeV $fm^5$. 
One can see that the SSRPA strengths are smaller than those of RPA, of the order of $\approx$10-20\%, because of the $2p-2h$ configuration mixing. The inclusion of the tensor force further reduces the strengths.

% The reduction is by about 11\% or 17\% with SAMi-T or SGII, respectively. The reduction is increased by adding the tensor force, being stronger in the SGII's parametrizations. 
% Small changes are introduced in the SAMi-T case, while in SGII+TUb they reduce the summed M1 strength further by about 12\% in SSRPA calculations, and produce a result close to the $(\gamma , n)$ experimental one. 
% As a general trend of the $2p-2h$ mixing and the tensor force with negative U term (SGII+TUb) produce a shift of the $M1$ strength towards higher energy, and the cumulative strength is quenched by about 30\%. These features are also consistent with the case of GT transitions discussed in Section \ref{Sec:Applications_SSRPA_GT}.
%The left side of Figure \ref{Fig:Magnetic_2} shows the unperturbed HF $1p-1h$ and RPA $M1$ excitation energies of SGII+TUb and SAMi-T without and with the tensor interactions for $^{48}$Ca (left panels) and $^{90}$Zr (right panels), respectively. At the HF level, the triplet-odd term acts uniquely on the energy splitting of neutron spin-orbit partners. The large negative $U$ value increases the spin-orbit splitting of like particles such that the tensor interaction of SGII+TUb with a larger negative $U$ gives a larger spin-orbit splitting in the two nuclei compared with those of SAMi-T, as seen in panels (a) and (c) of the right side of Figure \ref{Fig:Magnetic_2}. The strong $U$ value also acts strongly on the RPA correlations to push upwards the excitation energies of $M1$ states in the two nuclei $^{48}$Ca and $^{90}$Zr, as shown in panels (b) and (d) of the right side of Figure \ref{Fig:Magnetic_2}.

\begin{figure}
	\includegraphics[width=.49\linewidth]{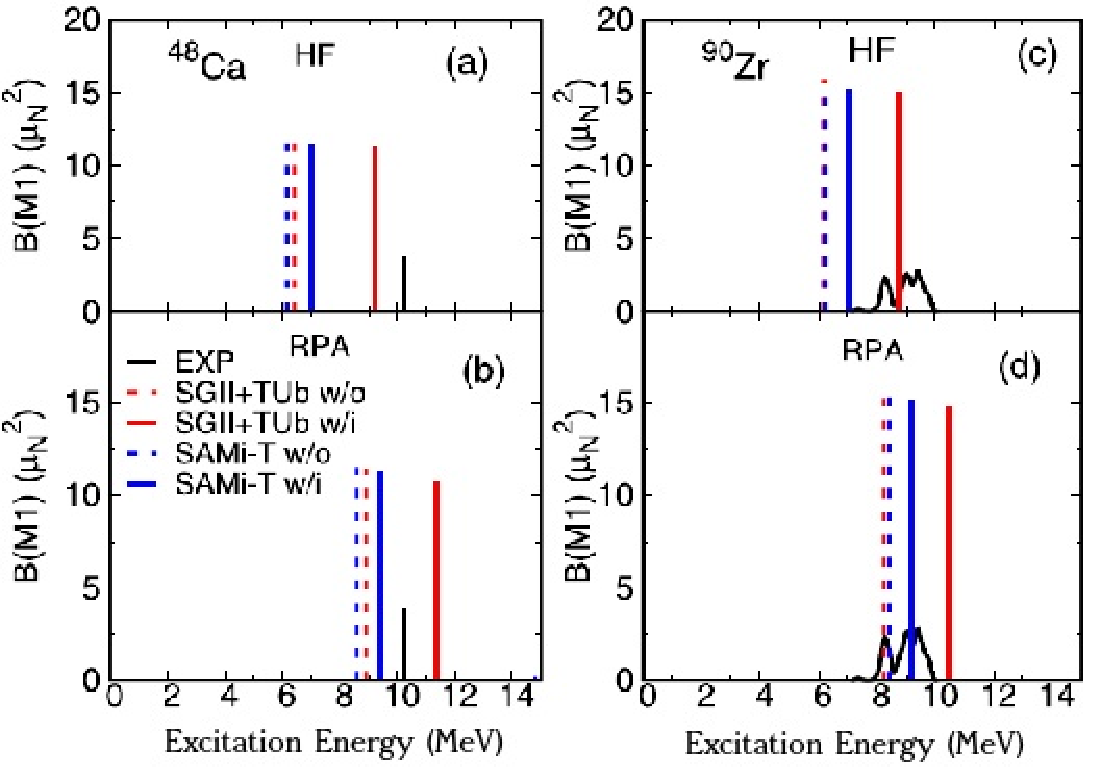}\hfill
	\includegraphics[width=.48\linewidth]{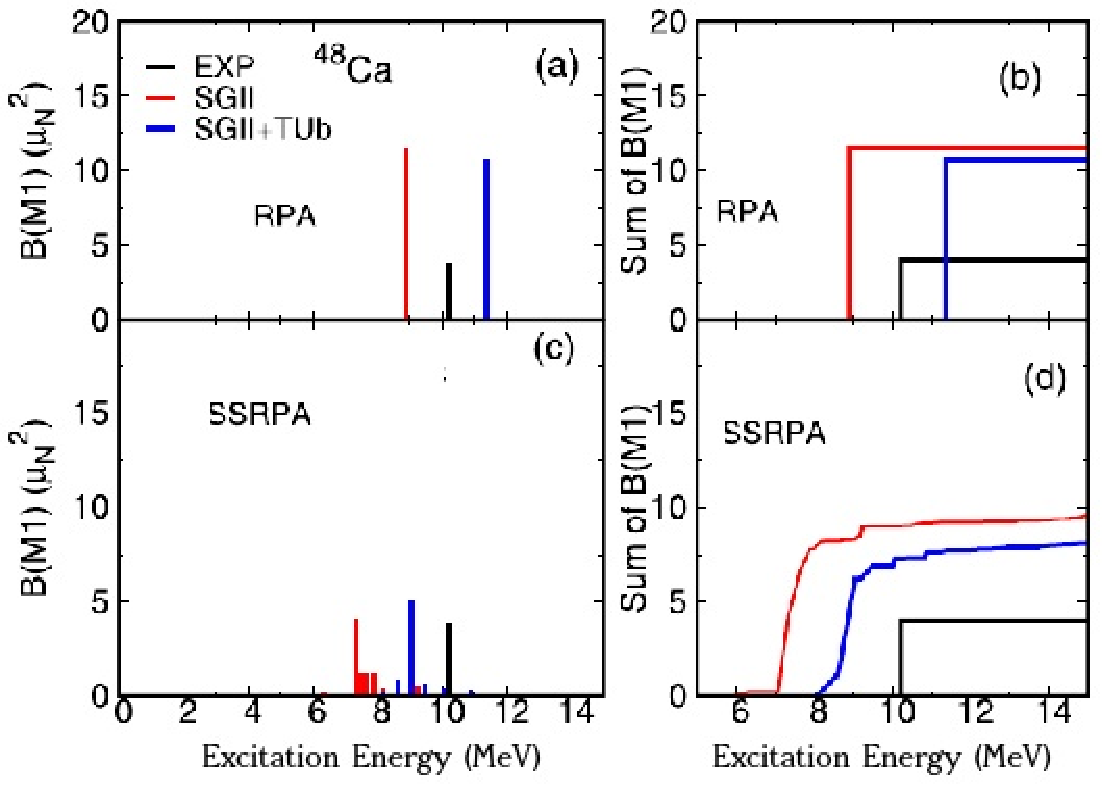}
	\caption{Left side: Unperturbed p-h and RPA strength distributions of $^{48}$Ca (left panels) and $^{90}$Zr (right panels) of M1 states calculated with the SGII+TUb and SAMi-T EDFs. The experimental data are from \cite{Birkhan2016}. Right side: $M1$ Strength distributions (left panels) and corresponding cumulative sums (right panels) for $^{48}$Ca calculated with the SGII, SGII+TUb EDFs in RPA (upper panels) and SSRPA (lower panels) compared with the experimental data \cite{Birkhan2016}. Adapted figures from Ref. \cite{Yang2024}.}
	\label{Fig:Magnetic_2}
\end{figure}

The unperturbed $\text{HF}$ $\text{1}p-\text{1}h$ and RPA $\text{M1}$ strengths, plotted on the left side of Figure \ref{Fig:Magnetic_2}, for $^{48}\text{Ca}$ (left panels) and $^{90}\text{Zr}$ (right panels), show a strong dependence on the $\text{U}$ value of the triplet-odd term in the SGII+TUb and SAMi-T interactions. At the $\text{HF}$ level, a large negative $\text{U}$ value increases the spin-orbit splitting for like particles. Thus, the SGII+TUb interaction, which has a larger negative $\text{U}$, yields a larger spin-orbit splitting in both nuclei compared to SAMi-T. This strong $\text{U}$ value also significantly influences RPA correlations, causing the $\text{M1}$ excitation energies to be pushed upwards in both $^{48}\text{Ca}$ and $^{90}\text{Zr}$ (panels (b) and (d)).
%%The right side of Figure \ref{Fig:Magnetic_2} shows the strength distributions and the corresponding cumulative sums of $M1$ transition in $^{48}$Ca calculated by RPA and SSRPA with SGII+TUb including tensor terms or not; SGII+TUb without tensor terms is actually SGII. As expected, the non-energy and inversely energy weighted sum rule moments $m_0$ and $m_{-1}$ are the same in RPA and SSRPA models, also for the $M1$ transition if the strength in the whole energy region is collected. The cumulative sums of $B(M1)$ are counted up to $E_{max} = 15$ MeV, to compare with the experimental measurement up to 15 MeV. The effects of the $2p-2h$ configuration mixings in SSRPA are clear in this figure. The RPA calculations either with or without tensor terms produce the main peaks with strengths larger than $10 \, \mu_N^2$, which is more than twice the experimental data. In SSRPA calculations, the $M1$ strengths are largely reduced, and the strength of the main peak ($\simeq 4.09 \, \mu_N^2$) becomes almost the same as the experimental one ($\simeq 4.0 \, \mu_N^2$). Besides the main peak, some states with small strength are distributed around the main peak in SSRPA. The tensor terms in SSRPA shift the main peak energy upwards by about 1.8 MeV, which is still almost 1 MeV lower than the experimental one. Panel (d) indicates that the cumulative sum of $B(M1)$ obtained by SSRPA up to 15 MeV is reduced by about 15\% of the total sum by the tensor correlations.
The $\text{M1}$ strength distributions and cumulative sums for $^{48}\text{Ca}$ are shown in the right side of Figure \ref{Fig:Magnetic_2}, using RPA and SSRPA with the $\text{SGII+TUb}$ interaction and without. The RPA calculations (with or without tensor terms) overestimate the main $\text{M1}$ peak strength, predicting values over $10 \, \mu_N^2$, which is more than double the experimental data.
 The SSRPA method significantly reduces the $\text{M1}$ strength of the main peak to $\simeq 4.09 \, \mu_N^2$, bringing it into good agreement with the experimental value ($\simeq 4.0 \, \mu_N^2$), while also distributing small strengths to other states. The tensor terms in SSRPA shift the main peak energy upwards by about $1.8\text{ MeV}$, though it remains $1\text{ MeV}$ below the experimental position. Overall, the tensor correlations in SSRPA reduce the cumulative $\text{B(M1)}$ sum up to $15\text{ MeV}$ by about $15\%$ of the total sum (panel (d)).

\subsection{Quadrupole response}
\label{Sec:Applications_SSRPA_Quadrupole}
\begin{figure}
	\centering
	\includegraphics[width=.5\linewidth]{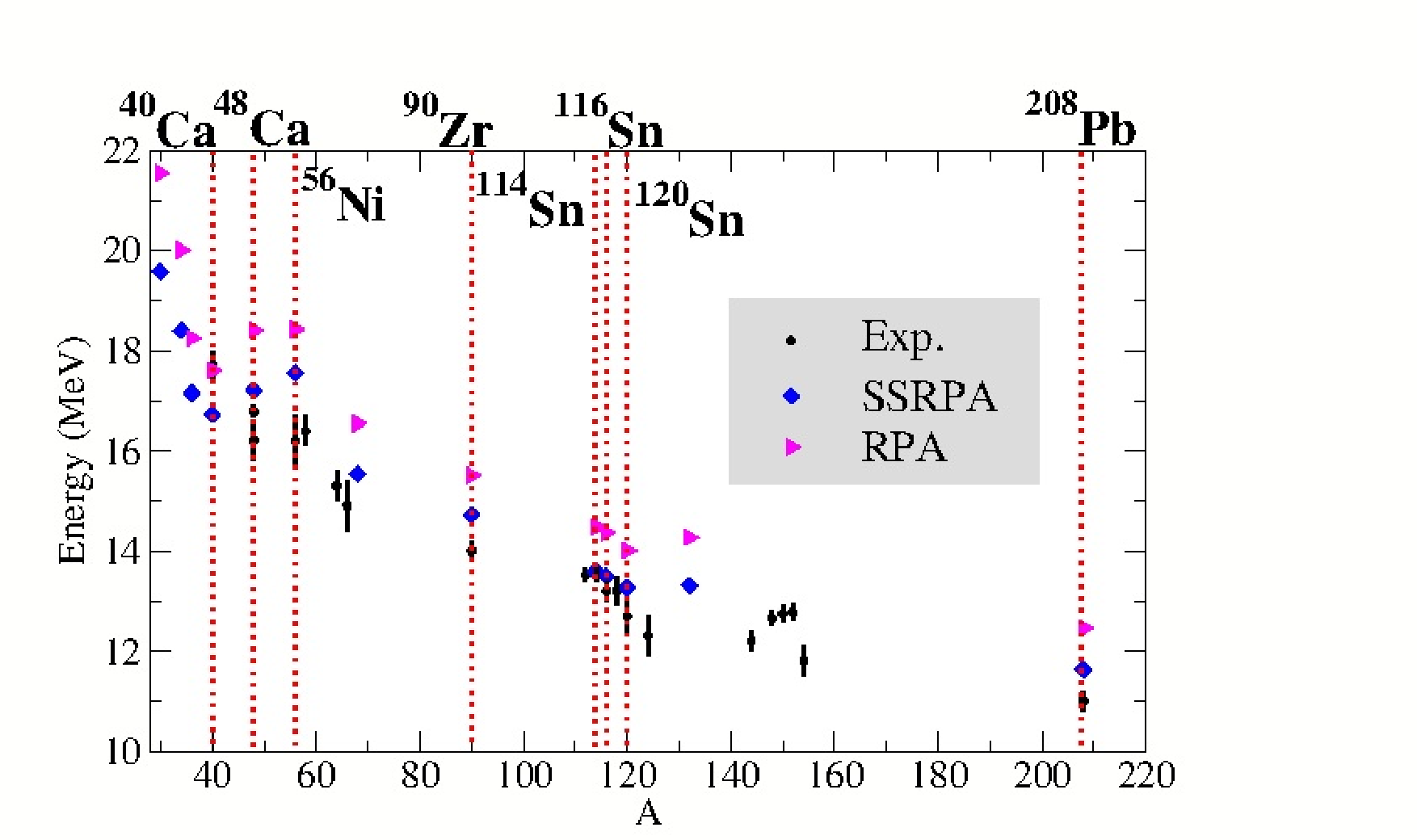}\hfill
	\includegraphics[width=.5\linewidth]{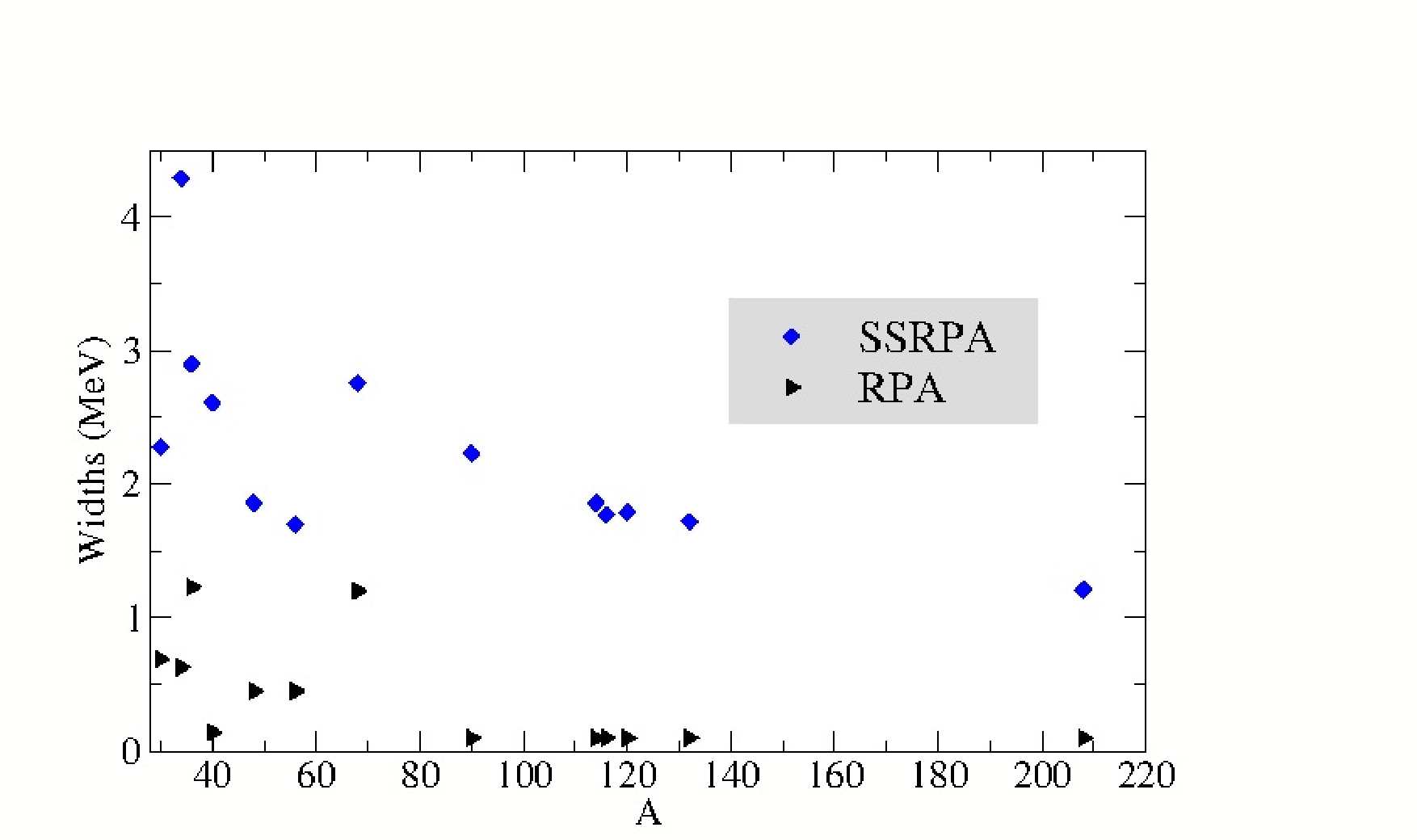}
	\caption{Left side: the experimental ISGQR centroids data are displayed as black circles (with their associated error bars) while the SSRPA (RPA) predictions are plotted as blue diamonds (magenta triangles). At $A=48$, there are two data points, corresponding to $^{48}$Ti and $^{48}$Ca. Right side: RPA and SSRPA theoretical widths. Adapted from Ref. \cite{Vasseur2018}.}
	\label{Fig:GQR_1}
\end{figure}

In Ref. \cite{Vasseur2018}, the SSRPA has been applied to the study of the ISGQR in several medium--mass and heavy closed-shell nuclei. The corresponding centroid energies are compared with the experimental one from Refs \cite{Lui2011,Youngblood1981,Li2010}. 
% While Ref. \cite{Lui2011} also provides Sn isotope IS GQR data with consistent centroid energies, its reported widths are notably larger compared to Refs. \cite{Youngblood1981,Li2010}. 
% 
 RPA and SSRPA calculations with the SLy4 parametrization \cite{SLY4} are presented below.  For each nucleus, the theoretical centroid energies $E_c$ and widths $\Gamma$ are obtained by fitting a Lorentzian distribution, as performed in Ref. \cite{Scamps2013}.
The left side of Figure \ref{Fig:GQR_1} illustrates the systematic trend of the centroids. The figure compares SSRPA results, shown as blue diamonds, with RPA results, shown as magenta triangles, and experimental data, represented by black circles with corresponding error bars. Vertical dotted red lines mark the nuclei for which a comparison between the theoretical predictions and experimental data is possible. The SSRPA centroids are observed to be consistently positioned at lower energies than the RPA values. % It is seen that the centroid energies of the ISGQR are significantly correlated with the effective mass, as detailed in Ref. \cite{Li2018}. A further discussion on this correlation and its impact on the effective mass beyond the mean-field approximation will be presented in Section \ref{Sec:BMF_effectivemass}. 
As a general trend, one can see that the SSRPA energies are in better agreement with the experimental values than the RPA centroids, which in general overestimate the data.
The SSRPA model is expected to significantly change the description of nuclear widths compared to the RPA approach. This change is due to the explicit inclusion of an additional spreading effect, through the coupling of $1p-1h$ and $2p-2h$ configurations, in addition to the Landau damping already present in RPA. The escape width is not accounted for. While the escape width is considered less influential than the spreading width, it could potentially modify some predictions, particularly for lighter nuclei. The right side of Figure \ref{Fig:GQR_1} displays the SSRPA and RPA widths. As expected, the SSRPA widths are consistently larger than the RPA widths. Furthermore, the figure reveals a general trend where both RPA and SSRPA exhibit a decrease in widths from lighter to heavier nuclei, indicating greater fragmentation in lighter nuclei. Because this effect is observed at the RPA level, it suggests that the increased fragmentation in lighter nuclei is due to stronger Landau damping, an effect considered in both RPA and SSRPA. While one can observe a general trend of Landau-damping attenuation increasing the nuclear mass, one can also note that $^{40}$Ca does not exhibit a significant Landau-damping effect, as its RPA width is notably small. For this specific nucleus, the effects arising from the mixing of $2p-2h$ configurations are particularly significant, resulting in a considerable increase in width from RPA to SSRPA.

\begin{figure}
	\includegraphics[width=.45\linewidth]{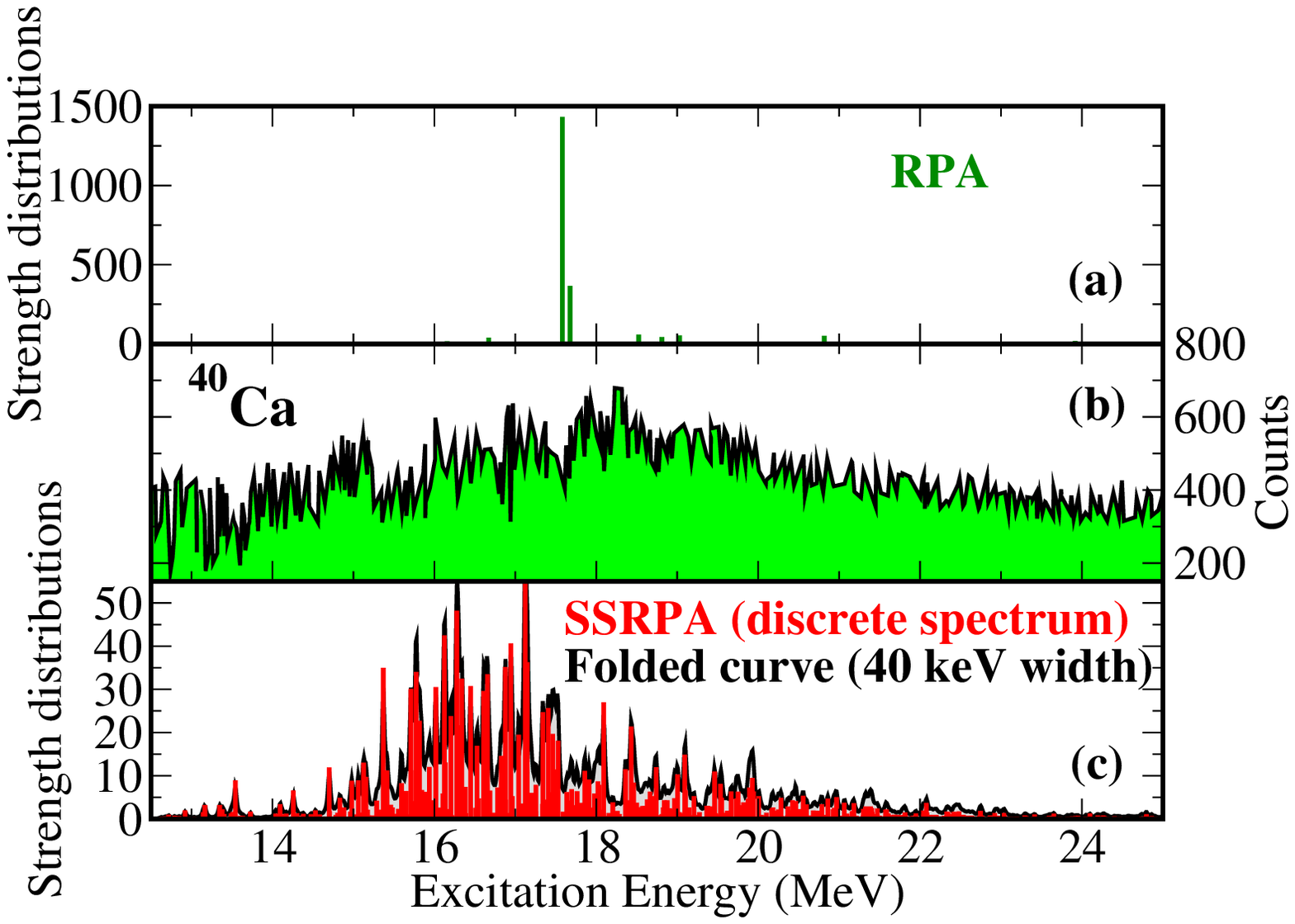}\hfill
	\includegraphics[width=.45\linewidth]{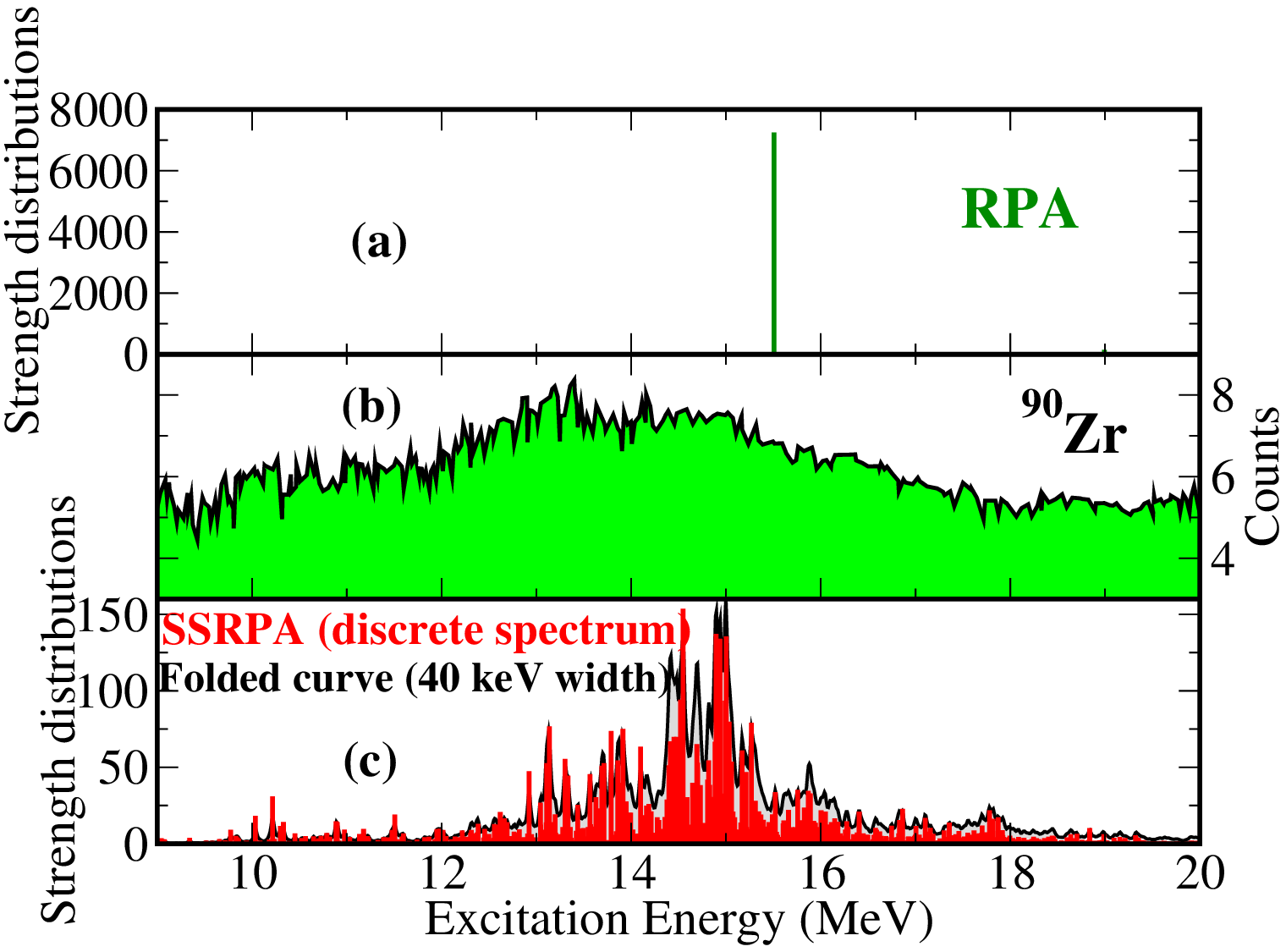}
	\caption{ ISGQR strength distributions obtained in RPA (panels (a)) and SSRPA (panels (c)) compared with the experimental spectrum (panels (b)) for $^{40}$Ca and $^{90}$Zr, on the left and righ side, respectively. The RPA and SSRPA discrete spectra are in units of $e^2$ fm$^4$. The SSRPA folded ones are in units of $e^2$ fm$^4$ MeV$^{-1}$. Adapted from Ref. \cite{Vasseur2018}.}
	\label{Fig:GQR_2}
\end{figure}

\begin{figure}
	\includegraphics[width=.45\linewidth]{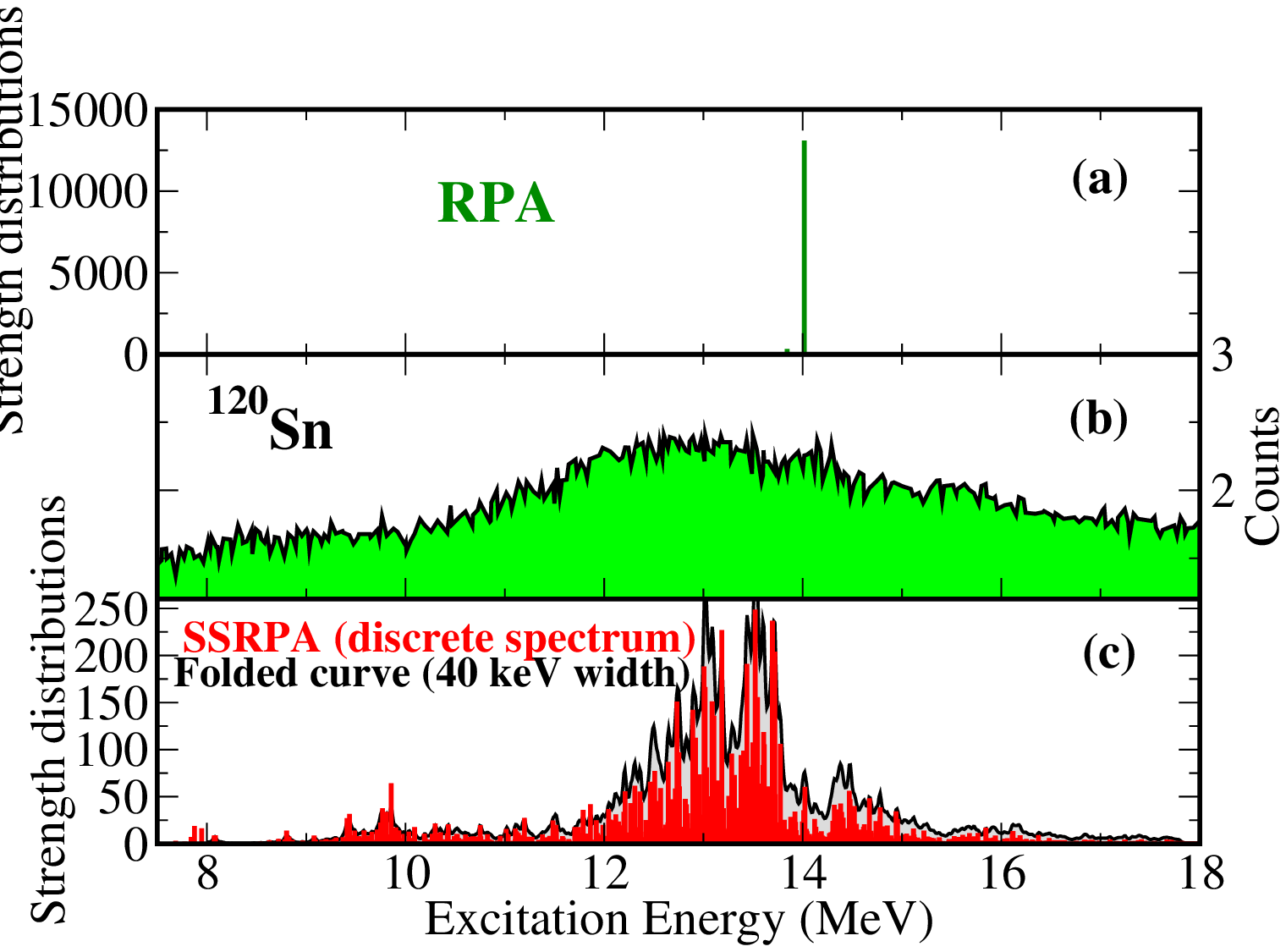}\hfill
	\includegraphics[width=.45\linewidth]{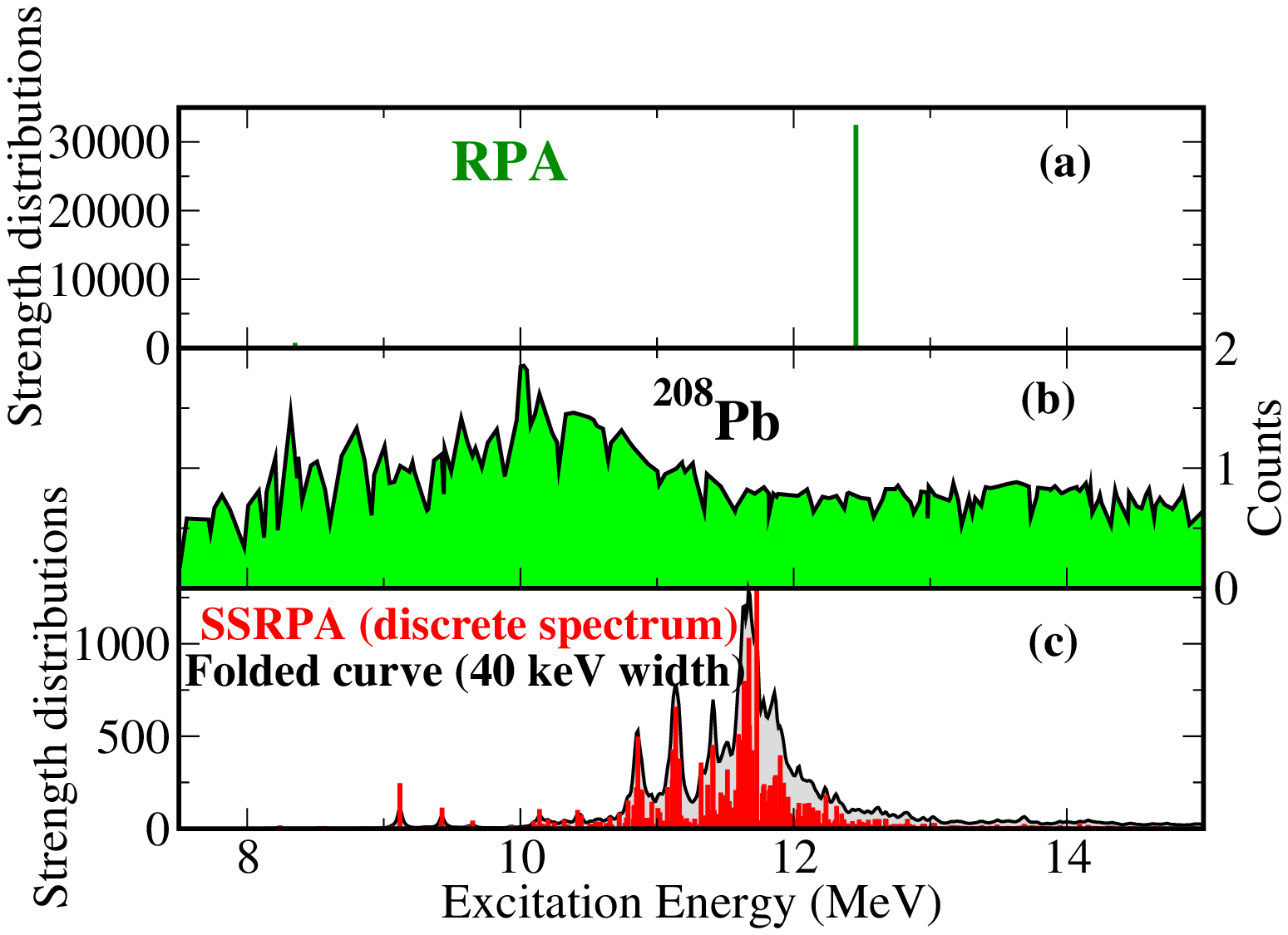}
	\caption{As in Figure \ref{Fig:GQR_2} but for $^{120}$Sn (left side) and $^{208}$Pb (right side). Adapted from Ref. \cite{Vasseur2018}.}
	\label{Fig:GQR_3}
\end{figure}

A detailed analysis of the  fine structure and fragmentation was performed with high-precision (p,p') studies \cite{Usman2011,Shevchenko2004}. The energy resolution of $\sim$40 keV can provide the experimental fine structure of the excitation spectra that is compared with the theoretical result in Figure \ref{Fig:GQR_2} for $^{40}$Ca and $^{90}$Zr, and Figure \ref{Fig:GQR_3} for $^{120}$Sn and $^{208}$Pb.
Panel (c) of the left side of Figure \ref{Fig:GQR_2} presents the SSRPA strength distribution (red bars) for the nucleus $^{40}$Ca. To facilitate comparison with the experimental spectrum (b), a folded curve (black solid line) is also shown, obtained with a Lorentzian of 40 keV width, corresponding to the experimental energy resolution. The folded curve accurately reflects the fine structure of the discrete spectrum. The $^{40}$Ca case is the only one where the SSRPA centroid energy is slightly underestimated compared to the experimental value (see left side of Figure \ref{Fig:GQR_1}), while the RPA centroid exhibits better agreement. However, one can clearly see that the SSRPA provides a closer description to data with respect to the RPA. The latter displays a single dominant peak (panel (a)), whereas the SSRPA strength distribution is considerably more fragmented and spans a broader energy range, aligning with the experimental data. The right side of Figure \ref{Fig:GQR_2} displays the case of $^{90}$Zr. In this case, the RPA centroid exceeds the experimental value by over 1 MeV (see left side of Figure \ref{Fig:GQR_1}). The SSRPA prediction, positioned at lower energies, shows improved agreement with data. Also in this case, a significant improvement in strength fragmentation is observed in SSRPA compared to the single dominant peak in RPA. Similar results  can be seen in Figure \ref{Fig:GQR_2} for $^{120}$Sn and $^{208}$Pb.

%Generally, all SSRPA centroids are slightly shifted downwards relative to RPA. In most cases, excluding $^{40}$Ca, this shift results in better agreement with experimental data. Regarding fragmentation and fine structure, SSRPA demonstrates a significant improvement over RPA, where the strength distributions for the four nuclei considered are characterized by a single dominant peak. The SSRPA strength distributions extend over a much broader energy window, encompassing the experimental response. The comparison with the experimental fine structure reveals a qualitative global agreement, as the model provides a fragmented response within the same energy region. However, the energy window of the experimental strength distribution is consistently broader than that of the SSRPA response, likely due to missing effects in the theoretical model, such as higher-order configurations (three-particle–three-hole, etc.) and spreading effects from coupling with the continuum.

By summarizing, the SSRPA shifts centroid energies slightly downward compared to RPA, leading to better agreement with experiment for most nuclei (excluding $^{40}$Ca).  The SSRPA offers a significant improvement in describing fragmentation and fine structure, replacing the single dominant RPA peak with a strength distribution that extends over a much broader energy window, thus qualitatively encompassing the experimental response. However, the experimental strength distribution is consistently broader than the SSRPA response, likely because the theoretical model currently omits higher-order configurations (like $3p-3h$) and coupling to the continuum (spreading effects).

\subsection{First applications in the relativistic point-coupling model }
\label{Sec:Applications_SSRPA_Relativistic}
\begin{figure}
	\includegraphics[width=.5\linewidth]{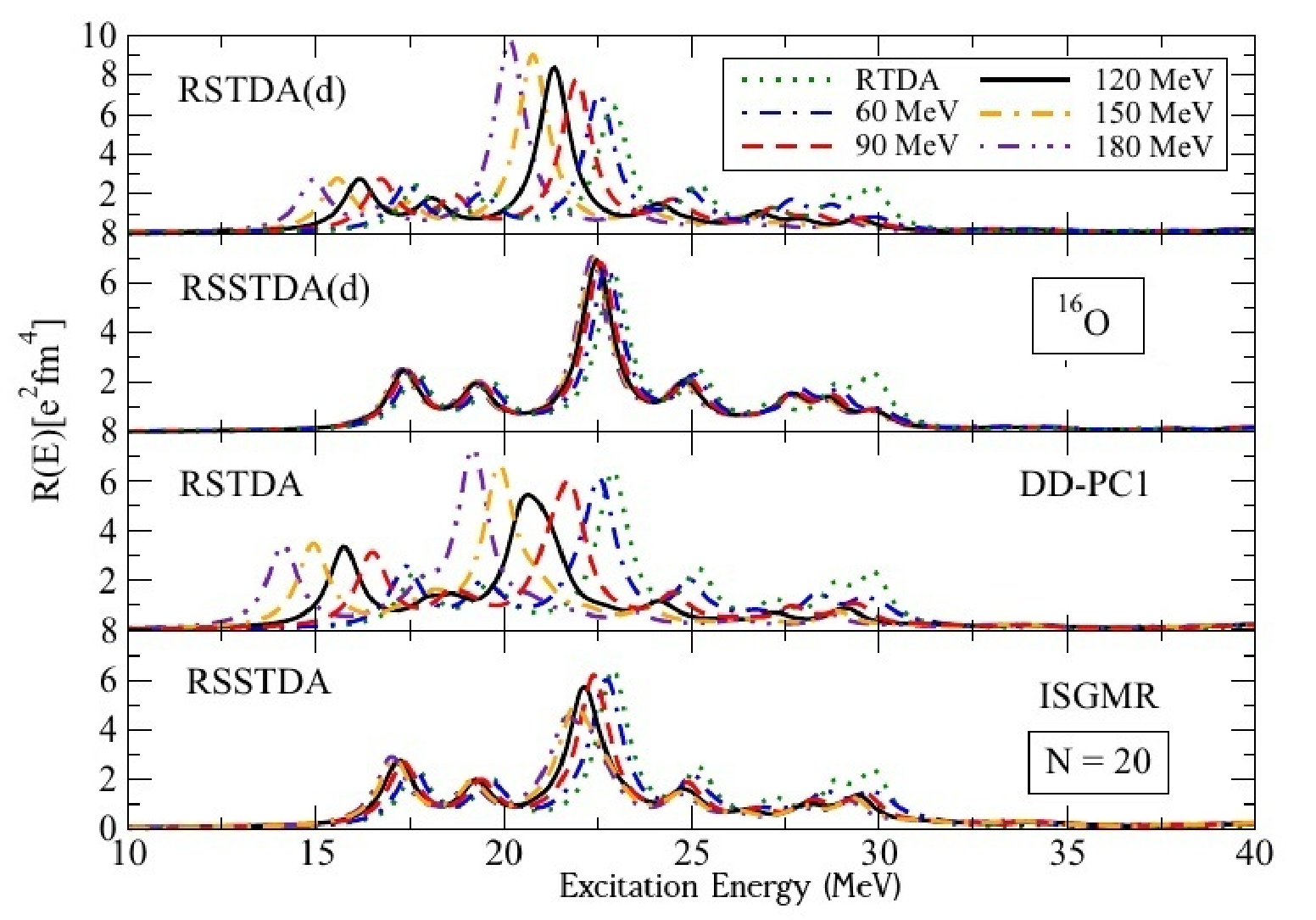}\hfill
	\includegraphics[width=.5\linewidth]{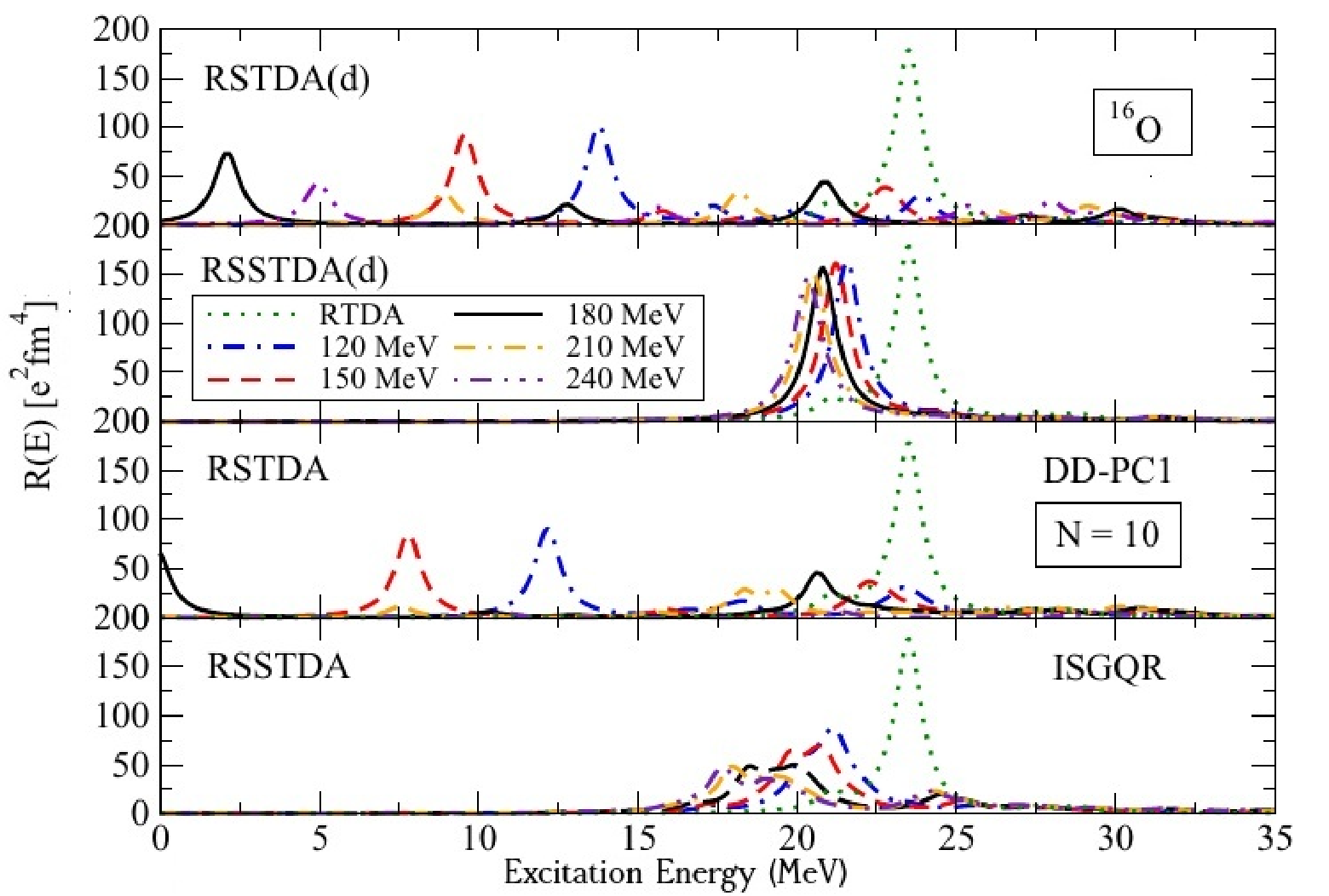}
	\caption{Strength distributions for ${}^{16}\text{O}$ using TDA, RSTDA, and RSSTDA with the full form and diagonal approximation (denoted as RSTDA(d) and RSSTD(d), respectively) as a function of excitation energy, for the ISGMR and ISGQR, on the left and right side, respectively. The results for different $2p-2h$ energy cutoffs are shown. Adapted from \cite{Vale2025}.}
	\label{Fig:SSRPA_Rel}
\end{figure}

The Relativistic STD Approximation (RSTDA) and the Subtracted RSTDA (RSSTDA), based on the relativistic nuclear energy density functional with point-coupling interaction (specifically the DD-PC1 parametrization \cite{DDPC1}) was presented for the first time in Ref. \cite{Vale2025}. Here we discuss the results for the ISGMR and ISGQR response in ${}^{16}\text{O}$. On the left side of Figure \ref{Fig:SSRPA_Rel}, the strength distribution is presented for the Relativistic Tamm-Dancoff Approximation (RTDA), the RSTDA, and its diagonal approximation RSTDA(d), along with the subtraction-based forms RSSTDA and RSSTDA(d) for the ISGMR case. Calculations are shown for increasing $2p-2h$ energy cutoffs ranging from $60 \text{ MeV}$ to $180 \text{ MeV}$. A clear trend is observed where the strength progressively shifts toward lower excitation energies as the cutoff energy increases. Conversely, the results based on the subtraction method demonstrate significantly faster convergence and more stable behavior with respect to the cutoff variation. On the right side of Figure \ref{Fig:SSRPA_Rel}, the strength distribution for the ISGQR is plotted, using $2p-2h$ energy cutoffs between $120 \text{ MeV}$ and $240 \text{ MeV}$. The ISGQR exhibits a behavior similar to the ISGMR concerning convergence, however, the differences when compared to the  RTDA results are more pronounced. %This greater deviation is likely attributable to the numerical approximations and basis restrictions employed in the ISGQR calculations.
Calculations are performed by using a spherical
harmonic oscillator basis, including 
states up to the maximum principal quantum number $N = 20$
in the case of ISGMR and $N = 10$ in the case of ISGQR, which might explain the greater deviation in this case.
The inclusion of $2p-2h$ configurations in both $\text{RSTDA}$ and $\text{RSSTDA}$ leads to the expected physical outcomes: peak fragmentation and shifts to lower excitation energies, qualitatively similar to results from non-relativistic SRPA and realistic interaction models. In the ISGMR case, the $\text{RSTDA}$ yields results closer to experimental values at very high energy cutoffs ($E_x \approx 120$ MeV). However, the $\text{RSSTDA}$, despite showing systematically higher eigenenergies than $\text{RSTDA}$, produces peak positions up to $\approx 1.5 \text{ MeV}$ lower than the RTDA result, better reproducing the observed fragmentation trends. In the case of the ISGQR, the $\text{RSSTDA}$ successfully reproduces the majority of experimentally observed low-lying states, capturing their main structural features with excitation energies deviating by up to $\approx 2 \text{ MeV}$. The deviations from experimental data are primarily attributed to two factors: the use of the $\text{DD-PC1}$ interaction, which was not optimized for light nuclei like ${}^{16}\text{O}$, and the use of the Relativistic Hartree \cite{Vale2025} approximation for the ground state, which omits explicit correlations known to influence the excitation spectrum significantly.

The RSTDA and RSSTDA results show qualitative similarities in the energy shifts and fragmentation distributions to those obtained with non-relativistic interactions discussed above. More systematic studies, overcoming the present limitations, towards a full self-consistent Second Relativistic RPA results are expected in the near future, and the comparison with the non-relativistic implementations will help clarifying its potential advantages.

\newpage
 \section{Applications of the SSRPA for charge-exchange excitations}\label{Sec:Applications_SSRPA_CE}
 
 \subsection{Introduction}
%\begin{comment}
%Charge‑exchange  excitations, and in particular the GT resonance, are collective spin–isospin oscillations that probe the spin–isospin component of the nuclear interaction and constrain nuclear effective interactions in that channel. GT excitations are relevant both two‑neutrino ($2\nu\beta\beta$) and neutrino-less double‐$\beta$ ($0\nu\beta\beta$) decay. Observation of $0\nu\beta\beta$ decay would confirm Majorana neutrinos and lepton‐number violation, with implications for physics beyond the Standard Model and the cosmic matter–antimatter asymmetry \cite{Avignone_2008,Engel_2017}. The decay rate scales with nuclear matrix elements dominated by the axial‐vector part of the GT operator. Current theoretical uncertainties  hinder experimental design and interpretation \cite{Engel_2017}. 
%	Beyond nuclear structure, GT modes are intrinsically connected to weak‑interaction processes of astrophysical interest. In stellar environments they govern electron‑capture and ordinary $\beta$‑decay rates, thus influencing core‑collapse supernovae and $r$‑process nucleosynthesis \cite{Langanke2003,Janka2007}. 
%\end{comment}	
GT excitations are spin-isospin excitations induced by the transition operators (\ref{Eq:Op-GT}).
%\begin{equation}
%	\hat{O}^{\pm}=\sum_{i=1}^{A} \sum_{\mu} \sigma_{\mu}(i) \tau^{\pm}(i),
%	\label{oper}
%\end{equation} 
%where $\tau^{\pm}$ are the 
%isospin raising ($+$) and lowering ($-$) operators, $\tau^{\pm}=t_x \pm it_y$, and $\sigma_{\mu}$ is the spin operator. 
The $T_{GT}^+$ operator generates the GT$^+$ strength (a neutron is added and a proton is removed), while the $T_{GT}^-$ operator produces the GT$^-$ strength (a neutron is removed and a proton is added). 
The Ikeda GT sum rule \cite{Ikeda}, relating the integrated strengths $S$ of the GT$^-$ and the GT$^+$ spectra to the number of neutrons $N$ and protons $Z$ of the nucleus, has the following expression: 
\begin{equation}
	S_{GT^-}-S_{GT^+}=3(N-Z).
	\label{ikeda}
\end{equation}
%This sum rule is model independent and can be easily deduced by using the properties of the isospin operators if the condition of completeness of states is fulfilled. 
In nuclei having a neutron excess, the strength $S_{GT^-}$ is dominant.

In the allowed GT approximation \cite{Feshbach_1974,Engel1999,Niu2018}
 the $\beta$ half-life is given by
\begin{equation}
	T_{1/2}=\frac{D}{g_A^2 \int_0^{Q_{\beta}}S(E) f(Z,\omega) dE},
	\label{niu1}
\end{equation}
where $E$ is the excitation energy referred to the ground state of the daughter nucleus, $	D=6163.4 \pm 3.8 s$, $Q_{\beta}$ is the $Q$ value: $Q_{\beta}= \Delta_{nH} - \Delta B$. $\Delta B $ is the difference of the binding energies of the mother and the daughter nuclei, $B_{mother}-B_{daughter}$, and $\Delta_{nH} = 0.78227 $ MeV is the mass difference between the neutron and the hydrogen. In Eq. (\ref{niu1}),
$g_A$ is the axial-vector coupling constant of the weak interaction. $S(E)$ is the GT strength function and
$f(Z,\omega)$ is the integrated phase volume or phase--space volume (which contains the lepton kinematics).

%Observation of $0\nu\beta\beta$ decay would establish the Majorana nature of neutrinos ($\nu$) and lepton‑number violation, with profound implications for physics beyond the Standard Model and the origin of the Universe’s matter‑antimatter asymmetry \cite{avignone2008,engel2017}. The $0\nu\beta\beta$ decay rate depends on nuclear matrix elements dominated by GT‑like operators $\propto g_A,\sigma\tau$; current theoretical predictions vary by factors of two to three, an uncertainty too large for guiding next‑generation experiments. Recent studies suggest that double‑CE measurements, which access correlated two‑GT operators, may help constrain $0\nu\beta\beta$ matrix elements \cite{cappuzzello,lenske,shimizu}.

Theoretical studies of the $\beta$-decay and GT excitations have been performed within the Shell Model approach \cite{Caurier1999,Yoshida2018,Saxena2018} and non-relativistic \cite{Engel1999,Bender2002,Fracasso2005,Fracasso2007,Bai2013,Martini2014,Bai2014,Sarriguren2018,Deloncle2017} and relativistic \cite{Paar2004,Niu2017,Liang2008,Ravlic2024,Robin2024}
approaches based on the RPA or QRPA methods. These models systematically over-predict the low‑energy GT strength (below $\sim$20 MeV), requiring phenomenological “quenching factors” to match experimental data \cite{Cao2019}. The over‑prediction is commonly attributed to missing physics beyond $1p-1h$ configurations or non-nucleonic degrees of freedom \cite{Engel_2017}. However, because the $\beta$-decay half-lives are only sensitive to the low-energy strength, the most challenging and intrinsic difficulty of the RPA method is in the underestimation of the strength obtained within the allowed window, typically leading to overestimated or even infinite half-lives. The role of BMF correlations, which are not taken into account in RPA, in describing $\beta$-decay half-lives has been explored in various theoretical works, both based on non-relativistic \cite{Niu2018,Niu2012,Mustonen2014,Niu2015,Niu2016,Liu2024} and relativistic \cite{Marketin2012,Litvinova2014,Robin2016,Robin2018,Robin2019} approaches. \emph{Ab initio} calculations based on chiral effective‑field‑theory interactions and consistent electroweak currents have recently demonstrated that two‑body weak currents play an important role in reducing the quenching problem \cite{Gysbers2019,Ekstrom2014,Coraggio2019}. 
\begin{figure}
	\includegraphics[width=.46\linewidth]{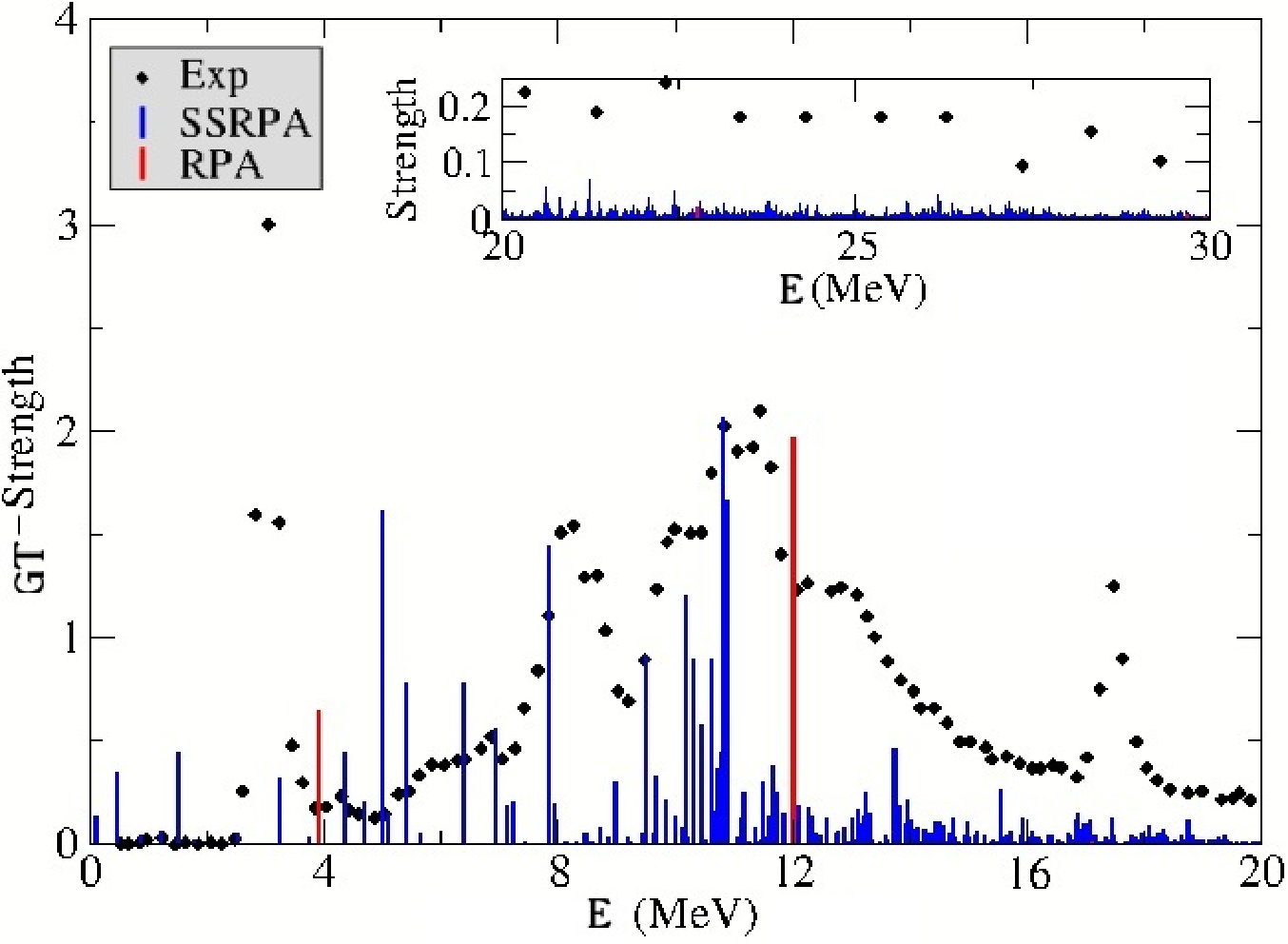
	}\hfill
	\includegraphics[width=.52\linewidth]{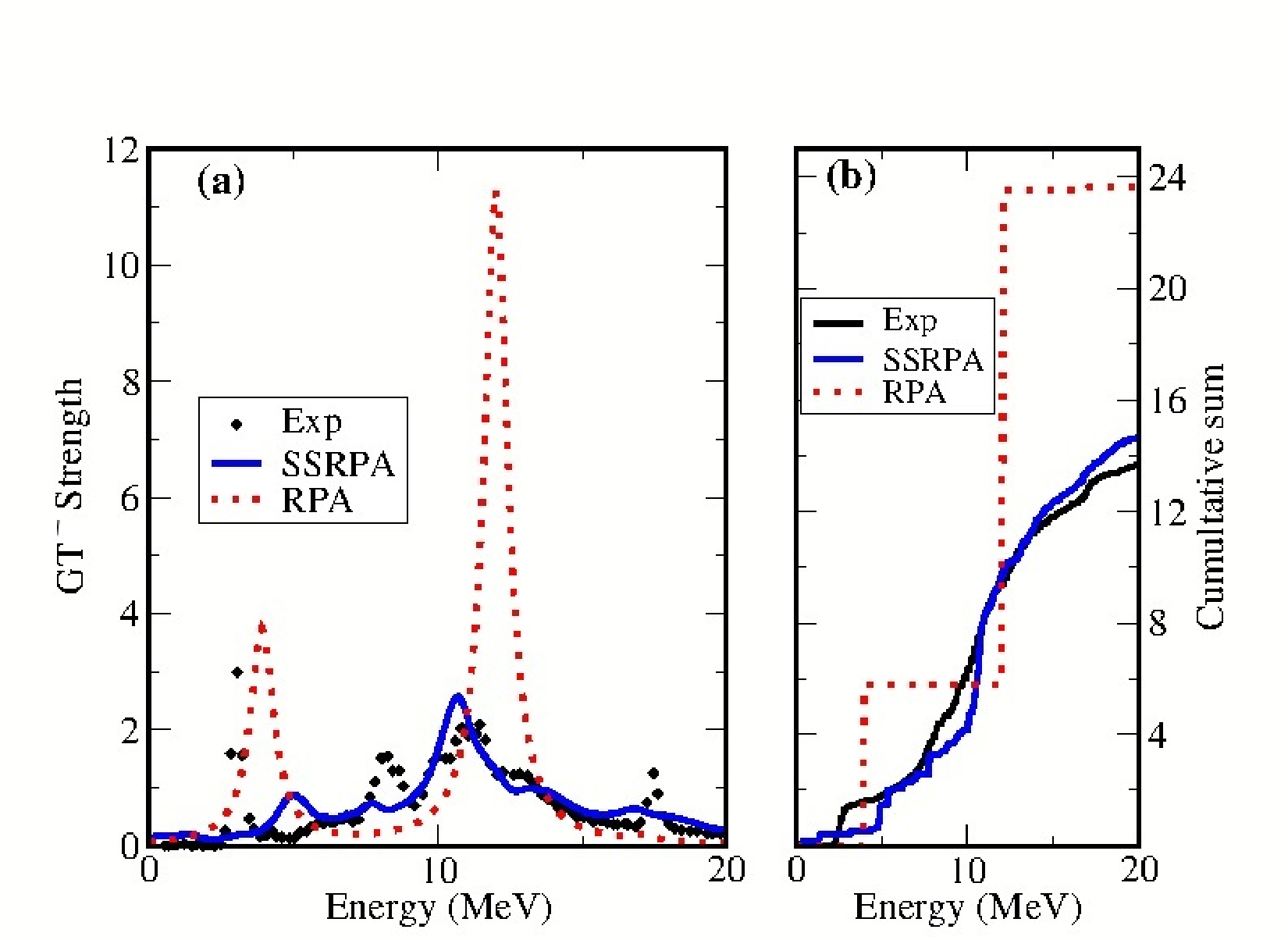}
	\caption{Left side: Experimental GT$^-$ strength in MeV$^{-1}$
		\cite{Yako2009} and discrete RPA and SSRPA strength distributions for $^{48}$Ca. The RPA strength has been divided by nine and the SSRPA strength by two so that the discrete distributions can be displayed on the same figure as the continuous experimental distribution (see text). The insert shows the energy region
		between 20 and 30 MeV. Right side: The RPA and SSRPA GT$^-$ responses for $^{48}$Ca folded with a Lorentzian having a
		width of 1 MeV and compared with experimental data (panel (a)). Cumulative strengths up to 30 MeV (panel (b)). The SGII force is used. 	Adapted from Ref.\cite{Gambacurta2020}.}
	\label{Fig:GT_3}
\end{figure}
%Starting from realistic nucleon-nucleon potentials, many-body perturbation theory has been utilized to derive effective shell-model operators, both for the Hamiltonian and the transition operators, that consistently incorporate correlations from outside the chosen shell-model space \cite{Coraggio2019} and showing that in such a way the quenching factors are strongly reduced if not unneeded. 

\subsection{GT excitations in SSRPA}
\label{Sec:Applications_SSRPA_GT}
The GT strength distributions in spherical closed-shell nuclei within the SSRPA approach have been recently studied \cite{Gambacurta2020, Gambacurta2022, Sagawa2022}.
% In Ref. \cite{Sagawa2022} the effect of the tensor force in the SSRPA has been also investigated. 
% The figures show that the strength distributions of the two calculations differ by not more than 0.1 MeV in the excitation energy. 
In Figure \ref{Fig:GT_3}, the experimental GT$^-$
strength \cite{Yako2009} below 30 MeV is shown (black points) and compared with the RPA and SSRPA results obtained with the SGII interaction \cite{SGII}. %Reference \cite{Yako2009} reports the results of a $^{48}$Ca(p, n) and $^{48}$Ti(n, p) experiments at a beam energy of 300 MeV performed at the Research Center for Nuclear Physics in Osaka. The integrated strength is only 64 $\pm$ 9\% of that given by the Ikeda sum rule. On the right panel, the RPA and SSRPA spectra are also shown.
On the left side, the discrete spectra are shown, where  the RPA and SSRPA strengths have been rescaled so that their respective highest peaks have approximately the same height as the corresponding experimental peak. This is done by dividing the RPA strength by nine and multiplying the SSRPA strength by two. 
%\textit{The SSRPA spectrum exhibits a pronounced fragmentation, especially in the $6\text{–}16\ \mathrm{MeV}$ window, where the strength clusters into three principal groups centered near $8$, $11$, and $14\ \mathrm{MeV}$, close to the experimental peak pattern. A further, much weaker assemblage appears around $17\ \mathrm{MeV}$, matching the highest‐energy feature seen experimentally. Moreover, the inset reveals a dense, extended tail of SSRPA strength between $20$ and $30\ \mathrm{MeV}$—a structure entirely missing in the RPA result, which consists of only a handful of isolated peaks and fails to reproduce the experiment’s intricate distribution.}
We notice that the SSRPA spectrum exhibits a pronounced fragmentation and a long high‐energy tail that accounts for the apparent deficit of strength at lower excitation energies. The $2p-2h$ configurations, whose density grows with excitation energy, produce an extended high‑energy tail by reshuffling strength away from lower-energy regions. 
% Consequently, the redistributed strength is dispersed over a wide energy range, rendering it difficult to distinguish from background from an experimental point of view.
At the very lowest end of the spectrum, SSRPA is somewhat less accurate, it predicts a dominant peak at $\sim5\ \mathrm{MeV}$, whereas the first experimental resonance lies at $3\ \mathrm{MeV}$. Nevertheless, SSRPA does generate some fragmented strength in the vicinity of $3\ \mathrm{MeV}$. In contrast, the RPA discrete spectrum shows only a single peak at $4\ \mathrm{MeV}$, with no appreciable fragmentation.

%To facilitate a direct comparison with experiment, the response functions were folded with a Lorentzian of width 1 MeV. 
In panel (a) of the right side of Figure  \ref{Fig:GT_3}, the folded RPA and SSRPA strength distributions are displayed together with the experimental data, while panel (b) shows the cumulative strength up to 30 MeV. Consistent with the discrete spectra, SSRPA reproduces the GT distribution very accurately, apart from a minor shift in the lowest‐energy peak, whereas RPA fails to capture the spectrum’s structure. 
The cumulative SSRPA strength is markedly lower than that of RPA and aligns closely with experiment even at low energies. The SSRPA curve is smooth and mirrors the experimental profile thanks to its realistic treatment of widths, fragmentation, and the subtraction procedure that correctly positions centroids. By contrast, the RPA curve exhibits pronounced steps due to its few isolated peaks. The ratio of experimental to theoretical integrated strength below 20 MeV is only 0.58 in RPA but rises to 0.93 in SSRPA, indicating that no quenching factor is required in the latter. The explicit inclusion of $2p-2h$ configurations in SSRPA efficiently generates the extended high‐energy tail that accounts for the “missing” strength. %When the strength is integrated up to 70 MeV, a total of 23.84 units is obtained; since the corresponding integrated $\beta^{+}$ strength is only 0.10, the Ikeda sum rule value of 24 is reproduced to within about 1 \%.

\begin{figure}
	\centering
	\includegraphics[width=.6\linewidth]{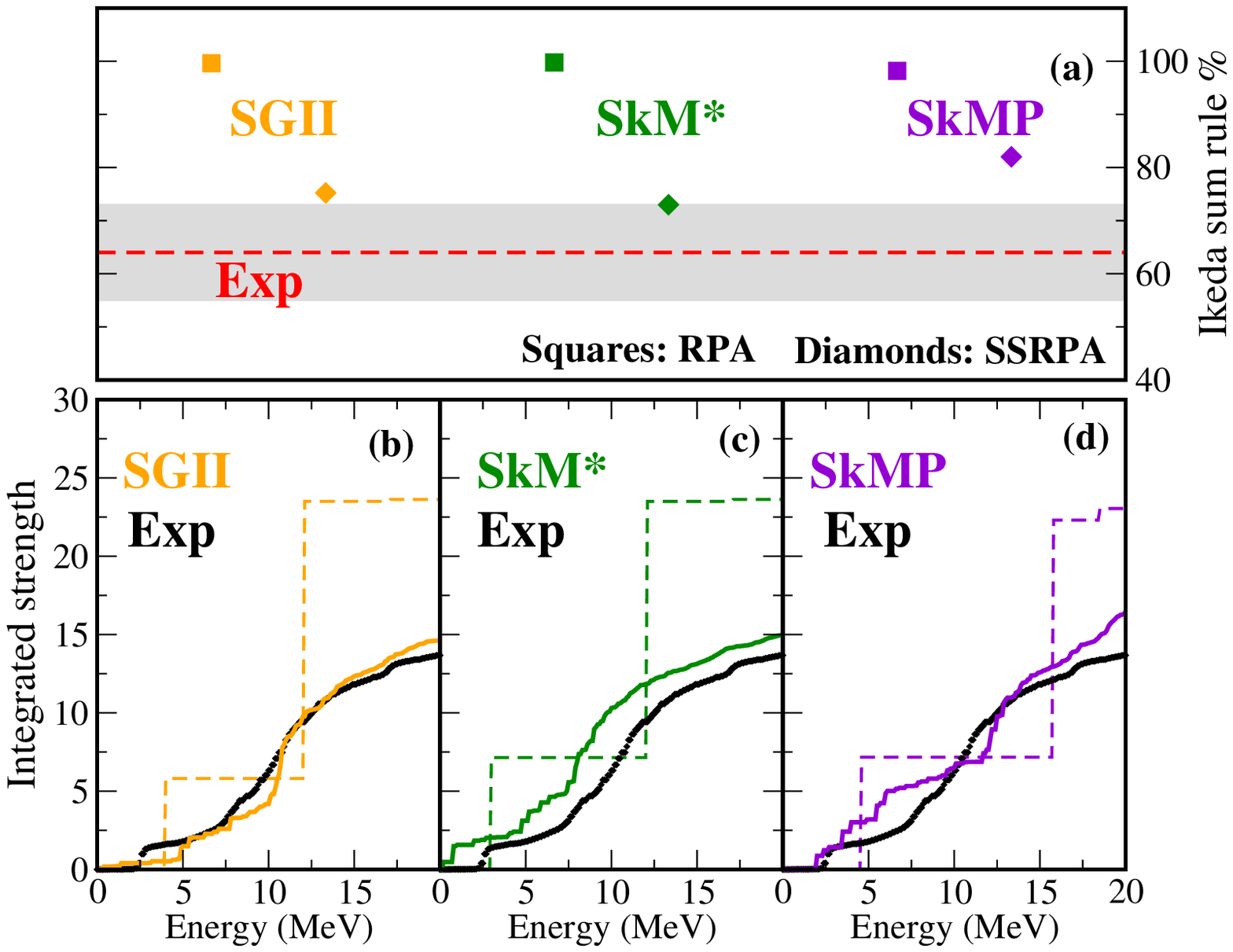}
	%	\hfill
	%	\includegraphics[width=.51\linewidth]{Fig4}
	\caption{ In panel (d) the experimental percentage of the Ikeda sum
		rule below 30 MeV extracted from Ref. \cite{Yako2009} (red dashed horizontal line),
		and its associated uncertainty (gray area), is compared with the RPA and SSRPA percentages obtained
		with the parameterizations SGII, SkM$^*$, and SkMP. Experimental data from Ref. \cite{Yako2009}. 
	Panels (b), (c), (d): SSRPA and RPA strengths integrated up to 20 MeV, in solid and dashed lines, respectively, obtained with the parameterizations SGII
	(b), SkM$^*$ (c), and SkMP (d)	
		Adapted from Ref. \cite{Gambacurta2020}.
	}
	\label{Fig:GT_4}
\end{figure}
To verify that the quenching of strength below 20 MeV is an intrinsic feature of SSRPA rather than an artifact of a specific single Skyrme interaction, Fig. \ref{Fig:GT_4} presents RPA and SSRPA results using three different parameterizations, the SGII \cite{SGII}, SkM$^*$ \cite{SKM}, and SkMP \cite{SKMP}—alongside the corresponding experimental data. In panels (b)–(d) the cumulative GT strength up to 20 MeV is plotted for each EDF, while panel (a) shows the fraction of the Ikeda sum rule exhausted by strength below 30 MeV (the experimental central value and its uncertainty are taken from Ref. \cite{Yako2009}). Across all three forces, SSRPA yields a systematically lower integrated strength than RPA in the low‐energy region, demonstrating that the quenching is genuinely introduced in the SSRPA.
% In particular, the SGII and SkM$^*$ interactions produce SSRPA percentages in panel (d) that closely match the experimental band. Furthermore, SGII most faithfully reproduces the fine details of the measured spectrum: in panel (a), the SSRPA curve follows the experimental profile almost exactly, confirming its capability to capture the observed fragmentation.

\begin{figure}
	\includegraphics[width=.52\linewidth]{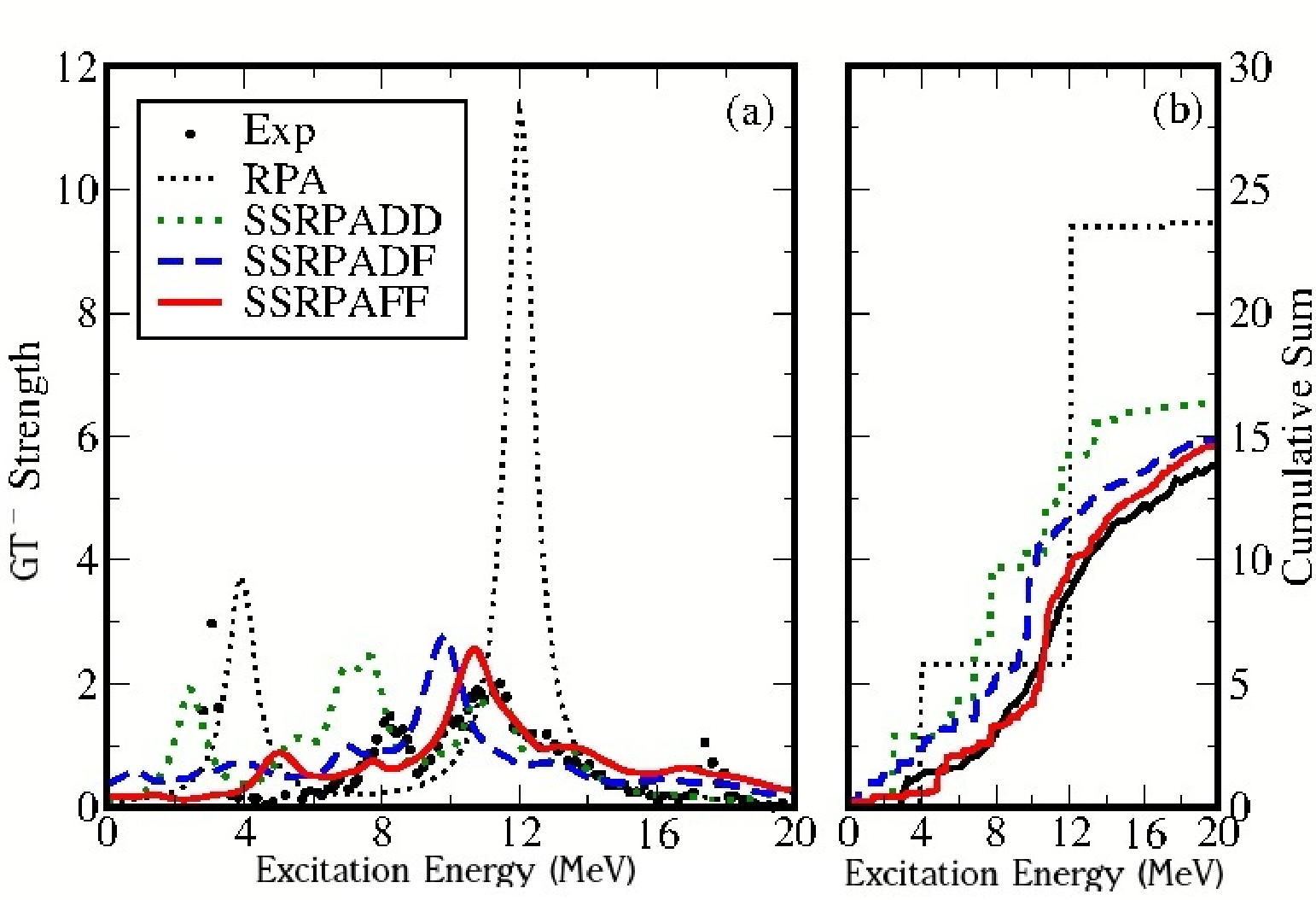}\hfill
	\includegraphics[width=.41\linewidth]{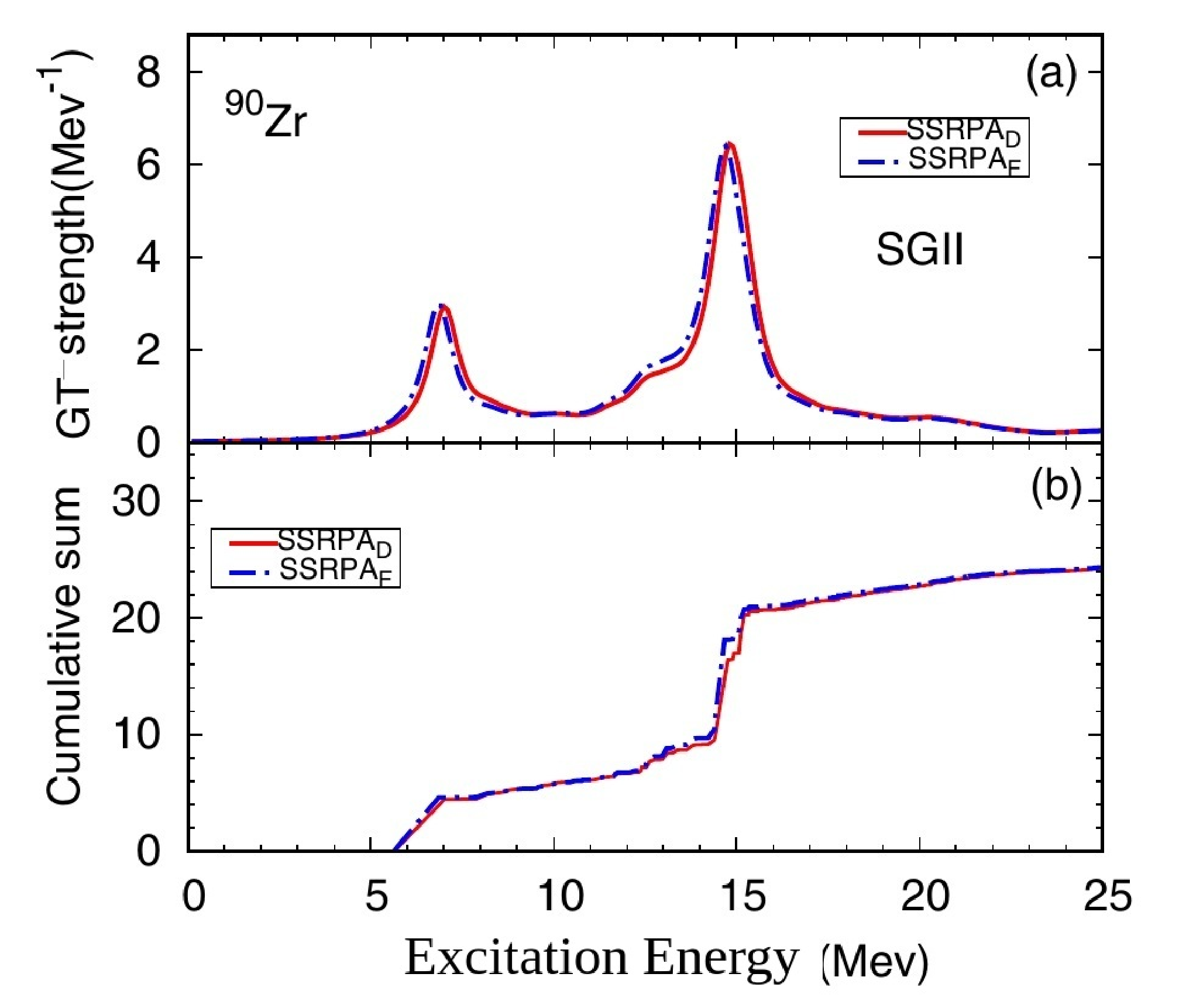}
	\caption{Left side: GT$^-$ strengths (panel (a)) obtained for $^{48}$Ca in MeV$^{-1}$ with the Skyrme interaction SGII and cumulative sums (panel (b)). Experimental data are extracted from Ref. \cite{Yako2009}. See the text for more details. Adapted from Ref. \cite{Gambacurta2022}.
		Right side: The GT$^-$ strength distributions (panel (a)) and corresponding cumulative
		sums (panel (b)) of $^{90}$Zr calculated in SSRPA with SGII force. The red line corresponds to the result obtained with the diagonal approximation used in the subtraction procedure, labeled SSRPAD , and the blue dashed line represents the full calculation, labeled SSRPAF. Adapted from Ref. \cite{Sagawa2022}.}
	\label{Fig:GT_1}
\end{figure}

On the left side of Figure \ref{Fig:GT_1}, the full GT$^-$ strength distribution for $^{48}$Ca, is shown in panel (a), while in panel (b), the corresponding cumulative sums are plotted. Three different kinds of SSRPA calculations are performed. The label SSRPADD refers to a calculation where the diagonal approximation is used twice both in the inversion and in the diagonalization procedures, while in SSRPAFF no approximation in used in either tasks. The SSRPDF indicates a calculation where the diagonal approximation for the matrix $A_{22}$ is adopted only in the subtraction procedure, where this matrix is inverted, whereas $A_{22}$ is fully treated in the diagonalization. The SSRPAFF case shows leads to a significant improvement of the results compared to RPA, as also shown in 
Ref. \cite{Gambacurta2020}. The SSRPADD provides a worse reproduction of the full spectrum.
An improvement is observed in the SSRPADF, yet having some differences with respect to the full calculation. Panel (b) of Figure \ref{Fig:GT_1} shows the corresponding cumulative sums, confirming that the SSRPADF scheme reproduces less well than the SSRPAFF one the experimental strength. At the same, a clear improvement with respect to the RPA is observed. This result indicates that, in the calculations done for $^{48}$Ca, neglecting the coupling between the $2p-2h$ configurations in the inversion of the matrix in the subtraction procedure has a non-negligible effect. However, as it has been shown in two set of independent calculations, see Refs \cite{Gambacurta2020} and \cite{Sagawa2022}, for heavier nuclei, the off-diagonal terms may be safely neglected in the matrix to invert without losing any important information in the excitation spectrum. As a consequence, the SSRPADF approximation becomes much more efficient. This can be seen on the right side of Figure \ref{Fig:GT_1}, where the GT strength distribution (and the corresponding cumulative sums) obtained in SSRPA for $^{90}Zr$ are shown in the upper (lower) panel. 
The SSRPA$_F$ results corresponds to the SSRPAFF case, while the SSRPA$_D$ results to the SSRPADF. One can see that in this case, the two sets of results are very close.

\begin{figure}
	\centering
	\includegraphics[width=.49\linewidth]{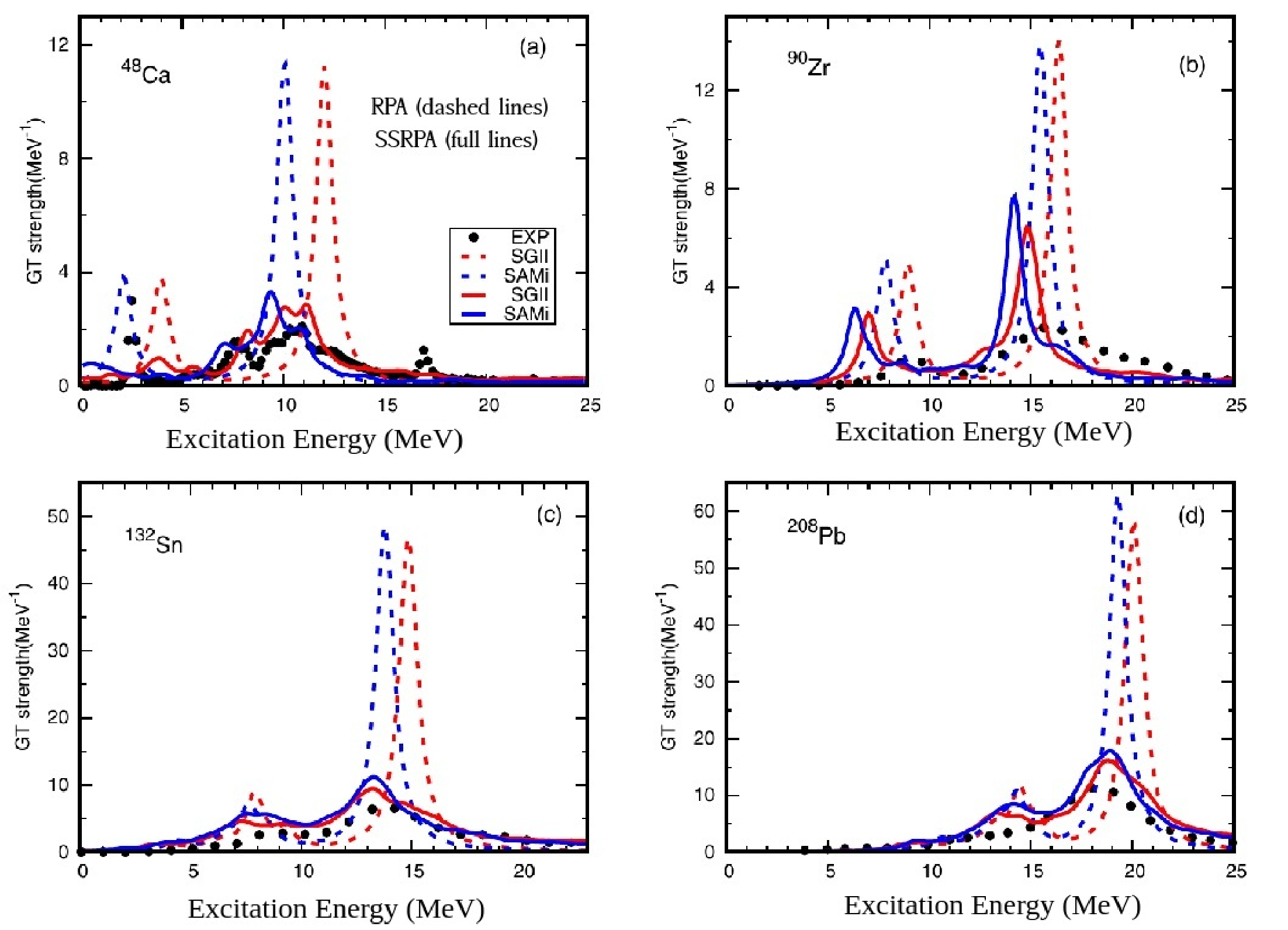}\hfill
	\includegraphics[width=.49\linewidth]{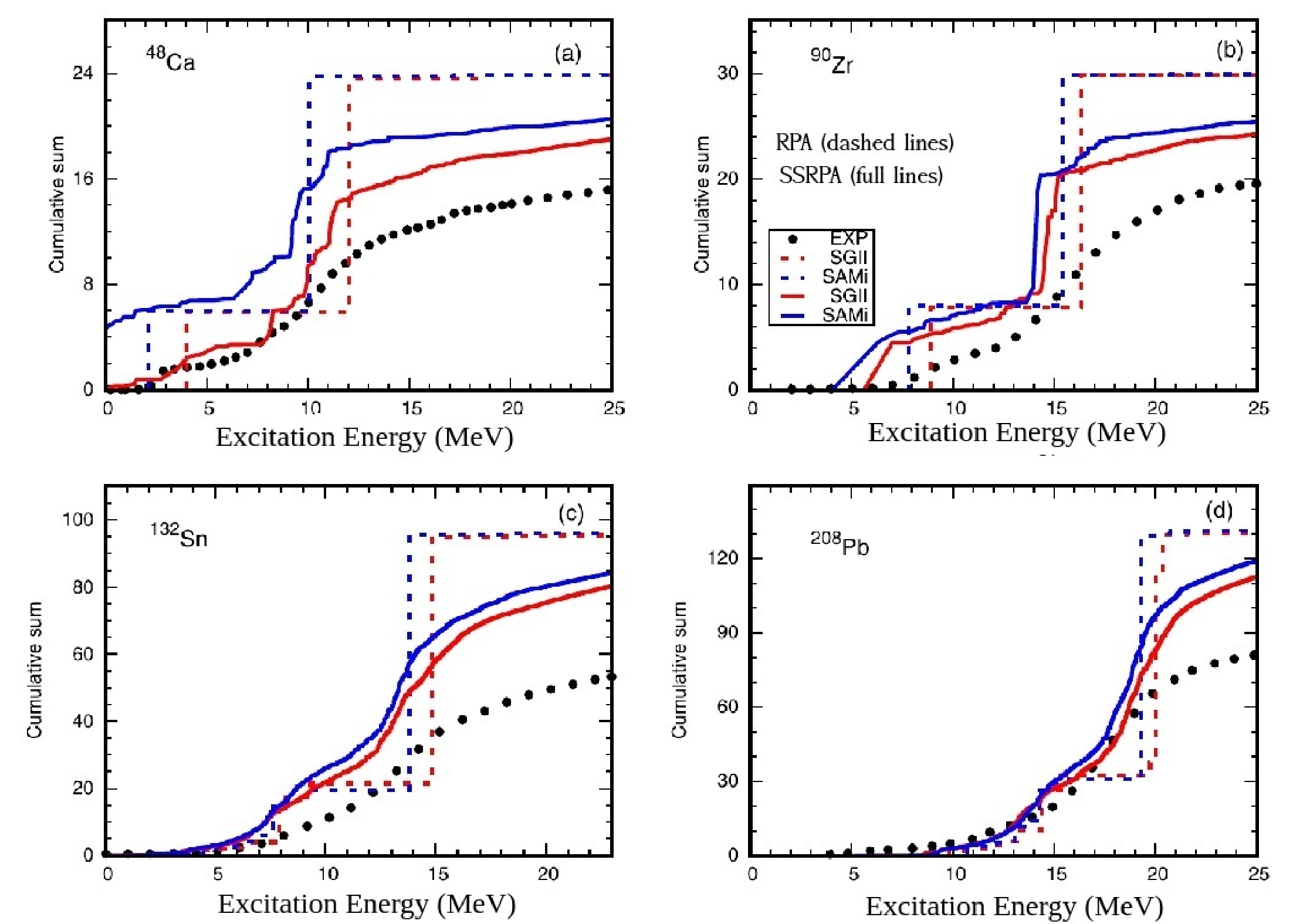}
	\caption{Left side: GT strength distributions for four nuclei: $^{48}$Ca (panel a), $^{90}$Zr (panel b), $^{132}$Sn (panel c), and $^{208}$Pb (panel d). The calculations were performed using  SGII (red lines) and SAMi (blue lines), within both the RPA (dashed lines) and the SSRPA (solid lines). For comparison, experimental GT strength data for each nucleus (from references \cite{Yako2009}, \cite{Wakasa1997}, \cite{Yasuda2018} and \cite{Wakasa2012} respectively) are shown as black filled circles. See the text for more details. Right side: corresponding cumulative sums. Adapted from Ref. \cite{Sagawa2022}. 	}
	\label{Fig:GT_5}
\end{figure}

Figure \ref{Fig:GT_5} illustrates the GT strength distributions for $^{48}$Ca (panel (a)), $^{90}$Zr (panel (b)), $^{132}$Sn (panel (c)), and $^{208}$Pb (panel (d)). These distributions were derived using the SGII (red lines) and SAMi (blue lines) EDFs within the RPA (dashed lines) and the SSRPA (solid lines).  The diagonal approximation was used for the nuclei with the exception of $^{48}$Ca for which the full subtraction was employed. Experimental data are also presented for comparison. One can see that RPA calculations employing the SAMI reproduce quite well the excitation energies of the dominant peaks. The SGII overestimates these energies by approximately 1–2 MeV relative to SAMi. Furthermore, RPA calculations significantly overestimate the strength of these main peaks, exhibiting values 5 to 7 times larger than the experimental observations. The SSRPA calculations show significant quantitative improvements in describing the GT strength distributions. The coupling of $2p-2h$ configurations leads to a spreading of the strength from the main peaks over a broader energy range. %Consequently, the strengths of the primary peaks in the SSRPA results are roughly 1.5 times the experimental values for $^{48}$Ca, $^{132}$Sn, and $^{208}$Pb, and approximately 3–4 times those for $^{90}$Zr.
Additionally, the SSRPA model induces a downward shift in the energy of the main peak. For the SGII, this downward shift is about 1 to 1.5 MeV across all four nuclei, leading to good agreement with the experimental peak positions. With the SAMI, the main peaks are shifted downwards by approximately 1 and 1.5 MeV in $^{48}$Ca and $^{90}$Zr, respectively, but remain largely unchanged in $^{132}$Sn and $^{208}$Pb.
The corresponding cumulative GT strength sums are shown in the right side of Figure \ref{Fig:GT_5}, where red and blue lines again correspond to SGII and SAMi results, respectively.
RPA quenching factors are of approximately 36.7\% for $^{48}$Ca ($E_{\text{max}} = 25$ MeV), 34.9\% for $^{90}$Zr ($E_{\text{max}} = 25$ MeV), 44.5\% for $^{132}$Sn ($E_{\text{max}} = 23$ MeV), and 38.6\% for $^{208}$Pb ($E_{\text{max}} = 25$ MeV). 
%These maximum energies were chosen based on the reliable limits of the available experimental data for each nucleus.
%The cumulative sums obtained from RPA calculations exhibit a rapid increase in the region of the main peak energies, approaching the sum rule limit.
 In contrast, the SSRPA cumulative sums show a more gradual increase. The corresponding quenching factors are approximately 20.7\% (SGII) and 14.4\% (SAMi) for $^{48}$Ca, 19.2\% (SGII) and 15.2\% (SAMi) for $^{90}$Zr, 16.4\% (SGII) and 12.5\% (SAMi) for $^{132}$Sn, and 14.7\% (SGII) and 10.0\% (SAMi) for $^{208}$Pb within the experimental energy ranges. For these nuclei, the SGII yields quenching factors about 5\% larger than those obtained with the SAMI. The SSRPA calculations with SGII achieve roughly half of the experimental quenching factor for $^{48}$Ca and $^{90}$Zr, and about 40\% for $^{132}$Sn and $^{208}$Pb. We conclude by noticing that, although both SSRPA calculations for $^{48}$Ca shown in Figures \ref{Fig:GT_1} and \ref{Fig:GT_5} employ the full subtraction procedure, they yield different cumulative sums. This discrepancy arises from the $J^2$ terms, which are accounted for only in Figure \ref{Fig:GT_5}. Because the SGII force was not fitted with these terms, their absence in the calculation actually might enhance the agreement with data,(see also the discussion at the end of Section IV of Ref. \cite{Sagawa2022}).

\begin{figure}
	\centering
	\includegraphics[width=.49\linewidth]{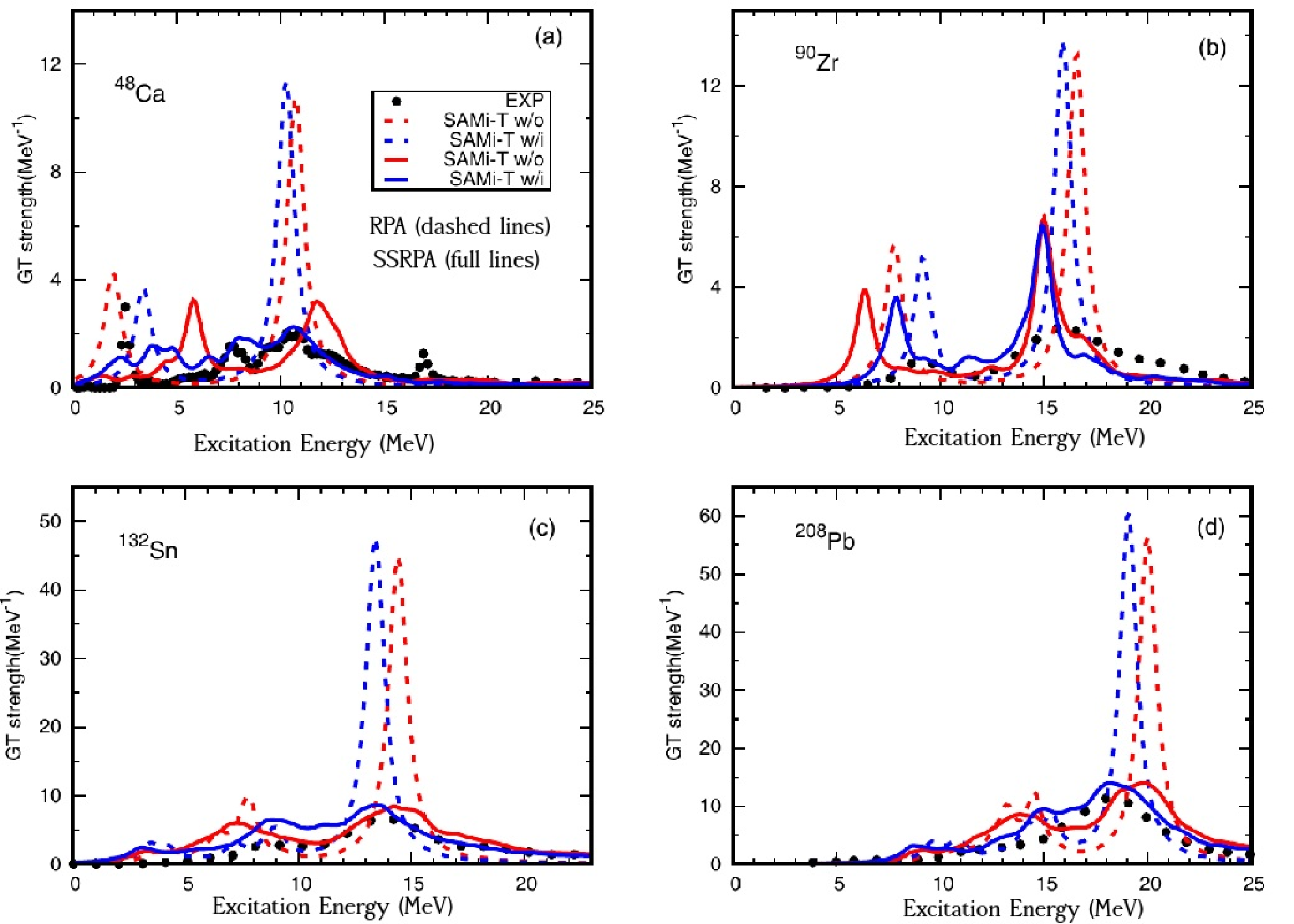}\hfill
	\includegraphics[width=.49\linewidth]{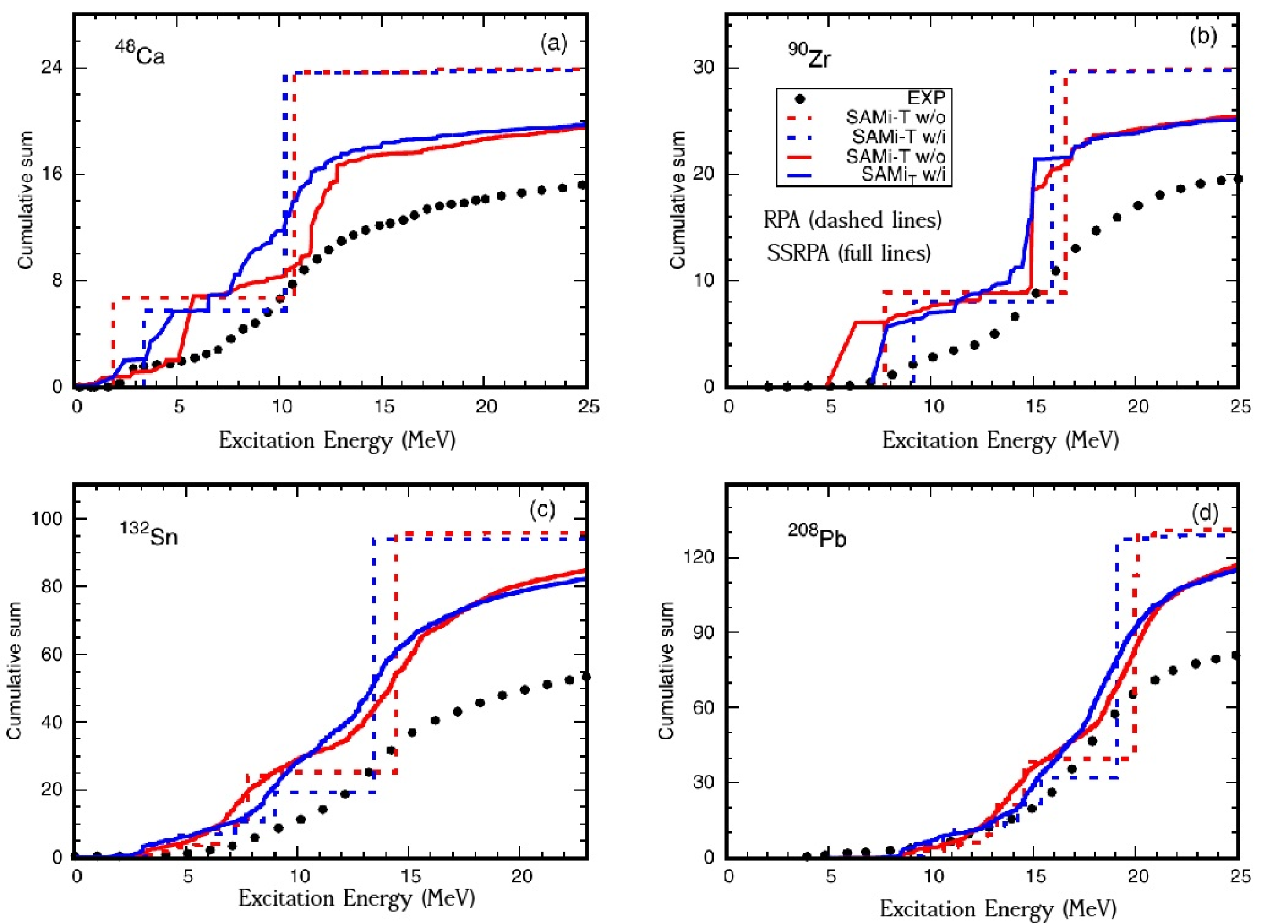}
	\caption{ Same as Figure \ref{Fig:GT_5}, but for the SAMi-T \cite{SAMIT} force with (label w/i ) or without ( labeled w/o) tensor terms. Adapted from Ref. \cite{Sagawa2022}.}
	\label{Fig:GT_7}
\end{figure}

\subsubsection{The role of the tensor force}
Tensor correlations have a significant influence on GT resonances within the RPA, see for example \cite{Bai2009,Cao2019}. It's role in the SSRPA for GT strength was studied in Ref. \cite{Sagawa2022}. Given that the tensor force in standard Skryme-EDFs is not optimally tuned for spin-isospin excitations, it is worth starting with Skryme-EDFs whose parameters, including tensor terms, were specifically fitted to reproduce a set of experimental observables, as for example the SAMi-T \cite{SAMIT} force. %These include SAMi-T \cite{SAMIT} and three EDFs from the TIJ \cite{Lesinski2007} family. After that, 
To further study the specific role of the tensor force on GT states, various choices of tensor terms added to existing Skryme-EDF parameterizations are also investigated. To this end, calculations with the   SGII + Te1, Te2, and Te3 were  performed \cite{Sagawa2022}. In these cases, the central part of the EDF remains unchanged, while the tensor forces are varied within a physically reasonable range to adequately describe GT and spin-dipole excitations, particularly concerning the excitation energies of the main peaks in several nuclei \cite{Bai2011}.

Left side of figure \ref{Fig:GT_7} presents the strength distributions calculated with the SAMi-T EDF, both with (blue lines) and without (red lines) tensor terms, within the RPA and SSRPA models. The RPA calculations without the tensor force exhibit main peak excitation energies that are approximately 1–2 MeV higher than those obtained with the tensor force across all considered nuclei. In SSRPA calculations, the inclusion of the tensor force leads to a downward shift of the main peaks by about 1 to 1.5 MeV, and the peak heights are reduced to values nearly matching the experimental data for $^{48}$Ca, $^{132}$Sn, and $^{208}$Pb. For $^{48}$Ca, the incorporation of tensor terms accurately reproduces not only the main peak at $E_x = 11$ MeV but also the shoulder structure at lower energy. Improved descriptions of the main peaks in terms of excitation energy and peak height are also observed for $^{132}$Sn and $^{208}$Pb. %Conversely, for $^{90}$Zr, the excitation energy of the main peak is almost unaffected by the tensor force, and agreement with experiment remains not so satisfactory. 
The corresponding cumulative sums are displayed in the right side of Figure \ref{Fig:GT_7} . The overall trends observed in RPA and SSRPA calculations are similar to those obtained with the SAMI (Figure \ref{Fig:GT_5}), showing only minor changes in the SSRPA case.

\begin{figure}
	\centering
	\includegraphics[width=.75\linewidth]{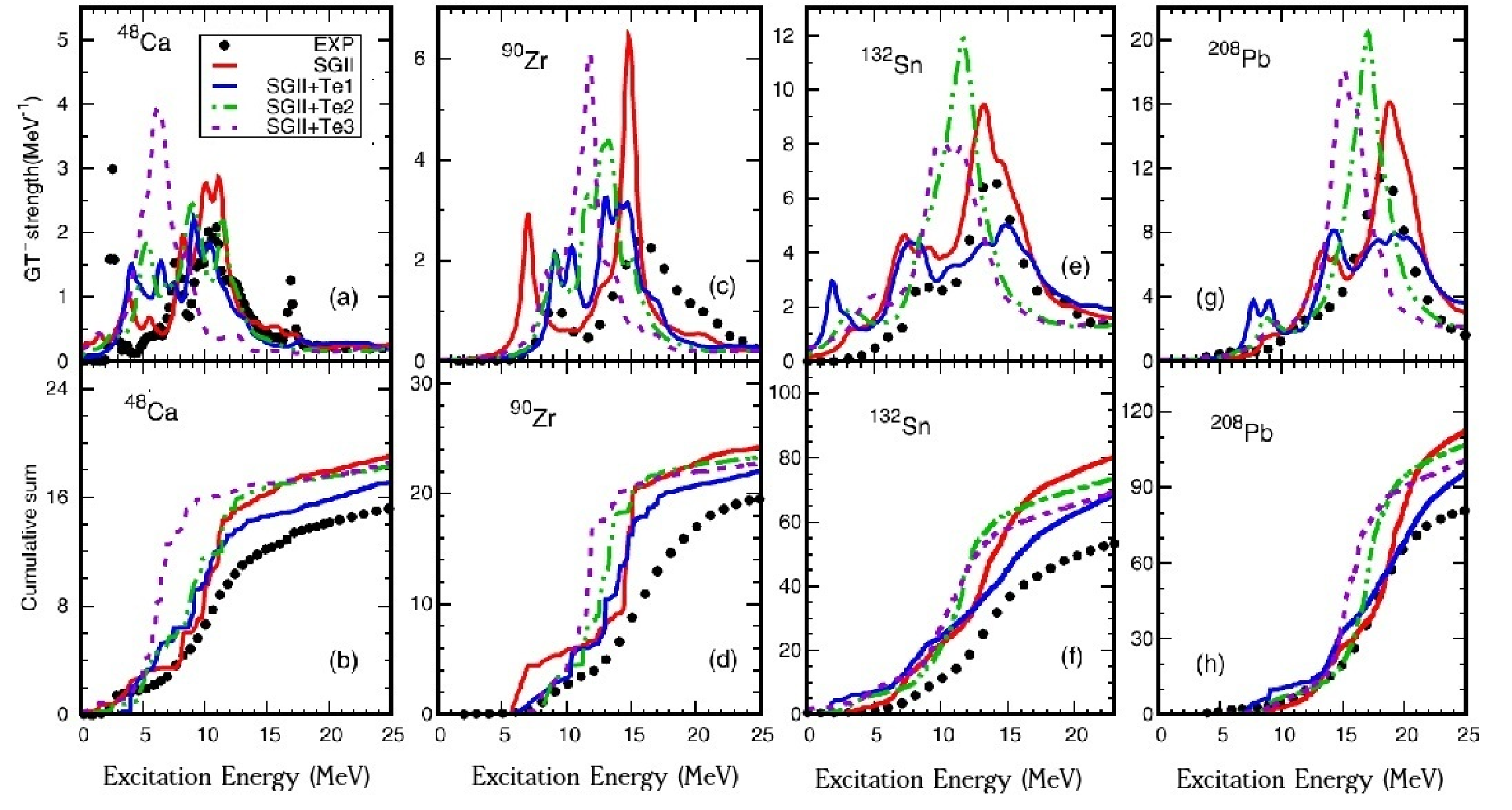}
	\caption{GT strength distributions for four nuclei (upper panels) and corresponding cumulative sums (lower panels) in: $^{48}$Ca (panel (a) and (b)), $^{90}$Zr (panel (c) and (d)), $^{132}$Sn (panel (e) and (f)), and $^{208}$Pb (panel (g) and (h)) but for the SGII, SGII +Te1, SGII +Te2, and SGII +Te3  shown by red solid lines, blue solid lines, green dash-dot-dot lines, and violet dashed lines, respectively. The experimental data are shown by the black filled circles. Adapted	from Ref. \cite{Sagawa2022}.
	}
	\label{Fig:GT_10}
\end{figure}
Figure \ref{Fig:GT_10} illustrates the strength distributions and their corresponding cumulative sums for $^{48}$Ca, $^{90}$Zr, $^{132}$Sn, and $^{208}$Pb, as computed using the SSRPA model with the SGII, SGII+Te1, SGII+Te2, and SGII+Te3 forces. The SGII+Te1 EDF leads to a reduction in the peak height of the primary GT resonance, without significantly changing the excitation energy compared to the SGII case. Consequently, the SGII+Te1 EDF provides a good reproduction of the main GT peaks in $^{48}$Ca, $^{132}$Sn, and $^{208}$Pb, accurately predicting both their excitation energies and peak heights. Nevertheless, it underestimates the excitation energy in $^{90}$Zr by approximately 1.5 MeV. The SGII+Te2 EDF yields a satisfactory description of the main peak energy in $^{48}$Ca. However, it underestimates the peak energies by about 2.5 MeV in $^{90}$Zr and $^{132}$Sn, and by roughly 1 MeV in $^{208}$Pb. The SGII+Te2 EDF overestimates the peak heights of the main GT resonances in $^{90}$Zr, $^{132}$Sn, and $^{208}$Pb. The SGII+Te3 EDF demonstrates poorer performance compared to the other two modified EDFs. It underestimates the excitation energies by approximately 5 MeV in $^{48}$Ca and $^{90}$Zr, and by about 3 MeV in $^{132}$Sn and $^{208}$Pb. 

\subsubsection{SSRPA in light nuclei and comparison with \textit{ab initio} calculations}
In  this section we consider the lighter systems $^{14}$C and $^{22}$O, for which \textit{ab initio} coupled-cluster (CC) calculations were performed in Ref. \cite{Ekstrom2014}. Although the Skryme-EDF approach is generally better suited for medium-mass and heavy nuclei, a qualitative comparisons with CC results can still be instructive.

\begin{wrapfigure}{l}{0.5\textwidth}
	\centering
	\includegraphics[width=0.9\linewidth]{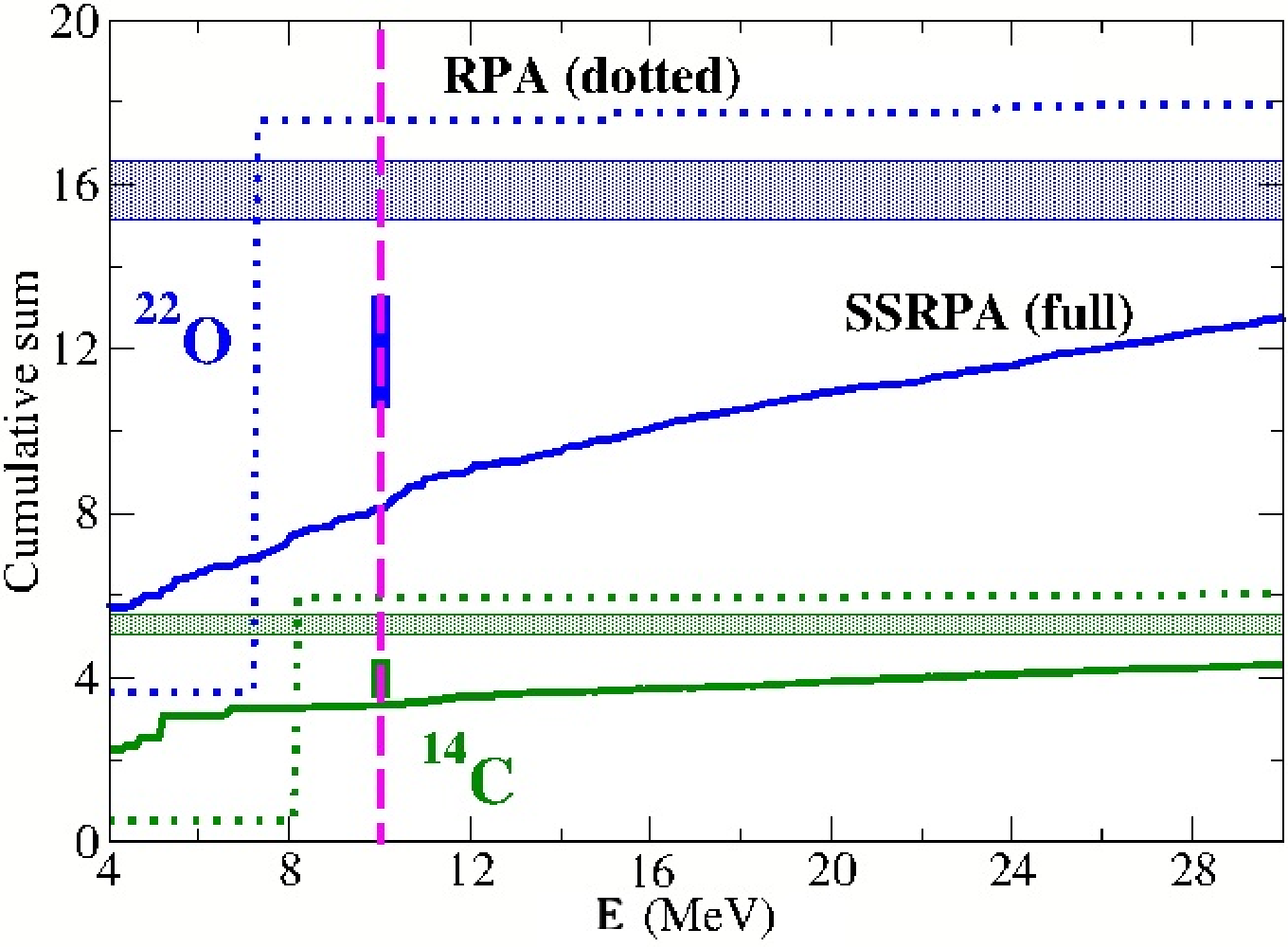}
	\caption{ Cumulative RPA and SSRPA $S_{GT^-}$ strengths for $^{22}$O (blue) and $^{14}$C (green). Horizontal shaded bands indicate the quenched total GT sum rule ($S_{GT^-} - S_{GT^+}$) predicted in Ref.\cite{Ekstrom2014} for each nucleus. The vertical dashed magenta line marks an excitation energy of 10 MeV, with the colored vertical bars representing the 70–80\% exhaustion of the quenched sum rule up to that energy, as reported in Ref. \cite
		{Ekstrom2014} (see text for details). Adapted from Ref. \cite{Gambacurta2022}.
	}
	\label{Fig:GT_12}
\end{wrapfigure}
The CC calculations in Ref. \cite{Ekstrom2014}, based on chiral effective field theory interactions, incorporate two-body (2B) currents into the GT transition operator in a manner consistent with the inclusion of three-body (3B) forces. This refinement significantly impacts the calculated GT strengths. By analyzing $^{14}$C, $^{22}$O, and $^{24}$O, it was shown that including 2B currents reduces the total GT strength compared to the nominal Ikeda value of $3(N-Z)$, yielding a quenched strength of approximately $q^2 \times 3(N-Z)$, with $q^2 \approx 0.84$–$0.92$. The unquenched value is recovered when the transition operator includes only its one-body component. This result supports the hypothesis that 2B currents can help resolve longstanding discrepancies between theoretical predictions and experimental GT strengths.

 Furthermore, it was shown that 70–80\% of the quenched strength is exhausted below 10 MeV of excitation energy. Similar conclusions were drawn in Ref. \cite{Gysbers2019} for $\beta$ decay in $^{100}$Sn, using the same theoretical framework. CC results are compared with RPA and SSRPA calculations using the SGII Skyrme force for $^{14}$C and $^{22}$O. 
%As in $^{48}$Ca, it is confirmed that a hybrid approximation (based on a diagonal treatment of the matrix inversion) is inadequate, underscoring the necessity of fully calculations.
 
 The cumulative $S_{GT^-}$ strengths from RPA and SSRPA are plotted in Figure \ref{Fig:GT_12} over the excitation energy range 4–30 MeV. As expected, $S_{GT^-}$ dominates over $S_{GT^+}$, with RPA cumulative sums nearly reaching the full Ikeda values of 6 and 18 for $^{14}$C and $^{22}$O, respectively. A vertical dashed magenta line in Fig.\ref{Fig:GT_12} marks 10 MeV, up to which 70–80\% of the quenched sum rule ($q^2 \times 3(N-Z)$) is exhausted. Horizontal bands in the figure indicate the quenched values for $^{14}$C and $^{22}$O, while vertical shaded intervals at 10 MeV show the 70–80\% region for each. The RPA cumulative sums exceed these intervals before 10 MeV, whereas SSRPA values remain below them and never reach the full quenched strength even up to 30 MeV. In particular, SSRPA results remain within the horizontal bands throughout, suggesting that the quenching mechanism in SSRPA, arising from the inclusion of $2p-2h$ configurations, is qualitatively consistent with that attributed to 2B currents in Ref. \cite{Ekstrom2014}.
Although a definitive comparison would require explicit inclusion of 2B operators in the present framework, the SSRPA results indicate that the inclusion $2p-2h$ configurations contributes significantly to the observed quenching.

\subsection{Beta-decay studies in SSRPA}
\label{Sec:Applications_SSRPA_Beta}
%The importance of BMF correlations in describing the beta half-life within RPA-based studies was shown in Ref. \cite{Niu2015} within the PVC model. 

\begin{wrapfigure}{l}{0.5\textwidth}
 	\vspace{-0.51cm}
 	\centering
 	\includegraphics[width=\linewidth]{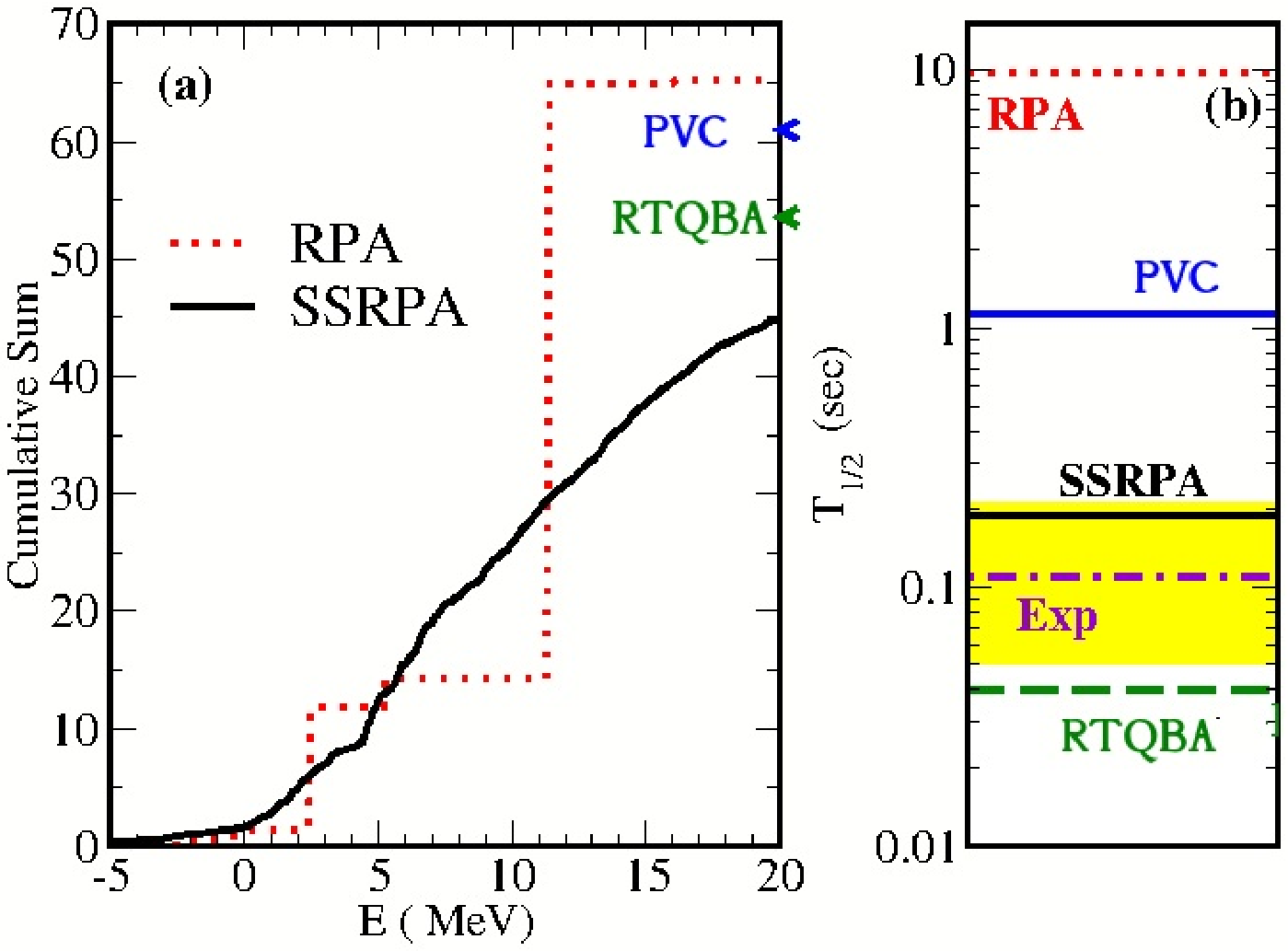}
 	\caption{ (a) Cumulative sum for different models (see legend and text) for the nucleus 
 		$^{78}$Ni in and (b) $\beta$-decay half-life for the same nucleus obtained in 
 		SSRPA, compared with predictions of other models and the experimental value \cite{Hosmer2005}. The yellow band represents the experimental uncertainty.
 		The RTBA \cite{Robin2018} and PVC \cite{Niu2015} results are shown for comparison. Adapted from Ref. \cite{Gambacurta2020}.
 	}
 	\label{Fig:Beta_0}
 \end{wrapfigure}

In Ref. \cite{Gambacurta2020}, the first evaluation of the beta half-life within the SSRPA was presented. In panel (a) of Figure \ref{Fig:Beta_0}, the cumulative sum of the GT strength for $^{78}$Ni obtained in SSRPA is compared to the RPA one, together with two other theoretical values, obtained within the RTQBA \cite{Robin2018} and PVC \cite{Niu2015}. The SSRPA integrated value is significantly lower than those of both the RPA and the other two BMF calculations. 
 
 Although the GT spectrum of $^{78}$Ni has not measured, the $\beta$-decay lifetime is known \cite{Hosmer2005} and it is shown in panel (b) together with the theoretical values for which the bare value $g_A=1.28$ for the weak axial-vector coupling constant has been used, instead of the usually adopted quenched value equal to $g_A=1.00$.  The SSRPA half-life is 0.19 seconds, very close to the experimental half-life. The RPA half-life, by contrast, is 9.51 seconds. 
 The RTQBA  \cite{Robin2018}  and PVC \cite{Niu2015} prediction are 0.04 and 0.69 seconds, respectively. The figure clearly shows the improvement induced by the BMF correlations within the three different models.
 
 More systematic calculations were performed afterwards in Refs \cite{Sagawa2023, Gambacurta2025}. 
In Ref. \cite{Sagawa2023}, $\beta$-decay half-lives of four semi-magic and magic nuclei, namely 
$^{34}$Si, $^{68,78}$Ni and $^{132}$Sn were investigated with different Skyrme interactions, studying also the effect of tensor interaction. The Skyrme forces SkM$^*$ \cite{SKM}, SIII \cite{SIII}, SLy5 \cite{SLY4}, SGII \cite{SGII}, SAMi \cite{SAMI}, and SAMi-T \cite{SAMIT} were employed.

\begin{wrapfigure}{r}{0.5\textwidth}
% 	\vspace{-0.5cm}
	\centering
	\includegraphics[width=0.859\linewidth]{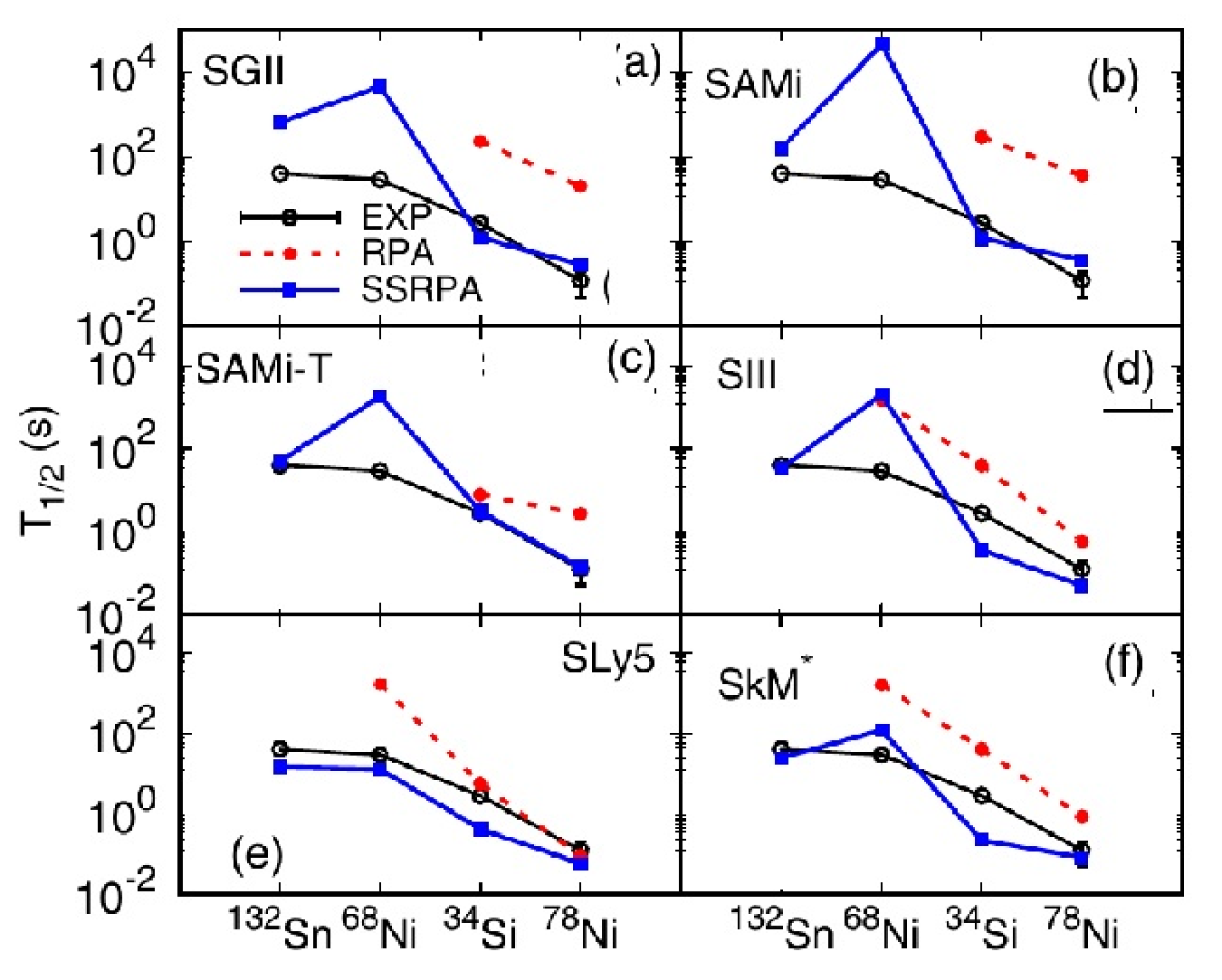}
	\caption{Experimental $\beta$-decay half-lives \cite{nndc} of $^{132}$Sn, $^{68}$Ni, $^{34}$Si, and $^{78}$Ni compared with the  RPA (red solid circles) and SSRPA ( blue solid
		squares) ones. 	Adapted from Ref. \cite{Sagawa2023}.	}
	\label{Fig:Beta_2}
\end{wrapfigure}
Figure \ref{Fig:Beta_2} presents a comparative analysis of $\beta$-decay half-lives for the selected set of nuclei, as predicted by the RPA and the SSRPA models, against experimental data. The RPA calculations generally exhibit a significant overestimation of the half-lives.
For $^{132}$Sn and $^{68}$Ni , the RPA predicts nuclear stability, resulting in infinite half-lives not shown in the figure. On the contrary, the SSRPA calculations yield finite half-life values for all considered nuclei, demonstrating a closer agreement with experimental observations. However, a discrepancy is observed between the SSRPA prediction and the experimental half-life for $^{68}$Ni.

SSRPA calculations employing the SLy5 and SkM* EDFs provide a better reproduction of the experimental half-lives for the four plotted nuclei compared to other EDFs. 
% Their prior SSRPA investigations [50] indicated that the SGII, SAMi, and SAMi-T EDFs consistently and accurately reproduce the giant GT (GT) strength distributions in the magic nuclei $^{48}$Ca, $^{90}$Zr, $^{132}$Sn, and $^{208}$Pb. In contrast, the SLy5 and SkM* EDFs yielded less satisfactory descriptions of these giant GT states.
As illustrated in panels (a) and (b) of Figure \ref{Fig:Beta_2}, the SSRPA calculations using the SGII and SAMI result in a reduction of approximately two orders of magnitude in the calculated half-lives of $^{34}$Si and $^{78}$Ni, leading to a good agreement with experimental values. Figure \ref{Fig:Beta_2}(c) highlights the inclusion of the tensor force in the SAMi-T EDF, demonstrating its significant impact in reducing the half-lives obtained from RPA calculations compared to the SAMI (Figure \ref{Fig:Beta_2}(b)), which lacks tensor terms.

 %Furthermore, a comparison of the SSRPA results obtained with SAMi and SAMi-T indicates that the tensor interaction noticeably shortens the half-lives of $^{132}$Sn and $^{68}$Ni by factors of approximately 4 and 25, respectively.
%While the SAMi and SAMi-T EDFs were optimized to the same experimental data set, some differences exist in the central part of the EDF due to the inclusion of tensor terms in SAMi-T.
\begin{figure}[h]
	\includegraphics[width=.5\linewidth]{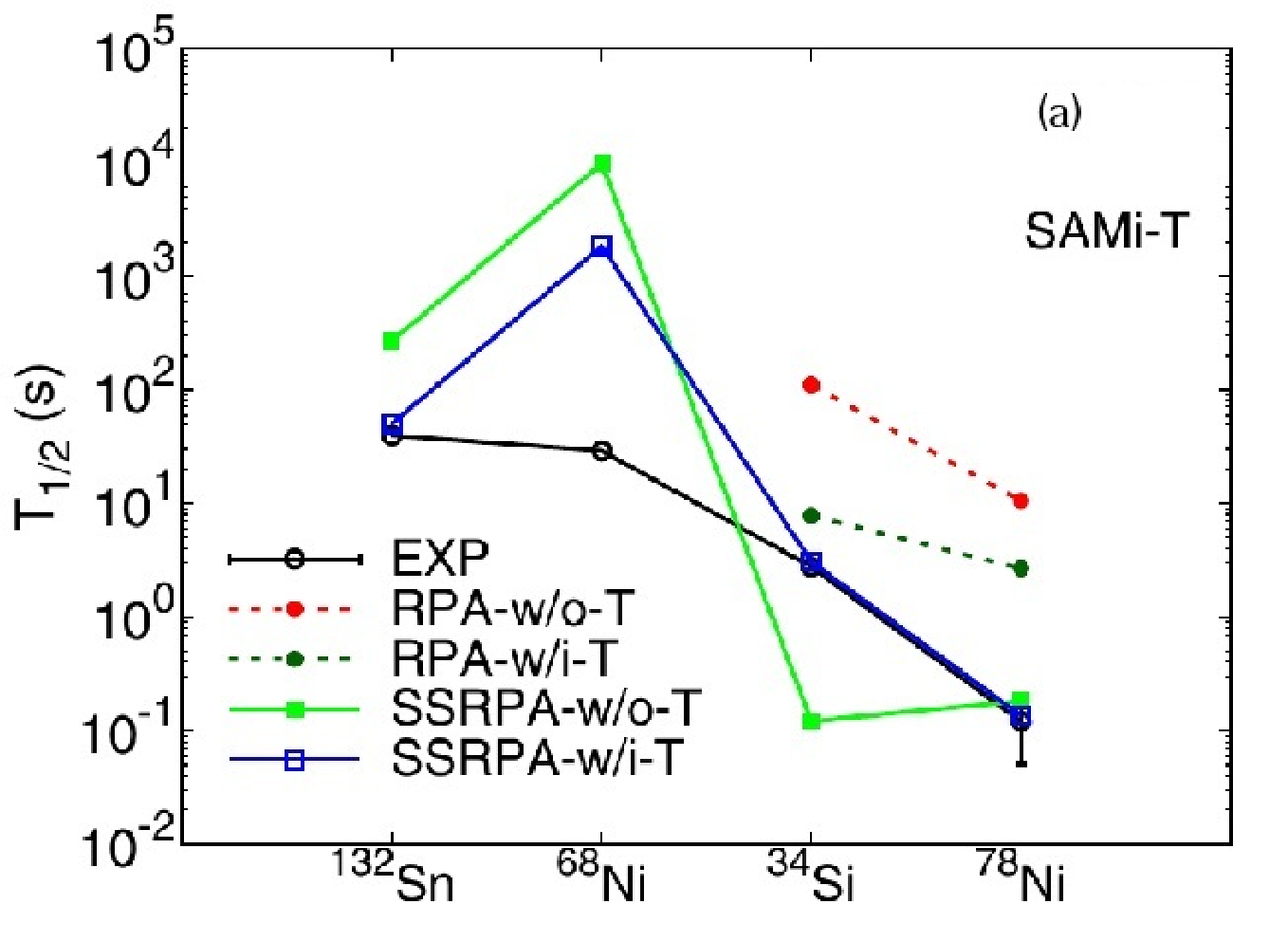}\hfill
	\includegraphics[width=.5\linewidth]{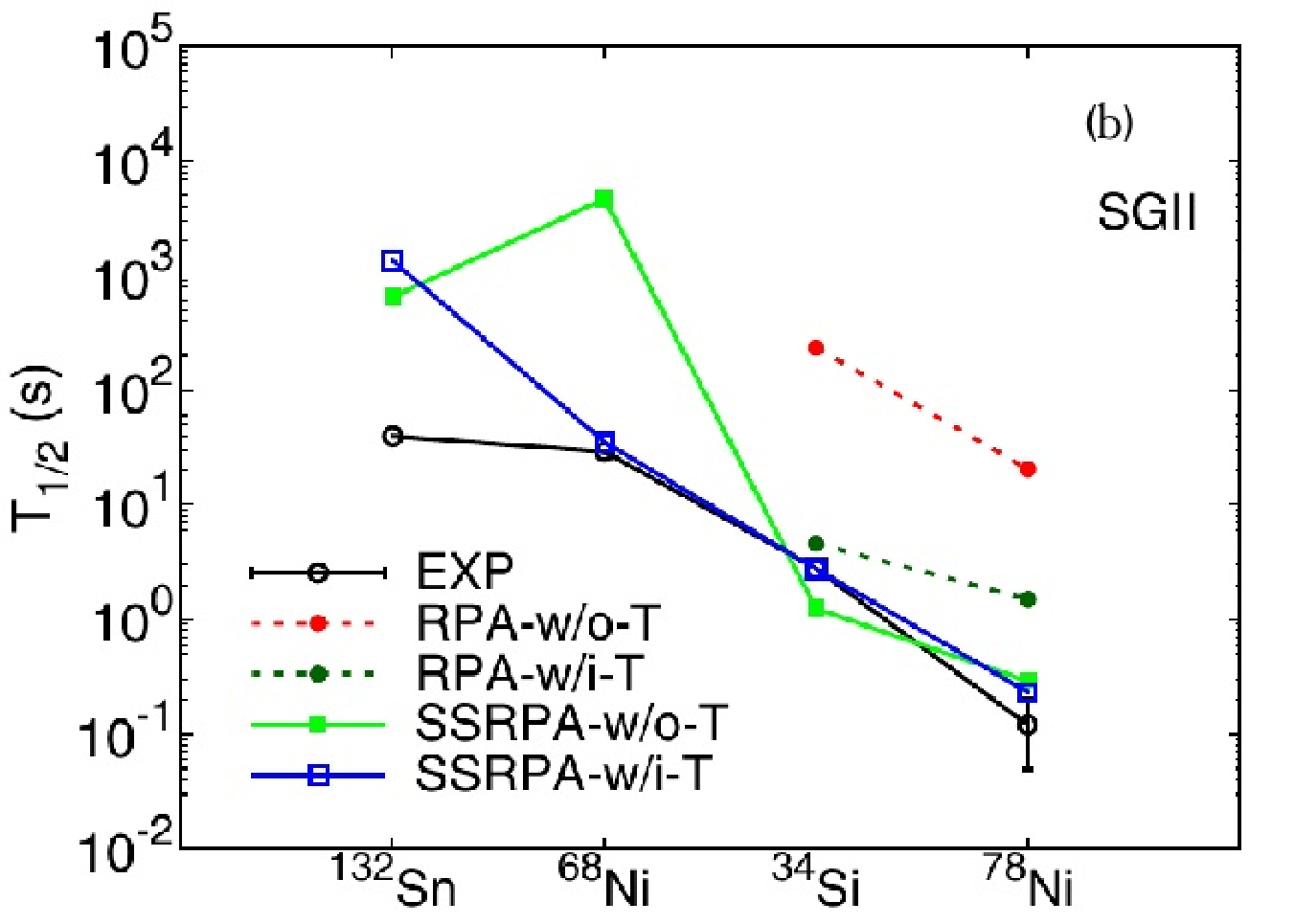}
	\caption{ $\beta$-decay half-lives of $^{132}$Sn, $^{68}$Ni, $^{34}$Si, and $^{78}$Ni calculated using the RPA and SSRPA models with the SAMi-T (left side) and SGII (right side) forces, both with ("w/i-T") and without ("w/o-T") the inclusion of tensor terms, compared to experimental data from Ref. \cite{nndc}. Experimental values are indicated by black empty circles.Left side: Right side: Figures from Ref. \cite{Sagawa2023}}
	\label{Fig:Beta_3}
\end{figure}
%To further investigate the specific effects of the tensor force, calculations were performed with the SAMi-T EDF both including and excluding the tensor terms. 
On the left side of Figure \ref{Fig:Beta_3} the $\beta$-decay half-lives are shown, calculated by RPA and SSRPA in two cases, labeled as "w/i-T" (with tensor) and "w/o-T" (without tensor). The inclusion of the tensor force at the RPA level significantly reduces the half-lives of $^{34}$Si and $^{78}$Ni. However, the calculated values remain larger than the experimental data. Infinite half-lives for $^{68}$Ni and $^{132}$Sn, even with the inclusion of tensor terms, are found. Consistent with the results obtained using the SAMI (Figure \ref{Fig:Beta_2}), the SSRPA calculation with the SAMi-T EDF excluding the tensor force reduces the half-lives of $^{34}$Si and $^{78}$Ni by more than two orders of magnitude compared to the corresponding RPA results. 
Nevertheless, these SSRPA calculations still overestimate the half-lives of $^{68}$Ni and $^{132}$Sn. The inclusion of the tensor force in the SSRPA calculations leads to a reduction in the half-lives of $^{132}$Sn and $^{68}$Ni by a factor greater than 5, while the half-life of $^{34}$Si is increased by approximately 30 times. Consequently, the SSRPA calculations incorporating the tensor force accurately reproduce the experimental half-lives of $^{132}$Sn, $^{78}$Ni, and $^{34}$Si, although the half-life of $^{68}$Ni remains overestimated.

%With the aim of optimizing the strengths of the tensor interaction, a series of calculations varying the parameters of the tensor force around the parameter set SGII+$T_{e1}$ were performed. This analysis revealed that the parametrization SGII+$T$(500, -280) provides the most realistic description for both investigated quantities. It is important to note that these optimized tensor strengths differ from those employed in the SAMi-T EDF, where $(T, U) = (415.5, -95.5)$. Specifically, the triplet-even component is similar in magnitude to that of SGII+$T$(500, -280), while the triplet-odd component is approximately three times weaker.

The right side of Figure \ref{Fig:Beta_3} illustrates the $\beta$-decay half-lives  calculated using the RPA and SSRPA models, both with and without the inclusion of the SGII augmented by the tensor terms $T$(500, -280). The influence of the tensor force on the RPA calculations exhibits trends analogous to those observed with the SAMi-T EDF. 
%Specifically, the tensor interaction significantly reduces the calculated half-lives of $^{34}$Si and $^{78}$Ni at the RPA level. However, it fails to produce finite half-life values for $^{68}$Ni and $^{132}$Sn within the RPA framework.
The inclusion of $2p-2h$ configurations leads to a reduction of approximately two orders of magnitude in the $\beta$-decay half-lives of $^{78}$Ni and $^{34}$Si, and importantly, yields finite half-life predictions for $^{132}$Sn and $^{68}$Ni. Within SSRPA, the tensor force has its strongest effect in $^{68}$Ni, where it induces a reduction in the half-life by two orders of magnitude. In contrast, the tensor terms for $^{34}$Si result in a slight enhancement of the half-life.

The effect of the $2p-2h$ configurations on the $\beta-$decay has been also studied in a subsequent independent work  \cite{Gambacurta2025} for four nuclei: $^{24}$O, $^{34}$Si, $^{78}$Ni, and $^{132}$Sn which can be safely considered as spherical non-superfluid nuclei where pairing correlations are not expected to play an important role. Indeed, as it can be seen from the previous figures, the nucleus $^{68}$Ni shows in general a more problematic behavior in reproducing the experimental data, also within the SSRPA. $^{68}$Ni is indeed super-fluid in the neutron channel, the inclusion of pairing correlations is therefore necessary. For this reason,  it was not included in Ref. \cite{Gambacurta2025}.

\begin{wrapfigure}{l}{0.5\textwidth}
	\centering
	\vspace{-0.5cm}
	\includegraphics[width=0.45\textwidth]{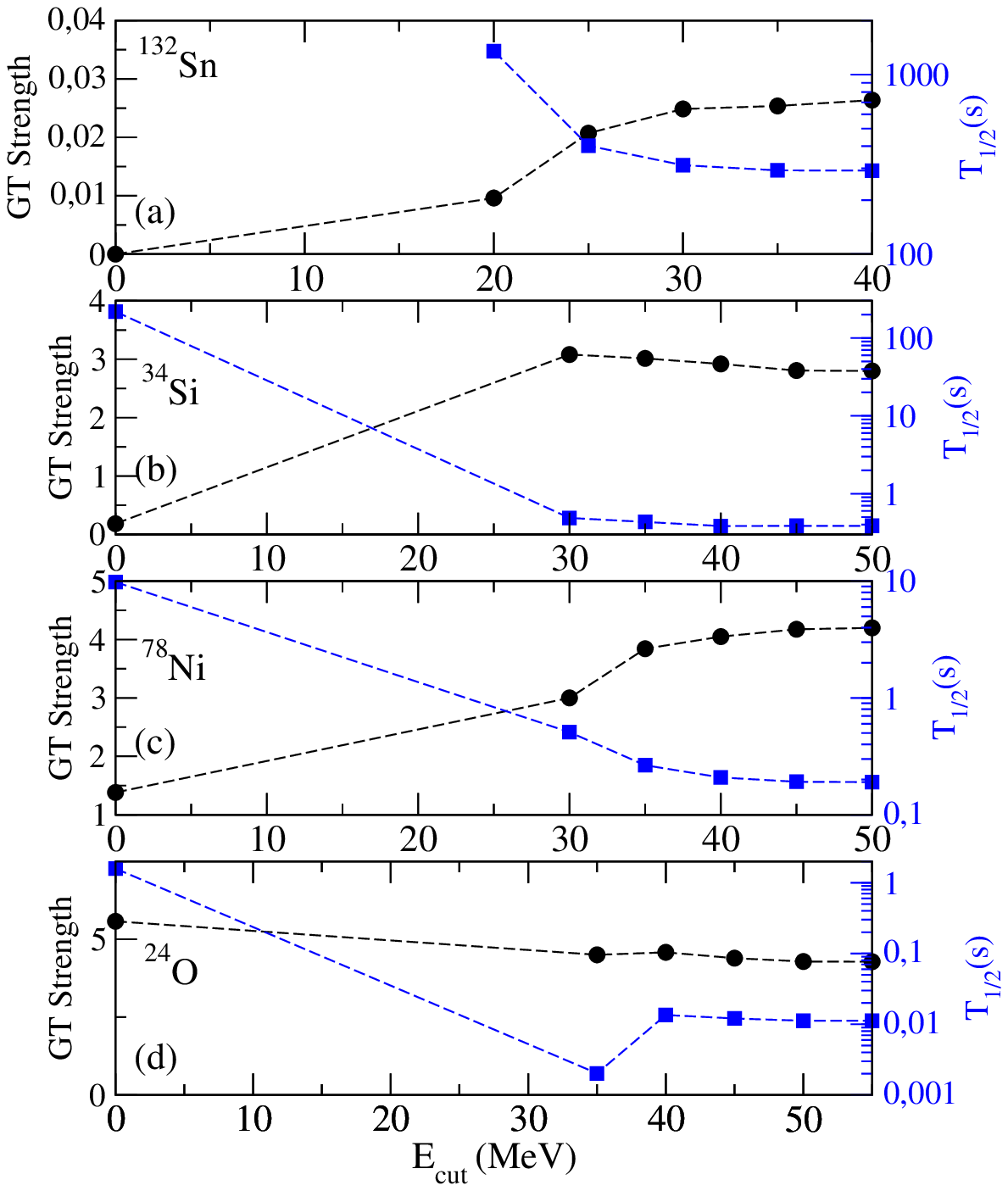}
	\vspace{-0.5 cm}
	\caption{SGII-SSRPA integrated GT strength in the $\beta$-window (black circle, right y-axis) and corresponding $\beta$-decay half-lives (blue square, left y-axis) for different nuclei
		%		$^{132}$Sn (panel (a)), $^{34}$Si (panel (b)), and $^{78}$Ni (panel (c)) and $^{24}$O (panel (d))
		for increasing values of the energy cutoff $E_{cut}$ on the $2p-2h$ configurations. The values for $E_{cut}=0$ MeV, corresponding to the RPA results, are shown as a reference. Adapted from Ref. \cite{Gambacurta2025}.}
	\label{Fig:Beta_4bis}
\end{wrapfigure}

 Calculations were performed by including all $1p-1h$ configurations with an unperturbed energy below 100 MeV were incorporated, accounting for over 99\% of the Ikeda sum rule across all the nuclei.
 
  To verify the convergence of our calculations with respect to the $2p-2h$ energy cutoff ($E_{cut}$), SSRPA calculations were performed for each nucleus and interaction, $E_{cut}$ was incrementally increased by 5 MeV in successive steps. This increment size was chosen to ensure a meaningful expansion of the $2p-2h$ configuration space for every nucleus studied. Convergence was considered achieved when the relative change in the $\beta$-decay half-life between two consecutive $E_{cut}$ evaluations dropped below 1\%.
While the precise $E_{cut}$ value required for convergence showed minor variations across different nuclei and interactions, a clear pattern emerged regarding the values that yielded stable results: 50-55 MeV for $^{24}$O, 40-45 MeV for $^{34}$Si and $^{78}$Ni, and 30-35 MeV for $^{132}$Sn. 
%The SGII interaction serves as a prime example of slower convergence, requiring higher cutoff values than other parametrizations to reach the stability.
Figure \ref{Fig:Beta_4bis} presents how the SSRPA integrated GT strength within the $\beta$-window (black circles, right y-axis) and the corresponding $\beta$-decay half-lives (blue squares, left y-axis) evolve with the energy cutoff $E_{cut}$. These plots are provided for $^{132}$Sn (panel (a)), $^{34}$Si (panel (b)), $^{78}$Ni (panel (c)), and $^{24}$O (panel (d)), and where the $E_{cut} = 0$ MeV results correspond to the RPA values, included for baseline comparison. A consistent observation across all the nuclei  is the reduction in $\beta$-decay half-life from RPA to SSRPA, with further decreases noted as $E_{cut}$ is progressively raised until convergence. Regarding the integrated strength, SSRPA generally yields higher values compared to RPA. The only exception is $^{24}$O, where a slight reduction is observed when moving from RPA to SSRPA (as also shown in panel (d) of Figure \ref{Fig:Beta_4bis}). Despite this minor anomaly for $^{24}$O, its SSRPA half-life remains shorter than its RPA counterpart. This outcome is primarily due to the energy-dependent phase-space volume $f(Z,\omega)$ (Eq. (\ref{niu1})), which assigns a specific weight to the strength of each state. This highlights a crucial point, the $\beta$-decay half-life is shaped by the distribution of strength within the $\beta$-window, not simply by the total integrated strength.
% It's worth noting that for $^{132}$Sn, the RPA half-life is theoretically infinite due to the complete absence of GT strength within the relevant energy window.

%Figure \ref{Fig:Beta_5} presents a comparison of $\beta$-decay half-lives calculated using the RPA and SSRPA theoretical frameworks with experimental data \cite{nndc}. Four Skyrme parametrizations were employed in this analysis: SAMI \cite{SAMI} (panel a), SGII \cite{SGII} (panel b), SkM$^*$ \cite{SKM} (panel c), and SLy5 \cite{SLY4} (panel d). The SSRPA calculations are done in the diagonal approximation in the subtraction procedure, a method validated for the scope of this study as detailed in Ref. \cite{Gambacurta2020}. 

%For certain nuclei, the RPA calculations failed to predict any state within the $\beta$-decay window, resulting in infinite half-lives. Notably, none of the adopted Skyrme forces yielded a finite $\beta$-decay half-life for $^{132}$Sn at the RPA level.

\begin{wrapfigure}{r}{0.5\textwidth}
	\centering
	%  	\vspace{-3.5mm}
	\includegraphics[width=0.49\textwidth]{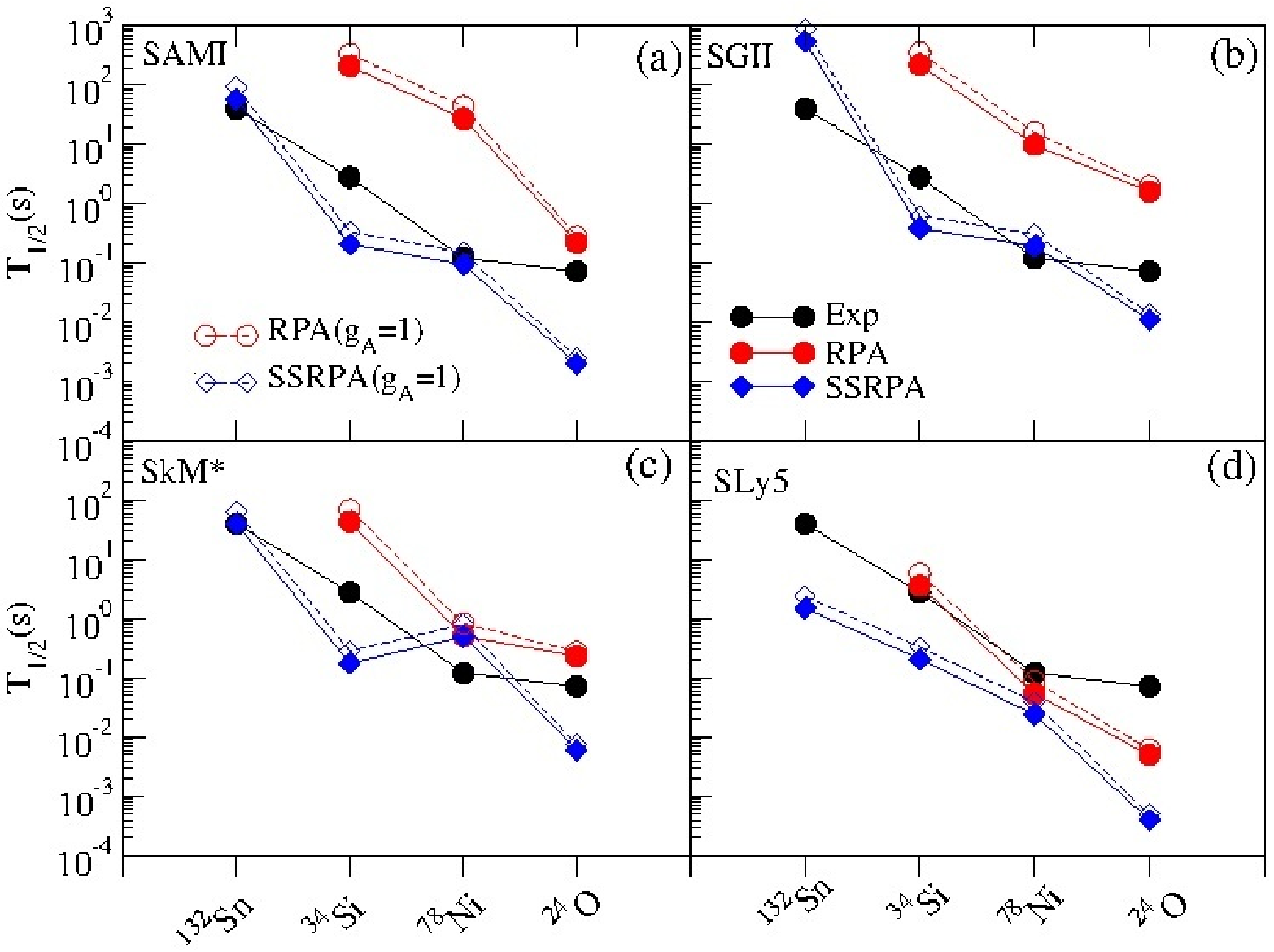}
	\caption{Experimental $\beta$-decay half-lives \cite{nndc} compared with those obtained within the RPA and SSRPA models. Full (empty) symbols are obtained by using the bare $g_A=1.26$ ($g_A=1.00$) constant. The RPA values that are not shown correspond to infinite half-life values. See the text for more details. Adapted from Ref. \cite{Gambacurta2025}.}
	\label{Fig:Beta_5}
\end{wrapfigure}

The resulting beta-decay half-lives are shown in Figure \ref{Fig:Beta_5}. In line with the previous discussed results, the RPA calculations significantly overestimated the experimental data, with the exception of the SLy5 force for $^{34}$Si and $^{78}$Ni.
 In contrast, the SSRPA calculations, for all Skyrme forces and nuclei, predicted excited states within the $\beta$-decay window, consequently providing finite half-lives. 
 %A consistent trend observed is that the inclusion of $2p-2h$ configurations in SSRPA, which leads to a richer and more fragmented spectrum compared to the RPA distribution, resulted in considerably lower half-lives, demonstrating an overall improved agreement with experimental data.
  Solid and dashed symbols represent results obtained using axial-vector coupling constants of $g_A=1.26$ and $g_A=1.00$, respectively. This comparison illustrates a genuine quenching effect within the SSRPA framework. The term "genuine quenching" emphasizes that this reduction arises microscopically from the inherent coupling with $2p-2h$ configurations within the SSRPA, unlike the \textit{ad hoc} quenching factors typically employed. This microscopically derived quenching effect is significantly stronger than the artificial one achieved by simply using an effective (reduced) $g_A=1.00$ value.
  
These results are consistent with those reported in Ref. \cite{Sagawa2023} and discussed above, with minor discrepancies potentially attributable to numerical details. In both sets of SSRPA calculations, the inclusion of $2p-2h$ configurations consistently lowered the $\beta$-decay half-lives, generally improving the agreement with experimental values compared to the RPA results. 
The general trend of the SSRPA results also aligns with findings obtained within the PVC model \cite{Niu2015}, where the effects of more complex configurations were introduced through coupling with collective phonons, a mechanism that appears more effective for $^{68}$Ni than the effects found in Ref. \cite{Sagawa2023}.

It is also noteworthy that, with the exception of $^{132}$Sn, for which all parametrizations yielded an infinite half-life at the RPA level, the SLy5 force provided RPA half-lives in good agreement with experimental data, particularly for $^{34}$Si and $^{78}$Ni. This force was specifically designed to improve the description of isospin degrees of freedom within the Skyrme functional, especially in nuclei far from stability, achieved partly by fitting to properties of asymmetric nuclear matter as pseudo-data. This protocol likely explains the good agreement observed with the SLy5 force at the RPA level. However, incorporating correlations beyond RPA, as done in SSRPA (see also Ref. \cite{Sagawa2023}) and PVC (Ref. \cite{Niu2018}) calculations, generally worsened the agreement with experimental data for this parametrization. Therefore, beyond-RPA approaches utilizing mean-field-optimized effective interactions need careful validation. While subtraction methods effectively address double-counting and restore the overall response, reproducing fine spectral details, such as strength in narrow energy regions as in the present study, may remain dependent on the specific effective interaction employed.

Given the strong sensitivity of $\beta$-decay half-lives to the low-energy strength distribution, the role of the $J^2$-terms\cite{Vautherin1972} deserves to be studied. These terms, arising from the central part of the nuclear interaction, contribute to the ground state energy through terms proportional to the square of the current density $J$ and are not consistently included in all parametrizations of Skyrme forces. While their contribution to the binding energy is often minimal, they may influence the low-lying distribution of the spectra, potentially competing with effects induced by the tensor force \cite{Bender2002}.

\begin{wrapfigure}{l}{0.4\textwidth}
	\centering
	%  	\vspace{-3.5mm}
	\includegraphics[width=.8\linewidth]{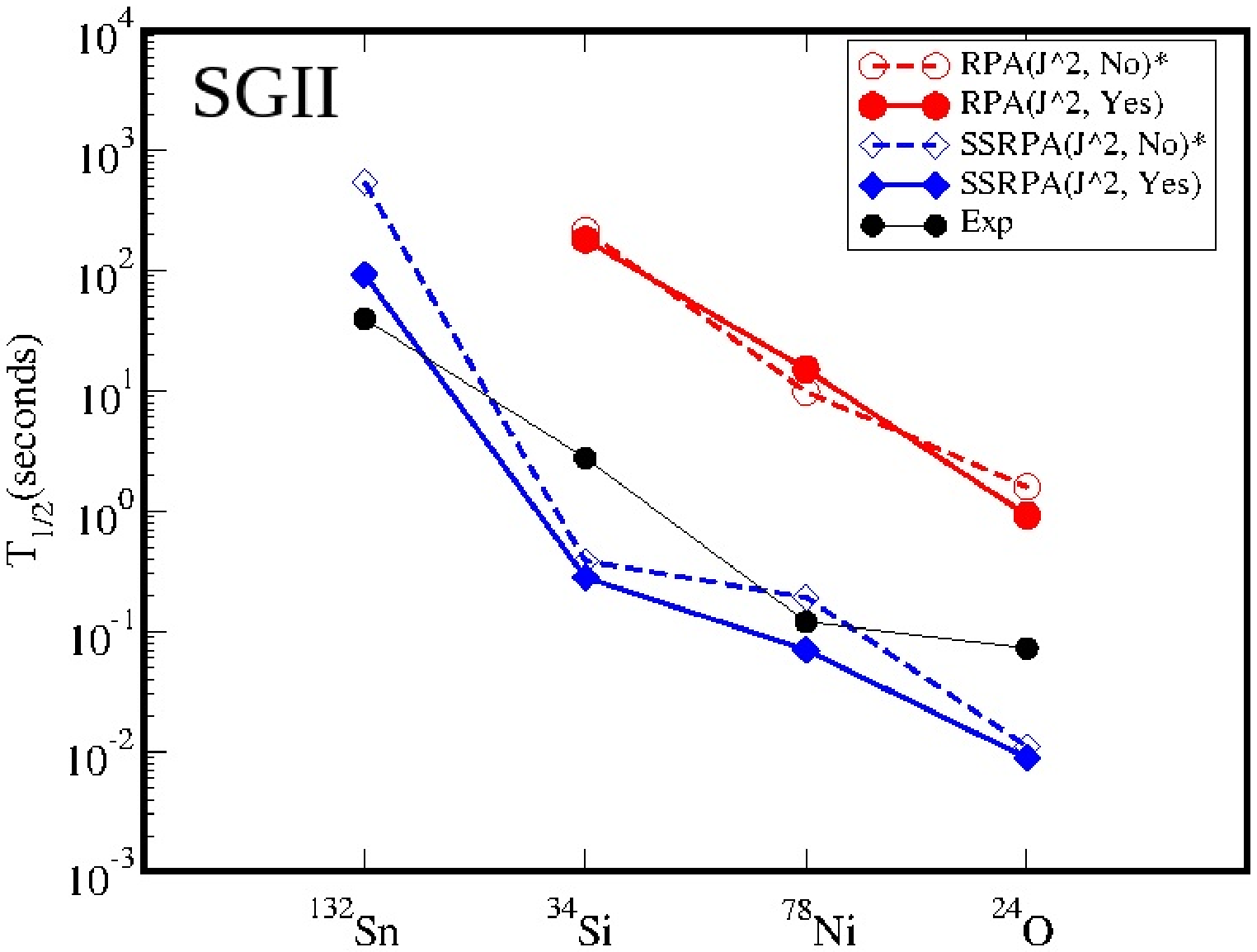}
	\hfill
% 	\vspace{-0.1 cm}
	\includegraphics[width=.8\linewidth]{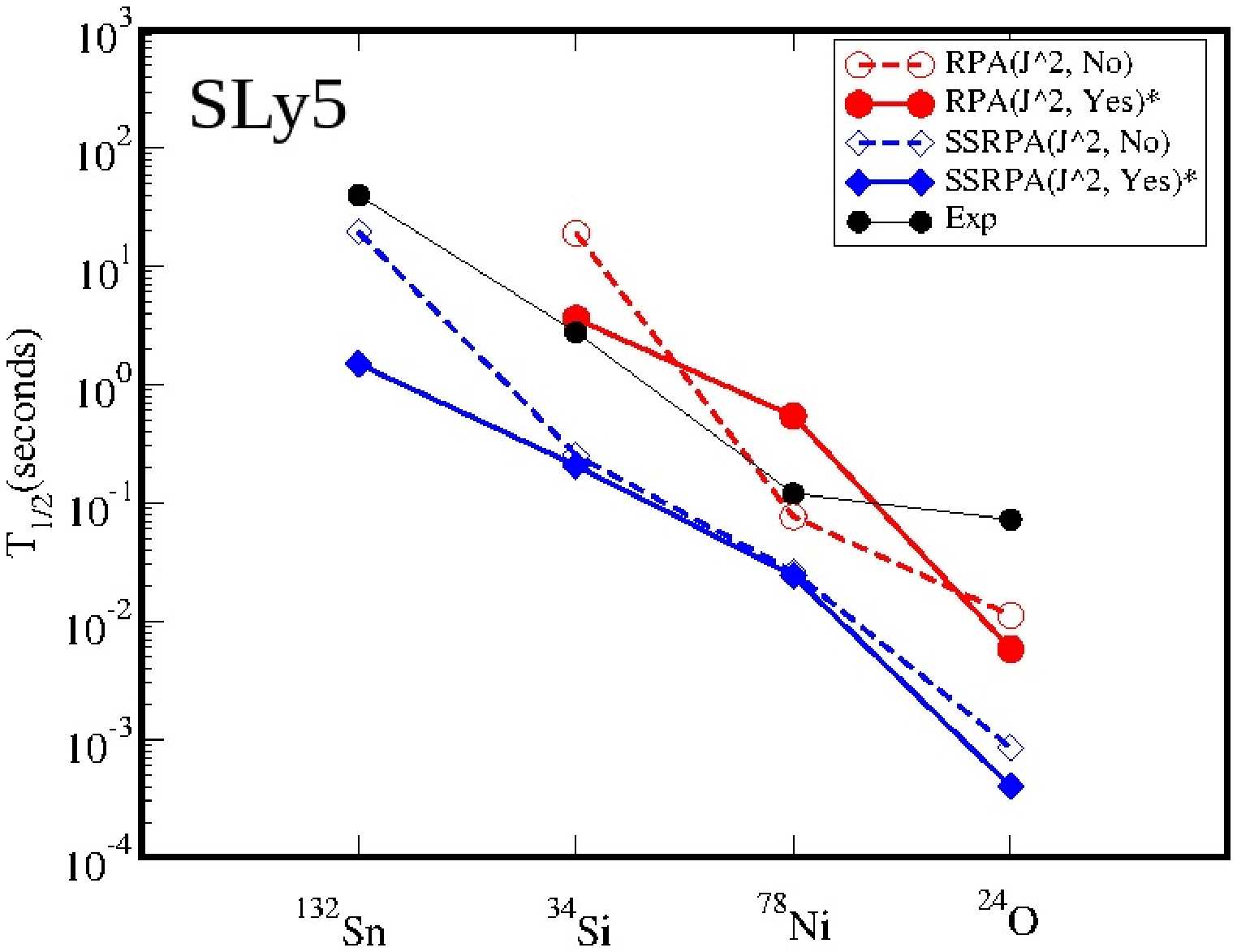}
	\caption{Comparison between the RPA and SSRPA with and without the $J^2$ terms. Top panel: the SGII interaction is used. The ``*'' symbol in the legend indicates that the SGII force has been derived without $J^2$ terms. Bottom side: The SLy5 interaction is used. The ``*'' symbol in the legend indicates that the SLy5 force has been derived with $J^2$ terms. Adapted from Ref. \cite{Gambacurta2025}.	}
	\label{Fig:Beta_6}
\end{wrapfigure}

To quantify the impact of the $J^2$-terms on $\beta$-decay half-lives, selected the SGII interaction, which was fitted without these terms, and and the SLy5 force which explicitly includes them, are considered in  Figure \ref{Fig:Beta_6}, respectively. For these two parametrizations, the effect of switching the $J^2$-terms on and off at both the RPA and SSRPA levels, is investigated. For the SGII interaction, we can see that, at the RPA level, the effect of the $J^2$-terms was found to be largely negligible for $^{34}$Si, with a slightly more noticeable impact for $^{78}$Ni, although their inclusion did not improve the agreement with experimental data in either case. However, at the SSRPA level, the $J^2$-terms exhibited a significant influence on the predicted half-life of $^{132}$Sn, with their inclusion further enhancing the agreement with experimental data. The effect was considerably weaker for the other nuclei, where the agreement was not substantially altered.

 For the SLy5 force, in the RPA calculations, the $J^2$-terms impacted the half-lives of $^{34}$Si and $^{78}$Ni, but in opposing directions; their exclusion worsened the agreement for $^{34}$Si while improving it for $^{78}$Ni. Conversely, within the SSRPA framework the results remained almost unchanged for $^{24}$O, $^{34}$Si, and $^{78}$Ni. Notably, for $^{132}$Sn, excluding the $J^2$-terms yielded results closer to the experimental data. These results suggest that the $J^2$-terms can play a significant role in determining $\beta$-decay half-lives, although in a manner that is dependent on both the specific nucleus and the Skyrme force employed. Furthermore, their effect appears to be comparable in magnitude to that of the tensor interaction above discussed. The competition between these two types of terms deserves careful consideration, particularly in the development of new energy density functionals aimed at achieving an improved description of $\beta$-decay processes and GT strength distributions.

%	\begin{wraptable}{r}{0.55\textwidth}
%		\small
%		\centering
%	\begin{tabular}{|ccccc|}
%		\hline
%		&&E$_{RPA}$= 0.310 MeV &&\\
%		\hline
%		$1p-1h$ conf. &$E_{1p-1h}$ &$N_{ph}$& &\\
%		$[\pi2s_{1/2},\nu2s_{1/2}]$& -1.379& 0.440 &&\\
%		% $[\pi1d_{3/2},\nu2s_{1/2}]$& 2.161& 0.035 &&\\
%		$[\pi1d_{3/2},\nu1d_{3/2}]$& -0.726& 0.513 &&\\
%		&$N_1=1.000$&$N_2=0.000 $ &&\\
%		\hline
%		$1p-1h$ conf. &$E_{1p-1h}$ &$A_{ph}$& $T_{ph}$&$b_{ph}$\\
%		\hline
%		$[\pi1d_{3/2},\nu1d_{5/2}]$& 6.038& -0.030& 3.094& -0.094 \\
%		$[\pi2s_{1/2},\nu2s_{1/2}]$& -1.379& 0.663& 2.443& 1.620 \\
%		$[\pi1d_{3/2},\nu1d_{3/2}]$& -0.726& 0.716& -1.548& -1.109 \\
%		\hline
%		$\sum b_{ph}$&& && 0.423\\
%		$P(\nu)$&& && 0.179\\
%		\hline
%	\end{tabular}
%	\caption {\label{tab-1} Particle-hole configurations which give the major contributions to the norm (upper part) and to the GT transition probability (lower part) for the RPA state located at 0.310 MeV. See upper panel of Figure \ref{Fig:Beta_7}. The superscripts $\pi$, $\nu$ refer to
%		proton and neutron states, respectively. See Eqs (\ref{norm}) and (\ref{prob}) for the definition of each quantity. The $E_{1p-1h}$ energies are given in MeV units. From Ref. \cite{Gambacurta2025}.
%		\textcolor{red}{here wrap}}
%	\label{tab-1}
%	\vspace{1cm}
%\end{wraptable}

\begin{wrapfigure}{l}{0.5\textwidth}
	\centering
	\includegraphics[width=\linewidth]{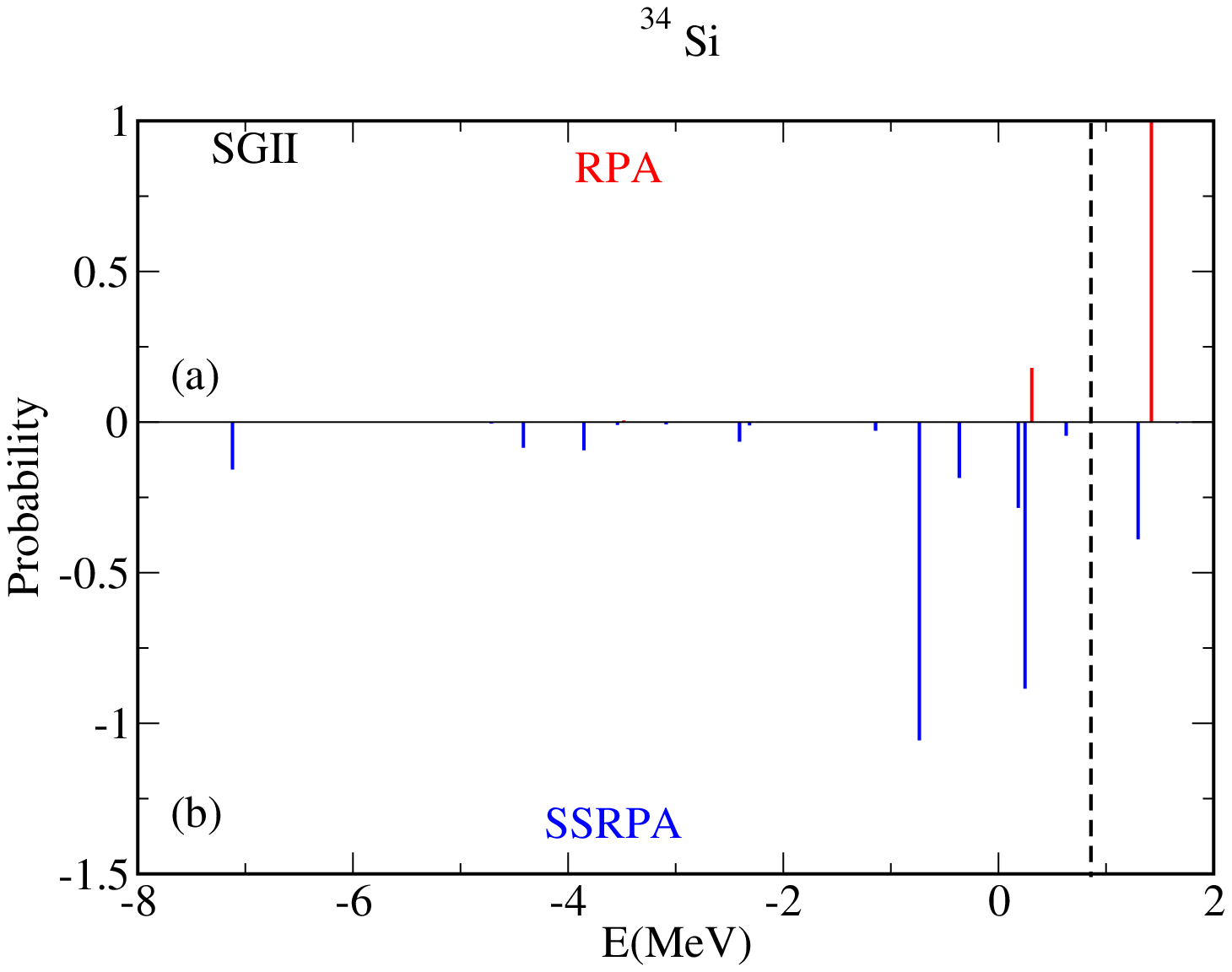}
	\caption{
		Low-lying GT strength distribution obtained in RPA and SSRPA for $^{34}$Si. Adapted from Ref. \cite{Gambacurta2025}.
	}
	\label{Fig:Beta_7}
\end{wrapfigure}

For the $^{34}$Si nucleus, a more detailed discussion analyzing the RPA and SSRPA wave functions of the most collective states contributing to the $\beta$ half-life is presented. In Figure \ref{Fig:Beta_7}, the GT strength distributions are shown as a function of the excitation energy with respect to the parent nucleus for the RPA (upper panel) and SSRPA (lower panel) with the SGII interaction. Only states with a transition probability larger than 0.1 and with energy lower than the $\Delta_{nH}$ value (black dashed line), relevant to the $\beta$-decay are considered.

We define \begin{equation}
	N_{ph}^{\nu} = |X^\nu_{ph}|^2 -|Y^\nu_{ph}|^2
	\label{norm}
\end{equation}
quantifying the amount of each $1p-1h$ configuration in the norm of each phonon state $\nu$, having that $\sum_{ph}N_{ph}^{\nu}=1$ in the RPA case, while in the SSRPA case, the normalizations reads as
\begin{equation}
	\sum_{ph}N_{ph}^{\nu}+	\sum_{php'h'}	N_{php'h'}^{\nu}=1
	\label{norm_tot}
\end{equation}
where

\begin{equation}
	N_{php'h'}^{\nu} = |X^\nu_{php'h'}|^2 -|Y^\nu_{php'h'}|^2
	\label{norm2}
\end{equation}
is the contribution of each $2p-2h$ configuration.

The probability to be excited $P(\nu)$ is given by the sum of each individual $1p-1h$ contribution $b_{ph}(\nu)$, that is the product of the ave-function amplitude $A_{ph}^\nu=X_{ph}^\nu - Y_{ph}^\nu$ multiplied by the matrix element of the transition operator $T_{ph}^\lambda$, having thus
\begin{equation}
	P(\nu) = |\sum_{ph} b_{ph}(\nu)|^2 =|\sum_{ph} (X_{ph}^\nu - Y_{ph}^\nu)
	T_{ph}^\lambda|^2=
	|\sum_{ph} A_{ph}^\nu
	T_{ph}^\lambda|^2.
	\label{prob}
\end{equation}

\begin {table}
\small
\centering
\resizebox{0.45\columnwidth}{!}{%
	\begin{tabular}{|ccccc|}
		\hline
		&&E$_{RPA}$= 0.310 MeV &&\\
		\hline
		$1p-1h$ conf. &$E_{1p-1h}$ &$N_{ph}$& &\\
		$[\pi2s_{1/2},\nu2s_{1/2}]$& -1.379& 0.440 &&\\
		% $[\pi1d_{3/2},\nu2s_{1/2}]$& 2.161& 0.035 &&\\
		$[\pi1d_{3/2},\nu1d_{3/2}]$& -0.726& 0.513 &&\\
		&$N_1=1.000$&$N_2=0.000 $ &&\\
		\hline
		$1p-1h$ conf. &$E_{1p-1h}$ &$A_{ph}$& $T_{ph}$&$b_{ph}$\\
		\hline
		$[\pi1d_{3/2},\nu1d_{5/2}]$& 6.038& -0.030& 3.094& -0.094 \\
		$[\pi2s_{1/2},\nu2s_{1/2}]$& -1.379& 0.663& 2.443& 1.620 \\
		$[\pi1d_{3/2},\nu1d_{3/2}]$& -0.726& 0.716& -1.548& -1.109 \\
		\hline
		$\sum b_{ph}$&& && 0.423\\
		$P(\nu)$&& && 0.179\\
		\hline
	\end{tabular}
}
\caption {\label{tab-1} Particle-hole configurations which give the major contributions to the norm (upper part) and to the GT transition probability (lower part) for the RPA state located at 0.310 MeV. See upper panel of Figure \ref{Fig:Beta_7}. The superscripts $\pi$, $\nu$ refer to
	proton and neutron states, respectively. See Eqs (\ref{norm}) and (\ref{prob}) for the definition of each quantity. The $E_{1p-1h}$ energies are given in MeV units. From Ref. \cite{Gambacurta2025}.}
	\end{table}

\begin{table}[ht] % Optional: Use the 'table' environment for floating/caption
	\centering
	\caption {\label{tab-2} Particle-hole configurations which give
		the major contributions to the norm (upper part) and to the GT transition probability (lower part) for the for the SSRPA states within the $\beta$ window located at -0.735 and, 0.247 MeV . See upper panel of Figure \ref{Fig:Beta_7}. The superscripts $\pi$, $\nu$ refer to
		proton and neutron states, respectively. See Eqs (\ref{norm}) and (\ref{prob}) for the definition of each quantity. The $E_{1p-1h}$ and $E_{2p-2h}$ energies are given in MeV units. From Ref. \cite{Gambacurta2025} .
	}
	\small
	\begin{minipage}[t]{0.48\textwidth} % Minipage for the first table (48% of text width)
 	\centering
%		\caption{Table A: First Dataset}
%		\label{tab:A}
\resizebox{\columnwidth}{!}{%
		\begin{tabular}{ccccc}
			\toprule
			%	\hline
			%	\hline
			&&E$_{SSRPA}$= -0.735 MeV &&\\
			\hline
			$1p-1h$ conf. &$E_{1p-1h}$ &$N_{ph}$& &\\
			$[\pi2s_{1/2},\nu2s_{1/2}]^{J=1^+}$& -1.379& 0.109 &&\\
			\\
			\\
			&&&$N_1= 0.125$&$N_2= 0.875 $ \\
			\hline
			$1p-1h$ conf. &$E_{1p-1h}$ &$A_{ph}$& $T_{ph}$&$b_{ph}$\\
			\hline
			$[\pi1d_{3/2},\nu1d_{5/2}]^{J=1^+}$& 6.038& 0.023& 3.094& 0.070 \\
			$[\pi2s_{1/2},\nu2s_{1/2}]^{J=1^+}$& -1.379& 0.330& 2.443& 0.807 \\
			$[\pi1d_{3/2},\nu1d_{3/2}]^{J=1^+}$& -0.726& -0.095& -1.548& 0.147 \\
			\hline
			$\sum b_{ph}$&& && 1.028\\
			$P(\nu)$&& && 1.056\\
			\midrule 
			%	\hline
			\multicolumn{3}{c}{$2p-2h$ conf. }&$E_{2p-2h}$ &$N_{php'h'}$\\
			\multicolumn{3}{c}{$\big[[\pi1d_{5/2}\nu2s_{1/2}]_{J_H=3}[\pi2s_{1/2}\pi1d_{3/2}]_{J_P=2}\big]_{J_{T}=1}$} & 9.407 & 0.353 \\
			\multicolumn{3}{c}{$\big[[\pi1d_{5/2}\nu1d_{3/2}]_{J_H=2}[\pi1d_{3/2}\pi1d_{3/2}]_{J_P=2}\big]_{J_{T}=1}$} & 12.946 & 0.087 \\
			\multicolumn{3}{c}{$\big[[\pi1d_{5/2}\nu1d_{3/2}]_{J_H=1}[\pi1d_{3/2}\pi1d_{3/2}]_{J_P=0}\big]_{J_{T}=1}$} & 12.946 & 0.071 \\
			\multicolumn{3}{c}{$\big[[\pi1d_{5/2}\nu1d_{5/2}]_{J_H=1}[\pi2s_{1/2}\pi2s_{1/2}]_{J_P=0}\big]_{J_{T}=1}$} & 5.867 & 0.057 \\
			\midrule 
		\end{tabular}
	}
	\end{minipage}% <--- IMPORTANT: The percent sign removes unwanted space!
	\hfill % Adds flexible horizontal space between the two tables
	\begin{minipage}[t]{0.48\textwidth} % Minipage for the second table
		\centering
%		\caption{Table B: Second Dataset}
%		\label{tab:B}
\resizebox{\columnwidth}{!}{%
			\begin{tabular}{ccccc}
			\toprule
			%\hline
			%\hline
			&&E$_{SSRPA}$= 0.247 MeV &&\\
			\hline
			$1p-1h$ conf. &$E_{1p-1h}$ &$N_{ph}$& &\\
			$[\pi2s_{1/2},\nu2s_{1/2}]^{J=1^+}$& -1.379& 0.092 &&\\
			$[\pi2s_{1/2},\nu1d_{3/2}]^{J=1^+}$& -4.266& 0.011 &&\\
			$[\pi1d_{3/2},\nu1d_{3/2}]^{J=1^+}$& -0.726& 0.024 &&\\
			&&&$N_1= 0.131$&$N_2= 0.869 $ \\
			\hline
			$1p-1h$ conf. &$E_{1p-1h}$ &$A_{ph}$& $T_{ph}$&$b_{ph}$\\
			\hline
			$[\pi1d_{3/2},\nu1d_{5/2}]^{J=1^+}$& 6.038& -0.013& 3.094& -0.039 \\
			$[\pi2s_{1/2},\nu2s_{1/2}]^{J=1^+}$& -1.379& 0.303& 2.443& 0.740 \\
			$[\pi1d_{3/2},\nu1d_{3/2}]^{J=1^+}$& -0.726& -0.154& -1.548& 0.238 \\
			\hline
			$\sum b_{ph}$&& && 0.941\\
			$P(\nu)$&& && 0.885\\
			\midrule 
			%\hline 
			\multicolumn{3}{c}{$2p-2h$ conf. }&$E_{2p-2h}$ &$N_{php'h'}$\\
			\multicolumn{3}{c}{$\big[[\pi1d_{5/2}\nu1d_{3/2}]_{J_H=3}[\pi1d_{3/2}\pi1d_{3/2}]_{J_P=2}\big]_{J_{T}=1}$} & 12.946 & 0.249 \\
			\multicolumn{3}{c}{$\big[[\pi1d_{5/2}\nu2s_{1/2}]_{J_H=2}[\pi2s_{1/2}\pi1d_{3/2}]_{J_P=2}\big]_{J_{T}=1}$} & 9.407 & 0.167 \\
			\multicolumn{3}{c}{$\big[[\pi1d_{5/2}\nu1d_{3/2}]_{J_H=1}[\pi1d_{3/2}\pi1d_{3/2}]_{J_P=0}\big]_{J_{T}=1}$} & 12.946 & 0.077 \\
			\multicolumn{3}{c}{$\big[[\pi1d_{5/2}\nu1d_{3/2}]_{J_H=3}[\pi2s_{1/2}\pi1d_{3/2}]_{J_P=2}\big]_{J_{T}=1}$} & 9.407 & 0.053 \\
			\midrule 
			%\hline
		\end{tabular}
	}
	\end{minipage}
\end{table}

Table \ref{tab-1} presents the dominant $1p-1h$ configurations contributing to the normalization (upper section) and the GT transition probability (lower section) for the RPA state located at 0.310 MeV. The superscripts $\pi$ and $\nu$ denote proton and neutron states, respectively, while $N_{ph}$ quantifies the contribution of each $1p-1h$ configuration to the norm (as defined in Eq. \ref{norm}), with corresponding unperturbed energies $E_{1p-1h}$ given in MeV units. 
%The state at 1.422 MeV, although its energy is close to the $\beta$-decay threshold, does not contribute to the $\beta$-decay lifetime and is included for completeness.
 The most significant configurations for both the norm and the transition probability are $[\pi1d_{3/2},\nu1d_{3/2}]$ and $[\pi2s_{1/2},\nu2s_{1/2}]$, both characterized by negative particle-hole excitation energies. A smaller contribution to the transition probability arises from the positive energy configuration $[\pi1d_{3/2},\nu1d_{5/2}]$ due to the strong matrix element of the transition operator $T_{ph}$. Indeed also for the state at 1.422 MeV which is outside the beta-window, the same configurations appear to be the most important ones. This observation suggests that the enhanced strength observed in SSRPA may be partially attributed to the energy shift towards lower values (or fragmentation) of this state as a consequence of the coupling with $2p-2h$ configurations. SSRPA calculations indeed a marked increase in the number of states within the $\beta$-decay window (as shown in Figure \ref{Fig:Beta_7}), leading to a shorter $\beta$-decay lifetime compared to RPA predictions. Tables \ref{tab-2} provide an analogous analysis to Table \ref{tab-1} for the two most collective SSRPA states. The norm of these states includes two components: $N_1$ representing the $1p-1h$ contribution and $N_2$ representing the $2p-2h$ contribution. The term $N_{php'h'}$ indicates the contribution of each $2p-2h$ configuration, with its unperturbed energy $E_{2p-2h}$ in MeV. Alongside the $1p-1h$ components (as in Table 1), the most significant $2p-2h$ components, defined as those contributing more than 0.05 to the norm, are also presented. The notation $\big[[hh']_{J_H}[pp']_{J_P}\big]_{J_{T}}$ is employed, where $J_H$ and $J_P$ are the total angular momenta of the two hole and two particle states, respectively, and $J_T$ is the total angular momentum. The analysis indicates that all the SSRPA states exhibit a dominant $2p-2h$ character (greater than 80\%), which is fragmented across multiple configurations. Furthermore, the largest contributions involve the $\pi1d_{5/2}, \nu1d_{3/2}, \nu2s_{1/2}$ hole states and the $\pi1d_{3/2}, \pi2s_{1/2}$ particle levels. This is consistent with these levels being closest to the Fermi energies for neutrons and protons.
  %Due to the QBA used also in the SSRPA framework, the transition probabilities are given by the $1p-1h$ configurations, as described by Eq. (\ref{prob}).
   The $[\pi2s_{1/2},\nu2s_{1/2}]$ component is observed to contribute to all the analyzed SSRPA states, both in terms of its $N_1$ content and its contribution to the transition probability. The $[\pi1d_{3/2},\nu1d_{3/2}]$ configuration, and to a lesser extent $[\pi1d_{3/2},\nu1d_{5/2}]$, also play significant roles in establishing the collectivity of these states. The introduction of $2p-2h$ configurations in the SSRPA framework results in an increased density of states within the $\beta$-decay window. These states are predominantly of $2p-2h$ nature, while their $1p-1h$ components are fragmented over the same types of configurations that appear in the RPA states presented in Table \ref{tab-1}.
\newpage

\section{BMF effects induced in SSRPA}
\label{Sec:BMF}
\subsection{Introduction}
The EoS is a key quantity in the description of nuclear systems, from atomic nuclei to nuclear compact massive objects \cite{Oertel2017,Rocamaza2018,Burgio2021,Agrawal2021}.
Expanding the EoS around saturation density and isospin asymmetry with a Taylor expansion one can get an expression in terms of some key parameters. For the present discussion, we recall here the nuclear incompressibility $K$, the symmetry energy at saturation ($J$) and its slope parameter ($L$), and the effective mass ($m^*$). These parameters are strongly linked to the properties of finite nuclear systems, such as the ISGMR centroid energy ($K_0$), neutron skin, low-lying dipole strength and polarizability, radii of neutron stars ($J$ and $L$), ISGQR centroid ($m^*$) \cite{Rocamaza2018,Agrawal2021}. EDF calculations, in addition to finite-nucleus observables (masses, radii, ... ), consistently provide model predictions for bulk parameters of the EoS. Many EDF parameters are constrained not only by using the properties of finite nuclei but also the EoS parameters which are used as pseudo data. EDF-based MF plus RPA calculations provide an economical and physically transparent mapping from experiment to EoS parameters. However, they do not include by construction dynamical correlations beyond the MF picture.  The study of their impact is two-fold. Besides the quantitative estimation of their role, it is also important to assess whether or not, the usual correlations between the EoS parameter and finite nuclear properties, which are usually derived within a MF approach, survive when BMF effects are introduced. For example, in the case of the nuclear response, the BMF models typically shift resonance centroids and produce a more fragmented strength.
That means that if one extract EoS parameters from MF + RPA calculations and ignore BMF corrections, one can get biased estimates and/or underestimated uncertainties. As a concrete example, ISGMR centroids in medium-heavy nuclei traditionally point to $K\approx 220\text{–}260$ MeV \cite{Garg2018} when analyzed with RPA-based fits. However, systematic differences across isotopes (Sn "softness" puzzle) \cite{Piekarewicz2007,Piekarewicz2010} suggested missing correlations. Recent studies \cite{Li2023} show that BMF correlations introduced within the quasiparticle Skyrme-PVC model shift monopole centroids by $\sim 0.5$ MeV (in Sn isotopes) while the shift being almost negligible in the $^{208}$Pb case. The inclusion of BMF effect reconcile thus ISGMR data across isotopic chains, implying that BMF shifts correspond to non-negligible corrections in extracted $K$. Similar conclusions were obtained also within a relativistic approach, based on the RQTBA calculations \cite{Litvinova2023}. In this Section, the role of BMF correlations introduced within the SSRPA and their impact on some EoS parameters are discussed.

\subsection{Soft modes and nuclear incompressibility}

\begin{figure}
	\includegraphics[width=.48\linewidth]{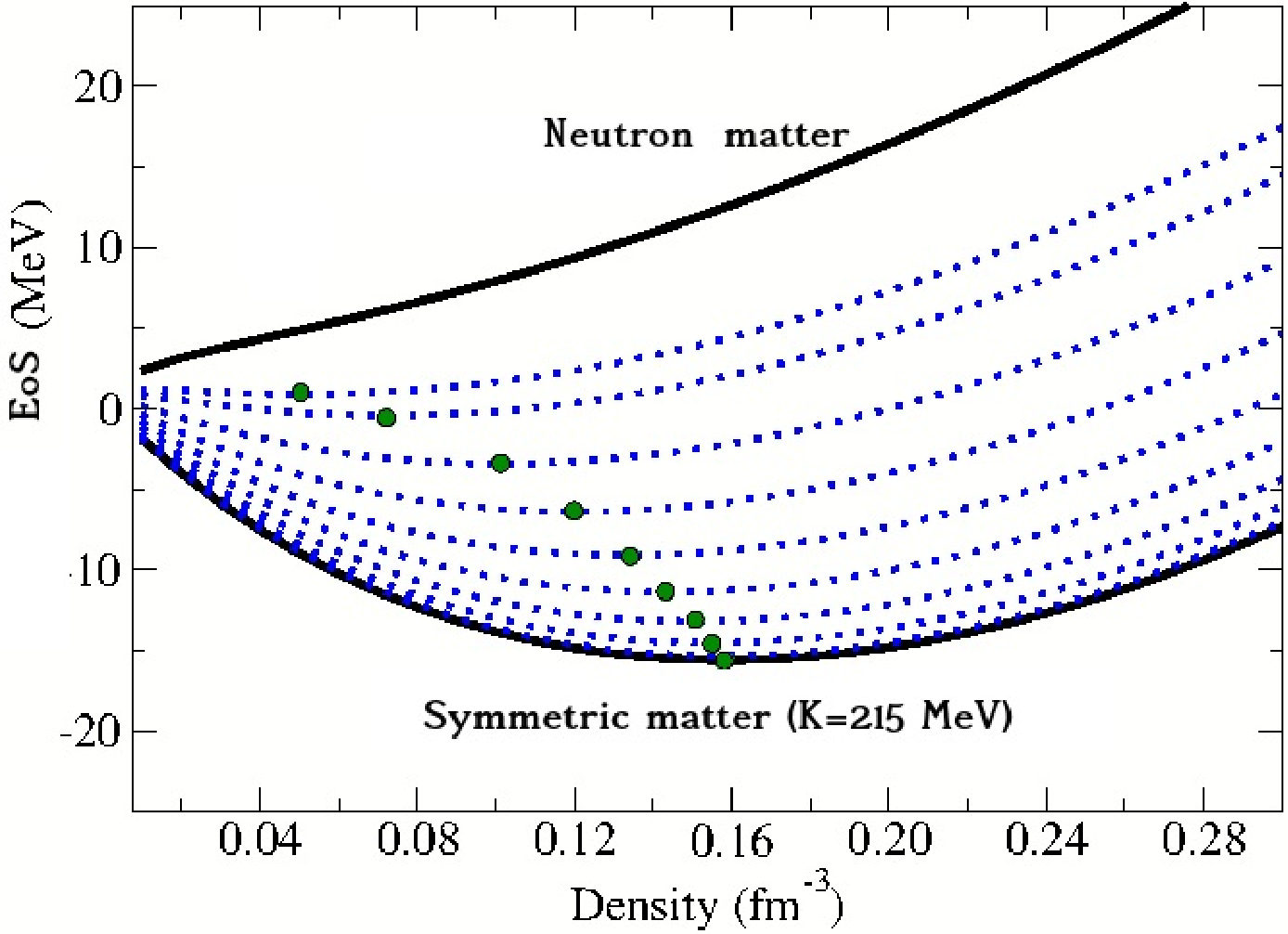}\hfill
	\includegraphics[width=.5\linewidth]{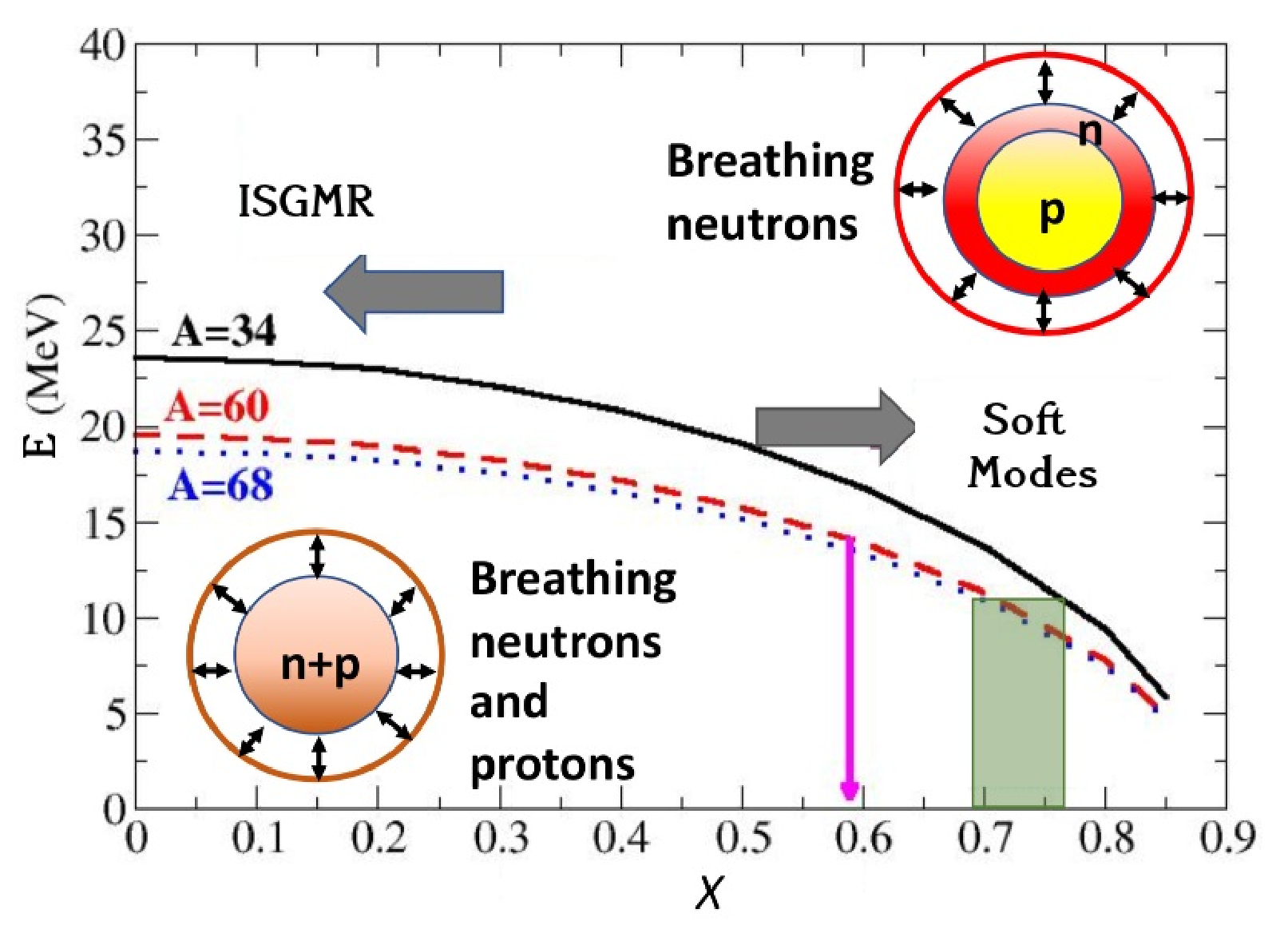}
	\caption{Left side: EoS for nuclear matter computed using the $\text{SGII}$ parametrization, ranging from symmetric nuclear matter ($\text{SNM}$) to pure neutron matter ($\text{PNM}$). Dotted lines represent the EoSs for asymmetric matter, defined by the asymmetry parameter $X$, increasing from $X=0.1$ (bottom curve) to $X=0.85$ (top curve). Green circles indicate the equilibrium points, at which the corresponding incompressibility coefficients $K_X$ are calculated (see the text for more details). The incompressibility value for $\text{SNM}$ is also shown. Right side: Centroid energy of the ISGMR as a function of the isospin asymmetry ($X$) for nuclei $A=34$, $60$, and $68$. The calculations are based on Eq. (\ref{eversusk2}). The $X=0$ intercept corresponds to the estimated $\text{GMR}$ centroid using the incompressibility of symmetric nuclear matter. The green shaded area highlights the range of isospin asymmetries predicted for the oscillating systems involved in the low-energy excitations observed at $11.075$ MeV (${}^{34}\text{Si}$), $8.838$ MeV (${}^{60}\text{Ca}$), and $11.021$ MeV (${}^{68}\text{Ni}$). The magenta arrow indicates the specific isospin asymmetry associated with the excitation mode at $14.087$ MeV in ${}^{60}\text{Ca}$. Adapted from \cite{Gambacurta2019}. }
	\label{Fig:Compressibility}
\end{figure}

The correlation between the centroid energies of ISGMR and the incompressibility modulus $K$ of symmetric nuclear matter has long been established~\cite{Blaizot1980,Garg2018}. The modulus for symmetric infinite matter is defined as
\begin{equation}
	K = 9\rho_0^2\left(\frac{\partial^2 E^{sym}/A}{\partial \rho^2}\right)_{\rho=\rho_0},
	\label{comsy}
\end{equation}
where $\rho_0=0.16$ fm$^{-3}$ and $E^{sym}/A$ denotes the EoS of symmetric matter. 
Within a liquid-drop model, the ISGMR centroid energy $E$ relates to $K$ as
\begin{equation}
	E = \sqrt{\frac{\hbar^2\pi^2}{15m}} \sqrt{\frac{K}{\eta_0^2}},
	\label{eversusk}
\end{equation}
and, with $\eta_0=r_0A^{1/3}$ ($r_0\simeq1$ fm),
\begin{equation}
	E \sim 5.22\,A^{-1/3}\sqrt{K}.
	\label{eversusk1}
\end{equation}

In Section \ref{Sec:Applications_SSRPA_Monopole}, the properties of the soft monopole modes in neutron-rich nuclei have been discussed. These states are mainly neutronic, involving only partial compression of the system. Thus, these modes should not be included in empirical extractions of $K$ from ISGMR data, which pertain to symmetric matter.

To describe such excitations, an incompressibility for asymmetric matter should be introduced. Defining the asymmetry parameter
\begin{equation}
	X = \frac{\rho_n-\rho_p}{\rho},
\end{equation}
the modulus for asymmetric matter reads
\begin{equation}
	K_X = 9\rho_{eq}^2\left(\frac{\partial^2 E^X/A}{\partial\rho^2}\right)_{\rho=\rho_{eq}},
	\label{com1}
\end{equation}
where $\rho_{eq}$ is the equilibrium density at given $X$.   In the left side of Figure~\ref{Fig:Compressibility}the dotted lines represents the EoS with several asymmetry values $X$ (from the bottom to the top) equal to 0.1 ($K_{X}=$ 211.59 MeV), 0.2 ($K_{X}=$ 203.54 MeV), 0.3 ($K_{X}=$ 187.33 MeV), 0.4 ($K_{X}=$ 166.40 MeV), 0.5 ($K_{X}=$ 139.90 MeV), 0.6 ($K_{X}=$ 108.30 MeV), 0.7 ($K_{X}=$ 72.35 MeV), 0.8 ($K_{X}=$ 33.74 MeV), 0.85 ($K_{X}=$ 13.32 MeV), with equilibrium points marked by green circles. Increasing $X$ shifts the minimum of the EoS toward lower densities and reduces its curvature, reflecting a decrease in $K_X$ with neutron excess.

Extending Eq.~(\ref{eversusk1}) yields
\begin{equation}
	E(X) \sim 5.22\,A^{-1/3}\sqrt{K_X},
	\label{eversusk2}
\end{equation}
linking excitation energy to the incompressibility of asymmetric matter. The right side of Figure~\ref{Fig:Compressibility} illustrates this qualitative dependence for selected nuclei ($^{34}$Si, $^{60}$Ca, $^{68}$Ni): as $X$ increases, $E(X)$ decreases, matching the microscopically predicted soft-mode energies.

In the case of excited states, the asymmetry of the oscillating subsystem is estimated using neutron and proton transition densities $\rho_\nu^n$ and $\rho_\nu^p$,
\begin{equation}
	X = \frac{X_N - X_P}{X_N + X_P}, ~\mbox{where}~~~
	X_N = 4\pi \int |\rho_\nu^n| r^2 dr, \quad
	X_P = 4\pi \int |\rho_\nu^p| r^2 dr.
	\label{xtot}
\end{equation}
The so-obtained values for the soft modes discusussed in Section \ref{Sec:Applications_SSRPA_Monopole} are shown in Table \ref{tab3}.
%The resulting $X$ values for selected excited states in $^{60}$Ca (see Figure \ref{Fig:Monopole_fig1}), $^{34}$Si (see Figure \ref{Fig:Monopole_fig4}) and $^{68}$Ni (see Figure \ref{Fig:Monopole_fig5}),  are listed in Table~\ref{tab3}. 

\begin{table}[h!]
	\centering
	\begin{tabular}{ccc}
		\hline
		Nucleus & $E$ (MeV) & $X$ \\
		\hline
		$^{34}$Si & 11.07 & 0.73 \\
		$^{68}$Ni & 11.02 & 0.78 \\
		$^{60}$Ca & 8.8 & 0.84 \\
		$^{60}$Ca & 14.1 & 0.73 \\
		\hline
	\end{tabular}
	\caption{Isospin asymmetry $X$ for selected monopole states. From Ref. \cite{Gambacurta2019}.}
	\label{tab3}
\end{table}

The analysis shows that the lowest-energy soft modes correspond to strongly asymmetric, neutron-dominated oscillations ($X\!\gtrsim\!0.7$), while higher-energy peaks, such as the 14.1 MeV mode in $^{60}$Ca, exhibit reduced asymmetry due to configuration mixing. Overall, Figures~\ref{Fig:Compressibility} highlight how the excitation energy decreases with increasing neutron excess, demonstrating a qualitative link between the soft monopole modes and the compressibility of neutron-rich matter.

This study is meant to be qualitative, primarily due to the following limiting factors. The approach combines MF values of the infinite-matter incompressibility modulus (K) with excitation energies predicted by a BMF model (e.g., SSRPA). The relationship connecting ISGMR centroids to the incompressibility $K$ is established using a simple Liquid Drop Model. A more microscopic and theoretically grounded relation would render such estimates more reliable. The low-lying "soft" modes discussed in Section \ref{Sec:Applications_SSRPA_Monopole} are predominately of a $1p-1h$ nature, despite exhibiting some small spreading into $2p-2h$ configurations. The link between nuclear-matter bulk properties and finite-nucleus observables is most reliable when the latter display strong collectivity.

\subsection{Symmetry energy and its slope $L$}
\label{Sec:BMF_symmetry}
\begin{figure}
	\includegraphics[width=.45\linewidth]{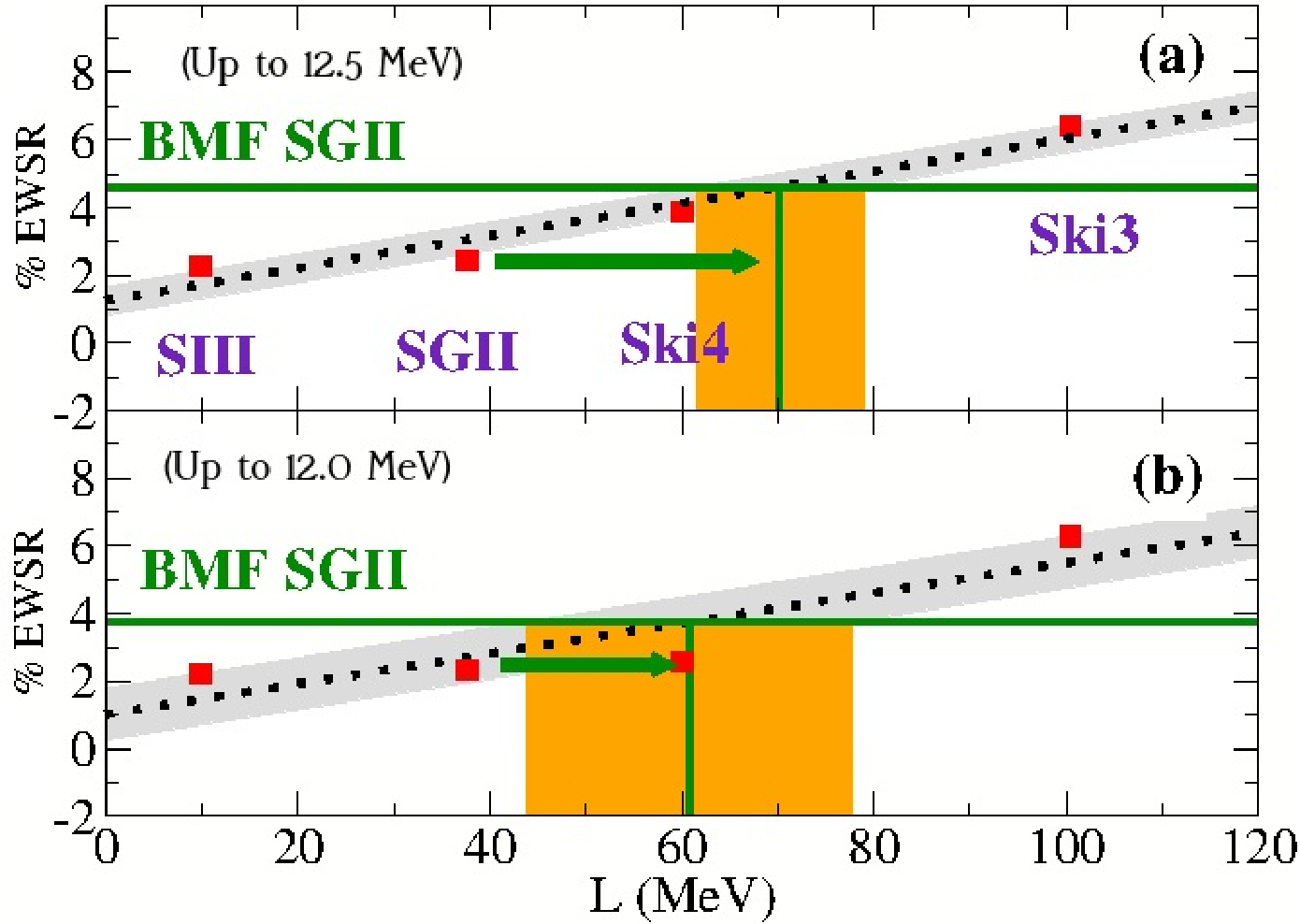}\hfill
	\includegraphics[width=.47\linewidth]{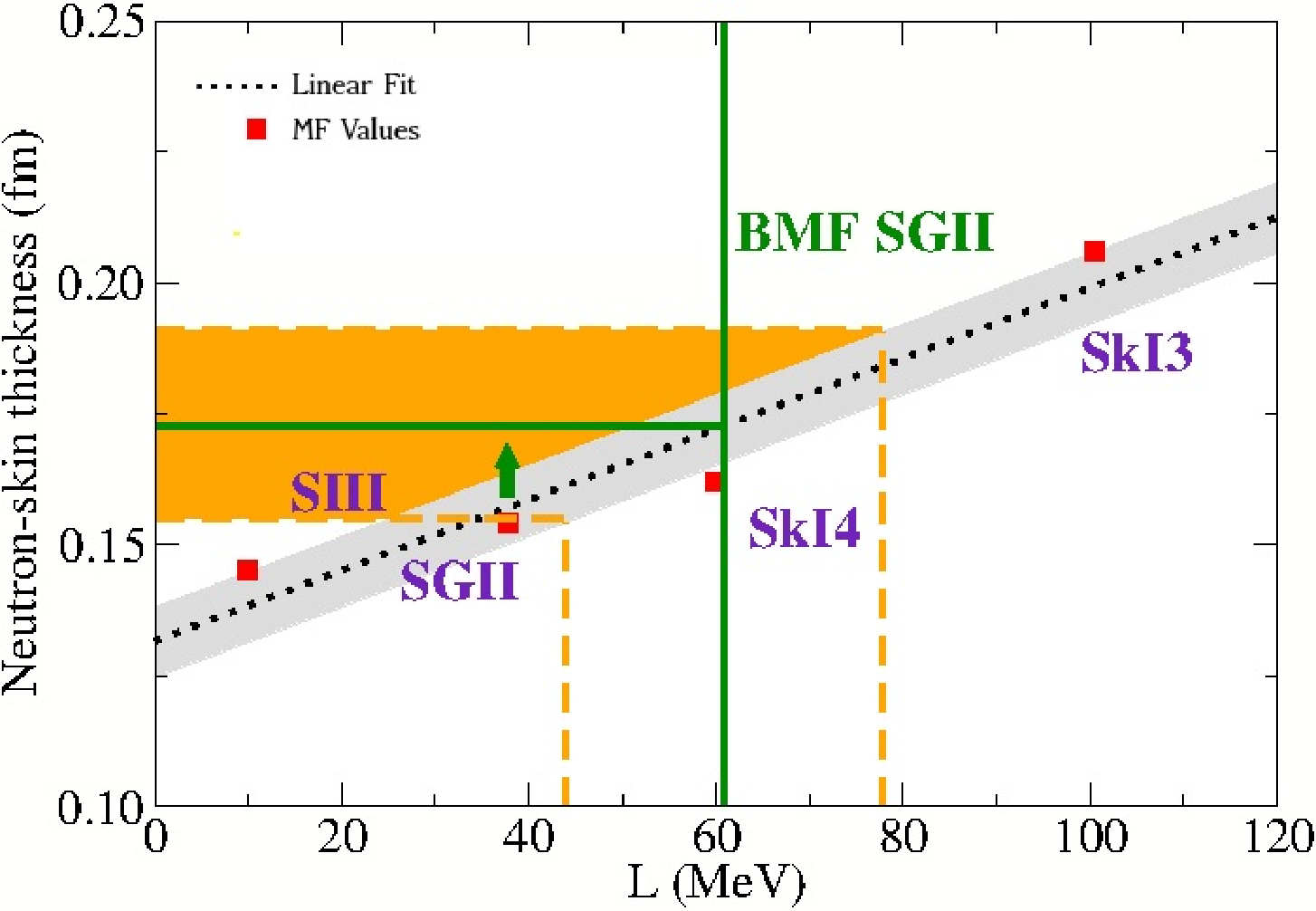}
	\caption{Left side: Percentage of the pygmy EWSR obtained in RPA for $^{68}$Ni with four Skyrme parametrizations  (red squares) as a function of their corresponding slope $L$ (MF values) of the symmetry energy. The dotted line indicates the linear fit of the four points with an associated uncertainty band (grey area). The horizontal green line indicates the SSRPA \% EWSR value for the case SGII and the BMF value of $L$ is correspondingly extracted (vertical green line) with an associated uncertainty (orange area). The percentage of EWSR is calculated up to 12.5 (12) MeV in panel (a) (panel (b)). 
		Right side: Neutron--skin thickness of $^{68}$Ni for the four Skyrme parametrizations SIII, SGII, SkI3, and SkI4 as a function of the slope $L$ (MF values) of the symmetry energy (red squares). The dotted line indicates the linear fit of the four points with an associated uncertainty band (grey area). The vertical green line indicates the BMF $L$ value for SGII with its uncertainty band (vertical orange dashed lines). The corresponding BMF value for $\Delta r_{np}$ is extracted (horizontal green line) with an associated uncertainty (orange area). Adapted from Ref. \cite{Grasso2020}.}
	\label{Fig:Ni68_Eos_fig1}
\end{figure}

The low-energy dipole strength of neutron rich nuclei, usually referred as PDR is strongly connected with the 
 neutron--skin thickness and to the $J$ and $L$ parameters of the EoS \cite{Bracco2019, Carbone2010, Piekarewicz2006, Klimkiewicz2007,Rocamaza2018,Typel2001,Yoshida2004}. Several correlations have been 
 pointed ou in different studies between the PDR strength, the neutron skin and the symmetry energy of infinite matter, and to its density dependence. For example,
 the neutron--skin thickness was found to be correlated with the symmetry energy calculated at the saturation density $J$ as well as to its slope \cite{Klimkiewicz2007,Furnstahl2002,Chen2005a,Centelles2009,Warda2009}, the slope being extremely important for example in heavy--ion collisions \cite{Chen2005a,Baran2005,Chen2005b,Li2008,Shetty2007,Famiano2006} and in nuclear astrophysics for the description of neutron stars \cite{Horowitz2001,Todd2005,Steiner2005,Lattimer2007}.
Correlations between the neutron--skin thickness and the product of the symmetry energy times the electric dipole polarizability were also investigated in Ref. \cite{Rocamaza2015}. 
However, all such correlations have been commonly investigated within a MF context. Therefore, it is interesting to investigate whether these correlations are robust or not, and also one can asses more quantitatively, though with some uncertainties, the impact of these BMF effects which are usually neglected.
Indeed, some of these correlations were first derived within quite general assumption, as for example those of the droplet models, suggesting that they might be general features of nuclei and nuclear matter and not a simple MF artifact. This is the case for example for the correlations found between the neutron--skin thickness of a nucleus of mass $A$ and the symmetry energy of matter $J$ minus the symmetry energy in the nucleus \cite{Centelles2009} or between the neutron--skin thickness and $J/Q$, where $Q$ is the so-called surface stiffness \cite{Warda2009}. Experimentally these correlations were investigated by means of different experimental constraints such as those coming from heavy--ion collisions, measurements of neutron skins, electric dipole polarizabilities, masses, IVGDR modes, isobaric analog states, as well as constraints coming from nuclear astrophysics (see, for instance, Refs. \cite{Tsang2012,Lattimer2013,Oertel2017,Li2013,kong2017,Lattimer2014}). 

In Ref. \cite{Grasso2020}, the low-lying dipole strength in $^{68}$Ni obtained in SSRPA and discussed in Section \ref{Sec:Applications_SSRPA_Dipole_Ni68}, was employed to study the impact of BMF correlations on the symmetry energy $J$ and its slope $L$. As a first step, four Skyrme parametrizations having progressively increasing values of $L$, namely, SIII \cite{SIII},	SGII \cite{SGII}, SkI4 \cite{SKI}, and SkI3 \cite{SKI} were chosen and the corresponding values of $J$ and $L$ are (28.16, 9.90), (26.83 , 37.70),(29.50 , 60.00) and (34.27 , 100.49), respectively, in MeV units. Such $J$ and $L$ values are associated with EoS computed at the MF level, corresponding to the leading order of the Dyson equation. 	It is then expected that, when BMF models are employed, these values might change because of higher order effects. One can  estimate indirectly such BMF effect, by comparing the RPA  and the SSRPA low--energy dipole strength.
%
%\begin {table} 
%\begin{center}
%	\begin{tabular}{ccc}
%		
%		\hline
%		\hline
%		Skyrme & $J$ (MeV) & $L$ (MeV) \\
%		\hline
%		SIII & 28.16 & 9.90 \\
%		SGII & 26.83 & 37.70 \\
%		SkI4 & 29.50 & 60.00 \\
%		SkI3 & 34.27 & 100.49 \\
%		\hline
%		\hline
%	\end{tabular}
%\end{center}
%\caption{MF values for the symmetry--energy coefficient computed at the saturation density $J$ and for the associated slope $L$ for the four Skyrme parametrizations indicated on the first column. From Ref. \cite{Grasso2020}.}
%\label{symL}
%\end {table} 

Starting from the RPA and SSRPA low-lying strength for the four mentioned forces, one can consider the correlation between the EWSR and slope of the symmetry energy $L$ whose results are shown on the left side of Figure \ref{Fig:Ni68_Eos_fig1}. The EWSR percentage for ${68}$Ni calculated up to 12.5 and 12 MeV, are shown in panel (a) and (b), respectively. By exploiting the correlation between low-lying EWSR percentages and $L$, the RPA results corresponding to four Skyrme parametrizations are plotted, showing a linear relationship. An uncertainty band was constructed by quantifying the average deviation of RPA data from the linear fit. Using the SSRPA-calculated EWSR percentage for SGII, a BMF $L$ value was extracted, revealing a significant increase compared to the MF value, thereby suggesting a stiffer EoS. This effect is even stronger with a 12.5 MeV cutoff, resulting in a larger BMF $L$ but a narrower uncertainty band. Transition density analysis confirmed the 12 MeV cutoff as optimal\cite{Grasso2020}, marking the transition from PDR to IVGDR tail characteristics. This analysis provides a qualitative assessment of BMF effects on $L$, demonstrating a consistent increase in $L$ when considering SSRPA compared to RPA. While the inclusion of additional functionals could broaden the uncertainty, the study demonstrates the qualitative influence of BMF effects on the symmetry energy slope.

\begin{figure}
	\includegraphics[width=.45\linewidth]{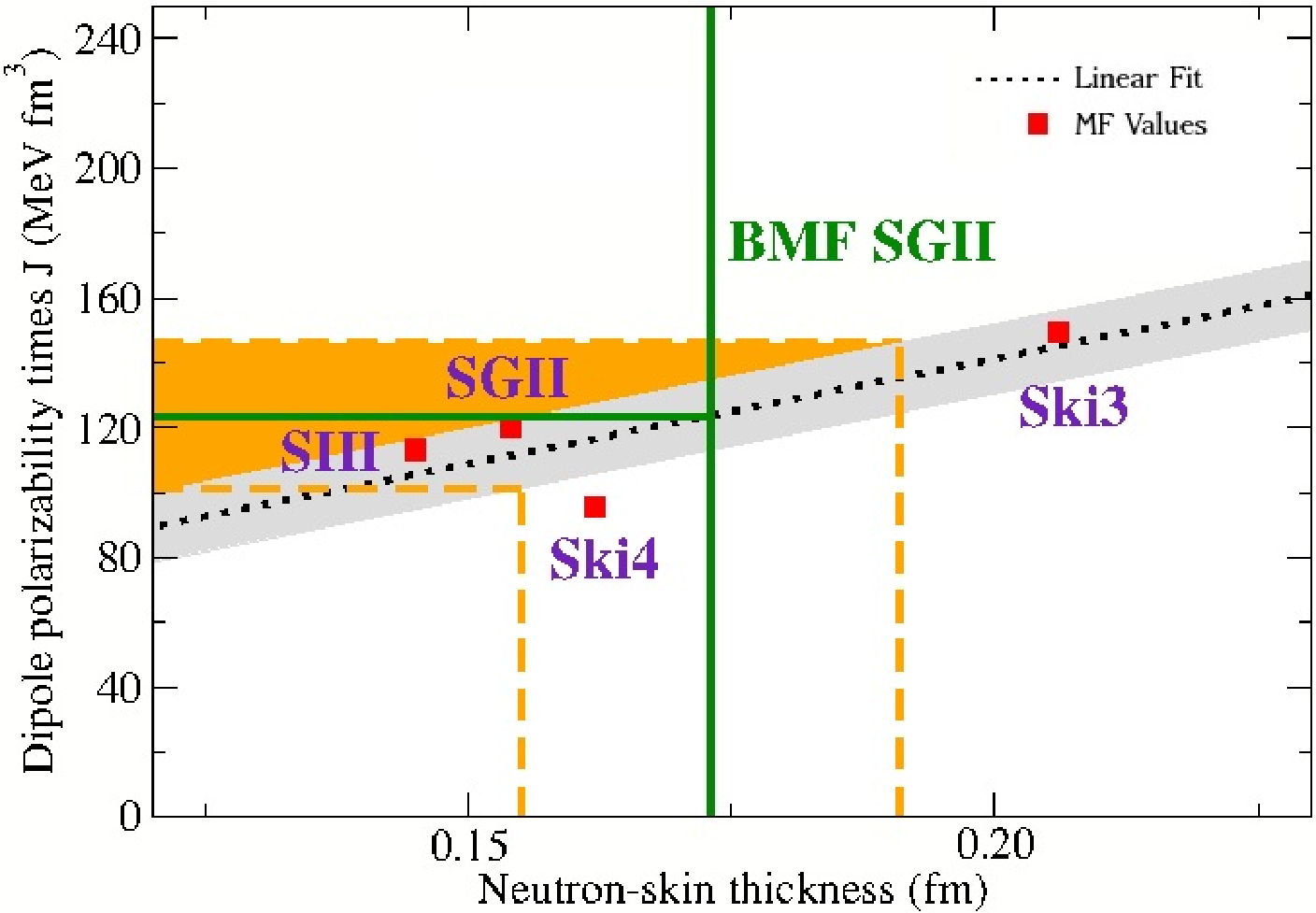}\hfill
	\includegraphics[width=.44\linewidth]{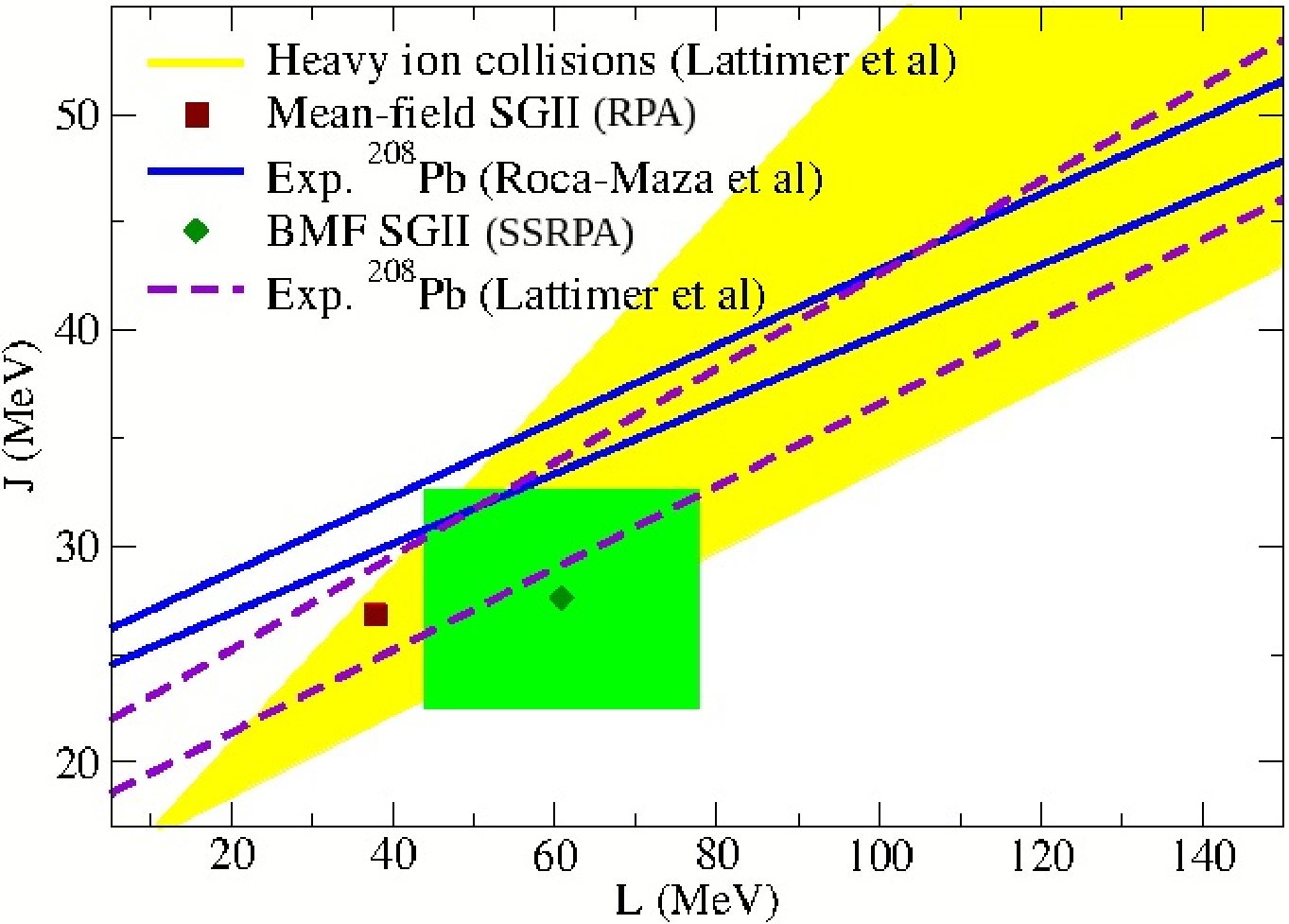}
	\caption{Left side: Electric dipole polarizability obtained in RPA for $^{68}$Ni times the MF value of $J$ as a function of the MF value of $\Delta r_{np}$ for the four Skyrme parametrizations (red squares). The dotted line represent a linear fit with its uncertainty band (gray area). The BMF value for $\Delta r_{np}$ is included for SGII (vertical green line) with its uncertainty band represented by vertical orange dashed lines. The corresponding BMF value of $\alpha J$ is extracted (horizontal green line) with its uncertainty (orange area).
		Right side: $J$ versus $L$ experimental constraints coming from different measurements together with the MF and the BMF points for the SGII force as a red square and a dark green diamond, respectively. The associated BMF uncertainty is indicated as a light green area. Adapted from Ref. \cite{Grasso2020}.
	}
	\label{Fig:Ni68_Eos_fig2}
\end{figure}
After that, one can consider the correlations between neutron skin thickness ($\Delta r_{np}$) and the symmetry energy slope ($L$) \cite{Centelles2009,Warda2009}. HF calculations were performed for $^{68}$Ni with four selected Skyrme parametrizations, and the resulting $\Delta r_{np}$ values were plotted against (MF) $L$ values in right side of Figure
\ref{Fig:Ni68_Eos_fig1}. A linear fit was generated, and an uncertainty band was estimated. This time, the BMF $\Delta r_{np}$ value was extracted using the BMF $L$ value just obtained in the previous analysis. For each (MF) $L$ value, the vertical distance between the HF neutron skin thickness and the corresponding point on the linear fit was computed. An average of these distances was then used to construct an uncertainty band with a vertical width equal to twice this average, represented by the gray area in the figure. The SGII BMF $L$ value, along with its uncertainty (represented by the vertical orange dashed lines), was then plotted on the figure. This allowed for the estimation of a BMF value for the neutron skin thickness, with an associated uncertainty (orange area). The (MF) $\Delta r_{np}$ value for the SGII parametrization was 0.154 fm, while the BMF value was 0.173 $\pm$ 0.018 fm. Qualitatively, one can concludes that BMF effects tend to increase the neutron skin thickness of the nucleus.

Reference \cite{Rocamaza2015} established a linear correlation between the product of electric dipole polarizability ($\alpha$) and symmetry energy at saturation density ($J$), and the neutron skin thickness ($\Delta r_{np}$) of a nucleus. This correlation was employed to determine a BMF value for $J$ based on the BMF $\Delta r_{np}$ value previously estimated. Due to the subtraction procedure, the dipole polarizability ($\alpha$) is the same in RPA and SSRPA, being equal to 4.48 fm$^{3}$ with the SGII force \cite{SGII}. 
The left side of Figure \ref{Fig:Ni68_Eos_fig2} illustrates the correlation between $\alpha J$ and $\Delta r_{np}$, derived from MF calculations using four Skyrme parametrizations. A linear fit was performed, and an uncertainty band was estimated. To extract a BMF value for $\alpha J$, the BMF $\Delta r_{np}$ value was used. For each of the four data points, the vertical distance between the (MF) $\alpha J$ value and the corresponding point on the linear fit was calculated. A vertical uncertainty band was defined as twice the average of these four distances. The BMF $\Delta r_{np}$ value, along with its uncertainty band, was then plotted, enabling the extraction of a BMF value for $J$. This BMF $J$ value, given that $\alpha$ is identical to the RPA value, was found to be 27.617 $\pm$ 5.004 MeV. BMF effects resulted in a slight increase in the symmetry energy at saturation density, although the MF value of 26.83 MeV was observed to fall within the uncertainty band.

The correlation previously discussed between the electric dipole polarizability multiplied by the symmetry energy and the neutron skin thickness was extended in Ref. \cite{RocaMaza2013} to a correlation between the dipole polarizability multiplied by the symmetry energy and its slope ($L$). This extension was specifically applied to the nucleus $^{208}$Pb. Using experimental measurements of the dipole polarizability, a relationship between the symmetry energy ($J$) and $L$ was later derived in Ref. \cite{Rocamaza2015}. Based on an experimental dipole polarizability value of 19.6 $\pm$ 0.6 fm$^{3}$, this relationship is represented on the right side of Figure
\ref{Fig:Ni68_Eos_fig2} as the region enclosed by the two blue solid lines. Conversely, utilizing a more recent dipole polarizability value of 20.1 $\pm$ 0.6 fm$^{3}$, as reported in Ref. \cite{Tami2011}, a slightly different constraint for $J$ and $L$  was derived in Ref. \cite{Lattimer2014}, which is shown as the region between the indigo dashed lines. The authors chose $^{208}$Pb as an illustrative example to emphasize the qualitative nature of such empirical constraints, as slight variations in measured dipole polarizability values can alter the constrained region for $J$ and $L$. The empirical constraint for $J$ and $L$ derived from heavy-ion collisions, also extracted from Ref. \cite{Lattimer2014}, is represented by the yellow area. The right side of Figure \ref{Fig:Ni68_Eos_fig2} also displays the MF values of $J$ and $L$ corresponding to the SGII parametrization, along with the BMF estimation and its associated uncertainty area.

The analysis revealed that the $L$ value is significantly more affected by BMF effects than the $J$ value. The MF point is located outside the region defined by the blue solid lines (from \cite{Rocamaza2015}), whereas the BMF area is compatible with this region when the uncertainty band is considered. Both the MF and BMF points are compatible with the empirical constraint represented by the indigo dashed lines (from Ref.  \cite{Lattimer2014}). The MF point is located within this area, while the BMF area remains compatible when the uncertainty region is considered. The yellow band, representing the empirical constraint from heavy-ion collisions (from Ref. \cite{Lattimer2014}), places the MF point at its boundary. The BMF estimation, however, favors a higher $L$ value, indicating a stiffer equation of state for pure neutron matter. The MF values and corresponding BMF estimations for $L$, $\Delta r_{np}$, and $J$ for the SGII parametrization are summarized in Table \ref{summa}.

\begin {table} 
\begin{center}
	\begin{tabular}{cccc}
		
		\hline
		\hline
		& $L$ (MeV) & $\Delta r_{np}$ (fm) & $J$ (MeV) \\
		\hline
		MF & 37.70 & 0.154 & 26.83 \\
		BMF & 60.815 $\pm$ 16.982 & 0.173 $\pm$ 0.018 & 27.617 $\pm$ 5.004 \\
		\hline
		\hline
	\end{tabular}
\end{center}
\caption{MF values and BMF estimations for the parametrization SGII for the slope of the symmetry--energy $L$, the neutron--skin thickness $\Delta r_{np}$, and the symmetry energy calculated at the saturation density $J$. Adapted from Ref. \cite{Gambacurta2025}.}
\label{summa}
\end {table} 

\subsection{Effective mass}
\label{Sec:BMF_effectivemass}
In Ref. \cite{Grasso2018}, the axial breathing mode has been investigated within the SSRPA to asses the modification of the nucleon effective mass due to the BMF effects. The analysis is focused in particular to the nuclei $^{48}$Ca and $^{90}$Zr. By examining the centroid energies of axial breathing modes, as determined by both the MF-based RPA and the SSRPA model, the modification induced is estimated. This estimation is achieved through the utilization of a relationship, derived from Landau's Fermi liquid theory, that connects the excitation frequency of axial modes to the square root of the ratio of the bare mass $m$ to the effective mass $m^*$. The observed enhancement of the effective mass is then discussed in relation to the renormalization of single-particle excitation energies, a process generated by the energy-dependent SSRPA self-energy correction. 

Landau's theory \cite{Pines,Abrikosov_1959} provides a powerful way to understand low-energy excitations in interacting Fermi systems, regardless of the specific system or energy scale. This theory simplifies the complex dynamics of interacting particles by introducing the concept of quasi-particles, which possess an effective mass, $m^*$, generated by inter-particle interactions. 
Consequently, the effective mass $m^*$ is a crucial quantity studied in many areas of many-body physics
and it is defined by the following equation 
\begin{equation}
	\frac{1}{m^*}=\frac{dE}{dk} \frac{1}{\hbar^2k} \,
	\label{defimstar}
\end{equation}
for a particle of energy $E$ and momentum $k$, with
\begin{equation}
	E=\frac{\hbar^2 k^2}{2m} + \Sigma_k + \Sigma_{k,E},
	\label{ener}
\end{equation}
where $ \Sigma_k + \Sigma_{k,E}$ is the self-energy, sum of the MF contribution $\Sigma_k$ (from the leading order of the Dyson equation in the perturbative many--body expansion) and of a BMF energy--dependent contribution $\Sigma_{k,E}$. 
In the MF approximation, the self--energy does not have any energy 
dependence and may have only a $k$ dependence. However, an explicit energy dependence is induced when contributions beyond the leading order of the Dyson expansion \cite{Fetter} are taken into account, overcoming the MF approximation. 
Using the definition of $m^*$ in Eq. (\ref{defimstar}), one can write
\begin{equation}
	\nonumber
	\frac{m^*}{m}=\left(1-\frac{\partial \Sigma_{k,E}}{\partial E}\right) \cdot \left(1+\frac{m}{\hbar^2k}\frac{\partial \Sigma_k}{\partial k}\right)^{-1} = \frac{m_E^*}{m} \cdot \frac{m_k^*}{m}, 
	\label{mstar}
\end{equation}
where the above expression defines the so--called $E$--mass $m_E^*/m$ and $k$--mass $m_k^*/m$. 
In cases where the MF self-energy does not depend on the momentum (for instance, with a zero--range interaction characterized only by a coupling constant, without any velocity--dependent terms) the $k$--mass is equal to 1. Therefore, an effective mass is generated only with second--order calculations.

Landau's Fermi liquid theory allows for a connection, within the local-density approximation, between the polaron's frequency, $\omega^*$, and its effective mass, $m^*$. Specifically, $\omega^*$ is proportional to the square root of the ratio of the bare mass, m, to the effective mass, $m^*$: $\omega^*$/$\omega =\sqrt{(1-A)m/m^*}$. Here, $\omega$ represents the trapping-potential frequency, and A quantifies the attraction between the impurity and surrounding atoms. A similar relationship holds in atomic nuclei \cite{Bohigas1979,Blaizot1980}, linking the centroid energies of ISGQRs, which correspond to nuclear axial breathing modes, to the effective mass in nuclear matter. The local-density approximation is applicable in this nuclear context because the effective mass varies smoothly with density. This relationship has been extensively employed in nuclear physics over the past decades to derive phenomenological constraints on the effective mass in nuclear matter from measurements of ISGQRs (see Ref. \cite{Li2018} and references therein).

Being the $E$-mass equal to 1 in the MF approximation, where $m^*=m^*_k$, any BMF effect produces a change of $m^*$ generated by the $E$-mass. Moreover, since the effective mass affects the density of states \cite{Erler2011}, BMF effects can produce a different single--particle spectrum, which is compressed if the effective mass is enhanced beyond the MF estimation. 
%These effects can be taken into account and assessed within the SSRPA model, in the framework of the energy--density--functional (EDF) theory \cite{Bender2003} with Skyrme forces\cite{Vautherin1972,Stone2007}. 
More precisely, the SSRPA energy--dependent $A$-type matrix elements may be written as 
\begin{equation}
 A_{1,1}^{SSRPA} (E) = \left[\epsilon_p - \epsilon_h \right]_{MF} 
	+ V_{phhp} \\ + \sum_{2,2'} \frac{A_{ph,2} A_{2',ph}}{E+i\eta-A_{2,2'}} 
	+ \sum_{2,2'} \frac{A_{ph,2} A_{2',ph}}{A_{2,2'}},
	\label{renosub}
\end{equation}
as discussed in Section \ref{Sec:Diagonal_SRPA}.
The energy--dependent self--energy correction provides a renormalization of the diagonal matrix elements 
$A_{1,1}$ and since these matrix elements contain the single--particle excitation energies, this renormalization induces a BMF renormalization of the $1p-1h$ single--particle spectrum.

\begin{figure}
	\includegraphics[width=.40\linewidth]{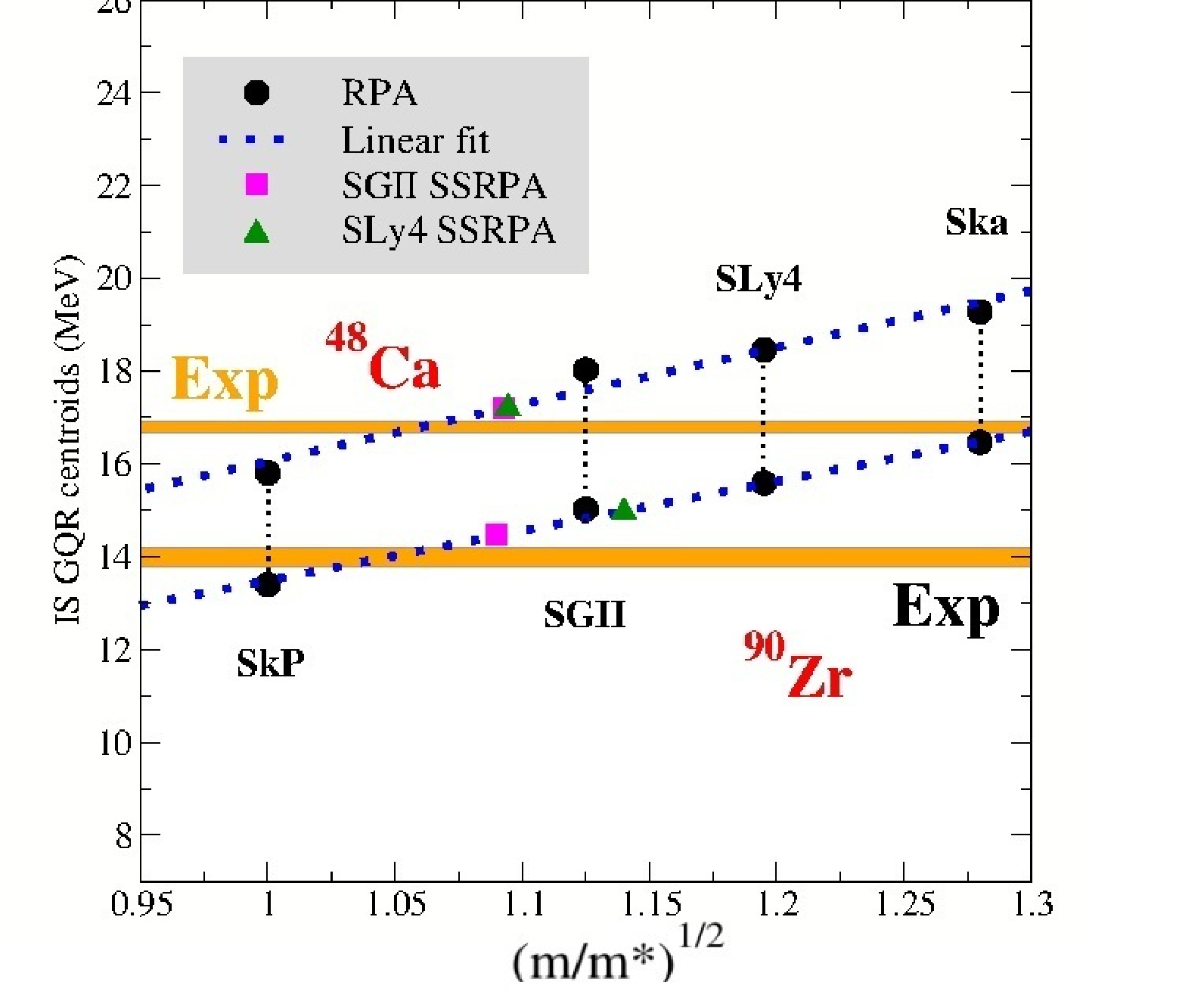}\hfill
	\includegraphics[width=.45\linewidth]{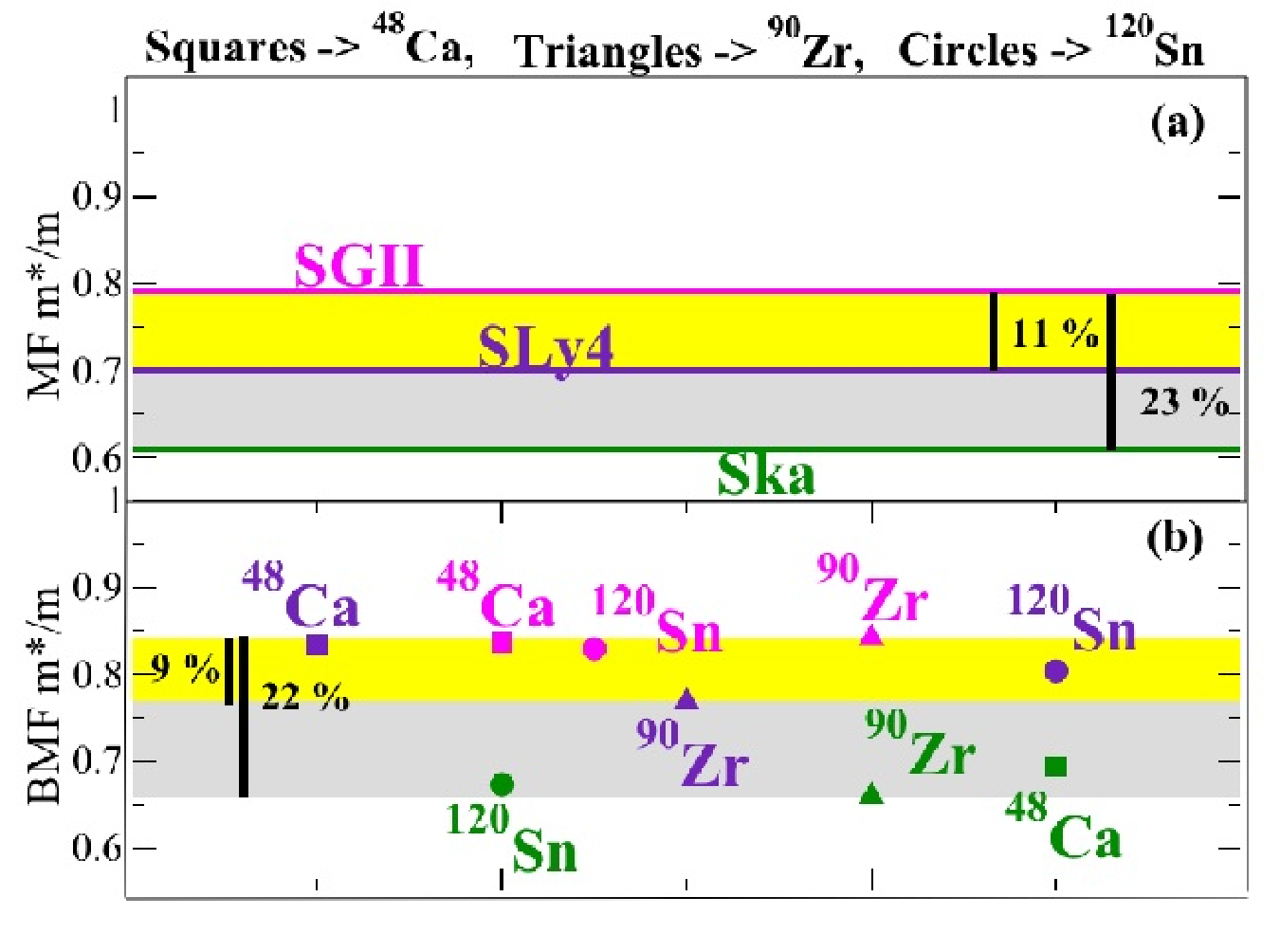}
	\caption{Left side: RPA Centroid energies of the ISGQRs for the nuclei $^{48}$Ca and $^{90}$Zr as a function of $\sqrt{m/m^*}$ for four Skyrme parametrizations. The corresponding linear fit is shown (blue dotted lines). The SSRPA-SLy4 and SSRPA-SGII centroids are reported, green triangles and magenta squares, respectively. The orange bands indicate the experimental values.
% 	Right side: (a) Theoretical error associated to the MF effective mass for nuclear matter induced by two Skyrme parametrizations, SLy4 and SGII (yellow band) and three Skyrme parametrizations, SLy4, SGII, and Ska (yellow plus gray band); (b) Same as in panel (a) but for the BMF effective mass. The four blue circles represent the calculations done with SLy4 and SGII for $^{48}$Ca and $^{90}$Zr whereas the red square represents the calculation done with Ska for $^{48}$Ca. Adapted from Ref. \cite{Grasso2018}.
	Right side: Theoretical error associated with the MF (panel (a)) and BMF (panel (b)) effective
mass for nuclear matter induced by two Skyrme parametrizations,
SLy4 (indigo) and SGII (magenta) (yellow band) and three Skyrme
parametrizations, SLy4, SGII, and Ska (green) (yellow plus grey
band).
Squares,
triangles, and circles represent the BMF effective masses extracted
from $^{48}$Ca, $^{90}$Zr, and $^{120}$Sn, respectively.  Adapted from Ref. \cite{Grasso2018}.
	}
	\label{Fig:EffectiveMass-1}
\end{figure}

On the left side of Figure \ref{Fig:EffectiveMass-1}, the RPA results reported for the medium-mass nucleus $^{48}$Ca and the heavier nucleus $^{90}$Zr using four Skyrme parametrizations: SkP \cite{SKP}, SGII \cite{SGII}, SLy4 \cite{SLY4}, and Ska \cite{SKA}, having MF effective masses of 1, 0.79, 0.7, and 0.61, respectively, in nuclear matter. The figure shows the RPA centroid energies for the ISGQR modes as a function of the square root of the ratio of the bare mass to the effective mass. Each centroid energy was associated with its corresponding MF effective mass in nuclear matter. A linear fit was performed on these four data points for each nucleus, represented by blue dotted lines, and the experimental values were also included as orange bands. In the same figure,
the SSRPA centroids for two selected parametrizations, SLy4 and SGII are also shown. 
The SSRPA yields centroid energies lower than the corresponding RPA values. This lowering of energies indicated an increase in the associated effective mass relative to the MF value. Consequently, it was deduced that, for $^{48}$Ca and $^{90}$Zr, the extracted effective mass for nuclear matter increased from 0.7 in the MF case to 0.834 and 0.769, respectively, for the BMF calculations of the ISGQR with SLy4. With SGII, the effective mass for nuclear matter increased from 0.79 to 0.837 and 0.842, respectively, for the calculations performed on $^{48}$Ca and $^{90}$Zr.

 The right side of Figure \ref{Fig:EffectiveMass-1} shows the theoretical uncertainty in effective mass ($m^*$) calculations. The  MF $m^*$ variations due to different Skyrme interactions yield an 11\% error (panel (a)). BMF calculations show a slightly reduced 9\% discrepancy (panel (b)), despite additional nucleus dependence. Including the Ska parametrization in BMF calculations increases the error to 21\%, comparable to the MF error (23\%). This indicates BMF $m^*$ extraction doesn't significantly amplify uncertainty beyond that already present in MF calculations due to interaction choice. E-mass calculations reveal BMF effects increase the E-mass by 6-16\%, with the largest change observed for $^{48}$Ca using the SLy4 parametrization. More precisely, from the left side of Figure \ref{Fig:EffectiveMass-1} , one can extract the average values of the $E$-mass, equal to 1.19 (1.06) with SLy4 (SGII) for $^{48}$Ca and to 1.10 (1.07) with SLy4 (SGII) for $^{90}$Zr.  A quantitatively similar estimate has been recently provided by including BMF effects within the PVC model \cite{Li2024}.

\newpage
	\section{The SSRPA future: short- and long-term perspectives}
	\label{Sec:Perspectives}
\subsection{Main applications}
Among the possible physics cases where the  SSRPA can be applied, we outline here a few special topics we believe correspond to the interests and timelines of current theoretical and experimental research efforts.

\textbf {Fine structure and total width}: The current interest in studying the fine structure and total width of GRs has been extensively discussed in the Introduction of this Review. In this context, the SSRPA can quantify how much of the total width of a GR originates from explicit ($2p-2h$) mixing (spreading) as opposed to Landau damping ($1p-1h$ fragmentation) and escape (continuum) effects. By comparing RPA and SSRPA calculations one can extract important information on the damping mechanisms. Another major contribution of the SSRPA is its ability to resolve the fine structure and fragmentation of GRs, and to extract the characteristic scales. The complexity of the $2p-2h$ space leads to the fragmentation of the collective GR strength, producing fine-structure patterns and characteristic scales that cannot be fully described by the standard RPA. Moreover, SSRPA provides not only energies and strengths but also microscopic transition densities that naturally include BMF effects. Comparing RPA and SSRPA transition densities reveals how the internal spatial structure of collective motion is modified by the coupling with the $2p-2h$ configurations. This is directly relevant for reaction-model calculations of inelastic cross sections, where SSRPA transition densities
can be easily used to compute form factors to be used in reaction codes. This is also particularly relevant when experimental studies are performed by using hadronic probes. Beyond GRs, the investigation of PDR modes within SSRPA is of special interest, aiming to clarify the nature and collectivity of these states.

\textbf{GR's Decay}: Studying GR's decay properties is as crucial as understanding their fine structure. Analysis of the GR decay products (e.g., particles or photons) can provide more detailed and selective information regarding the microscopic wave-function of the resonance. The experimentally accessible total width of a GR includes several components corresponding to distinct decay channels. Each decay channel is characterized by a specific branching ratio. Of particular interest is the gamma decay width, which is generally a minor component of the total width. Its significance stems from the fact that it is governed solely by the electromagnetic interaction, allowing for a model-independent interpretation of the underlying nuclear structure. GR decay is not restricted to the ground state, transitions to low-energy excited states are also possible. These channels offer a unique window into the microscopic nature of the GR, revealing, for instance, multi-phonon components within its wave function. From a theoretical perspective, such transitions involving excited states are fundamentally beyond the capabilities of the standard RPA, which describes nuclear excitations as one-phonon states built on the ground state. Therefore, describing transitions between two excited states requires theoretical frameworks that go beyond the RPA. In this context, SSRPA studies would be valuable for assessing their capability in accurately describing decay channels and branching ratios.
Beyond achieving agreement with experimental data, a comprehensive comparison with other advanced theoretical models, such as the PVC model or RTQBA, would give crucial information. Such comparisons will help elucidate whether the inclusion of many-body correlations, specifically through collective phonons or $2p-2h$ configurations, yields equivalent or different descriptions of GR decay properties.

\textbf{Beta decays} are of key interest in the study of CP violations (or equivalently T), providing access to the Vud element of the CKM matrix \cite{Hayen2024}. Beta decay studies in nuclei require more extensive and demanding calculations, but the corresponding experimental uncertainties are typically much smaller than other cases (e.g., for example compared to neutrons and pions). Both experimentally and theoretically, Fermi and GT transitions need to be accurately determined. The main challenges for current theories consist of reducing errors related to nuclear structure modeling. GT transitions play a crucial role, and the inclusion of complex configurations, such as the coupling with low-lying collective phonons \cite{Niu2015,Robin2019}, $2p-2h$ excitations \cite{Gambacurta2020} and two-body currents \cite{Gysbers2019,Ekstrom2014}, has been shown to significantly improve the description of the corresponding half-lives calculations. In this respect, systematic SSRPA studies, extended also to superfluid nuclei are of primary importance to confirm the key-role of the $2p-2h$ configurations, in particular in addressing the quenching problem.

\textbf{Neutrino-less double-beta}: The experimental observations of the neutrino-less double-beta decay would indicate that the neutrino is a Majorana particle and would be clear evidence of physics beyond the Standard Model. Moreover, it would allow the extraction of the neutrino’s mass, provided that the nuclear matrix elements describing the transition are sufficiently well predicted by theoretical models. Unfortunately, current predictions based on state-of-the-art calculations show a strong model dependence (Fig.5 of Ref. \cite{Engel_2017}). Theoretical approaches fall into two main categories. The first category includes RPA(QRPA) based methods, having the advantage of using a large single particle basis which guarantees the satisfaction of sum rules, but they include only 1p-1h (2qp) excitations. The second one includes those where many-body configurations can be taken into account in a given single particle basis, which for numerical reasons it must be severely truncated (e.g., Interacting Shell Model, Coupled-Cluster, IBM, GCM, Projected GCM, etc). The SRPA (and Second QRPA, see below) can be considered as an intermediate step providing insight into the discrepancies. Moreover, the results on the single beta decay obtained within the SRPA are quite encouraging and promising.

\subsection{Theoretical extensions and computational developments}
A number of important theoretical extensions and computational developments
remain to be addressed in the coming years related to the SRPA and SSRPA framework. 
\subsubsection{Theoretical extensions}

\textbf {Second Quasi-particle RPA}: A major challenge for future work is the generalization of SRPA to superfluid
and open-shell nuclei through a fully consistent Second Quasi-particle RPA (SQRPA). The latter would be obtained by expressing the excitation operators in terms of quasi-particle operators. 	Starting from the standard QRPA operators, the SQRPA ones would also contain 4-quasi-particle creation and annihilation terms. Beside the formal derivation of SQRPA equations, which at best of our knowledge have never been derived, the main challenge would consist in the computational cost, as the matrix dimensions are expected to increase with respect to the SRPA case. Such an extension would allow to include pairing correlations and to study the majority of nuclei on the nuclear chart.

\textbf{Self-consistent framework:} Of great importance is the further development of fully self-consistent implementations based on modern Skyrme, Gogny, or covariant energy density
functionals, including all the terms of the residual interaction at HF, RPA, SRPA and SSRPA levels. This ensures the full theory is internally consistent and improves the robustness of transition strengths and sum rules.
In this context, deriving effective interactions whose parameters are tuned at a BMF level presents an alternative path. Currently, we are constrained to using effective interactions fitted only at the MF level (e.g., HF and HF+RPA). The use of the subtraction procedure is then employed to avoid double counting issues. Ideally, fitting the force directly at the SRPA level would avoid these issues entirely. While the high computational cost of SRPA currently makes a direct fit unfeasible in the near future, the development and use of emulators could potentially make this goal achievable.

\textbf{SSRPA based on ab initio rooted EDF:} A powerful avenue is to merge SSRPA with \textit{ab initio} based energy density functionals, where the use of phenomenological tuning is avoided or strictly limited. Also in this case, the use of the subtraction procedure might be not necessary. In this way, a more robust and predictive framework would be obtained, with a strong link with the underlying and more fundamental physics. 

\textbf{Continuum SSRPA:} Another interesting attempt would be the introduction of the coupling to the 	continuum within the SSRPA formalism. A continuum SSRPA would enable a consistent description of
escape and spreading widths. This would be of particular importance especially in the case of weakly bound systems and exotic nuclei near the
drip lines. One possible strategy would consist in implementing it in the coordinate-space representation \cite{Mizuyama2009,DeDonno2011,Mizuyama2012} to incorporate the continuum  correctly. 

\textbf{Hybrid approaches:} From a conceptual standpoint, deeper connections between the SSRPA and
energy-dependent Green's function or PVC frameworks
should be further investigated. Specific studies and applications where the methods are compared would be useful and would help clarify the complementary
roles of static and dynamical correlations in the different approaches. A better understanding would also open to the possibility to develop extended combined approaches. BMF correlations are introduced via the coupling with collective phonons and $2p-2h$ excitations in the PVC and SRPA, respectively. The first strategy is particular effective in describing the spreading width, the second one in the description of the fine structure of the strength distributions. One could thus envisage to develop an \textit{hybrid} approach where the phonon operators are expressed in terms of both collective phonons (especially low-lying ones) and $2p-2h$ excitations. The orthogonality of the two spaces would be guaranteed by the norm matrix in the corresponding equations that would have a generalized eigenvalue form. This extension could provide a
unified microscopic framework capable of describing both fragmentation and damping over a wide energy range.

\textbf{Finite-temperature SSRPA:} Generalizations of the SSRPA to finite-temperature regimes represent also an interesting direction to be explored. A finite-temperature SSRPA could be used
to study hot nuclei (dependence of the width with the excitation energy for example) and astrophysical environments (beta decay studies for example). The extension would not represent a significant further effort both formally and computationally, as the main difference would be in the appearance of the temperature-dependent occupation numbers in the SRPA matrix elements.

\subsubsection{Computational and numerical advances}
From a computational point of view, the most demanding operations required by SRPA and SSRPA calculations are:\\
a) the evaluation of the matrix elements to be diagonalized and their storing;\\
c) in the case of the SSRPA, the inversion of the $A_{22}$ matrix;\\
b) the diagonalization of SRPA matrix  to obtain the eigenvalues and eigenfunctions.\\
The numerical complexity of the first two operations depends on the dimension of the involved matrices. The
diagonalization cost depends also on the number of requested eigenvalues. In this respect, some possible strategies are listed below.
%\\
%\textbf{Parallel implementations} on modern
%high-performance computing architectures, including GPU acceleration, can
%further extend the reach of SSRPA toward heavy nuclei and large-scale calculations. \\ 
%The development of \textbf{iterative algorithms},
%Krylov-subspace solvers, and energy-dependent reduction schemes will be crucial
%to make systematic studies feasible.\\
%Another breakthrough would be the use of \textbf{eigenvalue solver} which can target specific region of the spectrum (avoiding the full diagonalization) and the implementation of GPU based large-scale eigenvalue solver (Invidia, ViellaCl, MAGAMA, ...).\\ 
%Another promising avenues might involve
%the use of \textbf{accelerator methods} such as the importance-truncation \cite{Roth2007,Stumpf2016}, the eigenvector continuation \cite{Frame2018, Konig2020,Companys2023} or tee reduced-basis method (RBM) (projection-based emulators) \cite{Bonilla2022,Duguet2024} specifically adapted to the SSRPA case to identify the most relevant $2p-2h$
%configurations and accelerate convergence.

\noindent
\begin{description}
	
	\item[-] Parallel implementations on modern
	high-performance computing architectures, including GPU acceleration, can
	further extend the reach of SSRPA toward heavy nuclei and large-scale calculations. 
	\item[-] The development of iterative algorithms,
	Krylov-subspace solvers, and energy-dependent reduction schemes will be crucial
	to make systematic studies feasible.
	\item[-] Another strong advantage would be the use of eigenvalue solver which can target specific region of the spectrum (avoiding the full diagonalization) and the implementation of GPU based large-scale eigenvalue solver (Invidia, ViellaCl, MAGAMA, ...). 
	\item[-] Another promising avenues might involve
	the use of methods such as the importance-truncation \cite{Roth2007,Stumpf2016}, the eigenvector continuation \cite{Frame2018, Konig2020,Companys2023} or the reduced-basis method  (projection-based emulators) \cite{Bonilla2022,Duguet2024} specifically adapted to the SSRPA case to identify the most relevant $2p-2h$
	configurations and accelerate convergence.
	\item[-] Considering the high-computational cost required by the SRPA, the use of emulators could be beneficial both for systematic calculations and in the attempt to derive new EDFs specifically tailored at SRPA level. In this regard, recent applications  of surrogate models \cite{Jin2025} to emulate the QRPA description and based on the  parametric matrix models \cite{Cook2025}, could be extended to the SRPA case.
\end{description}

	\newpage
	\section{Summary and Outlook}\label{Sec:Summary}
%	\subsection{Summary}
In this review, we have presented and discussed in detail the most recent developments of the SRPA and SSRPA methods, also in connection with some of the main current experimental and theoretical challenges in the study of nuclear collective excitations, some of which have been briefly outlined in the introductory Section \ref{Sec:Intro}.

	In Section \ref{Sec:FormalPart}, the formal properties of the SRPA were derived within the method of the equations of motion and discussed. %Additional formal properties, such as the behavior of the moments and sum rules, were also examined. 
	The conceptual and mathematical motivations for introducing the subtraction procedure within the EDF framework were then discussed. The subtraction procedure  acts as a regularization method to remove double-counting issues when effective interactions are employed and to cure ultraviolet divergences appearing in the case of zero-range interactions. The SSRPA equations are then introduced, and some of its  properties and approximations discussed. 
	
	Section \ref{Sec:Applications_SRPA} is entirely devoted to large-scale SRPA applications, performed over the past fifteen years. These implementations have become feasible without  severe truncations of the model space, as it was necessary in earlier works, and without resorting to approximations in the evaluation of matrix elements beyond the standard RPA level. The section illustrates a variety of results obtained with different types of effective interactions, ranging from microscopic to phenomenological ones. Regardless of the specific interaction or type of excitation considered, SRPA calculations consistently exhibit two distinctive features:
	(i) a pronounced fragmentation of the strength distribution, and
	(ii) a systematic downward shift of the centroid energies with respect to RPA predictions.
	This energy shift, typically amounting to several MeV, is somewhat pathological, especially within the EDF framework, where the RPA usually reproduces the experimental GR centroid energies within 1–2 MeV. These findings clearly demonstrate the need to incorporate the subtraction procedure within the EDF-based SRPA to properly regularize the theory and increase its predictive power. The Section concludes discussing SRPA instabilities, further emphasizing the need of advancing beyond the standard SRPA scheme.
	
	Section \ref{Sec:Applications_SSRPA_CC} thus covers the most recent applications of the SSRPA, with the Skyrme and Gogny interactions and in the Relativistic point-coupling model, for charge-conserving excitations. The very first implementation of the SSRPA is discussed to show to what extent it is able to cure the pathological red-shift observed in SRPA. After that, several applications are presented for the study of the monopole, dipole and quadrupole response, underlining the advantages of the SSRPA in describing key features of collective excitations, such as strength distributions, their centroids and spreading width. The Section concludes with the very recent application of the relativistic SSTDA scheme, where similar results to the non-relativistic case were found. Similarly, Section \ref{Sec:Applications_SSRPA_CE} presents the SSRPA results for charge-exchange excitations, in particular related to the GT channel and beta-decay. It is shown how the inclusion of the $2p-2h$ configurations within the SSRPA  significantly improves the reproduction of the GT resonance centroid and widths, and the strength distribution within the beta window, resulting in a better description of the corresponding beta decay life time.
	
	In Section \ref{Sec:BMF}, the BMF effects induced by the SSRPA on key properties of the nuclear equation of state are discussed. In particular, we examine their impact on the symmetry energy and its slope, the effective mass, and the incompressibility, and provide qualitative estimates of these effects. 
	
		Finally, in Section \ref{Sec:Perspectives}, the future perspectives of the SSRPA, regarding its possible applications, and formal or computational extensions have been discussed. Concerning the applications, those related to the single-beta and double-beta decay are perhaps the ones that fit best the current experimental interests and the most recent SSRPA applications. Among the extensions, the development of the SQRPA to include pairing correlations is  mandatory to enlarge the applicability of the SRPA/ SSRPA framework. Another possible avenue is the development of the SSRPA in the coordinate space representation to  include the coupling to the continuum and thus describe the escape and spreading width consistently.
		 
% 
	
%	\bibliography{mybibfile}
	%Please use Bib\TeX\ to generate your bibliography and include DOIs whenever available. Example of bib file: 
	
	%%%%%%%%%%%%%%%%%%%%%%%%%%%%%%%%%%%%%%%%%%%%%%%%%%%%%%%%%%%%%%%%%%%
	% Encoding: ISO-8859-1

	%@Article{Eichmann:2016yit,
	%author = {Eichmann, Gernot and Sanchis-Alepuz, Helios and Williams, Richard and Alkofer, Reinhard and Fischer, Christian S.},
	%title = {{Baryons as relativistic three-quark bound states}},
	%journal = {Prog. Part. Nucl. Phys.},
	%year = {2016},
	%volume = {91},
	%pages = {1-100},
	%archiveprefix = {arXiv},
	%doi = {10.1016/j.ppnp.2016.07.001},
	%eprint = {1606.09602},
	%owner = {chfi},
	%primaryclass = {hep-ph},
	%slaccitation = {%%CITATION = ARXIV:1606.09602;%%},
	%timestamp = {2018.08.02},
	%}

	%@Comment{jabref-meta: databaseType:bibtex;}
	%%%%%%%%%%%%%%%%%%%%%%%%%%%%%%%%%%%%%%%%%%%%%%%%%%%%%%%%%%%%%%%%%%%

	\newpage
	\appendix
	\renewcommand*{\thesection}{\Alph{section}}
	
	\section{RPA and SRPA matrices in the m-scheme}
	\label{Sec:AppMScheme}
	
	Considering the Hamiltonian $H = \sum_i \epsilon_i a^\dagger_i a_i + \frac{1}{4} \sum_{tqrs} V_{tqrs} a^\dagger_t a^\dagger_q a_s a_r$, where $\epsilon_i$ are the single-particle energies obtained from the HF calculation and $V_{tqrs}$ are the antisymmetrized matrix elements ($V_{tqrs}=\langle tq\mid \hat{V}\mid rs \rangle-\langle tq\mid \hat{V}\mid sr \rangle$) of the residual two-body interaction $\hat{V}$, one gets for the RPA matrix (\ref{Eq:RPAmatA_M}) and (\ref{Eq:RPAmatB_M}) the following expressions:
	\begin{equation}
	\mathcal{A}_{11} \equiv	\mathcal{A}_{ph, p'h'} = (\epsilon_p - \epsilon_h) \delta_{pp'} \delta_{hh'} + {V}_{ph'hp'}
		%\langle ph' | V_{res} | p'h' \rangle
	\end{equation}	
	\begin{equation}
	\mathcal{B}_{11} \equiv	\mathcal{B}_{ph, p'h'} = {V}_{pp'hh'}.
		%\langle pp' | V_{res} | hh' \rangle	
	\end{equation}
	%where ${V}_{ijkl}={V}_{ijkl}-{V}_{ijlk}$.
	Considering the SRPA matrices, as the QBA is still used, and the Hamiltonian contains only a one-body and (no density-dependent) two-body terms, it can be shown 
	that $\mathcal{B}_{12}$, $\mathcal{B}_{21}$ and $\mathcal{B}_{22}$ are zero. The other matrix elements (\ref{a12}) and (\ref{a22}) are equal to: 
	
	\begin{eqnarray}\label{a12-expr}
		\mathcal{A}_{12}=\mathcal{A}_{ph,p_1p_2h_1h_2}&=&\big\langle HF |
		\big[a_{h}^{\dag}a_{p},[H,a_{p_1}^{\dag}a_{p_2}^{\dag}a_{h_2}a_{h_1}
		]\big]| HF \big\rangle \nonumber\\
		&=&\hat{U}_{h_1 h_2}{V}_{h_1pp_1p_2}\delta_{hh_2}-\hat{U}_{p_1 p_2}{V}_{h_1h_1p_1h}\delta_{pp_2}
	\end{eqnarray}
	\begin{eqnarray}
		\label{a22-expr}
		\mathcal{A}_{22}=\mathcal{A}_{p_1h_1p_2h_2,p'_1h'_1p'_2h'_2}&=&\big\langle HF |\big[a_{h_1}^{\dag}a_{h_2}^{\dag}a_{p_1}a_{p_2},[H,
		a_{p'_2}^{\dag}a_{p'_1}^{\dag}a_{h'_2}a_{h'_1} ]\big]| HF \big\rangle=\nonumber \\
		&=&(\epsilon_{p_1}+\epsilon_{p_2}-\epsilon_{h_1}-\epsilon_{h_2})\hat{U}_{p_1 p_2}\hat{U}_{h_1h_2}\delta_{h_1h'_1}\delta_{p_1p'_1}\delta_{h_2h'_2}\delta_{p_2p'_2}
		\nonumber \\
		&+&
		\hat{U}_{h_1h_2}{V}_{p_1p_2p'_1p'_2}\delta_{h_1h'_1}\delta_{h_2h'_2}\nonumber\\
		&+&
		\hat{U}_{p_1p_2}{V}_{h_1h_2h'_1h'_2}\delta_{p_1p'_1}\delta_{p_2p'_2}\nonumber\\
		&+&
		\hat{U}_{p_1 p_2}\hat{U}_{h_1 h_2}\hat{U}_{p'_1p'_2}\hat{U}_{h'_1 h'_2}{V}_{p_1h'_1h_1p'_1}\delta_{h_2h'_2}\delta_{p_2p'_2}
	\end{eqnarray}
	where $\hat{U}_{ij}$ is the antisymmetrizer for the indices $i$, $j$.
\section{RPA and SRPA matrices in the J-scheme}
\label{Sec:AppJScheme}
The RPA matrices in the J-coupled scheme are:
\begin{eqnarray} \label{Eq:RPA_A_J}
	\mathcal{A}_{11'}\equiv \mathcal{A}_{ph;p'h'} 
	&=& (e_p-e_h)\delta_{pp'}\delta_{hh'} 
	% \nonumber \\
	% &&
	+ \langle ph^{-1};J | V | p'{h'}^{-1};J\rangle
	\nonumber \\ 
	&=& (e_p-e_h)\delta_{pp'}\delta_{hh'} 
	% \nonumber \\
	% &&
	+ \sum_{J_1}(-1)^{j_{h}+j_{p'}-J_1} (2J_1+1)
	% \nonumber \\ &&
	\times
	\left\{ \begin{array}{ccc}j_{p} & j_{h'} & J_1\\ j_{p'} & j_{h} & J \end{array} \right\} 
	\langle ph';J_1 | V | h p';J_1\rangle 
	\label{E:Amat} 
\end{eqnarray} 
\begin{eqnarray} \label{Eq:RPA_B_J}
	\mathcal{B}_{11'} \equiv	\mathcal{B}_{ph;p'h'} 
	&=& \langle (ph^{-1};J) (p'{h'}^{-1};J) | V | 0\rangle 
	\nonumber \\ 
	&=& \sum_{J_1}(-1)^{j_{h}+j_{p'}+J-J_1} (2J_1+1) 
	% \nonumber \\ &&
	\times
	\left\{ \begin{array}{ccc}j_{p} & j_{p'} & J_1\\ j_{h} & j_{h'} & J \end{array} \right\} 
	(1+\delta_{pp'})^{1/2} 
	(1+\delta_{hh'})^{1/2} 
	\nonumber \\ & &\times 
	\langle pp';J_1 | V | h h';J_1\rangle 
	. 
	\label{E:Bmat} 
\end{eqnarray} 
where $e_i$ are the HF single particle energies and $V$ the residual interaction.

The $\mathcal{A}_{12}$  and $\mathcal{A}_{22}$ matrices are 
%describes the coupling between $1p-1h$ and $2p-2h$ states:
\begin{eqnarray} 
\mathcal{A}_{12}\equiv 	\lefteqn{
		\mathcal{A}_{ph;p_1p_2h_1h_2J_pJ_h} 
		=\langle ph^{-1};J | V | (p_1p_2;J_p)(h_1h_2;J_h)^{-1};J \rangle} && \nonumber \\
	% && \langle ph^{-1};J | V | (p_1p_2;J_p)(h_1h_2;J_h)^{-1};J \rangle
	% \nonumber \\
	&=& [1-(-1)^{j_{h_1}+j_{h_2}-J_h} \hat{U}_{h_1h_2}]
	\frac{\delta_{h_1h}}{(1+\delta_{h_1h_2})^{1/2}}
	% \nonumber \\
	%\times
	(-1)^{j_{p}+j_{h_2}+J+J_h} 
	\hat{J}_p\hat{J}_h 
	% \nonumber \\ & &
	%\times
	\left\{ \begin{array}{ccc}J_p & J & J_h\\ j_{h_1} & j_{h_2} & j_p \end{array} \right\} 
	%\nonumber \\ & \times & 
	\langle p_1p_2;J_p | V | ph_2;J_p\rangle 
	\nonumber \\ & - &
	[1-(-1)^{j_{p_1}+j_{p_2}-J_p} \hat{U}_{p_1p_2}]
	\frac{\delta_{p_1p} }{(1+\delta_{p_1p_2})^{1/2} }
	% \nonumber \\ & &
	%\times
	(-1)^{j_{p_1}+j_{p_2}+J+J_h} 
	\hat{J}_p\hat{J}_h 
	% \nonumber \\ & &
	%\times
	\left\{ \begin{array}{ccc}J_h & J & J_p\\ j_{p_1} & j_{p_2} & j_h \end{array} \right\} 
	%\nonumber \\ & & \times 
	\langle hp_2;J_h | V | h_1h_2;J_h\rangle 
	, 
	\label{Eq:SRPA_A12_J} 
\end{eqnarray} 
\begin{eqnarray} 
\mathcal{A}_{22'}\equiv 	\lefteqn{
		\mathcal{A}_{p_1p_2h_1h_2J_pJ_h;p_1'p_2'h_1'h_2'J_{p'}J_{h'}} 
		= } \nonumber \\
	&& \delta_{p_1p_1'}\delta_{h_1h_1'}\delta_{p_2p_2'}\delta_{h_2h_2'}
	\delta_{J_pJ_{p'}}\delta_{J_hJ_{h'}}(e_{p_1}+e_{p_2}-e_{h_1}-e_{h_2}) 
	\nonumber \\ &+&
	\langle
	(p_1p_2;J_p)(h_1h_2;J_h)^{-1};J 
	| V | 
	(p_1'p_2';J_{p'})({h_1'}{h_2'};J_{h'})^{-1};J 
	\rangle
	\nonumber \\
	&=& \delta_{p_1p_1'}\delta_{h_1h_1'}\delta_{p_2p_2'}\delta_{h_2h_2'}
	(e_{p_1}+e_{p_2}-e_{h_1}-e_{h_2}) 
	\nonumber \\ &+& 
	\delta_{p_1p_1'}\delta_{p_2p_2'} 
	\delta_{J_pJ_{p'}}\delta_{J_hJ_{h'}} [1+(-1)^{J_p}\delta_{p_1p_2} ] (1+\delta_{p_1p_2})^{-1} 
	% \nonumber \\ & &
	\times
	\langle h_1'h_2';J_h | V | h_1h_2;J_h\rangle 
	\nonumber \\ &+&
	+\delta_{h_1h_1'}\delta_{h_2h_2'}
	\delta_{J_pJ_{p'}}\delta_{J_hJ_{h'}} [1+(-1)^{J_h}\delta_{h_1h_2} ] (1+\delta_{h_1h_2})^{-1} 
	% \nonumber \\ & &
	\times
	\langle p_1p_2;J_p | V | p_1'p_2';J_p\rangle 
	\nonumber \\ &+& 
	[1-(-1)^{j_{p_1}+j_{p_2}-J_p} \hat{U}_{p_1p_2}] (1+\delta_{p_1p_2})^{-1/2}
	% \nonumber \\ &&
	\times
	[1-(-1)^{j_{h_1}+j_{h_2}-J_h} \hat{U}_{h_1h_2}] (1+\delta_{h_1h_2})^{-1/2}
	\nonumber \\ &&
	\times
	[1-(-1)^{j_{p_1'}+j_{p_2'}-J_p} \hat{U}_{p_1'p_2'}] (1+\delta_{p_1'p_2'})^{-1/2}
	% \nonumber \\ &&
	\times
	[1-(-1)^{j_{h_1'}+j_{h_2'}-J_h} \hat{U}_{h_1',h_2'}] (1+\delta_{h_1'h_2'})^{-1/2}
	\nonumber \\ && \times
	\delta_{h_2h_2'}\delta_{p_2p_2'} (-1)^{1+j_{p_1}+j_{p_2}+j_{h_1}+j_{h_2}} 
	\hat{J}_p \hat{J}_{p'} \hat{J}_h \hat{J}_{h'} 
	\nonumber \\ && \times 
	\sum_{L}(-1)^{J_h-J_{h'}+J-L} (2L+1)
	\left\{ \begin{array}{ccc}J_p & J_{p'} & L\\ J_{h'} & J_h & J \end{array} \right\} 
	% \nonumber \\ &&
	\times
	\left\{ \begin{array}{ccc}J_p & J_{p'} & L\\ j_{p_1'} & j_{p_1} & j_{p_2} \end{array} \right\} 
	\left\{ \begin{array}{ccc}J_h & J_{h'} & L\\ j_{h_1'} & j_{h_1} & j_{h_2} \end{array} \right\} 
	\nonumber \\ &&
	\times
	\sum_{J_1}(-1)^{j_{h_1}+j_{p_1'}-J_1} (2J_1+1)
	% \nonumber \\ &&
	\times
	\left\{ \begin{array}{ccc}j_{p_1} & j_{h_1'} & J_1\\ j_{h_1} & j_{p_1'} & L \end{array} \right\} 
	\langle p_1h_1';J_1 | V | p_1'h_1;J_1\rangle 
	. 
	\label{Eq:SRPA_A22_J} 
\end{eqnarray} 
	where $\hat{U}_{ij}$ is the antisymmetrizer with respect to the indices $ij$. 
\section{One-body transition operators}
%Transition operators empl along the text.
%\subsection{Monopole and Quadrupole operators}
\textbf{Monopole and Quadrupole operators}\\
For the monopole ( $\lambda=0$) and quadrupole  ($\lambda=2$) case, the transition operators used are
\begin{equation}\label{Eq:Op-J02-isos}
	T_{\lambda}^{IS}=\sum_{i=1}^{A} r_i^2 Y_{\lambda 0}(\hat{r}_i)~~~
\end{equation}
\begin{equation}\label{Eq:Op-J02-isov}
	T_{\lambda}^{IV}=\sum_{i=1}^{A} r_i^2 Y_{\lambda 0}(\hat{r}_i)\tau_z(i)
\end{equation}
in the isoscalar and isovector channel, respectively.
%, where $n=\lambda$ except for $\lambda=0$ where $n=2$.
% and $\tau_z$ is the isospin .
\\
%\subsection{Dipole case}
%\vspace{2cm}
\textbf{Dipole electric (E1) and magnetic (M1) operators}\\
In the dipole case, the following operators are employed:
\begin{equation}\label{Eq:Op-J1-em}
	T_{E1}^{(em)} = \sum_{p} r_{p}Y_{10}(\hat{r}_{p})%\equiv O^{1}_{E1}
\end{equation}
\begin{equation}\label{Eq:Op-J1-isov}
	T^{IV}_{E1}=\sum_{p=1}^{Z} r_p Y_{1 0}(\hat{r}_p) -\sum_{n=1}^{N} r_n Y_{1 0}(\hat{r}_n)
\end{equation}
\begin{equation}\label{Eq:Op-J1-isos}
	T^{IS}_{E1}=\sum_{p=1}^{Z} r_p Y_{1 0}(\hat{r}_p) +\sum_{n=1}^{N} r_n Y_{1 0}(\hat{r}_n)%\equiv O^{sp}
\end{equation}
being the pure electromagnetic, isovector and isoscalar operators respectively.

The intrinsic dipole isovector operator, containing the factors for the center of mass correction, is
\begin{equation}\label{Eq:Op-J1-IV-CM}
	T^{IVD}_{E1}=\sum_{p=1}^{Z} \frac{N}{A} r_p Y_{1 0}(\hat{r}_p) -\sum_{n=1}^{N} \frac{Z}{A} r_n Y_{1 0}(\hat{r}_n).%\equiv O^{2}_{E1}
\end{equation}
%having that
%\begin{equation}
%	T^{IVD}_{E1}=T_{E1}^{(em)}-\frac{Z}{A}	T^{IS}_{E1}.
%\end{equation}

%
%
%
%\begin{equation}
%	T^{\mathrm{sp}}=\frac{A}{Z} (T_{E1}^{(em)}-T_{E1}^{(IVD)})
%\end{equation}
%
%
%
%one and the latter will be used to being the generator of the spurious $1^-$ state.

In the isoscalar channel, we also consider the second order operators \cite{Harakeh_book, BM85} 
\begin{equation}\label{Eq:Op-J1-isos-r3}
	T_{\mathrm{E1}}^{IS-2} = \sum_{i=1}^A r_i^3 Y_{10}(\hat{r}_i)
\end{equation}
and the intrinsic one  with center of mass correction \cite{SGII}
\begin{equation}\label{Eq:Op-J1-isos-r3-corrected}
	T_{\mathrm{E1}}^{IS-2,cm} = \sum_{i=1}^A (r_i^3 - \frac{5}{3}\langle r^2\rangle r_i ) Y_{10}(\hat{r}_i).
\end{equation}
%where $\eta = \frac{5}{3}\langle r^2\rangle$~\cite{SGII},

The magnetic dipole operator reads as 
\begin{equation}\label{Eq:Op-J1-magnetic}
	T_{M1}=\frac{3}{4\pi}\sum_i^A\big[g_i^l \mathbf{l_i}+ g_i^s \mathbf{s_i}\big] \mu_N
%	\label{Eq:M1op}
\end{equation}
where $\mathbf{l_i}$ and $\mathbf{s_i}$ are the orbital and spin angular momenta, respectively, 
$g_l$ and $g_s$ the corresponding gyromagnetic factors of nucleons and $\mu_N$ is the Bohr magneton of a nucleon \cite{Fujita2011}.\\
%\vspace{2cm}
%\subsection{GT operator}
\textbf{Gamow-Teller (GT) operators}

GT excitations are induced by the transition operators
\begin{equation}
	T_{GT}^{\pm}=\sum_{i=1}^{A} \sum_{\mu} \sigma_{\mu}(i) \tau^{\pm}(i),
	\label{Eq:Op-GT}
\end{equation} 
where $\tau^{\pm}$ are the 
isospin raising ($+$) and lowering ($-$) operators, $\tau^{\pm}=t_x \pm it_y$, and $\sigma_{\mu}$ is the spin operator. 

\section{Formal derivation of the rearrangement terms in SRPA}
\label{Sec:Rear}
Let's start by expressing the ground state $|\Psi >$ as
\begin{eqnarray}
	|\Psi > = e^{S} |\Phi >,
	\label{eq1}
\end{eqnarray}
where $|\Phi>$ is the HF ground state and $S$ is a linear superposition of 
$1p-1h$ and $2p-2h$ configurations, 
\begin{eqnarray}
	S = \sum_{ph} C_{ph} a^{\dag}_p a_h + \frac {1} {2} 
	\sum_{pp'hh'} C_{pp'hh'} a^{\dag}_p a^{\dag}_{p'} a_{h'} a_h~.
	\label{eq2}
\end{eqnarray}

Let us denote with $h$, $i$, $j$, $k$ and $l$ hole states and with 
$m$, $n$, $p$ and $q$ particle states, while we use Greek letters for a generic particle states. We introduce the anti-symmetrized coefficients
\begin{eqnarray}
	\hat{C}_{\alpha \beta \gamma \delta} = C_{\alpha \beta \gamma \delta} 
	- C_{\alpha \beta \delta \gamma}~.
	\label{eq3}
\end{eqnarray}

Moreover, the Hamiltonian contains a density-dependent two-body term:
\begin{eqnarray}
	H=\sum_{\alpha \beta} t_{\alpha \beta} a^{\dag}_{\alpha} a_{\beta} + 
	\frac {1} {4} \sum_{\alpha \beta \gamma \delta}
	\hat{V}_{\alpha \beta \gamma \delta} (\rho) a^{\dag}_{\alpha} 
	a^{\dag}_{\beta} 
	a_{\delta} a_{\gamma},
	\label{H}
\end{eqnarray}
where
\begin{eqnarray}
	\hat{V}_{\alpha \beta \gamma \delta} (\rho) = 
	V_{\alpha \beta \gamma \delta} (\rho) - 
	V_{\alpha \beta \delta \gamma} (\rho)~.
	\label{eq5}
\end{eqnarray}

In the present formulation, the coefficients $C_{ph}$ and $C_{p'p'',h'h''}$ associated with the $1p-1h$ and $2p-2h$ ground state configurations, respectively, are assumed to be infinitesimally small and are treated as variational parameters. This assumption facilitates the expansion of the expectation values of one- and two-body operators as a power series in these coefficients. The expansion is then truncated at the second order, meaning that the expectation values will contain terms up to quadratic order in $C_{ph}$ and $C_{p'p'',h'h''}$.

By expanding then the one-body density matrix $\rho$ around the HF density $\rho^{(0)}$ up to the second order in $S$, we have

\begin{eqnarray}
	\rho_{\alpha \beta} = 
	<\Psi| a^{\dag}_{\beta} a_{\alpha} |\Psi> = <\Phi| 
	e^{S^{\dag}} a^{\dag}_{\beta} a_{\alpha} e^S |\Phi>
	= <\Phi| (1+S^{\dag} + \frac {1} {2} S^{\dag 2}+... ) 
	a^{\dag}_{\beta} a_{\alpha} (1+ S + \frac {1} {2} S^2 +...) | \Phi> 
	\nonumber \\
	\sim \rho_{\alpha \beta}^{(0)} + 
	<\Phi| a^{\dag}_{\beta} a_{\alpha} S + S^{\dag} 
	a^{\dag}_{\beta} a_{\alpha} |\Phi> + 
	<\Phi| \frac {1} {2} a^{\dag}_{\beta} a_{\alpha} S^2 + 
	S^{\dag} a^{\dag}_{\beta} a_{\alpha} S + 
	\frac{1} {2} S^{\dag 2} a^{\dag}_{\beta} a_{\alpha}|\Phi> =
	\rho_{\alpha \beta}^{(0)} + \delta \rho_{\alpha \beta}.
	\label{eq7}
\end{eqnarray}
The variation of the density $\delta \rho_{\alpha \beta}$ is given by the sum of
linear $\delta \rho^{(1)}$ 
\begin{eqnarray}
	\delta \rho^{(1)}_{\alpha \beta} = 
	<\Phi| a^{\dag}_{\beta} a_{\alpha} S + S^{\dag} 
	a^{\dag}_{\beta} a_{\alpha} |\Phi>,
	\label{eq8}
\end{eqnarray}
and quadratic 	$\delta \rho^{(2)}$ contributions
\begin{eqnarray}
	\delta \rho^{(2)}_{\alpha \beta} = 
	<\Phi| \frac {1} {2} a^{\dag}_{\beta} a_{\alpha} S^2 + S^{\dag} a^{\dag}_{\beta} a_{\alpha} S + 
	\frac{1} {2} S^{\dag 2} a^{\dag}_{\beta} a_{\alpha}|\Phi>.
	\label{eq8_1}
\end{eqnarray}

More specifically, we have that 
\begin{eqnarray}
	\delta \rho^{(1)}_{hh'}= \delta \rho^{(1)}_{pp'} =0,
	\label{eq9}
	\delta \rho^{(1)}_{ph}= C_{ph},
	\label{eq10}
	\delta \rho^{(1)}_{hp} = C^*_{ph},
	\label{eq11}
\end{eqnarray}
\begin{eqnarray}
	\delta \rho^{(2)}_{ph}= \sum_{mi}C^*_{mi} \hat{C}_{pmhi},
	\label{eq12}
	\delta \rho^{(2)}_{hp} = \sum_{mi} C_{mi} \hat{C}^*_{pmhi},
\end{eqnarray}
\begin{eqnarray}
	\delta \rho^{(2)}_{hh'} = -\sum_{m} C^*_{mh} C_{mh'} - \frac {1} {2} 
	\sum_{mni} \hat{C}^*_{mnih} \hat{C}_{mnih'},
	\label{eq14}
\end{eqnarray} 
\begin{eqnarray}
	\delta \rho^{(2)}_{pp'} = \sum_{i} C^*_{p'i} C_{pi} + \frac {1} {2} 
	\sum_{mij} \hat{C}^*_{p'mij} \hat{C}_{pmij}~.
	\label{eq15}
\end{eqnarray} 

The mean value of the Hamiltonian in $|\Psi>$, $<H>$, can be also expanded as 
\begin{eqnarray}
	<H>= <\Phi|H|\Phi> + \sum_{mi} (C^{*}_{mi} \lambda_{mi} (\rho) 
	+ C_{mi} \lambda_{im} (\rho)) 
	+ \sum_{i<j,m<n} (\hat{C}^{*}_{mnij} \hat{V}_{mnij} (\rho) + 
	\hat{C}_{mnij} \hat{V}_{ijmn} 
	(\rho)) + F^{(2)},
	\label{meanH}
\end{eqnarray} 
where
\begin{eqnarray} 
	\lambda_{ab}(\rho) = t_{ab} + \sum_k \hat{V}_{akbk}(\rho),
	\label{eq17}
\end{eqnarray}

which, in cases where the interaction does not depend on the density, 
define the single-particle energies. and $F^{(2)}$ is the sum of the
quadratic contributions in $C$ and $C^*$.

The matrices $A$ appearing in the SRPA equations are defined as 
\begin{eqnarray}
	A_{mi,pk } = \left[\frac {\delta^2 <H> } 
	{\delta C^*_{mi} \delta C_{pk}}\right]_{C=C^*=0}\equiv A_{11},
	\label{A11}
\end{eqnarray}

\begin{eqnarray}
	A_{mi,pqkl } = \left[\frac {\delta^2 <H> } {\delta C^*_{mi} 
		\delta \hat{C}_{pqkl}}\right]_{C=C^*=0} \equiv A_{12},
	\label{A12}
\end{eqnarray}

\begin{eqnarray}
	A_{mnij,pqkl } = \left[\frac {\delta^2 <H> } 
	{\delta \hat{C}^*_{mnij} \delta \hat{C}_{pqkl}}\right]_{C=C^*=0} \equiv A_{22},
	\label{A22}
\end{eqnarray}
where $A_{11}$ is the usual RPA matrix. 
The corresponding matrices $B$ are obtained by replacing in the 
previous equations the derivative with respect to $C$ and $\hat{C}$ with the
derivative with respect to
$C^{*}$ and $\hat{C}^{*}$, respectively.
%	For example, the RPA matrix $B$ is written as 
%	\begin{eqnarray}
	%		B_{mi,pk } = \left[\frac {\delta^2 <H> } {\delta C^*_{mi} 
		%			\delta C^*_{pk}}\right]_{C=C^*=0}\equiv B_{11}.
	%		\label{B11}
	%	\end{eqnarray}

When the interaction is density-independent, only the $F^{(2)}$ term of Eq. (\ref{meanH}) contributes, as the other terms are not quadratic in the C coefficients. 	This recovers the SRPA matrices derived by Providencia (see Eqs. (9)-(12) of Ref. \cite{Providencia1965}). If the interaction is density-dependent, rearrangement terms emerge, generated by the first three terms of Eq. (\ref{meanH}). 	 The focus here is on the rearrangement terms that must be incorporated into the SRPA matrices of type '12' ($A_{12}$ and $B_{12}$) and '22' ($A_{22}$ and $B_{22}$).
Expanding now $\hat{V}$ around the HF density 
$\rho^{(0)}$ up to second order one gets: 
\begin{eqnarray}
	&& \hat{V}_{\alpha \beta \gamma \delta} (\rho) \sim 
	\hat{V}_{\alpha \beta \gamma \delta} (\rho^{(0)}) + 
	\sum_{a b} \left[ \frac {\delta \hat{V}_{\alpha \beta \gamma \delta}} 
	{\delta \rho_{ab}}\right]_{\rho=\rho^{(0)}} \delta \rho_{ab} + \frac {1} {2} 
	\sum_{a b c d } \left[ \frac {\delta^2 \hat{V}_{\alpha \beta \gamma \delta} } 
	{\delta \rho_{ab} \delta\rho_{cd}}\right]_{\rho=\rho^{(0)}} 
	\delta \rho_{ab} \delta \rho_{cd} ,
	\label{eq6}
\end{eqnarray}
where 
\begin{eqnarray}
	\delta \rho_{\alpha \beta}=\delta \rho_{\alpha \beta}^{(1)}+\delta \rho_{\alpha \beta}^{(2)}.
\end{eqnarray}
By employing this expansion for the interaction, it can be demonstrated that the standard RPA rearrangement terms are recovered. For instance, the RPA matrix $A_{11}$ defined in Eq. (\ref{A11}) gives:
\begin{equation}
	A_{mi,pk } = \delta_{ki}\epsilon_{mp}-\delta_{mp}
	\epsilon_{ki}+\mathscr{V}_{mkip},
	\label{A11-RPA}
\end{equation}
with the single-particle energies given by
\begin{equation}
	\epsilon_{ab} =\lambda_{ab}(\rho^{(0)})+\frac{1}{2}\sum_{kk'} 
	\left[\frac {\delta \hat{V}_{k k' k k'}} 
	{\delta \rho_{ba}} \right]_{\rho=\rho^{(0)}} \rho_{ba}, %\rho_{kk'}^{(0)}
	\label{spenergy}
\end{equation} 
where $\rho_{ba}=\psi_b \psi^*_a$ and $\psi$ are the HF single-particle wave 
functions. 
The rearrangement contribution in Eq. (\ref{spenergy}) comes out from the 
derivative of the first term appearing on the right-hand side of Eq. (\ref{meanH}).
%		in the derivation of Eq. (\ref{spenergy}), we have used that $\rho_{hh'}^{(0)}=\delta_{hh'}$.
In Eq. (\ref{A11-RPA}), the residual interaction is 
\begin{eqnarray}
	\mathscr{V}_{mkip}=\hat{V}_{mkip} (\rho^{(0)}) + 
	\sum_{h} \left[ \frac {\delta \hat{V}_{m h i h}} 
	{\delta \rho_{kp}}\right]_{\rho=\rho^{(0)}} \rho_{kp} 
	+ \sum_{h } \left[ \frac {\delta \hat{V}_{h k h p}} 
	{\delta \rho_{mi}}\right]_{\rho=\rho^{(0)}} \rho_{mi} 
	+ \frac {1} {2} 
	\sum_{h h' } \left[ \frac {\delta^2 \hat{V}_{h h' h h'} } 
	{\delta \rho_{mi} \delta\rho_{kp}}\right]_{\rho=\rho^{(0)}} 
	\rho_{mi} \rho_{kp},
	\label{vres-a11}
\end{eqnarray}
where the first term is the original interaction while the others are the 
rearrangement terms. 
%	The second and third terms come out from the 
%	quantities $\lambda$ 
%	appearing in Eq. (\ref{meanH}) and the last term derives 
%	from $<\Phi |H| \Phi>$ in Eq. (\ref{meanH}). 

Let us consider now the rearrangement terms of the matrix elements beyond the usual RPA.
First, it comes out that no rearrangement terms appear in the matrix 
$A_{22}$, Eq. (\ref{A22}), except for the contributions
\begin{eqnarray}
	A_{mnij,pqkl}^{(rearr)}=
	\frac{1}{2}\delta_{nq}\delta_{ik}\delta_{jl}\sum_{kk'} 
	\left[\frac {\delta \hat{V}_{k k' k k'}} 
	{\delta \rho_{pm}}\right]_{\rho=\rho^{(0)}} \rho_{pm} 
	+\frac{1}{2}\delta_{mp}\delta_{ik}\delta_{jl}\sum_{kk'} 
	\left[\frac {\delta \hat{V}_{k k' k k'}} 
	{\delta \rho_{qn}}\right]_{\rho=\rho^{(0)}} \rho_{qn} \nonumber \\
	-\frac{1}{2}\delta_{mp}\delta_{nq}\delta_{jl}\sum_{kk'} 
	\left[\frac {\delta \hat{V}_{k k' k k'}} 
	{\delta \rho_{ki}}\right]_{\rho=\rho^{(0)}} \rho_{ki} 
	-\frac{1}{2}\delta_{mp}\delta_{nq}\delta_{ik}\sum_{kk'} 
	\left[\frac {\delta \hat{V}_{k k' k k'}} 
	{\delta \rho_{lj}}\right]_{\rho=\rho^{(0)}} \rho_{lj}
	\label{a22-r}
\end{eqnarray}
generated by the first term of Eq. (\ref{meanH}). 
These terms,together with the quantities $\lambda$, correctly yield the single-particle energies for a density-dependent interaction. This ensures complete consistency with the single-particle energies that appear in the RPA matrix $A_{11}$, 	Eqs. (\ref{A11-RPA}) 
The other terms do not provide rearrangement terms in the residual interaction of the matrix $A_{22}$.
No rearrangement terms appear in the matrix $B_{22}$. 
%	The fact that rearrangement terms are not found in 
%	the residual interaction of the matrices '22' introduces a kind of 
%	asymmetry between the matrices '11' and 
%	'22'. This asymmetry can be attributed to two primary reasons: first, the linear term of the density variation ($\delta \rho $) does not depend on the two-body coefficients; and second, the interaction itself depends exclusively on the one-body density.

Concerning the matrices coupling $1p-1h$ with $2p-2h$ 
configurations, the same kind of contributions like in Eq. (\ref{a22-r}) is found which gives a vanishing term owing to the HF condition (see for example Eq. (5.35) of \cite{RS.80}):
\begin{equation}
	\epsilon_{kp}=\lambda_{kp}(\rho^{(0)})+\frac{1}{2}\sum_{kk'} 
	\left[\frac {\delta \hat{V}_{k k' k k'}} 
	{\delta \rho_{pk}}\right]_{\rho=\rho^{(0)}} \rho_{pk} =0 .
	\label{spea12}
\end{equation}

In addition to that, we have only one type of rearrangement terms in the 
residual interaction, given by

\begin{eqnarray}
	A_{mi,pqkl }^{(rearr)} = \left[ \frac {\delta \hat{V}_{klpq}} 
	{\delta \rho_{im}}\right]_{\rho=\rho^{(0)}}\rho_{im }.
	\label{A12-r}
\end{eqnarray}

and 
\begin{eqnarray}
	B_{mi,pqkl }^{(rearr)}=\left[ \frac {\delta \hat{V}_{klpq}} 
	{\delta \rho_{mi}}\right]_{\rho=\rho^{(0)}}\rho_{mi }.
	\label{B12-r}
\end{eqnarray}
for the $B_{12}$ matrix.	
We recall that, in the case of a two-body interaction that does not 
depend on the density, 
the matrices 
$B_{22}$ and $B_{12}$ are equal to zero. In the present 
case, $B_{12}$ 
has a non vanishing contribution coming from the rearrangement terms. %Applications of the SRPA with the so-derived rearrangement terms are discussed in Section \ref{Sec:ApplicationPart_SRPA_Rear}.

\newpage	
	\section{List of Acronyms} % Use \chapter* or \section* as appropriate
%	\end{tabular}
%\end{table}
\begin{minipage}[c]{0.5\textwidth}
	\centering
	\begin{tabular}{|l|l|}
		\hline
		Acronym & Meaning \\
		\hline
		 1p-1h   & One-Particle--One-Hole \\
		1h-1p   & One-Hole--One-Particle \\
		2p-2h   & Two-Particle--Two-Hole \\
		2h-2p   & Two-Hole--Two-Particle \\
		2qp      & Two Quasiparticles \\
		%BCS      & Bardeen--Cooper--Schrieffer \\
		BMF      & Beyond-Mean-Field \\
		CC       & Coupled Cluster \\
		CWT      & Continuous Wavelet Transform \\
		EDF      & Energy Density Functional \\
		EoM      & Equations of Motion \\
		EoS      & Equation of State \\
		EWSR     & Energy-Weighted Sum Rule \\
		GR       & Giant Resonance \\
%		GDR      & Giant Dipole Resonance \\
%		GMR      & Giant Monopole Resonance \\
%		GQR      & Giant Quadrupole Resonance \\
		GT       &Gamow--Teller \\
		HF       & Hartree--Fock \\
%		HFB      & Hartree--Fock--Bogoliubov \\
		ISGMR    & Isoscalar Giant Monopole Resonance \\
		ISGQR    & Isoscalar Giant Quadrupole Resonance \\
		IVGDR    & Isovector Giant Dipole Resonance \\
		MF       & Mean-Field \\
		PDR      & Pygmy Dipole Resonance \\
		PVC      & Particle--Vibration Coupling \\
		QBA      & Quasi-Boson Approximation \\
		QPM      & Quasiparticle--Phonon Model \\
		QRPA     & Quasiparticle Random-Phase Approximation \\
		RPA      & Random-Phase Approximation \\
    	RQTBA    & Relativistic Quasiparticle Time-Blocking\\
		RRPA     & Relativistic Random-Phase Approximation \\
		RSTD     & Relativistic Second Tamm--Dancoff \\
		RSSTD    & Relativistic Subtracted Second Tamm--Dancoff  \\
		SQRPA    & Second Quasiparticle Random-Phase Approximation \\
		SRPA     & Second Random-Phase Approximation \\
		SSRPA    & Subtracted Second Random-Phase Approximation \\
		STD      & Second Tamm--Dancoff Approximation \\
		SSTD     & Subtracted Second Tamm--Dancoff Approximation \\
		TD       & Tamm--Dancoff \\
		UCOM     & Unitary Correlation Operator Method	\\
		\hline
%		\hline
	\end{tabular}
\end{minipage}
%\begin{minipage}[c]{0.5\textwidth}
%	\centering
%	\begin{tabular}{|ll}
%		\hline
%		Acronym & Meaning \\
%		\hline
%		RSTD     & Relativistic Second Tamm--Dancoff \\
%		RSSTD    & Relativistic Subtracted Second Tamm--Dancoff  \\
%		SQRPA    & Second Quasiparticle Random-Phase Approximation \\
%	SRPA     & Second Random-Phase Approximation \\
%	SSRPA    & Subtracted Second Random-Phase Approximation \\
%	STD      & Second Tamm--Dancoff Approximation \\
%	SSTD     & Subtracted Second Tamm--Dancoff Approximation \\
%	TD       & Tamm--Dancoff \\
%	UCOM     & Unitary Correlation Operator Method	\\
%	\hline
%	&\\
%	&\\
%	&\\
%	&\\
%	&\\
%	&\\
%	&\\
%	&\\
%	&\\
%	&\\
%	&\\
%	&\\
%	&\\
%	&\\
%	&\\
%	&\\
%	&\\
%	&\\
%	&\\
%	&\\
%	&\\
%	&\\
%	&
%	
%	\end{tabular}
%\end{minipage}
\newpage

	\bibliography{mybibfile}	
\end{document}